\documentclass[prd,preprint,amsmath,amssymb,nofootinbib,eqsecnum]{revtex4-1}
\usepackage{latexsym,graphics,color,epsfig,mathrsfs}

\newcounter{Diagrams}
\Alph{Diagrams}
\newtheorem{Diag}{}[Diagrams]

\newcommand{\ie}{{i.e.}}
\newcommand{\cf}{{cf.}}
\newcommand{\eg}{{e.g.}}
\newcommand{\aka}{{a.k.a.}}

\newcommand{\viz}{{viz.}}
\newcommand{\wrt}{with respect to}
\newcommand{\lhs}{left-hand side}

\newcommand{\rhs}{right-hand side}

\newcommand{\naive}{na\"{\i}ve}
\newcommand{\naively}{na\"{\i}vely}

\newcommand{\be}{\begin{equation}}
\newcommand{\ee}{\end{equation}}
\newcommand{\bea}{\begin{eqnarray}}
\newcommand{\eea}{\end{eqnarray}}
\newcommand{\beas}{\begin{eqnarray*}}
\newcommand{\eeas}{\end{eqnarray*}}

\newcommand{\bear}{\begin{array}{l}}
\newcommand{\eear}{\end{array}}

\newcommand{\bcf}{\begin{center}\begin{figure}}
\newcommand{\ecf}{\end{figure}\end{center}}

\newcommand{\bct}{\begin{center}\begin{table}}
\newcommand{\ect}{\end{table}\end{center}}

\newcommand{\ds}{\displaystyle}

\newcommand{\eq}[1]{(\ref{eq:#1})}
\newcommand{\eqn}[1]{equation~(\ref{eq:#1})}
\newcommand{\Eqn}[1]{Equation~(\ref{eq:#1})}

\newcommand{\eqs}[2]{(\ref{eq:#1}) and~(\ref{eq:#2})}

\newcommand{\eqns}[2]{equations~(\ref{eq:#1}) and~(\ref{eq:#2})}
\newcommand{\Eqns}[2]{Equations~(\ref{eq:#1}) and~(\ref{eq:#2})}

\newcommand{\sect}[1]{section~\ref{sec:#1}}
\newcommand{\Sect}[1]{Section~\ref{sec:#1}}
\newcommand{\sects}[2]{sections~\ref{sec:#1} and~\ref{sec:#2}}

\newcommand{\fig}[1]{figure~\ref{fig:#1}}

\newcommand{\app}[1]{appendix~\ref{app:#1}}

\newcommand{\D}{d}

\newcommand{\integral}[1]{\int \! d #1 \,}
\newcommand{\Nintegral}[1]{\int \! d^N #1 \,}
\newcommand{\Int}[1]{\int \!\! d^{\D} \! #1 \,}
\newcommand{\FourInt}[1]{\int \!\! d^4 \! #1 \,}

\newcommand{\MomInt}[2]{\int \!\! \frac{d^{#1} #2}{(2\pi)^{#1}} \, }

\newcommand{\volume}[1]{d^{\D} \! #1 \,}

\newcommand{\Fint}[1]{\int \mathcal{D} #1 \,}

\newcommand{\DD}[1]{\delta^{(#1)}}

\newcommand{\SU}{\mathrm{SU}}

\newcommand{\der}[2]{\frac{d #1}{d #2}}
\newcommand{\pder}[2]{\frac{\partial #1}{\partial #2}}

\newcommand{\fder}[2]{\frac{\delta #1}{\delta #2}}
\newcommand{\dfder}[3]{\frac{\delta^2 #1}{\delta #2 \delta #3}}
\newcommand{\Or}{\mathrm{O}}

\newcommand{\order}[1]{\Or \bigl( #1 \bigr)}
\newcommand{\hf}{\frac{1}{2}}

\newcommand{\mom}{\mathrm{mom}}
\newcommand{\Omegasl}[1]{\not{\!\Omega}_{#1}}
\newcommand{\emc}{\gamma_{\mathrm{EM}}}

\newcommand{\tr}{\mathrm{tr}\,}
\newcommand{\Tr}{\mathrm{Tr}\,}

\newcommand{\nCr}[2]{
		\Bigl(
		\begin{matrix}
			#1
		\\[-1.5ex]
			#2
		\end{matrix}
		\Bigr)
}

\newcommand{\abs}[1]{
	\left\vert #1 \right\vert	
}

\newcommand{\norm}[1]{
	\lVert #1 \rVert	
}

\newcommand{\inner}[2]{
	\langle #1,#2 \rangle
}

\newcommand{\eval}[1]{
	\langle #1 \rangle
}

\newcommand{\DiracD}[2]{
	\delta^{#1}(#2)
}

\newcommand{\deltahat}[1]{
	\hat{\delta}(#1)
}

\newcommand{\pf}{\mathcal{Z}}

\newcommand{\cutoff}{K}

\newcommand{\ep}{C}
\newcommand{\dd}{\dot{\ep}}
\newcommand{\knl}[1]{\cdot {#1}\cdot}
\newcommand{\ctp}{\mathrm{C}^{-1}}

\newcommand{\DummyKernel}{\ensuremath{\stackrel{\bullet}{\mbox{\rule{1cm}{.2mm}}}}}

\newcommand{\critexp}{\lambda}

\newcommand{\flow}{\Lambda \partial_\Lambda}
\newcommand{\flowlam}{\flow\bigr\vert_{\lambda}}
\newcommand{\totalflow}{\Lambda \der{}{\Lambda}}
\newcommand{\op}{\mathcal{Y}}
\newcommand{\Count}{\Delta}

\newcommand{\Sint}{S^{\mathrm{I}}}
\newcommand{\Stilde}{\tilde{S}}

\newcommand{\Siv}[1]{S^{\mathrm{I}(#1)}}
\newcommand{\hS}{\hat{S}}
\newcommand{\hSvert}[1]{\hat{S}^{(#1)}}
\newcommand{\Svert}[1]{{S}^{(#1)}}
\newcommand{\hSint}{\hat{S}^{\mathrm{I}}}
\newcommand{\Sigint}{\Sigma^{\mathrm{I}}}
\newcommand{\Swil}{S_{\mathrm{W}}}
\newcommand{\Tact}{T}

\newcommand{\coupled}{\mathcal{P}}
\newcommand{\fpop}{\mathscr{I}}

\newcommand{\dual}{\mathcal{D}}
\newcommand{\dualv}[1]{\mathcal{D}^{(#1)}}
\newcommand{\dopi}{\overline{\mathcal{D}}}
\newcommand{\dopiv}[1]{\overline{\mathcal{D}}^{(#1)}}
\newcommand{\hDopi}{\hat{\Pi}}
\newcommand{\dualshift}{\mathcal{E}}

\newcommand{\homog}{\mathcal{H}}
\newcommand{\h}{h}

\newcommand{\dep}{\overline{\ep}}

\newcommand{\dimc}[1]{\left[ #1 \right]_\mathrm{c}}

\newcommand{\classical}[3]{\fder{#1}{\phi} \knl{#2} \fder{#3}{\phi}} 
\newcommand{\quantum}[2]{\fder{}{\phi} \knl{#1} \fder{#2}{\phi}}

\newcommand{\eop}{\mathcal{O}}
\newcommand{\marginal}{\eop_\mathrm{mar}}

\newcommand{\jop}{\mathscr{O}}

\newcommand{\intconst}{B}
\newcommand{\const}{c}

\newcommand{\proj}{\mathcal{P}}

\newcommand{\normalization}{\mathcal{N}}
\newcommand{\adjop}{\hat{L}}
\newcommand{\adjopb}{\hat{M}_\star}
\newcommand{\weight}{G}
\newcommand{\normconst}{b}

\newcommand{\classifier}{\hat{\mathcal{M}}_\star}
\newcommand{\classifierJ}{\hat{\mathcal{M}}_\star^J}

\newcommand{\diagnorm}{\Upsilon}
\newcommand{\sym}{\mathcal{S}}
\newcommand{\conn}{\mathrm{c}}

\newcommand{\gibbs}{u}

\newlength{\LabLength}

\newcommand{\dec}[3][0]{\left[ #2 \hspace{#1em} \right]^{#3}}

\newcommand{\hepth}[1]{hep-th/#1}
\newcommand{\hepph}[1]{hep-ph/#1}
\newcommand{\heplat}[1]{hep-lat/#1}
\newcommand{\condmat}[1]{cond-mat/#1}

\newcommand{\nuclth}[1]{nucl-th/#1}

\begin{document}

\title{Fundamentals of the Exact Renormalization Group}

\author{Oliver~J.~Rosten}

\affiliation{Department of Physics and Astronomy, University of Sussex, Brighton, BN1 9QH, U.K.}
\email{O.J.Rosten@Sussex.ac.uk}

\begin{abstract}

Various aspects of the Exact Renormalization Group (ERG) are explored,
starting with a review of the concepts underpinning
the framework and the circumstances under which it
is expected to be useful. A particular emphasis is placed on the intuitive picture provided for both renormalization in quantum field theory and universality associated with second order phase transitions.
A qualitative discussion of triviality, asymptotic freedom and asymptotic safety is presented.

Focusing on scalar field theory, the construction of assorted flow equations is considered using a general approach, whereby different ERGs follow from field redefinitions. It is recalled that Polchinski's equation can be cast as a heat equation, which provides intuition and computational techniques for what follows. The analysis of properties of exact solutions to flow equations includes a
proof that the spectrum of the anomalous dimension at critical fixed-points is quantized.

Two alternative methods for computing the $\beta$-function in $\lambda \phi^4$ theory
are considered. For one of these it is found that all explicit dependence on 
the non-universal differences between a family of ERGs cancels out, exactly.
The Wilson-Fisher fixed-point is rediscovered in a rather novel way.

The discussion of nonperturbative approximation schemes focuses on the derivative expansion,
and includes a refinement of the arguments that, at the lowest order in this approximation, a function can be constructed which decreases monotonically along the flow.

A new perspective is provided on the relationship between the renormalizability of the 
Wilsonian effective action and of correlation functions, following which the construction of manifestly gauge invariant ERGs is sketched, and some new insights are given. Drawing these strands together
suggests a new approach to quantum field theory.

\end{abstract}

\maketitle

\tableofcontents

\section{Introduction}

The physical intuition which underpins the Exact Renormalization Group%
\footnote{%
The ERG is also commonly referred to as the Functional Renormalization Group, the Nonperturbative Renormalization Group and,
occasionally, the Continuous Renormalization Group.}
(ERG)
derives from an observation which is so familiar as to be considered almost mundane: namely that the natural description of physics generally changes with the scale at which observations are made. 
Crudely speaking, this is no more high-minded a statement than saying that the world around us looks rather different when viewed through a microscope. More precisely, our parametrization of some system in terms of both the degrees of freedom and an action specifying how they interact generally change with scale.
In essence, the ERG is a mathematical formulation of this idea. 

As pointed out in~\cite{Wetterich-Rev}---and rather more entertainingly in~\cite{Cardy-Book}---a useful way to view the ERG is like a microscope of varying resolving power (but where this microscope is abstract in the sense that it operates on the action, rather than on physical samples). Starting from a description of physics at some short distance scale, the ERG allows us to go (in principle) step by step to a long distance description. Working in position space, we can envisage each of these steps as constituting some sort of averaging procedure over local patches of the system. In momentum space, this process of iteratively `coarse-graining' degrees of freedom starts by taking account of high energy fluctuations (either quantum or statistical) and gradually includes those of lower and lower energy.
As this coarse-graining procedure is performed, we thus expect to see the microscopic description
of the system under analysis transmogrifying into a description more appropriate to the
macroscopic behaviour.

The central ingredient of the ERG is the Wilsonian effective action. Let us suppose that we have modelled some system by providing a description at a high energy scale, the `bare scale', $\Lambda_0$. This description is provided by the bare action, $S_{\Lambda_0}$,
which encodes the types and strengths of the various interactions 
 (we will later discuss, at much greater length, precisely what is meant by the bare action). Now, following the above philosophy, we integrate out degrees of freedom between the bare scale and a lower, effective scale, $\Lambda$. In general, the action will change during this procedure, resulting in a  Wilsonian effective action, $S_\Lambda$, that is usually different from the
bare action. Roughly speaking, one can consider the Wilsonian effective action to provide the appropriate description of physics at the effective scale.

It is the ERG equation, \aka\ flow equation, which governs the behaviour of the Wilsonian
effective action under infinitesimal changes of the effective scale. For some set of fields, $\varphi$, this equation (which actually has many guises) takes the basic form
\[
	-\flow S_\Lambda [\varphi] = \ldots,
\]
where the derivative is performed at constant $\varphi$.

Whilst we will work in the continuum for most of this review, for the qualitative discussions in this section and the next we will frequently discuss models formulated on a lattice, due to the extra intuition that they provide. In this context, we will consider discrete, rather than infinitesimal changes of the scale. Strictly speaking, we are no longer dealing with the ERG, as its alternative name `the Continuous Renormalization Group' suggests. However, since we will learn lessons that are pertinent to the ERG, proper, and since our real concern in this paper is infinitesimal changes of scale in continuum systems, we will not be too fussy about this distinction. Where it matters, we will use the term `Wilsonian Renormalization Group' for the discrete case.

A natural and pertinent question to ask is when the ERG approach is useful. One can always
attempt to construct an ERG, though there are many cases where this is perhaps an
academic, rather than practical, exercise. 
As particularly emphasised in the celebrated review of Wilson and Kogut~\cite{Wilson},
the diagnostic for when the ERG comes into its own is the number of degrees
of freedom within a correlation length, $\xi$. Let us suppose that this number is small
compared to the total number of degrees of freedom in whatever system we happen
to be considering. Then we can see that there is at least some level of simplification,
since the properties of the entire system are expected to be essentially the same as a much
smaller subsystem whose characteristic dimension is $\xi$. Nevertheless, this might
not be of much help. For example, a piece of ferromagnetic material
could have $\order{10^{23}}$ degrees of freedom. If it turns out that
there are `only' $\order{10^{10}}$ degrees of freedom within one correlation length
then the problem of understanding the system is not really any easier.

However, in favourable circumstances, the number of degrees of freedom within
a correlation length is just a few or, in the optimal case, only one. In such a scenario we can make
real progress, since the task of understanding the bulk properties of the system has
been reduced to a problem which we might have some hope of solving.%
\footnote{%
Though even a cluster of as little as three atoms requires further approximations
to render it analytically soluble.
}
It is in this regime that the ERG has, perhaps, little to offer. Rather, it is in the opposite
regime---where there are many degrees of freedom (anywhere from
hundreds to infinity) per correlation length---that the
formalism has become an indispensable tool.

The reason why the ERG can be expected to be useful
in such situations boils down to the coarse-graining procedure, \emph{together with
an assumed locality of the interactions} in the system under analysis. 
If the interactions are local with a range $\order{L_0}$, then the idea is to break
the system up into small patches of this characteristic size. In an ideal situation, each
patch will contain  just a few degrees of freedom. So far, this sounds
similar to what we do when the correlation length is small, where we have
no need for the ERG. The difference, of course, is that since the correlation
length is large, we cannot expect to deduce the bulk properties of the system
directly from these small subsystems. However, suppose that we now
coarse-grain over patches with characteristic size $2L_0$ (for argument's sake). Since the interactions
are local (and, ideally, the number of degrees of freedom we have to deal with is small), 
we can hope to figure out the results of this procedure, even though
$\xi \gg L_0$. What we will find is a description of the system with fewer degrees
of freedom but a range of interaction which has roughly doubled. (In other words, starting
from the bare action we compute a Wilsonian effective action appropriate to the coarse-grained
system.)

But have we really gained anything? This procedure is most tractable when there
happens to be a small number of degrees
of freedom within a patch of characteristic size $L_0$.
But this means that the coarse-graining procedure does not
reduce the number of degrees of freedom very much. So, if  there were a large number 
of degrees of freedom per correlation
to start with, then this is still true after the first coarse-graining. 
But here is the crucial point:
the procedure can be iterated. At each stage,
we need only understand how to coarse-grain over neighbouring
patches.
And if we iterate the procedure enough times, then we arrive at
a description of the physics appropriate to scales of order the 
correlation length. This is at the heart of why the ERG is so useful.

There are many systems for which the ERG approach is profitable.
In this review, we will focus on relativistic  Quantum Field Theories (QFTs)
and statistical systems in the vicinity of a critical (\aka\ continuous or second order)
phase transition. Of the others, it is worth mentioning, in passing, the Kondo problem~\cite{Kondo}
(a magnetic impurity in a metal), due to the role this played in the development of the ERG~\cite{Wilson-Kondo}.

In the context of QFT, where any finite region contains an infinite number of
degrees of freedom, we might wonder how the ERG can be expected to
be of any use. However, there is hope because the interactions are
point-like. Indeed, considering continuum QFT as the limit of a lattice model
should make it clear that the density of degrees of freedom can 
be compensated by locality of the interactions. Further insight is provided
by working in momentum space. In the continuum case, 
each ERG step corresponds to integrating over an infinitesimal 
momentum shell. Thus, we attempt to take account of the modes
in the path integral gradually, rather than all at once. Of course, this by no means
guarantees that each coarse-graining step can be done in an analytically
controlled way; indeed, we expect this to be true only in special circumstances,
such as when there is a small parameter available. 

Nevertheless, one of the great strengths of
the ERG is that, although the flow equation cannot be exactly solved in general, various approximation schemes have been developed which are nonperturbative in essence 
(as will be outlined in \sect{Truncations}). (It should be borne in mind that the flow equation amounts to an exact reformulation of the path integral and, as such, contains the complete nonperturbative information of the theory at hand.)
Whilst these approximation schemes have errors which are hard to assess, their very existence
provides a method for attacking some exceedingly difficult problems. Examples include the strongly coupled regime of Quantum ChromoDynamics (QCD) and the nonperturbative renormalization of quantum gravity. (References can in \sect{Other-Overview}.)

Irrespective of the practical details of attempting 
quantitative calculations within the ERG, its other great use
is providing a qualitative---and profoundly physical---understanding
of two intimately related phenomena: the behaviour of statistical systems near 
to a critical phase transition and the nonperturbative
renormalizability of QFTs.

As it turns out, to most conveniently understand both renormalization and critical phenomena,
we must add a second ingredient to the ERG transformation (on top of the coarse-graining): a rescaling. With the above points in mind, 
we can quickly see what this amounts to by working on a lattice, with a spin at each site. Let us suppose that we coarse-grain over squares of $n\times n$
lattice sites. This means that $n\times n$ groups of spins are replaced by a single `blocked' spin and so the 
distance between blocked spins is $n$ times the original lattice spacing (as we will explicitly illustrate in the next section). Now, if we wish to compare the descriptions of the original system and the coarse-grained system,
we should rescale the lattice spacing to its original size.%
\footnote{For the continuous RG, this step can be most conveniently achieved by measuring all dimensionful quantities in terms of the effective scale, as will be described later.}
Taking the ERG transformation to include both the coarse-graining and rescaling steps,
it is the fixed-points of this transformation that are instrumental to understanding both renormalizability in QFT and critical phenomena.

At an intuitive level, the reason for this is that these fixed-points correspond to scale-invariant theories: the description of the system after coarse-graining and then rescaling does not change. 
From the point of view of statistical mechanics, it is precisely such theories that we expect to describe the long-distance dynamics of systems at criticality: for so long as one is looking at scales appreciably higher than the absolute cutoff (which might be \eg\ the molecular spacing), then the theory appears to be scale-invariant. Perhaps the canonical example of this is a ferromagnet for which (having set any external magnetic field to zero) the temperature is adjusted to bring the system to its critical point.%
\footnote{%
One further phenomena which is too beautiful to resist mentioning, at least in passing, is that of critical opalescence. A fluid which is otherwise transparent to visible light is, through tuning external parameters, brought towards a critical phase transition. 
Approaching criticality, the size at which structure is present increases, eventually encompassing the length scale of visible light, causing the sample to become opaque (so long as there is a difference in the refractive index of the two phases).}%

In a simple model, one can visualize this system as a lattice of little magnets (or spins), oriented either up or down. Assuming no  external magnetic field is present then, above the critical point, one finds a jumble of essentially uncorrelated spins. Below the critical point, the sample is magnetized, and there is a preponderance of either ups or downs. 
However, precisely at criticality, the net magnetization is zero and the correlation length is infinite. At this point, the system is scale-invariant in the precise sense that the long range dynamics encoded in the bare action correspond to those of the appropriate fixed-point theory. Interestingly, as vigorously emphasised in~\cite{Delamotte-Rev}, it does not follow from this that the popular picture of scale invariance being manifest in the physical structure of clusters of spins is correct. This false picture posits that if we identify 
 a cluster of mostly ups then, zooming in, it appears that this cluster is itself made up of clusters of mostly ups or downs, which in turn are made up of clusters of mostly ups or downs, and so on and so on. Compelling as it is, this \naive\ picture is wrong.

Moving on, critical fixed-points also form the basis for constructing nonperturbatively renormalizable QFTs. Ignoring the largely uninteresting non-critical fixed-points (which we will return to in \sects{Gen-TP}{CorrFns-Ball}), fixed-points correspond to massless, scale-invariant theories. As such, there  cannot be any dependence on a bare scale, which is just another way of saying that the theory can be renormalized. Moreover, one can construct scale-\emph{dependent} renormalizable theories by considering theories whose ultraviolet (UV) dynamics are governed by a critical fixed-point.

Further developing and refining this discussion of renormalization forms an important part of this review.
Indeed, the main aims of this paper are to:
\begin{enumerate}
	\item Elucidate the very physical picture of renormalization encapsulated by the ERG;
	
	\item Describe the construction of various flow equations;
	
	\item Recall some exact statements pertaining to the solutions of particular flow
	equations and derive some new ones;
	
	\item Describe methods for performing actual calculations with the ERG, both
	perturbative and otherwise;
	
	\item Present a new insight into the relationship between the renormalizability of the
	Wilsonian effective action and the renormalizability of correlation functions.
\end{enumerate}
As such, it is hoped that this review will, on the one hand, provide a thorough grounding in the basic ideas of the ERG approach, with the presentation being complementary to that of the existing reviews~\cite{Wilson,Wegner_CS,Fisher-Rev,Aoki-Rev,TRM-Elements,B+B,Wetterich-Rev,Polonyi-Rev,Canet-Rev,Mouhanna-Rev,Delamotte-Rev,JMP-Review,Gies-Rev,Kopietz-Book}. (For Wilson's personal perspective on the early development of the subject of renormalization and critical phenomena, as a whole, see~\cite{Wilson-RMP-83}.)
On the other hand, a number of new results/methodologies will be presented.
 Since applications are not the main focus of this paper, a comprehensive review of the associated literature will not be found here. That said, for applications which are mentioned (the focus being on high energy physics), the original literature is cited, pointers to appropriate reviews are given (including more specialist reviews than the ones just mentioned), and an effort is made to mention recent important work.

The rest of this paper is structured as follows.
Rather than immediately introducing specific forms of
the flow equation, in the next section we will discuss qualitative aspects pertaining to both the construction and application of the formalism.  
Various flow equations are presented in \sect{ERG} for scalar field theory.  
The focus is on so-called generalized flow equations, in contrast to many recent reviews~\cite{Wetterich-Rev,Canet-Rev,JMP-Review,Gies-Rev}, which 
deal exclusively with the `effective average action' formalism (the effective average action is discussed in \sect{EEA}).
It is recalled in \sect{heat} that certain flow equations can be written in the form of a heat equation.
This observation is useful for much of the subsequent analysis, providing both some extra intuition
and useful tools.

Aspects of exact solutions of the flow equation are analysed in \sect{solve}, and in some sense this is the heart of the quantitative side of this paper. 
The discussion begins with an analysis of fixed-point solutions. Many of the general considerations of
\sect{GenCon} are illustrated with a discussion of the Gaussian fixed-point in \sect{GFP}. Inspired by some of the technology of \sect{heat}, in \sect{FP-dual} a number of new results are derived, including  a
proof that the spectrum of the anomalous dimension at critical fixed-points
is quantized (equivalently, discrete). Moving on to scale-dependent solutions,
a refinement of the arguments pertaining to the nonperturbative renormalizability of theories sitting on a renormalized trajectory is given in \sect{renorm}. Finally, in \sect{self-similar}, a loose end pertaining to the linearization of the flow equation in the vicinity of a fixed-point is tied up.
 
 \Sect{beta} is devoted to discussing the $\beta$-function in $\lambda \phi^4$ theory. Two different methods of computation are presented in \sects{Canonical}{SFM}, based on different definitions of the coupling. For one of these it is found that all explicit dependence on 
the non-universal differences between a family of ERGs cancels out, exactly.
Finally, in \sect{WF}, the Wilson-Fisher fixed-point is uncovered, in a rather novel manner. 

One of the strengths of the ERG is that it supports
intrinsically nonperturbative approximation schemes, as discussed further in 
\sect{Truncations}. In terms of techniques, the main focus is on the `derivative expansion'---discussed in \sect{Trunc-DE}---in
which the interactions in the Wilsonian effective action are ordered according
to the number of powers of momenta they contain. Amongst other things, 
at lowest order in this approximation scheme, the argument that
a function can be constructed which decreases monotonically along the flow is
recalled and further developed. \Sect{Trunc-Opt} is devoted to the optimization
of truncation schemes and some associated issues.

\Sect{CorrFns} deals with the computation of correlation functions in the ERG.
The relationship between renormalizability of the Wilsonian effective action and the renormalizability of correlation functions is fleshed out, as is the realization of dilatation covariance at a critical fixed-point. A deep insight into the difference between critical and non-critical fixed-points is also presented.
A sketch of how the generalized approach to ERGs can be applied to theories with non-scalar field content is given in \sect{other-theories}. 
Most of the exposition deals with gauge theory, and it is recalled---quite remarkably---that the
generalized approach to ERGs admits a \emph{manifestly} gauge invariant
formulation: no gauge fixing is ever performed. Some new insights into this formalism are presented.
References
to work done using the alternative, effective average action approach can also be found
in this section.

The conclusion summarizes the compelling picture of QFT uncovered by the ERG and elucidates
some of the potentially exciting consequence of the fresh point of view provided by \Sect{CorrFns}.

\section{Qualitative Aspects}
\label{sec:Qualitative}

\subsection{Blocking}

As emphasised in the introduction, the central techniques behind the ERG are the coarse-graining of degrees of freedom and a
rescaling which restores the cutoff to its original value. We now flesh out the illustrative example given in the introduction (which is, strictly, in the context of the WRG) by taking a two dimensional system in which we have a lattice of spins, $s$, each of which we take to point either up or down (equivalently, $s=\pm1$).
A particular configuration of this system is shown in the first panel of \fig{block}.
 In fact, we suppose that the full lattice is much bigger than we can show. The coarse-graining procedure amounts to choosing blocks of spins and averaging over them to give new spins, $s'$. 
This is essentially the celebrated blocking procedure of Kadanoff~\cite{Kadanoff}.
Note, though, that the coarse-graining procedure is performed `under the partition function' rather than on physical realizations of the system. With this in mind, the only restrictions that we will place on this procedure are that it is performed only over local patches and that the partition function does not change.
 These points will be discussed further in \sects{local}{General_ERG}.
For definiteness---and as indicated---we have chosen $3\times3$ blocks. Our averaging procedure is
such that if there are more ups than downs, then $s'$ is up (corresponding to $s'=+1$: the magnitude of the spins does not change in this example), and vice-versa. As can be easily checked, this does indeed preserve the partition function, as shown explicitly in~\cite{Cardy-Book}.

The second panel in \fig{block} indicates the result of averaging over the spins. Notice that the lattice spacing (\ie\ cutoff) has increased by a factor of three, as anticipated in the introduction. Now we rescale, to reduce the lattice spacing back to its original size. This has the effect of sucking into our picture parts of the lattice which were previously off the page. The block with which we started now occupies only a small part of the visible portion of the lattice, as indicated by the dashed boundary.
\bcf[h]
	\[
	\resizebox{\textwidth}{!}{\ensuremath{\begin{array}{c}\begin{picture}(0,0)%
\epsfig{file=pstex/Blocking_3x3-a.pstex}%
\end{picture}%
\setlength{\unitlength}{3947sp}%
\begingroup\makeatletter\ifx\SetFigFont\undefined%
\gdef\SetFigFont#1#2#3#4#5{%
  \reset@font\fontsize{#1}{#2pt}%
  \fontfamily{#3}\fontseries{#4}\fontshape{#5}%
  \selectfont}%
\fi\endgroup%
\begin{picture}(28912,11086)(2057,-12310)
\end{picture}%
 \end{array}}}
	\]
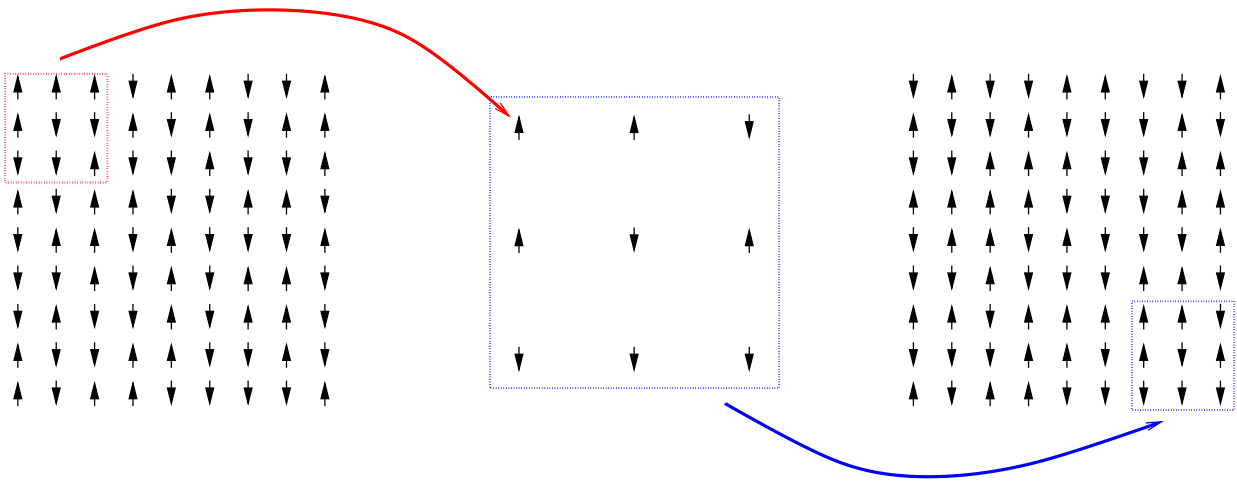
\caption{Block-spinning: starting from a microscopic description, $3\times 3$ blocks of spins are averaged over, using the `majority rules' prescription. Next, the system is rescaled to restore the lattice
spacing to its original value.}
\label{fig:block}
\ecf

An obvious question to ask concerns the effect of this procedure. Let us start by supposing that, for argument's sake, before any coarse-graining takes place the spins interact only with their nearest neighbours (the Ising model).  We emphasise that this is a choice we are making, amounting to the choice of bare action (we will discuss in \sect{renormalizability} the important issue of the extent to which we can choose the bare action in various circumstances ). Now, what interactions are exhibited by the blocked spins? In general, the blocked spins exhibit \emph{all possible interactions}. In other words, in addition to nearest neighbour interactions, there will be next-to nearest neighbour interactions, next-to-next-to nearest neighbour interactions and so forth. However, let us emphasise that this does not spoil the locality we prized so highly in the introduction. Deferring a  precise discussion of locality to \sect{local}, we note that changes to the longer-range interactions induced by the blocking procedure are suppressed.

In general, the result of iterating this procedure is that the various strengths of all the interactions change at each step. 
This suggests an intuitive way to visualize what is going on. Let us consider `theory space': the space of all possible interactions. Thus, we consider one axis to be labelled by the strength of the nearest neighbour interaction, one to be labelled by the strength of the next-to nearest neighbour interaction and so forth. Points in this space represent particular Wilsonian effective actions. Since we expect this action to change with the RG procedure, we hop around in theory space. Perhaps the most important qualitative feature of theory space is that it can have fixed-points under the RG procedure (it should be emphasised that both the blocking and rescaling steps are included when we talk about the RG procedure).

In \fig{flow} we show a qualitative picture of what the various RG flows might look like in the vicinity of some critical fixed-point. For the case of discrete blocking transformations, like the one we have been considering,  we have joined the dots, to give the smooth lines in the picture. Later in this review, we will focus on the case of continuum models and will consider infinitesimal changes in the scale, in which
case the flows are anyway smooth.
\bcf[h]
	\[
	\resizebox{\textwidth}{!}{\ensuremath{\begin{array}{c}\input{pstex/Flow.pstex_t} \end{array}}}
	\]
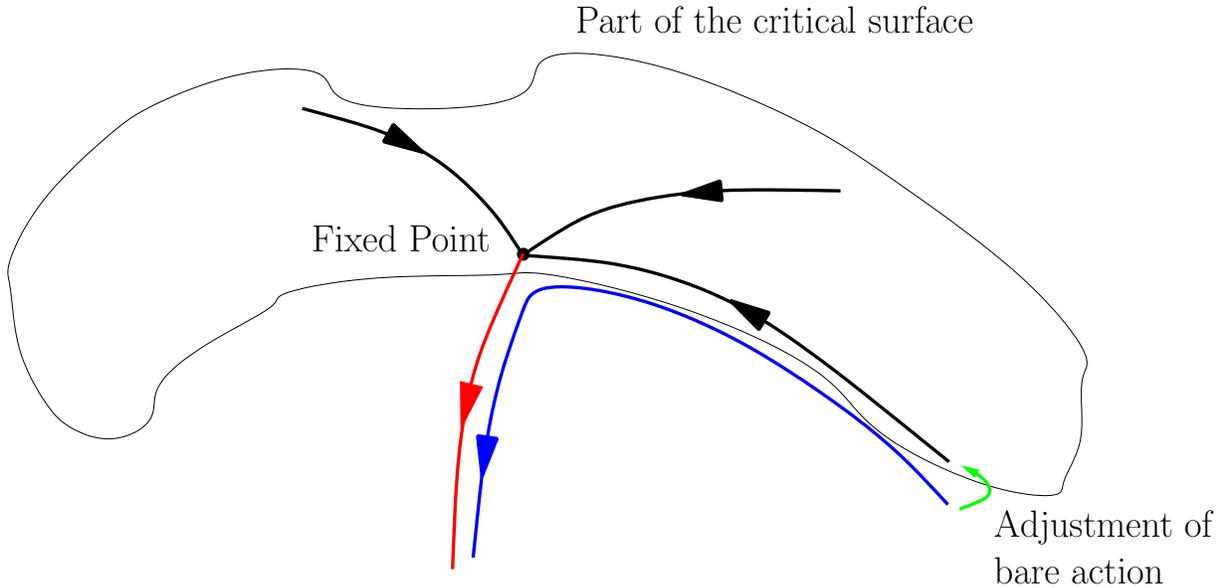
\caption{Renormalization group flows (from ultraviolet to infrared) in the vicinity of a fixed-point. The thick black lines
represent flows within the critical surface, only part of which is shown. The red line emanating
from the fixed-point is called a renormalized trajectory. The blue line shows a flow which
starts just off the critical surface. By adjusting the bare action, this flow can be tuned
towards the  critical surface.}
\label{fig:flow}
\ecf

Given a critical fixed-point, we can consider the surface constructed by demanding that all actions
on the surface flow into the fixed-point under the RG procedure. This defines the critical surface of the fixed-point \emph{under consideration}. We emphasise this last point because theory space might support several fixed-points, each of which will have its own critical surface. The portion of the critical surface in the infinitesimal neighbourhood of the fixed-point is spanned by the so-called irrelevant operators.%
\footnote{%
In this context, `operators' are actually commuting functionals of the fields; at a notational level, we will distinguish these from derivative operators by decorating the latter with a hat, whenever confusion is likely.} 
These operators are called irrelevant simply because their coefficients in the action decrease to zero as the fixed-point is approached \ie\ as we descend into the infrared (IR).

Conversely, the relevant operators are those whose coefficients grow as we flow towards the IR.%
\footnote{Marginal operators---to be discussed in detail later---are those which, to leading order in a perturbation about a fixed-point, are neither relevant nor irrelevant. When this property is spoilt at higher orders, we generally lump such marginally (ir)relevant operators together with the other (ir)relevant operators, unless there is some particular reason to consider them separately. Some operators exist which are exactly marginal and one in particular will play an important role in \sect{solve}.}
 Thus, if we consider a bare action slightly displaced from the critical surface, then the flow will start by driving it towards the fixed-point (the blue line in \fig{flow}). At some stage, however, a relevant operator will have grown to such a size as to become important and will then drive the action away from the fixed-point. With this simple picture, we can already gain a qualitative understanding of universality in critical phenomena. 

Let us start by imagining that we have a sample of some material which can be described by an action in a certain theory space (\ie\ the space consisting of all theories with a particular field content, possibly with some symmetry constraints). An example might be a lump of ferromagnet which we model as above. Now, experimentally, we know that to approach the ferromagnetic phase transition we must adjust two quantities: we must set the external magnetic field to zero (as it happens) and must careful tune the temperature to its critical value.
 Thus, temperature and magnetic field constitute the relevant directions of this system\footnote{Of these
 two relevant directions, the magnetic one is symmetry breaking, since it defines a preferred orientation
 for the spins, whereas the temperature direction is symmetry conserving. The case of a single symmetry preserving relevant direction is the canonical example of a critical system. Those systems with additional symmetry preserving relevant directions are often referred to as `multicritical'.
 }: by tuning them to their critical values we draw our initial bare action on to the critical surface, as indicated by the green arrow in \fig{flow}. Note that this is not an RG flow: here we are adjusting external parameters to change the bare action.

Having made this adjustment, \emph{now} we consider the effects of the RG flow: this tells us that the IR dynamics of the system are those of the fixed-point if we are strictly on the critical surface. 
Clearly, this picture will be repeated wherever we start on the critical surface. With this in mind, suppose that there exists some system with a wildly different microscopic description from our model of a ferromagnet which, nevertheless, can be modelled as a bare action in the same theory space. Although this action will be very different from the one corresponding to the ferromagnet, if we tune the relevant parameters such that it too is drawn towards the critical surface, then its IR dynamics will also be described by the fixed-point.
Systems which exhibit the same IR dynamics, in this way, are said to be in the same `universality class'.

For a system with $n$ relevant directions, Cardy~\cite{Cardy-Book} provides a  typically nice piece of
imagery: as an experimentalist trying to induce such a system to undergo a second order phase
transition, one must carefully dial to the correct position $n$ knobs which control the physical values of the associated parameters.

We can also ask what happens if we are just away from criticality \ie\ suppose that the relevant parameters have been adjusted such that the action almost, but not quite, touches the critical surface. Now the dynamics at some range of low energies are dominated by the fixed-point, whereas those at lower energies still are determined by the flow away from the fixed-point along the relevant direction(s). The structure of the rest of theory space---particularly whether or not there are any other fixed-points---will determine how sensitive the far end of such trajectories are on the boundary conditions. 

To conclude this section, we will expand on the point made in the introduction that not all fixed-points are critical. For example, sticking with the theory space appropriate to the two-dimensional Ising model, we can flow away from the critical fixed-point along the relevant temperature direction, ultimately hitting the `high-temperature fixed-point' at infinite temperature. This terminology is occasionally (and confusingly) used in zero-temperature QFT, along with `infinite-mass fixed-point'. 
We will have more to say about non-critical fixed-points in \sect{Gen-TP} and, particularly, \sect{CorrFns-Ball}.

\subsection{Renormalizability}
\label{sec:renormalizability}

With just a little extra effort, we can get a feeling for what is meant by renormalizability in
the nonperturbative sense (we will give a quantitative treatment in \sect{renorm} which, like the one given here, is based on that of Morris~\cite{TRM-Elements}). For the purposes of doing
so, we shall suppose that the usual notion of renormalizability---\ie\ renormalizability of the Green's functions---can be identified with renormalizability of the Wilsonian effective action. This is actually a more subtle point than is usually indicated, as we will discuss in \sect{CorrFns}. Ignoring this for the time being, let us work in (Euclidean) momentum space, recalling that the bare scale is denoted by $\Lambda_0$. Now imagine flowing down to the effective scale, $\Lambda$, arriving at an effective action which depends on both $\Lambda$ and $\Lambda_0$. At this stage, we pose the question: are there any such effective actions for which $\Lambda_0$ can be safely sent to infinity? By `safely' we mean that any divergences can be absorbed into a finite number of (renormalized) couplings. Note that the process of sending $\Lambda_0 \rightarrow \infty$ is often called `taking the continuum limit'.

The first observation to make is that fixed-point theories are, trivially, renormalizable! Since fixed-point theories are independent of scale, they are necessarily independent of $\Lambda_0$, which can thus be trivially sent to infinity. To see this in a little bit more detail, let us follow convention and introduce the `RG-time', $t\equiv\ln\mu/\Lambda$, where $\mu$ is an arbitrary scale, so that  $-\flow$ can just be replaced by $\partial_t$. This `time' runs from $-\infty$ in the UV to $+\infty$ in the IR. We also now indicate the typical dependencies of the \rhs\ of a certain class of flow equations:
\be
	\partial_t S_t[\varphi] = \mathcal{F}
	\left(
		S_t[\varphi], \fder{S_t[\varphi]}{\varphi}, \dfder{S_t[\varphi]}{\varphi}{\varphi}
	\right).
\label{eq:template}
\ee
Throughout this paper, we will use a $\star$ to denote fixed-point quantities. So, a fixed-point action is defined by
\be
	\partial_t S_\star[\varphi] = 0.
\label{eq:fp-criterion}
\ee
Now, does this really imply independence on $\Lambda_0$? Why, for example, could we not have dependence on (say) the ratio of a bare mass to the bare scale, \viz\ $m_0/\Lambda_0$? The point is as follows. Since we have rescaled to dimensionless variables, all couplings, $g$, in the action are dimensionless. From the solutions of~\eq{template}, it is apparent that these couplings will depend on $t$. Additional scales could creep in via a boundary condition $g(t=t_0) = g_0$. However, at a fixed-point, the couplings are independent of $t$, so new scales cannot appear in this way and the fixed-point action really is scale-invariant. The only way this could be violated is if an additional scale explicitly appears on the \rhs\ of~\eq{template}. This is not the case for the theories considered in this paper, though it can happen. For example, in noncommutative theories (for reviews see~\cite{Riv-Review,Szabo-Rev,Douglas:2001ba}), the dimensionful noncommutativity parameter, $\theta$, does indeed explicitly appear in the flow equation. In this case, one must carefully reconsider the criteria for nonperturbative renormalizability~\cite{RG+OJR}.%
\footnote{Given the big deal that has been made about locality in the introduction, one might wonder what point there is in constructing an ERG for noncommutative theories. Interestingly, such theories can be reformulated in terms of infinite dimensional matrices~\cite{MatrixBase}, and a cutoff can be implemented
by smoothly suppressing those rows and columns beyond a certain point. Constructing a flow equation
in this `matrix base'~\cite{G+W-PC,G+W-2D,G+W-4D,RG+OJR} has proven very profitable.}

Having discussed scale-invariant renormalizable theories, we should now ask whether it is possible to find
scale \emph{dependent} renormalizable theories? The answer is, of course, yes. To do so, we perturb a fixed-point action along one (or more) of the associated relevant directions. The resulting trajectories which emanate from the fixed-point are Wilson's `renormalized trajectories' (\eg\ the red line in \fig{flow}). As the name suggests, such actions are nonperturbatively renormalizable, the proof of which will be recalled in \sect{renorm}. Intuitively, it is perhaps obvious, since the UV dynamics is controlled by a fixed-point and we know that fixed-point theories are renormalizable. 

The actions along a renormalized trajectory are sometimes called `perfect actions'~\cite{Perfect}. Presuming that all quantities have been rendered dimensionless via an appropriate rescaling with
$\Lambda$, a crucial feature that renormalized trajectories exhibit is, as emphasised by Morris, \emph{self-similarity}~\cite{Shirkov-SelfSim}. Given some set of fields, $\varphi$, self-similarity means that all scale dependence is carried through the renormalized couplings, $g_i$, and the anomalous dimensions of the fields, $\eta_j$:
\be
	S_t[\varphi] = S(g_i(t),\eta_j(t))[\varphi].
\label{eq:self-similar}
\ee
Let us now stress a very important point, which can be a source of confusion. Renormalized trajectories are spawned by perturbing a fixed-point in some finite number of relevant directions. However, a finite distance along the flow the action generally receives contributions from all possible operators, including the irrelevant ones. The point is that the couplings of these latter operators---whose contribution to the action vanishes as we trace our way back into the UV---depend entirely on the $g_i(t)$. Of course, computing this dependence is the difficult bit!
[The perceptive reader might wonder why we need more than one coupling to specify the scale dependence in~\eq{self-similar}. The point is that each of the couplings carries information about an integration constant which forms part of the boundary condition for the flow. The anomalous dimensions come along for the ride because, as will see in \sect{renorm}, they require their own 
renormalization conditions.]

Returning to the question of renormalizability it is apparent that, nonperturbatively, this
boils down to the existence of fixed-points in theory space, and the renormalized trajectories that such fixed-points support.\footnote{We are ignoring the existence of limit cycles or other exotic RG behaviour~\cite{Wilson-RG+Strong,Bedaque-3body,Bernard+LeClair,Glazek+Wilson-LC-I, Braaten-IRQCD-LC,Russian-Doll-I,Russian-Doll-II,Cycles-Log-Periodic,Glazek+Wilson-LC-II,LeClair-RG-LC}. 
For renormalizable theories which are unitary upon continuation to Minkowski space this is justified in two dimensions on the basis of Zamolodchikov's c-theorem~\cite{Zam-cfn}. We will have more to say about this in \sect{cfn}.} Note that this suggests a rather different way of looking at field theory than is perhaps the norm. A standard approach would be to write down an action, understood as a bare action, and then to perform a (perturbative) analysis of the renormalizability of its correlation functions. In the ERG approach, we start by \emph{solving} the ERG equation to ascertain the spectrum of fixed-points.%
\footnote{This is much easier said than done, as we will discuss in \sect{Truncations}.}
If we find a fixed-point, then we linearize the ERG equation about the fixed-point to determine whether the various operators are relevant, irrelevant or marginal.

When we linearize about a fixed-point, the flow equation can be separated in $t$ and $\varphi$.%
\footnote{Actually, this not the general solution to the linearized flow equation. We will see in \sect{self-similar} why we nevertheless focus on these solutions. Given this choice, it will become apparent in
\sect{FP} that demanding locality (in the sense of \sect{local}) of the eigenperturbations quantizes
the $\lambda_i$.
\label{foo:self-similar}
}
\be
	S_t[\varphi] = S_\star[\varphi]  + \sum_i \alpha_i e^{\lambda_i t} \eop_i[\varphi],
\label{eq:perturb}
\ee
where the $\alpha_i$ are integration constants, the $\lambda_i$ are the RG-eigenvalues\footnote{The symbol $\lambda$ will also be used for the four-point coupling in scalar field theory.} and the 
$\eop_i[\varphi]$ are the eigenperturbations (\aka\ eigenoperators or just operators).
Substituting this into the flow equation, and working to linear order in the perturbation yields something of the form
\be
	\classifier \eop_i[\varphi] = \lambda_i \eop_i[\varphi],
\label{eq:eigen}
\ee
where $\classifier$ is a differential operator, the form of which depends on the choice of flow equation; a specific realization will be given in \sect{GenCon}. This equation can, in principle, be solved to yield both 
the $\lambda_i$ and the $\eop_i[\varphi]$. Those operators for which $\lambda_i >0 $ are relevant, since these increase in importance with increasing $t$. Conversely, those operators for which $\lambda_i <0$ are irrelevant. In the special case that $\lambda=0$, the operator is called marginal. One must go to the next order in the perturbation (and maybe beyond this) to determine whether an operator is marginally relevant [\ie\ relevant but growing only as $t$ (or slower still), rather than $e^t$], marginally irrelevant, or exactly marginal.\footnote{Loosely speaking, a finite perturbation along
an exactly marginal operator will not induce a flow. Whilst this encapsulates the basic idea, things are a little bit more subtle than this, as we will discuss in \sect{GenCon}.}

Before continuing with the main theme of our exposition, we pause to give context to a subtlety which will play an important role later. In addition to the classifications just mentioned, operators can be additionally divided up into whether they are `scaling operators' or `redundant operators'.\footnote{In the literature on asymptotic safety in quantum gravity, the couplings associated to these operators are often referred to as essential and inessential, respectively.} 
Redundant operators are associated with local field redefinitions and so carry no physics. For the rest of this section, we shall suppose that we are just considering the scaling operators.

It is the spectrum of relevant operators (including those which are marginally relevant) that determines the renormalized trajectories. If we decide that we would like to consider theories on renormalized trajectories emanating from a particular fixed-point, then the freedom we have amounts to choosing 
the integration constants, $\alpha_1, \ldots,\alpha_n$, associated with the relevant operators.

With this picture in mind, let us now revisit precisely what is meant by a bare action. Away from a renormalized trajectory, it is clear: the bare action is the boundary condition to our flow, being as it is the form of the action specified at some short distance scale. But along a renormalized trajectory, the boundary condition amounts to integration constants associated with the relevant operators. At some point near the top end of the trajectory, we could decide to call the action the bare action, but this choice of scale is arbitrary. For this reason, it is perhaps more illuminating to replace the notion of a bare action in this context with the notion of the perfect action in the vicinity of the UV fixed-point. To emphasise one last time: perfect actions are \emph{solved for}, 
given a choice of integration constants, and not chosen outright.

Before moving on, it is worth addressing the question of whether it makes sense to refer to fixed-points as UV fixed-points or IR fixed-points. For critical fixed-points, such a distinction only makes sense once something is said about the RG trajectories under consideration. If a critical fixed-point is considered, just in its own right, then it makes no sense to ascribe to it any notion of UV or IR since a fixed-point is, by definition, scale-independent. Of course, if we now say that we are considering RG trajectories flowing into a fixed-point then, \emph{for these trajectories}, the fixed-point governs the IR behaviour. But we might instead consider flows along the relevant directions of the very same critical fixed-point, in which case it can act as a UV fixed-point. Thus, context is everything. Note that non-critical fixed-points do not support relevant directions and so are sinks for RG trajectories~\cite{Wegner_CS}. Consequently, they can be unambiguously referred to as IR fixed-points.

\subsection{Asymptotic Safety and all that}
\label{sec:AS}

In this section we enumerate the various types of scale-dependent renormalizable theories that can be supported by fixed-points. First of all, let us consider a Gaussian fixed-point, and suppose that it has no interacting relevant directions. If this is the only fixed-point in theory space, then there are no non-trivial theories which are renormalizable beyond perturbation theory. This is illustrated in the first panel of \fig{behaviour}, where it is supposed that the Gaussian fixed-point has just a relevant mass direction, as would be the case in scalar field theory for $\D \geq 4$. In this situation, \emph{theory space} (rather than one particular trajectory) is said to suffer from the triviality problem, meaning that there are no non-trivial bare actions for which the bare scale can be removed. (See~\cite{Callaway} for a detailed discussion of various aspects of triviality.)
\bcf[h]
	\[
	\resizebox{\textwidth}{!}{\ensuremath{\begin{array}{c}\begin{picture}(0,0)%
\epsfig{file=pstex/Trivial.pstex}%
\end{picture}%
\setlength{\unitlength}{3947sp}%
\begingroup\makeatletter\ifx\SetFigFont\undefined%
\gdef\SetFigFont#1#2#3#4#5{%
  \reset@font\fontsize{#1}{#2pt}%
  \fontfamily{#3}\fontseries{#4}\fontshape{#5}%
  \selectfont}%
\fi\endgroup%
\begin{picture}(10195,4444)(1659,-6402)
\end{picture}%
 \end{array}}}
	\]
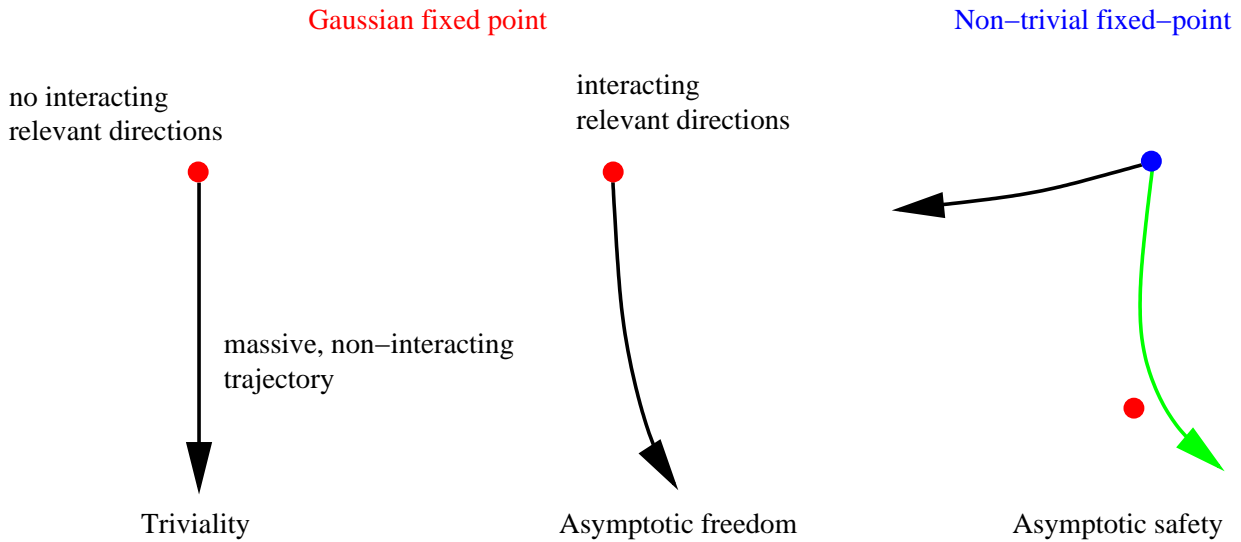
\caption{A cartoon depicting triviality, asymptotic freedom and asymptotic safety. Along a massive, non-interacting trajectory,  interesting interactions are never generated, which is illustrated by the straight line in the first panel (even in this case, the strengths of various two-point interactions do actually vary, but this is hidden by the choice of subspace on to which we have projected). The curved lines in the other panels are supposed to indicate more interesting RG flows.}
\label{fig:behaviour}
\ecf

More interesting is the case where the Gaussian fixed-point has \emph{interacting} relevant directions, as is the case for \eg\ QCD or scalar field theory in $\D <4$. Now the Gaussian fixed-point supports non-trivial renormalized trajectories, as indicated in the second panel of \fig{behaviour}. Such trajectories exhibit the celebrated asymptotic freedom. (Note the distinction between an asymptotically free \emph{trajectory} and a \emph{theory space} afflicted by triviality.)

The final case is where there exists a non-trivial fixed-point which supports renormalized trajectories, as shown in the third panel of \fig{behaviour}. In this case, the theory is said to be asymptotically safe,
a term coined by Weinberg~\cite{Weinberg-Erice,Weinberg-AS}.

Let us now consider a special case: an asymptotically free theory which supports a renormalized trajectory which just so happens to pass close to the Gaussian fixed-point, as depicted by the green line in \fig{behaviour}. The reason this is interesting to consider is because one can do perturbation theory in the vicinity of the Gaussian fixed-point. What would one conclude about the renormalizability of the theory based on such a perturbative analysis? That the theory is non-renormalizable, since it does not lie on a trajectory emanating from the Gaussian fixed-point! Of course, the problem with this analysis is that it is being done about the `wrong' fixed-point. The renormalizability of  this theory is determined by the fixed-point up in the UV.

To look at this another way is to say that, just because a perturbative analysis of some bare action 
in the vicinity of the Gaussian fixed-point indicates that it is non-renormalizable, does not mean that such an action does not lie close to (or on, but one would have to be mighty lucky to guess that right!) a renormalized trajectory emanating from some non-trivial fixed-point. This is the motivation behind some current and intense work into quantum gravity (see the end of \sect{other-theories} for references).

So, what do these considerations tell us about some familiar quantum field theories? As mentioned above, QCD is renormalizable nonperturbatively, being as it is asymptotically free. However, for scalar field theory in $\D\geq4$, the Gaussian fixed-point does not have any interacting relevant directions: only the mass is relevant. (In $\D=4$, the marginal four-point coupling is irrelevant by virtue of the positive coefficient of the one-loop $\beta$-function.) Moreover, in~\cite{Trivial} it was argued that the Gaussian fixed-point is the only physically acceptable critical fixed-point%
\footnote{By this we mean that the fixed-point is suitably local (in the sense of \sect{local}) and that the theory is unitary upon continuation to Minkowski space. In fact, as we will recall in \sect{Gen-TP}, there is an infinite family of non-interacting fixed-point theories which violate the latter constraint. Moreover, the possibility of interacting theories of this type has not been ruled out. Whilst their discovery would be interesting from the point of view of understanding theory space, such theories would not offer a physical solution to the triviality problem.}%
, 
adding weight to the general expectation that scalar field theory in $\D\geq4$ suffers from the triviality problem. (Of course, in this context, we understand scalar field theory to be a shorthand for the theory space of all scalar field theories.)

An obvious question is how this picture is reconciled with the very well known \emph{perturbative} renormalizability of the $\lambda \phi^4$ theory in $\D=4$.%
\footnote{It is almost a perversity that a particularly efficient proof of the \emph{perturbative} renormalizability 
of this theory---namely the refinement of Polchinski's proof~\cite{pol} by
Keller, Kopper and Salmhofer~\cite{Salmhofer}---uses the ERG which, as we have been discussing at length, provides a deep understanding of precisely why this theory is not renormalizable!
In a series of papers~\cite{K+K-QED_1,K+K-Comp_I,K+K-Comp_II,K+K-massless,K+K-QED_2,K+K-Minkowski}, Keller and Kopper have further developed
the flow equation approach to perturbative renormalizability.
See also~\cite{Bonini-PertRenorm,Muller-PertRenorm,Ball+Thorne}.}
 The resolution to this apparent paradox resides in the fact that the standard perturbative analysis involves a sleight of hand. Let us suppose that we specify a $\lambda \phi^4$ bare action and now integrate out degrees of freedom down to the effective scale, yielding an effective action $S_{\Lambda,\Lambda_0}$. For small coupling, we can write the result of doing this as a perturbative series plus nonperturbative power corrections, which we can write \emph{schematically} as:
\be
	S_{\Lambda,\Lambda_0}[\phi] = \sum_{i=0}^\infty \lambda^{i-1} S_i[\phi]
	+ \order{\Lambda/\Lambda_0}.
\label{eq:Action-pert}
\ee
If we now send $\Lambda_0 \rightarrow \infty$, then what remains is an expression for
the action written in self-similar form [$S_\Lambda = S(\lambda)$] and so we might be tempted to conclude that the theory is renormalizable.

However, taking the limit $\Lambda_0\rightarrow \infty$ is a formal and, strictly, illegal operation
since the remaining perturbative series is in fact ambiguous, as a consequence of UV renormalons.
Let us unpick this statement by first recalling some features of perturbative series in QFT, 
following Beneke~\cite{BenekeReview}.

To begin, consider some function of a parameter $\alpha$, $R(\alpha)$, for which
there is a power series,
\be
	R(\alpha) \sim \sum_{n=0}^\infty r_n \alpha^{n+1},
\label{eq:series}
\ee
assumed to be divergent.
If the perturbative coefficients, $r_n$, grow factorially with $n$, then one can attempt to assign
a value to the divergent sum via the Borel transform:
\[
	B[R](s) = \sum_{n=0}^\infty \frac{r_n}{n!} s^n.
\]
Should the following integral exist, then one can use the Borel transform to construct a function
with the same power series as $R$:
\be
	\tilde{R} = \int_0^\infty ds e^{-s/\alpha} B[R](s).
\label{eq:Borel-int}
\ee
In certain circumstances~\cite{Sokal} $\tilde{R}$ and $R$ coincide, but in general they may differ
by terms exponentially small in the coupling, \ie\ of the form $e^{-\mathrm{const}/\alpha}$.
Anyhow, this subtlety is not of importance for our concerns, and we will just suppose for simplicity that
$\tilde{R}$ and $R$ are the same.

Now, the Borel integral~\eq{Borel-int} will exist only if (i) the integrand dies off sufficiently rapidly
for large $s$; (ii) there are no poles along the positive real axis.
In the case that there \emph{are}
poles along the real axis, one can of course deform the contour of integration around the poles,
but there is an ambiguity about how to do so. As we have written things, \eq{series} tells us nothing
about which prescription should be adopted; but that is down to us being sloppy. In such cases,
we expect that $R$ would look something like
\be
	R(\alpha) =  \sum_{n=0; \, \pm}^\infty r_n \alpha^{n+1}
	+ \order{e^{-1/\alpha}}_{\mp},
\label{eq:proper-exp}
\ee
where the $\pm$ on the asymptotic series tells us whether to evaluate the Borel  integral in the
upper or lower complex plane. The crucial point is that this prescription is correlated
with a prescription for evaluating the $\order{e^{-1/\alpha}}$ terms. 

Beneke~\cite{BenekeReview} gives a very instructive example of how this works in practice. Denoting the
logarithmic derivative of the $\Gamma$ function by $\Psi$, the following function is analytic
in the entire complex plane except at $\alpha=0$:
\[
	R(\alpha) \equiv \sum_{n=0}^\infty (-1)^n \frac{\Psi(n)}{n!\alpha^n}.
\] 
For $\alpha>0$, this can be re-expressed as
\[
	R(\alpha) = -\sum_{n=0; \, \pm}^\infty n! \alpha^{n+1} + e^{-1/\alpha} (-\ln \alpha \mp i\pi).
\]
Taking both the perturbative series, and the exponentially small terms, \emph{and a consistent prescription for evaluating both}, a unique function can be reconstructed.

So how is all of this relevant to the renormalizability of $\lambda \phi^4$ in $\D=4$? In this
case we do not know the full function $S(\lambda)$ and so we do not have the luxury of being able to make absolute statements. However, we do expect there to be poles
along the positive real axis of the Borel plane, arising from UV renormalons. UV/IR renormalons refer to poles in the Borel plane arising from large/small loop momenta in certain types of Feynman diagram.
Poles in the Borel plane can have other origins (such as instantons in appropriate theories) but, in the current context, it is sufficient to recognize that there are renormalon contributions, at the very least.

The presence of these poles tells us that the (divergent) perturbative series in~\eq{Action-pert} is, by itself, ambiguous and that
in order to reconstruct $S_{\Lambda,\Lambda_0}$ we must keep the $ \order{\Lambda/\Lambda_0}$ terms. Consequently, we do not expect the limit $\Lambda_0\rightarrow\infty$ to exist, in the strict sense. But if we keep the $ \order{\Lambda/\Lambda_0}$ terms then self-similarity---and hence renormalizability---is manifestly destroyed by the presence of the scale $\Lambda_0$. 
The relationship between 
the $\Lambda/\Lambda_0$ `power corrections'  and terms which are exponentially small in the coupling can be made clear  by noticing that, to one-loop order,
\be
	\frac{\Lambda}{\Lambda_0} = 
	\exp\left[-\frac{1}{\beta_1 \lambda(\Lambda)} + \frac{1}{\beta_1 \lambda(\Lambda_0)}\right],
\label{eq:power-corrections}
\ee
where, as usual, $\beta_1$ is the one-loop coefficient of $\beta = \Lambda d \lambda /d\Lambda$.

Let us mention that in the constructive approach to QFT~\cite{Riv-Book} it is the presence of a Landau pole that is identified as the impediment to removing the bare cutoff. If the Landau pole is indeed present (as opposed to an artefact of perturbation theory) then it does, of course, destroy self-similarity.

As a final point, it is worth contrasting the above to what happens in a strictly renormalizable theory.
First of all, the type of diagrams which previously gave the UV renormalon problem still produce
poles in the Borel plane, but they now appear on the negative axis and so are harmless. Consequently, self-similarity is not spoiled by the explicit appearance of a UV scale. Nevertheless, it might well be
that there are still poles on the positive axis coming from some other source (for example, in QCD IR renormalons produce poles along the positive real axis). There is nothing wrong with this: there is no reason why perturbation theory should be Borel resummable in a strictly renormalizable theory. The point is that the exponentially small corrections must now occur in strictly self-similar form. This means that the power corrections are of the type $\mu/\Lambda = e^{t}$. (Recall that $\mu$ is an arbitrary scale. We can, of course, \emph{choose} to set $\mu$ to some value and, in QCD, it might be that this value is what we have decided to call $\Lambda_{\mathrm{QCD}}$. But this does not violate self-similarity: there is nothing fundamental about such a choice, and what we call $\Lambda_{\mathrm{QCD}}$ is anyway down to definition. On the other hand, the presence of a definite scale where a theory breaks down---\ie\ a Landau pole---is a different kettle of fish.)

\section{Flow Equations For Scalar Field Theory}
\label{sec:ERG}

In this section we will discuss the construction of flow equations in a very general
context. Following the excellent examples of Wegner~\cite{Wegner_CS} and Bagnuls and Bervillier~\cite{B+B}, the next subsection will be devoted to fixing notation and recalling a few
elementary facts. \Sect{local} deals with the issue of locality and, with this behind us, 
we turn to the construction of a large family of flow equations in \sect{General_ERG}, focusing particularly on those with a structure similar to Polchinski's~\cite{pol}. In \sect{EEA} we introduce the `effective average action', the flow of which can be derived from Polchinski's equation via a Legendre transform. 
\Sect{rescale} is devoted to the matter of transferring to dimensionless variables, allowing
us to arrive at the flow equation which will be used for much of the rest of the paper.
Some insight into the structure of flow equations is provided by their diagrammatic representation,
discussed in \sect{Diagrammatics}. Finally, some other ERGs are briefly mentioned in \sect{otherflow}.

\subsection{Notation \& Conventions}
\label{sec:Notation}

Throughout this paper we work in $\D$ Euclidean dimensions.
Euclidean space is the natural setting for the ERG, since it allows an easy separation
of modes into high/low energies 
(the indefinite signature of Minkowski space means that high energy states can
have small or vanishing invariant masses, which presents difficulties).
For simplicity (and, in some instances, tractability), most of our work will focus on theories of a single scalar
field,  $\phi$. The symbol $\varphi$ will be used to denote some collection of fields, which need
not be restricted to just scalars (but could represent just $\phi$). As we will see in \sect{General_ERG}, our blocking procedure acts on the fields and so, generally speaking, they depend on $\Lambda$. However, only in situations where this dependence is important will we bother to indicate it explicitly.

The Euclidean coordinate vector will be denoted by  $x$, and the momentum by
$p$. As is commonly the case in the literature, the same symbol will be used for 
the norm, with the meaning being clear by the context: if $x$ or $p$ appears as 
an argument, \eg\ $\phi(x)$ or $\phi_x$, then it is understood as the coordinate vector 
(explicitly, $x^\mu$). The scalar product of two vectors is denoted using a dot, \viz\
$p\cdot x$. If a coordinate appears squared, then obviously the norm is meant
\eg\ by $p^2$ we mean just $p \cdot p$. 

The Fourier transform of $\phi(x)$ is:
\be
	\phi(p) = \Int{x} \phi(x) e^{-i p\cdot x},
	\qquad
	\phi(x) = \MomInt{\D}{p} \phi(p) e^{i p\cdot x}.
\label{eq:FT}
\ee
Notice that we are (to borrow from programming terminology) using an `object-oriented' notation for
$\phi$: the same symbol is used for $\phi(x)$ and its Fourier transform, with the argument telling
us how $\phi$ should be interpreted [but we will not go as far as writing \eg\ $\phi.x()$!]. As usual, letters at the end of the alphabet $x,y$ will stand for
position-space coordinates, whereas letters closely following $p$ will be understood as momenta.
In this vein, we will use an object-oriented, compact notation for various integrals:
\[
	\int_x \equiv \Int{x}, \qquad \int_p \equiv \MomInt{\D}{p}.
\]
The Dirac $\delta$-function---which is, of course, not really a function but a distribution---will be
loosely understood as
\be
	\DiracD{\D}{x} = \int_p e^{i p\cdot x}.
\label{eq:DD}
\ee

The functional derivative \wrt\ $\phi(x)$ will be denoted, as usual, by $\delta /\delta \phi(x)$
and satisfies
\be
	\fder{\phi(y)}{\phi(x)} = \DiracD{\D}{y-x}.
\label{eq:dphidphi-pos}
\ee
The functional derivative \wrt\ $\phi(p)$ is defined via Fourier transform:
\be
	\fder{}{\phi(p)} \equiv \Int{x} e^{i p\cdot x} \fder{}{\phi(x)}.
\ee
Using this equation, together with~\eqs{FT}{DD}, we see that
\be
	\fder{\phi(p)}{\phi(q)} = \Int{x} e^{i(q-p)\cdot x} = (2\pi)^{\D} \DiracD{\D}{p-q}
	\equiv \deltahat{p-q}.
\label{eq:dphidphi}
\ee

In addition to being used for the scalar product between two vectors, a dot will also be used
to denote integrals over functions of the coordinates \eg\
\be
\begin{split}
	A \cdot B &\equiv  \int_p A(p) B(-p) = \int_{x} A(x) B(x), 
\\
	A \cdot K \cdot B & \equiv  \int_p A(p) \cutoff(p,-p) B(-p) = \int_{x,y} A(x) \cutoff(x-y) B(y),
\end{split}
\label{eq:dot-notation}
\ee
where%
\footnote{As always, it is translational invariance that allows us to extract the momentum conserving $\delta$-function: its presence follows from the automatic invariance of the integral on the \rhs\ under the change of variables $x_\mu \mapsto x_\mu + a_\mu, \ y_\mu \mapsto y_\mu + a_\mu$, together with
invariance of $\cutoff(x-y)$ under the same shift.
}
\[
	\cutoff(p,q)\deltahat{p+q} = \int_{x,y} \cutoff(x-y) e^{i (p\cdot x + q\cdot y)};
	\qquad
	\mathrm{with} \ \cutoff(p^2) \equiv \cutoff(p,-p).
\]
Similar notation to~\eq{dot-notation} is used in the cases where either $A$, $B$ or both are functional derivatives, though care must be taken with the momentum space arguments when expanding out the shorthand. For example, $\phi \cdot \delta / \delta \phi = \int_p \varphi(p) \, \delta / \delta \phi(+p)$. Whilst easy to check explicitly, the intuitive reason for this result can be seen by allowing this operator to act on $\phi \cdot \phi = \int_q \phi(q) \phi(-q)$: the $ \delta / \delta \phi(+p)$ eats a field leaving behind $ \phi(-p)$.

Notice from~\eq{dot-notation} that we will always interpret things like $A \cdot \cutoff \cdot B$ in momentum space first and then transfer to position space if required. This will enable us to use simple notation. For example, we will regularly encounter an object $\cutoff'(p^2)$, where a prime denotes a derivative \wrt\ the argument. If we take our object-oriented notation too seriously, then in position space this would be $\cutoff'\bigl((x-y)^2\bigr)$ but where now the prime should not be interpreted as a derivative \wrt\ the argument! Using the same symbol for things like $\phi(x)$ and $\phi(p)$ on the one hand, but on the other interpreting more complex expressions first in momentum space, enables us to keep notational clutter to a minimum.

We conclude this section by discussing the dimensionality of the various objects introduced.
The canonical (\aka\ engineering) dimension of some quantity, $X$, will be denoted by $\dimc{X}$.
Lengths, $L$, have dimension $-1$ whereas energies have dimension $+1$:
\[
	\dimc{L} = -1, \qquad \dimc{\Lambda} = +1.
\]
The canonical dimension of the scalar field, $\phi(x)$, follows from inspection of the standard kinetic term
$\int_x \partial_\mu \phi(x) \partial_\mu \phi(x)$. Since this is a contribution to the action, it must
be dimensionless and we therefore conclude that
\[
	\dimc{\phi(x)} = \frac{\D-2}{2}, \qquad \dimc{\phi(p)} = -\frac{\D+2}{2},
\]
where the dimensionality of $\phi(p)$ follows from that of $\phi(x)$, given their relationship via Fourier transform, \eq{FT}.
The canonical dimensions of the various other objects that we have introduced are:
\be
	\dimc{\DiracD{\D}{x}} = \D,\qquad \dimc{\deltahat{p}} = -\D,
	\qquad
	\dimc{\fder{}{\phi(x)}} = \frac{\D+2}{2}, \qquad 
	\dimc{\fder{}{\phi(p)}} = \frac{2-\D}{2}.
\label{eq:various}
\ee

Of course, one of the things which makes quantum field theory so rich
is that quantum fields can acquire anomalous dimensions, essentially meaning that the scaling dimension of the field is not equal to the canonical dimension. 
In the context of the ERG, we will see in \sect{rescale} that this is a subtle point.

As a final point, we anticipate that we will find it useful to render
the field dimensionless using appropriate powers of $\Lambda$.
Taking the field to have canonical dimension (the following is essentially
unchanged in the presence of anomalous scaling)
we introduce new variables
\be
	\tilde{\phi}(\tilde{x})= \tilde{\phi}(x,\Lambda) =   \phi(x) / \Lambda^{(\D-2)/2},
	\qquad
	\tilde{\phi}(\tilde{p})= \tilde{\phi}(p,\Lambda) =   \phi(p) \Lambda^{(\D+2)/2},
\label{eq:dimless-fields}
\ee
where
\be
	\tilde{x} \equiv x\Lambda, \qquad \tilde{p} \equiv p/\Lambda.
\label{eq:dimless-coords}
\ee
Notice that [as we could have anticipated from~\eq{various}]
\be
	\fder{}{\phi(x)} = 
	\int_{\tilde{y}}
	\fder{\tilde{\phi}(\tilde{y})}{\phi(x)} \fder{}{\tilde{\phi}(\tilde{y})}
	=
	\Lambda^{\D} \int_y \frac{1}{\Lambda^{(\D-2)/2}} \DD{\D}(y-x) \fder{}{\tilde{\phi}(\tilde{y})}
	=
	\Lambda^{(\D+2)/2} 
	\fder{}{\tilde{\phi}(\tilde{x})},
\label{eq:derphi-chain}
\ee
from which it follows that
\be
	\fder{\tilde{\phi}(\tilde{y})}{\tilde{\phi}(\tilde{x})} =  \DD{\D}(\tilde{y}-\tilde{x}),
	\qquad
	\fder{\tilde{\phi}(\tilde{p})}{\tilde{\phi}(\tilde{q})} =  \deltahat{\tilde{p}-\tilde{q}}.
\ee

\subsection{Locality}
\label{sec:local}

In the introduction, the importance of locality in the intuitive framework underpinning the
early works on the ERG (and WRG) was stressed. Roughly speaking, we might
imagine a scenario where, in the UV, we start off with a local action. Iterating
the ERG procedure, the Wilsonian effective action remains local at all finite
intermediate scales, $\Lambda$. However, in the limit $\Lambda \rightarrow 0$,
we might expect non-localities to emerge in certain cases; after all, an infinite
number of steps $\Lambda \mapsto \Lambda- \delta \Lambda$ have been
performed.

To sharpen this discussion, there are several different
notions of (non)locality that must be delineated. In particular, and as we will see
in the next section, the flow equation actually introduces non-localities in to the Wilsonian
effective action, even at non-zero values of $\Lambda$, for theories we might
expect to be strictly local. However, such non-localities are of a 
very particular, `soft' type. 

For example, we will see that a typical two-point contribution
to the action takes the form
\[
	\hf
	\Int{x} \Int{y} \phi(x) X_\Lambda(x-y) \phi(y)=
	\MomInt{\D}{p} \phi(-p) X_\Lambda(p^2) \phi(p),
\]
where $X_\Lambda(x-y)$ is some kernel which, whilst possibly having a local component
which goes as $\DiracD{\D}{x-y}$, has other components which do not. 
If we simply accept for the moment that
this is what we find, then it is clear that there is some degree of
non-locality present, with the scale being set by $\Lambda$. 
The soft non-locality mentioned a moment ago is
often referred to as `quasi-locality' and, in the current context,
would be the requirement that $X_\Lambda(p^2)$ has an all-orders 
Taylor expansion for small $p^2/\Lambda^2$. Equivalently, in position space, the
above contribution to the action exhibits an all-orders derivative expansion. 
Note that quasi-locality forbids, for example, contributions to the action like
\[
	\Int{x} \phi(x) \Int{y} \phi(y).
\]
It is easy to generalize these considerations to the full Wilsonian effective action. Working in position space,  a quasi-local action exhibits a derivative expansion%
\footnote{%
In \sect{Trunc-DE} we will describe an approximation scheme
based on this expansion.
}:
\be
	S_\Lambda[\phi] \sim
	\Int{x}
	\Bigl[
		V_\Lambda(\phi) + W_\Lambda(\phi) \partial_\mu \phi \partial_\mu \phi + \order{\partial^4}
	\Bigr],
\label{eq:FullAction-DE}
\ee
where $V$ and $W$ do not contain derivatives but are otherwise arbitrary. To transfer to momentum space, let us suppose that the action can be expanded in powers
of the field:
\be
\begin{split}
	S_\Lambda[\phi] & = \sum_n \int_{x_1,\ldots,x_n}
	\frac{1}{n!}
	\Svert{n}_\Lambda(x_1,\ldots,x_n) \phi(x_1)\cdots \phi(x_n)
\\
	& =
	 \sum_n \int_{p_1,\ldots,p_n}
	\frac{1}{n!}
	\Svert{n}_\Lambda(p_1,\ldots,p_n) \phi(p_1)\cdots \phi(p_n)\deltahat{p_1+\cdots+p_n},
\end{split}
\label{eq:action-exp}
\ee
where, in the second line,  we have assumed translation invariance of the vertices so that
\be
	\Svert{n}_\Lambda(p_1,\ldots,p_n) \deltahat{p_1+\cdots+p_n}
	=
	\int_{x_1,\ldots,x_n} \Svert{n}_\Lambda(x_1,\ldots,x_n) 
	e^{i (p_1\cdot x_1 + \cdots + p_n \cdot x_n)}.
\ee
Again, we have used an object oriented notation for the vertices, $\Svert{n}$. 
 Let us also take the opportunity to
 introduce the following shorthand:
 \be
 	\Svert{2}_\Lambda(p^2) \equiv \Svert{2}_\Lambda(p,-p).
 \ee

Quasi-locality requires that the $\Svert{n}_\Lambda(p_1,\ldots,p_n)$ can be Taylor expanded in the $p_i/\Lambda$. It is thus apparent that a quasi-local
theory becomes strictly local in the limit $\Lambda \rightarrow \infty$. It is worth pointing out that, since
this limit can only be taken for nonperturbatively renormalizable theories, theories defined by a bare action away from a renormalized trajectory have some irreducible non-locality present at the scale
of the bare cutoff. 

With this in mind we will, nevertheless, henceforth loosely take non-locality to
refer only to those
functions which (with the extraction of a \emph{single} momentum conserving $\delta$-function, if appropriate) have non-analytic dependence on momenta. (For the rest of this paper, we will have
no need to distinguish such theories from quasi-local theories for which the limit $\Lambda \rightarrow \infty$ does not exist.)

In this paper, we shall display a preference for UV actions which are quasi-local. 
This is, of course, in accord with the discussion in the introduction of the circumstances
under which the ERG is expected to be useful. 
Moreover, 
this restriction is apparently necessary in order for cluster decomposition
to be realized by a QFT~\cite{WeinbergI}. 
Nevertheless, this prejudice for quasi-locality is inflicted at the level of \emph{solutions} to the flow equation; there is nothing to stop one investigating non-local solutions, should we so desire.
Indeed,
in \sects{GFP}{Gen-TP} we will use a sufficiently
simple example to do precisely this. However, without further restrictions, we will see that there are an uncountable infinity of fixed-points, with a continuous spectrum of RG eigenvalues and it is not clear how to make sense of this.

However, whilst we are free to relax the restriction to quasi-local solutions of the flow equation, we strictly adhere to the demand that all \emph{inputs} to the flow equation are quasi-local, at least for 
$\Lambda>0$. This is necessary in order that blocking is performed only over local patches~\cite{aprop} and
ensures that, if we start from a quasi-local action, this property will be realized
all the way along the flow, at least for $\Lambda>0$. At $\Lambda=0$,  it is quite legitimate for non-local interactions to arise from a quasi-local action since, although blocking is only over local patches, an
infinite number of RG steps have been performed. Note, though, that
this is not to say that the action in the $\Lambda\rightarrow 0$ limit is necessarily non-local, merely that such non-locality is a possibility.

\subsection{Generalized ERGs for Scalar Field Theory}
\label{sec:General_ERG}

In this section, we give a derivation of several flow equations for scalar field theory, using general principles. The flow equations that we will discuss have a structure similar to Polchinski's~\cite{pol}. It should be pointed out that, for the Polchinski equation at any rate, there are alternative derivations. In particular, a much more mathematically minded approach is given in~\cite{Salmhofer}. 

It is always important to remember that the ERG is really an auxiliary construction in QFT: by this it is meant that the physics is contained in the partition function, coupled to operators via various sources, and that the ERG is just one particular way (with its own strengths and weaknesses) of extracting the physics. 
Indeed, universal quantities know nothing about the introduction of an effective scale as a computational device. But part of the point is that the converse is not true; the Wilsonian effective action \emph{does} know about universal quantities and can be useful in their evaluation.

As such, it is a fundamental requirement of the ERG that the partition function is left invariant under the flow (otherwise it would be the actual physics, rather than our \emph{description} of the physics that would change under the RG procedure). Consequently, but rather abstractly, this means that a family of ERG equations follows by taking~\cite{TRM+JL,mgierg1,ym1}
\be
-\flow e^{-S_\Lambda[\phi]} =  \int_x \fder{}{\phi(x)} \left(\Psi_\Lambda(x) e^{-S_\Lambda[\phi]}\right),
\label{eq:blocked}
\ee
where the $\Lambda$-derivative is taken at constant $\phi$.
Invariance of the partition function, $\pf = \Fint{\phi} \, e^{-S_\Lambda[\phi]}$, formally follows from the
total derivative on the \rhs\ of~\eq{blocked}.\footnote{We are not going to take any particular care over the measure and, indeed, will generally discard constant contributions to the action being as they are unimportant for our considerations.}
The object $\Psi$ (which in general is both a function and a functional of $\phi$)  parametrizes the continuum analogue of a Kadanoff blocking (the precise link will be made below). The only definite requirements on $\Psi$ are that~\cite{aprop}:
\begin{enumerate}
	\item It does indeed correspond to a (continuum) blocking procedure, where 
	the blocking is performed only over local patches;

	\item  It ensures UV regularization of the flow equation, which can be achieved by including a 
	(suitably strong) UV cutoff in $\Psi$.
\end{enumerate}

To make all of this more concrete~\cite{Wetterich-AA+RGE,mgierg1}, let us explicitly relate $\Psi$ to the blocking procedure. Just as in the discrete case, the effective field is written as some average over the bare field:
$\phi(x) = b_\Lambda[\phi_0](x)$. To implement locality,  we demand that the blocking procedure
is suitably local. For example, given a kernel $f_\Lambda(z)$ which is steeply decaying
for $z\Lambda>1$, we could choose $b_\Lambda[\phi_0](x) = \int_y f_\Lambda(x-y)\phi_0(y)$. Note, though, that there are many other choices we could make and that there is no need for $b_\Lambda[\phi_0]$ to be linear in field.

Using the blocking functional, we can write the effective action in terms of the 
bare action as follows:
\be
	e^{-S_\Lambda[\phi]} = \Fint{\phi_0} \delta\bigl[ \phi - b_\Lambda[\phi_0]\bigr]
	e^{-S_{\Lambda_0}[\phi_0]}.
\label{eq:blocking}
\ee
Integrating over $\mathcal{D} \phi$ on both sides, it is clear that (formally) the partition function is left invariant under this procedure.
We can now relate $\Psi$ to $b_\Lambda$ by recognizing that if we choose
\be
	\Psi_\Lambda(x) e^{-S_\Lambda[\phi]}
	=
	 \Fint{\phi_0} \delta\bigl[ \phi - b_\Lambda[\phi_0]\bigr]
	 \Lambda \pder{b_\Lambda[\phi_0](x)}{\Lambda}
	e^{-S_{\Lambda_0}[\phi_0]},
\label{eq:Psi-explicit}
\ee
then~\eq{blocked} follows from~\eq{blocking}. Note that this form of $\Psi$ is consistent with Wegner's observation~\cite{WegnerInv} that $\Psi$ should depend on $S_\Lambda$ (a fact which makes the flow equation non-linear).

The flow equation corresponding to $\Psi$ follows directly from~\eq{blocked}:
\be
	-\flow S_\Lambda[\phi]
	=
	\int_x \fder{S_\Lambda}{\phi(x)} \Psi_\Lambda(x) - \int_x \fder{\Psi_\Lambda(x)}{\phi(x)}.
\label{eq:primitive-flow}
\ee
The two terms on the \rhs\ are often called the classical and quantum terms, respectively. The reason for this nomenclature is clear from a diagrammatic point of view, since the first term generates tree-like diagrams whereas the second generates loop diagrams, as we will see explicitly in \sect{Diagrammatics}. However, it must be borne in mind that the classical diagrams have vertices which incorporate quantum fluctuations down to the effective scale and so this classical interpretation needs to be taken with a pinch of salt.

Before moving on, it is well worth noting that the flow equation~\eq{primitive-flow} 
follows
from the infinitesimal field redefinition~\cite{Wegner_CS,WegnerInv}
\be
	\phi'(x) = \phi(x) - \delta t  \, \Psi_\Lambda(x),
\ee
where $\delta t = - \delta \Lambda /\Lambda$.
Under the path integral, this change of variables induces a change to the action and a non-trivial
Jacobian given, respectively, by
\begin{align*}
	S_\Lambda[\phi'] & = S_\Lambda[\phi] 
	- \delta t \int_x \Psi_\Lambda(x) \delta S_\Lambda [\phi]/ \delta \phi(x) + \order{(\delta t)^2}
\\
	\abs{\fder{\phi'}{\phi}} & = 1  - \delta t \int_x \delta \Psi_\Lambda(x) /\delta \phi(x) 
	+ \order{(\delta t)^2}.
\end{align*}
This implies that
\be
	\pf = \Fint{\phi'} e^{-S_\Lambda[\phi']} 
	= \Fint{\phi}
	e^{-S_\Lambda[\phi] + \delta t \, \mathcal{G}_\mathrm{tra}[\Psi] S_\Lambda[\phi]} +\order{(\delta t)^2},
\label{eq:ChangeVariables}
\ee 
where, using Wegner's notation~\cite{WegnerInv,Wegner_CS},
\be
	\mathcal{G}_\mathrm{tra}[\Psi] S_\Lambda[\phi]= 
	\int_x
	\left\{
	 	\Psi_\Lambda(x) \fder{S_\Lambda[\phi]}{\phi(x)}
		- \fder{\Psi_\Lambda(x)}{\phi(x)}
	\right\}.
\label{eq:G_tra}
\ee
The `tra' stands for `transformation of variables'. Equating $S_{\Lambda - \delta \Lambda}[\phi]$ with $S_\Lambda[\phi] - \delta t \mathcal{G}_\mathrm{tra}[\Psi] S_\Lambda[\phi]$
(up to higher order terms) reproduces the flow equation~\eq{primitive-flow} in the limit $\delta \Lambda \rightarrow 0$. Viewing the flow equation as coming from a change of variables has been thoroughly explored in~\cite{TRM+JL,TRM+JL-conf} (see also~\cite{Aoki-Scheme}).

For the rest of this paper we shall almost exclusively work with those $\Psi$s which yield flow equations with the 
same basic structure as Polchinski's~\cite{pol}. To this end, we need to introduce two new objects,
the `ERG kernel', $\dd_\Lambda(x-y)$---which incorporates the UV regularization---and the `seed action'~\cite{aprop,scalar2,mgierg1,mgierg2,qcd}, $\hS_\Lambda$.
Momentarily suppressing our curiosity about both of these objects we take
\be
\label{eq:Psi}
	\Psi_\Lambda(x) = \hf \dd_\Lambda(x-y) \fder{\Sigma_\Lambda}{\phi(y)},
\ee
where
\be
	\Sigma_\Lambda \equiv S_\Lambda - 2\hS_\Lambda.
\label{eq:Sigma}
\ee
Let us emphasise that~\eq{Psi} corresponds to a \emph{choice} for $\Psi$ that we are not compelled to make.

Resolutely refusing to say any more about $\dd$ or the seed action for
a moment longer, we substitute~\eq{Sigma} into~\eq{primitive-flow} to
yield:
\be
	-\flow  S 
	= \hf \fder{S}{\phi} \cdot \dd \cdot \fder{\Sigma}{\phi} 
	- \hf \fder{}{\phi} \cdot \dd \cdot \fder{\Sigma}{\phi}
\label{eq:ProtoFlow}
\ee
where we have dropped the various subscripted $\Lambda$s, for brevity, and employ the shorthand
introduced in~\eq{dot-notation}. The form of this equation tells us some important things
about $\dd$.

First of all, let us note that since the Wilsonian effective action is
dimensionless, the same must be true of the object
\[
	\classical{}{\dd}{}.
\]
Therefore, the dimensionality of $\dd$ is related to that of $\phi$.
We will proceed by supposing that $\phi$
has canonical scaling dimension.
This sounds like it might be too restrictive. However, as we will
discuss further in \sect{rescale}, in this approach the anomalous dimension
(typically) appears via the usual modification of the kinetic term by
the field strength renormalization.
Anyhow, recalling~\eq{various} we thus conclude
that $\dd$ has mass dimension $-2$; in addition we know that
$\dd$ is quasi-local and incorporates UV regularization.

To construct a $\dd$ that satisfies all of these criteria
let us introduce an object which looks like a UV regularized
propagator:
\be
	\ep_\Lambda(p^2) \equiv \ep(p^2;\Lambda) = \frac{\cutoff (p^2/\Lambda^2)}{p^2},
\label{eq:sensible}
\ee
where $\cutoff (p^2/\Lambda^2)$ is a UV cutoff function, which exhibits a derivative expansion, and which we choose to normalize such that
$\cutoff(0)=1$. The cutoff function decreases monotonically, 
decaying fast enough for large momenta  (how fast depends on what we are trying to achieve:
it may be possible to regularize theories on particular RG trajectories with power law decay but to ensure, for example,
that all eigenperturbations of the Gaussian fixed-point are finite requires decay faster than any power, as we will see in \sect{Gen-TP}). 
The point of all this is that we can use $\ep$ to construct a suitable $\dd$ by taking
\be
	\dd_\Lambda(p^2) \equiv \dd(p^2;\Lambda) 
	= -\totalflow \ep_\Lambda(p^2) = \frac{2\cutoff'(p^2/\Lambda^2)}{\Lambda^2},
\label{eq:dd}
\ee
where here the prime means a derivative \wrt\ the argument of the associated object.%
\footnote{A prime will be used to denote several different things
throughout this paper, with the meaning hopefully being clear from the context.}

Before moving on, let us say a few more things about $\ep$.
Using object-oriented notation, we have
\be
	\ep_\Lambda(x-y) = \int_p \ep_\Lambda(p^2) e^{ip\cdot(x-y)}.
\ee
We will frequently refer to $\ep$ as an effective propagator.
In the literature, the symbol $c$ is sometimes used for the cutoff function (our $\cutoff$), with
$\ep$ represented by $\Delta$. However, we will reserve $\Delta$ for later use.  
In the more mathematical literature, 
one often finds the $\ep_\Lambda(p^2)$ of~\eq{sensible}
referred to as a `covariance' and, moreover, that $e^{-\hf \phi \cdot \ep^{-1} \cdot \phi}$ is absorbed into the measure of the functional integral. Finally, we will often omit writing the explicit $\Lambda$-dependence and so just write $\ep(p^2)$.

At this stage, the only object in our flow equation~\eq{ProtoFlow} that we are yet to discuss
is the seed action, the interpretation of which is as follows. Fixing $\Psi$ to take the form~\eq{Psi} represents a constraint on the allowed blocking functionals, the residual freedom of which is carried by the form of the ERG kernel and the seed action. In principle, the seed action can be an arbitrarily complicated functional of the field, so long as it has a derivative expansion. 
[Note, though, that we cannot make the tempting choice $2\hS = S$, since then the flow equation is
linear in the action and so, recalling the discussion around~\eq{Psi-explicit}, does not implement a blocking procedure.]

Unlike the Wilsonian effective action---for which we solve---the seed action is an input to the flow equation. Generally speaking, universal quantities must come out independent of the choice of seed action and so, in this sense, it does not matter how it is chosen. Indeed, it is often instructive to leave it unspecified in scalar field theory as one finds, without too much work, that it often cancels out of many quantities of interest.%
\footnote{%
Actually, this used to be a lot of work~\cite{scalar2}, but in the present paper the old analysis is radically simplified.
}
We will see this explicitly for the the $\beta$-function of $\lambda \phi^4$ theory in \sect{Canonical} and for the correlation functions in \sect{comp-corr-fns}.
Indeed, in an ideal world, we would always leave the seed action as general as possible. However,
we will encounter examples in this paper where this makes life too hard 
(for the moment---hopefully this will change in the future)
and so instead make
the simplest choice.
In scalar field theory, at any rate, this amounts to setting the interactions of the seed action to zero, which yields Polchinski's equation (the complications arising in gauge theory will be discussed in \sect{other-theories}).

To obtain Polchinski's equation we split the Wilsonian effective action and seed action according to
\be
	S[\phi] =  \hf \phi \cdot \ep^{-1} \cdot \phi + \Sint[\phi],
	\qquad
	\hS[\phi] = \hf \phi \cdot \ep^{-1} \cdot \phi + \hSint[\phi],
\label{eq:split}
\ee
and set $\hSint = 0$. There are a number of comments to make. Let us start by analysing what this splitting means for the Wilsonian effective action. At first sight, since our choice of $\ep$ seems to  correspond to a massless propagator, we might suppose that our splitting corresponds to a massless action with interactions carried by $\Sint$. But this is not the right way of looking at things: it is quite permissible for $\Sint$ to contain a mass term. Indeed, it is even permissible for $\Sint$ to contain a term which subtracts off some or even all of the regularized kinetic term! Presumably, the resulting theory would not be unitary upon continuation to Minkowski space, but that is a secondary consideration: first and foremost, we are interested in solutions of our ERG equation; their interpretation can come later. Indeed, we will recover in \sect{Gen-TP} a class of solutions found by Wegner~\cite{Wegner_CS} which correspond precisely to $\Sint$ removing the $\order{p^2}$ piece of the kinetic term. So, from this point of view, calling
$\ep_\Lambda(p^2) =\cutoff(p^2/\Lambda^2)/p^2$ a regularized propagator is in some sense
putting the coach before the horse: having solved our ERG equation, it might be that the propagator
actually turns out to go like $1/p^4$. Either way---and this is important---the cutoff function
does not itself introduce new poles into whatever the propagator ends up being.

So much for the splitting of the Wilsonian effective action. As for the seed action, the choice $\hSint = 0$ is the simplest. One might suppose that the simplest choice is $\hS = 0$ but, given our choice of $\Psi$ and $\dd$, we can now see why this is not so.  First of all, let us look at the quantum term. Up to a (divergent) vacuum energy term,  which we discard, this term can be obtained simply by replacing $\Sigma$ with $\Sint$ (for $\hSint=0$). Actually, this does not tell us much at all since, up to a different vacuum energy term, we could make the same replacement for $\hS=0$.
But what about the classical term? Now we can see the point of the previously mysterious factor of two in front of the $\hS$ contribution in~\eq{Sigma}. We have that
\[
	S[\phi] =  \hf \phi \cdot \ep^{-1} \cdot \phi + \Sint_\Lambda[\phi],
	\qquad
	\Sigma[\phi] = - \hf \phi \cdot \ep^{-1} \cdot \phi + \Sint_\Lambda[\phi].
\]
Consequently, in the classical term---which is bilinear in $S$ and $\Sigma$---the cross-terms cancel.
Recognizing that
\[
	-\flow  \hf \phi \cdot \ep^{-1} \cdot \phi = 
	- \hf \phi \cdot \ep^{-1}\, \dd \, \ep^{-1} \cdot \phi,
\]
we thus see that the flow equation does indeed  reduce to Polchinski's, which is written entirely in terms of $\Sint$:
\be
	-\flow \Sint = \hf \classical{\Sint}{\dd}{\Sint} - \hf \quantum{\dd}{\Sint}.
\label{eq:Pol}
\ee

It will now be very profitable to unpick how much of what we have just done depends on the various choices we have made. \Eqns{Psi}{Sigma} are low level choices, that will be employed almost exclusively throughout this entire paper, from which the flow equation~\eq{ProtoFlow} follows directly. This flow equation is often referred to as a generalized ERG equation~\cite{scalar2,aprop,mgierg1,qcd}.
The choice~\eq{dd} is a valid one so long as we take the field to have canonical 
dimension (which we emphasise does not prohibit the appearance of a field strength
renormalization in the action, as will be properly discussed in \sect{rescale}).
Integrating up~\eq{dd} yields~\eq{sensible}. Given our pre-existing knowledge of QFT,
we interpret $\ep$ as a UV regularized propagator. But let us emphasise once again
that this interpretation can be misleading: it might be that, after solving the flow equation
for the Wilsonian effective action, it does not even have a standard kinetic term!
Nevertheless, even if this is true, we are always free to make the splittings~\eq{split}, which
we can understand as definitions for $\Sint$ and $\hSint$.

Leaving $\hS$ unspecified, the generalized flow equation can be rewritten as
\be
	-\flow \Sint = \hf \fder{\Sint}{\phi} \cdot \dd \cdot \fder{\Sigint}{\phi} 
	- \hf \fder{}{\phi} \cdot \dd \cdot \fder{\Sigint}{\phi}
	- \phi \cdot \ep^{-1} \dd \cdot \fder{\hSint}{\phi},
\label{eq:flow}
\ee
where we take the obvious definition $\Sigint \equiv \Sint - 2\hSint$.
Trivially, \eq{flow} reduces to the Polchinski equation if we set $\hSint=0$.

Let us conclude this section by mentioning that it is easy to extend the
flow equation to $N$ scalar fields: we just include a classical and quantum term on the \rhs\ for each of the new fields and take the effective action to depend on the complete set, which we will denote by $\varphi_i$. Thus we introduce a set of kernels, $\dd_{ij}$, a sensible choice for which is
\be
	\dd_{ij}(p^2) = \dd(p^2) \delta_{ij}.
\ee
The generalized flow equation~\eq{ProtoFlow} becomes:
\be
	-\flow  S 
	= \hf \fder{S}{\varphi_i} \cdot \dd_{ij} \cdot \fder{\Sigma}{\varphi_j} 
	- \hf \fder{}{\varphi_i} \cdot \dd_{ij}  \cdot \fder{\Sigma}{\varphi_j},
\label{eq:flow-N}
\ee
where a sum over repeated indices is understood.
[It would be entirely reasonable to remove the indices entirely, 
allowing the dots sandwiched between the functional derivatives
and the kernels to stand both for an integral over momentum and
a sum over (suppressed) indices.] Whilst this flow equation and its cousins can
be used to study completely general theories of $N$ scalar
fields, they are more commonly used to study O$(N)$ 
scalar field theory,
by restricting the action to be invariant under global O$(N)$
transformations.

In \sect{other-theories} we will
consider flow equations for theories containing fields
other than scalars.

\subsection{The Effective Average Action}
\label{sec:EEA}

Currently, by far and away the most popular flow equation for performing practical calculations is the flow equation for the `effective average action', $\Gamma_\Lambda$, the \emph{IR} regulated 
generator of one-particle irreducible (1PI) diagrams. That the regularization is IR and not UV is perhaps surprising but there is an intuitive explanation~\cite{TRM-Elements}. As ever, consider integrating out degrees of freedom between $\Lambda_0$ and $\Lambda$. For the remaining unintegrated modes, $\Lambda$ acts as a UV cutoff; this is the picture we have been employing up until now. Contrariwise, for the \emph{integrated} modes, $\Lambda$ acts as an IR cutoff. From this perspective, it is not so surprising that one can flip between the two viewpoints. Indeed, the flow equation for the effective average action is actually related to the Polchinski equation by a Legendre transform~\cite{TRM-ApproxSolns,Ellwanger-1PI}.

There are a number of different ways of deriving the flow equation for $\Gamma_\Lambda$~\cite{Wetterich-1PI,Bonini-1PI,Ellwanger-1PI,TRM-ApproxSolns}. We will follow a recent, elegant method due to Osborn and Twigg~\cite{HO-Remarks}. First of all, we simply \emph{define} and object $G_\Lambda$ via a Legendre transform relation:
\be
G_\Lambda[\Phi,D] \equiv \Sint_\Lambda[\phi]
	-
	\hf
	\bigl(
		\Phi  -\phi
	\bigr)
	\cdot
		D_\Lambda
	\cdot
	\bigl(
		\Phi  - \phi
	\bigr),
\label{eq:map-Pol}
\ee
where we will leave $D_\Lambda(p^2)$ undetermined, for the moment. We understand $\Phi$ to be defined via the relationship
\be
	\fder{\Sint_\Lambda[\phi]}{\phi(p)} = -D_\Lambda(p^2)
	\bigl[
		\Phi(-p)  - \phi(-p)
	\bigr]
	.
\label{eq:dS/dphi}
\ee
Using the complementary relationship
\be
	\fder{G_\Lambda[\Phi]}{\Phi(p)} = 
	- D_\Lambda(p^2)
	\bigl[
		\Phi(-p)  - \phi(-p)
	\bigr]
\label{eq:Gamma_Phi}
\ee
it follows that
\be
	\flow \bigr\vert_\phi \Sint_\Lambda[\phi] =
	\flow \bigr\vert_\Phi G_\Lambda[\Phi] 
	- \hf 
	\bigl(
		\Phi  -\phi
	\bigr)
	\cdot
		\dot{D}_\Lambda
	\cdot
	\bigl(
		\Phi  - \phi
	\bigr)
\label{eq:flows}
\ee
with
$
	\dot{D}_\Lambda(p^2) \equiv -\flow D_\Lambda(p^2).
$
Now, substituting~\eq{dS/dphi} into~\eq{flows} it is apparent that if we choose
\be
	\bigl[D_\Lambda(p^2)\bigr]^{-2} \dot{D}_\Lambda(p^2) = \dd_\Lambda(p^2)
\label{eq:D-eq}
\ee
then we find that
\be
	-\flow \Bigr\vert_\Phi G_\Lambda[\Phi] = - \hf \quantum{\dd_\Lambda}{\Sint_\Lambda}.
\ee
To re-express the \rhs\ in terms of $G$, we functionally differentiate both sides of~\eq{dS/dphi} \wrt\ $\phi$, and both sides of~\eq{Gamma_Phi} \wrt\ $\Phi$ from which we conclude that
\begin{multline}
	\int_q
	\biggl\{
		\dfder{S_\Lambda[\phi]}{\phi(p)}{\phi(q)} - D_\Lambda(p^2) \deltahat{p+q}
	\biggr\}
	\biggl\{
		\dfder{\Gamma_\Lambda[\Phi]}{\Phi(-q)}{\Phi(-p')} + D_\Lambda(q^2) \deltahat{p'+q}
	\biggr\}
\\
	=
	-\bigl[D_\Lambda(p^2)\bigr]^2
	\deltahat{p-p'}.
\end{multline}
Defining
$G^{(2)}_\Lambda \equiv \delta^2 G_\Lambda / \delta \Phi \delta \Phi$ and discarding a vacuum energy term, we arrive at the following flow equation:
\be
	-\flow G_\Lambda[\Phi]
	=
	\hf \Tr
	\Bigl\{
	 \dot{D}_\Lambda
	 \Bigl[
	 	D_\Lambda + G_\Lambda^{(2)}
	 \Bigr]^{-1}
	 \Bigr\}.
\label{eq:EEA-flow}
\ee
Let us now return to~\eq{D-eq}. Any solution to this equation which is quasi-local is legitimate; we will investigate two choices. First of all, let us take
\be
	D_\Lambda(p^2) = D^{\Lambda_0}_\Lambda(p^2) 
	= \frac{1}{\ep_{\Lambda_0}(p^2) - \ep_\Lambda(p^2)}
	\qquad
	\mathrm{with}
	\qquad
	\Gamma_\Lambda[\Phi] \equiv G[\Phi,D^{\Lambda_0}_\Lambda]
.
\label{eq:Choice-TRM}
\ee
In this case, \eq{EEA-flow} becomes the flow equation written down in~\cite{Bonini-1PI,Ellwanger-1PI,TRM-ApproxSolns}:
\be
	-\flow \Gamma_\Lambda[\Phi]
	=
	\hf \Tr
	\Bigl\{
	 \dot{D}^{\Lambda_0}_\Lambda
	 \Bigl[
	 	D^{\Lambda_0}_\Lambda + \Gamma_\Lambda^{(2)}
	 \Bigr]^{-1}
	 \Bigr\}.
\label{eq:Morris-1PI}
\ee
 As shown by Morris~\cite{TRM-Elements}, $\Gamma_\Lambda$ is an IR regularized generator of 1PI Green's functions and reduces to the standard effective action in the limit $\Lambda \rightarrow 0$. Referring back to~\eq{map-Pol}, note that~\eq{Choice-TRM} represents the unique choice for which the Wilsonian effective action and effective average action coincide at the bare scale.

Next, let us make the choice
\be
	D_\Lambda(p^2) = D^{\infty}_\Lambda(p^2) = \frac{1}{1 - \ep_\Lambda(p^2)}
	\qquad
	\mathrm{with}
	\qquad
	\Gamma'_\Lambda[\Phi] \equiv G[\Phi,D^{\Lambda_0}_\Lambda]
,
\label{eq:Choice-CW}
\ee
in which case the flow equation~\eq{EEA-flow} reduces to the one written down by Wetterich~\cite{Wetterich-1PI}, which in its standard form follows from the following changes of variables: we write $\Lambda$ as $k$, shift $\Gamma'_k[\Phi] \rightarrow \Gamma'_k[\Phi] + \hf \int_p \Phi(p) \Phi(-p) p^2$ and identify $R_k(p^2) = D^{\infty}_\Lambda(p^2) + p^2$. Notice that $D^{\infty}_\Lambda(p^2) = \lim_{\Lambda_0 \rightarrow \infty} D^{\Lambda_0}_\Lambda(p^2)$. However, removing the bare scale from the regulator in this way does not compromise the UV regularization of the flow equation since (as can be readily checked) $\dot{D}^\infty_\Lambda$ dies off rapidly in the UV.

Let us now investigate the difference between $\Gamma_\Lambda$ and $\Gamma'_\Lambda$. First of all, suppose that we are dealing with a theory which sits either at a fixed-point or on a renormalized trajectory. In this case, we can take the limit $\Lambda_0 \rightarrow \infty$ on both sides of~\eq{Morris-1PI}, after which (modulo the trivial changes mentioned above) the equation takes precisely the same form as Wetterich's and so we can identify $\Gamma'_\Lambda$ with $\Gamma_\Lambda$. However, for a non-renormalizable theory, we cannot remove the bare scale in the way. In this case, whilst both perfectly well defined objects, $\Gamma_\Lambda$ and $\Gamma'_\Lambda$ are not quite the same. Since the former reduces to the standard effective action in the limit $\Lambda \rightarrow 0$, the same cannot be true of the latter. Note, though, that for RG trajectories which lie on or close to the critical surface of some fixed-point, universality means that differences between $\Gamma_\Lambda$ and $\Gamma'_\Lambda$ will be suppressed by powers of $\Lambda_0$.

Let us conclude this section by noting that of all the derivations of the flow equation for the effective average action, Wetterich's differs most in spirit from the above (see also~\cite{Gies-Rev} for a clear discussion).
The starting point can again be traced back to the partition function but with several differences to the generalized flows of~\sect{General_ERG}. First of all, whilst UV regularization is assumed to be present (to make subsequent steps well defined) it is not made explicit. Secondly, the partition function is \emph{modified} via the inclusion of an \emph{additive} IR cutoff, which can be thought of as a momentum-dependent mass term. In this sense, the lineage of Wetterich's equation arguably begins with a paper by Symanzik~\cite{Symanzik-Power}. In this work, a mass term---albeit a momentum-independent one (meaning, amongst other things, that the resulting flow equation is not UV regularized)---is added to the action and the effects of varying this addition considered. However, the power of flow equations like~\eq{Morris-1PI}---for which a potted history can be found in the `note added' at the end of~\cite{TRM-ApproxSolns}---derives from their Wilsonian heritage.

\subsection{Rescalings}
\label{sec:rescale}

As mentioned in the introduction, the classic ERG procedure consists of two steps:
a coarse-graining, followed by a rescaling. Traditionally~\cite{Wilson,WegnerInv,Wegner_CS}, 
this latter operation is performed by considering
an explicit
dilatation and computing its effect on the effective action. Equivalently,
as noted by Morris~\cite{TRM-Deriv}, we can instead rescale
all quantities to dimensionless ones using the effective scale, $\Lambda$. 

However, there is a subtlety concerning precisely what we mean by 
dimensionless. Recall that we have formulated our flow equation in terms
of a field with canonical scaling dimension. Therefore, we can reduce
things to dimensionless variables by performing the change of variables~\eqs{dimless-fields}{dimless-coords}. Dropping all the tildes, we can equivalently view this change of variables as inducing the
shifts
\be
	\phi(x) \mapsto \Lambda^{(\D-2)/2} \phi(x),
	\quad
	x \mapsto x /\Lambda;
	\qquad
	\phi(p) \mapsto \Lambda^{-(\D+2)/2} \phi(p),
	\quad
	p \mapsto p\Lambda.
\label{eq:canonical-rescaling}
\ee
Nevertheless, we might well suspect that this is not the end of 
the story, since there is no mention here of any anomalous dimension.

We can get a feeling for what is going on by supposing, to begin
with, that the full bare action possesses a standard kinetic term. 
Along the flow, we expect this piece of the action will become modified by a 
scale-dependent factor, which we will denote by $1/Z_\Lambda$ and identify with the
field strength renormalization, viz
\[
	\frac{1}{2 Z_\Lambda}  \MomInt{\D}{p} \phi(-p,\Lambda) p^2 \phi(p,\Lambda).
\]
Moreover, let us define all the other couplings in the action such that a factor
of $1/\sqrt{Z}$ is extracted for each power of the field. For example, the momentum-independent four-point term would read:
\[
	\frac{\lambda}{4! Z_\Lambda^2} 
	\int_{p_1,\ldots,p_4} \phi(p_1,\Lambda) \cdots \phi(p_4,\Lambda)
	\hat{\delta}(p_1+\cdots+p_4).
\]

Now consider an RG step $\Lambda \rightarrow \Lambda -\delta \Lambda$. 
Recalling~\eq{G_tra}, it is apparent that the change induced in the action due to the
change of $Z_\Lambda$ can be undone by a quasi-local field redefinition (actually, a
strictly local redefinition, in this case). Specifically, if in this particular case the anomalous dimension is identified according to
\be
	\eta = \Lambda \der{\ln Z_\Lambda}{\Lambda},
\label{eq:eta}
\ee
then the necessary change to the field is
\be
\phi \mapsto \phi \Bigl(1 -  \frac{\eta}{2} \frac{\delta \Lambda}{\Lambda} \Bigr). 
\label{eq:AnomalousPhiShift}
\ee
In this example, we have identified $Z_\Lambda$ as a redundant
(or inessential) coupling.%
\footnote{Strictly speaking, we have not really identified $Z$ as a redundant coupling in the right way. Redundant couplings should be identified by first finding a fixed-point and then finding the associated operators \cf~\eq{perturb}. Of these, we then identify the subset which are redundant, thereby determining the redundant couplings in the vicinity of this \emph{particular} fixed-point. It is an important point that operators which are redundant at one fixed-point are not necessarily redundant at another.
}
Furthermore, by performing this rescaling after every RG step, we can ensure that
the coefficient of the standard kinetic term never flows.
Note, though, that unless otherwise specified, we will not insist on a canonically normalized
kinetic term. In this case we identify $Z$ as the field strength renormalization only
up to a scale-independent constant. At the level of the flow equation, the redefinition of the
field can be achieved by shifting $\Psi \mapsto \Psi - \eta /2 \, \phi$ so that~\eq{ProtoFlow} 
becomes:
\be
	\left(-\flow + \frac{\eta}{2} \Count_\phi\right) S[\phi]
	=
	\hf \classical{S}{\dd}{\Sigma} - \hf \quantum{\dd}{\Sigma},
\label{eq:flow-Ball-a}
\ee
where $\Count_\phi \equiv \phi \cdot \delta/\delta\phi$ is the `$\phi$-counting operator'.

The question is, though, why perform this additional rescaling, given
that~\eq{canonical-rescaling} is sufficient to reduce everything to
dimensionless form?
Recall that our motivation for rescaling is to conveniently uncover fixed-points, which govern the critical behaviour of physical systems. Now, the equivalence theorem (see~\cite{Ball+Thorne} for an excellent discussion in the context of the ERG) tells us that infinitesimal quasi-local field redefinitions leave the S-matrix---equivalently physics---invariant. So, if a coupling can be removed from the action by a redefinition such as~\eq{AnomalousPhiShift}, then there is no need for it to stop flowing at what is, for the remaining couplings, a fixed-point. Consequently, in order that the criterion~\eq{fp-criterion} should be physically useful, it is clear that we should apply it to the flow equation for which $Z_\Lambda$ has been removed by the appropriate rescaling of the field. (This discussion has assumed the presence of a standard kinetic term, but the lack of such an object
is not an impediment. In such
a case we can still perform a rescaling so as to remove the scale dependence
associated with the normalization of the field; prescriptions for doing this  will be
discussed in \sect{renorm}.)

However, this analysis begs a further question: if, to uncover fixed-points, we should remove $Z$ from the action, why do we not do the same for all the other redundant couplings? Indeed, precisely such a scheme is advocated in Weinberg's seminal paper on asymptotic safety~\cite{Weinberg-AS}, a point of view which is adopted by some subsequent works, see in particular~\cite{Percacci-NewtonConstant,Percacci-RG+Units,Percacci-Review}. Let us emphasise that there is nothing wrong with removing all redundant couplings from the action; however, it is unlikely that this procedure will reveal any new fixed-points. 

To understand the reason why, 
recall that the anomalous dimension can be taken into account in the flow equation by performing the field redefinition~\eq{AnomalousPhiShift}. This introduces a new term on the \rhs\ of the flow equation, $-\eta/2 \, \phi \cdot \delta S_\Lambda /\delta \phi$. The anomalous dimension, $\eta$, obtains some universal value, $\eta_\star$, at a given critical fixed-point.
With this in mind, consider performing additional field redefinitions, each of which we agree to associate with its own anomalous dimension, $\gamma_i$. Now, the spectrum of critical fixed-points clearly includes all of those found before, corresponding simply to $\gamma_{i\star}=0$. Is it, then, not reasonable to suppose that there might be additional fixed-points for which one or more of the $\gamma_{i\star}$ are non-vanishing? The point is that for a genuinely new fixed-point to exist---\ie\ one describing different physics from all others---it is not simply enough for a fixed-point to be found with one or more of the $\gamma_{i\star} \neq 0$: in addition, the spectrum of these anomalous dimensions must be quantized (\ie\ discrete). To see why this is the case, consider the following example. Suppose that a fixed-point exists not just for some $\gamma_{j \star} = 0$, but also for a continuous range of values in the neighbourhood of zero. Then these `new' fixed-points can be reached by a succession of infinitesimal, quasi-local field redefinitions, starting from the original fixed-point with $\gamma_{j \star}=0$. Being as they are related in this way, these fixed-points must describe the same physics (the fixed-points are equivalent, to use the standard lingo).

This leads us to consider the question as to whether the spectra of any the $\gamma_{i\star}$ can be quantized.
Before addressing this directly, let us note that precisely the same arguments can be applied to $\eta_\star$. In particular, for fixed-points with differing values of $\eta_\star$ to be genuinely different (in the sense of not describing the same physics) it must not be possible to go from one to the other via infinitesimal quasi-local field redefinitions. This suggests that the spectrum of $\eta_\star$ is quantized---and indeed it is, as we will see in \sect{FP-dual}. Now, Wegner pointed out that if the spectrum of $\eta_\star$ is quantized then there necessarily exists a marginal, redundant direction~\cite{Wegner_CS}.
His proof will be recalled in \sect{GenCon}, where it will become apparent that quantization of any of the $\gamma_{i\star}$ also implies the existence of a marginal, redundant direction. From the perspective of \sect{FP-dual}, it will be seen to be likely that each quantized anomalous dimension must come with its own marginal, redundant direction (for this not to be true, the direction in theory space associated with a quantized $\gamma_{i\star}$ would have to have a very particular, non-zero projection on to the direction associated with $\eta_\star$). Consequently, it is probably the case that, for there to be any \emph{necessity} to use flow equations possessing extra terms which take the $\gamma_{i\star}$ into account, the fixed-points of these flow equations possess more than one marginal, redundant direction. Obviously, since fixed-points are known to exist for all $\gamma_{i\star} = 0$, it suffices to check whether extra marginal, redundant operators exist for the standard flow equation.  The existence of additional directions of this type seems rather unlikely (they are certainly not present at the Gaussian fixed-point), though it would be nice to have a proof, one way or the other.

We have seen above that the anomalous dimension of the field can be taken into account in the flow equation by including in the blocking functional a linear, infinitesimal field redefinition which depends on $\eta$. It is instructive to see what happens if we instead perform the finite field redefinition
\be
	\phi'(x,\Lambda) = \phi(x,\Lambda) Z^{-1/2},
\label{eq:field-rescale}
\ee
as in~\cite{scalar1,scalar2}.
(Here we are taking a prime to denote a new variable, rather than a derivative.)
Accompanying the change of field variable is a change to the action, so that
\be
	S[\phi] = S'[\phi'].
\ee
Moreover, since the field redefinition is linear, the Jacobian present under the path integral is just an uninteresting constant which we ignore. Consequently, we can perform the redefinition~\eq{field-rescale} directly at the level of the flow equation. Indeed,
using the chain rule and~\eq{field-rescale} we have that
\begin{align}
	-\flow\bigr\vert_\phi S[\phi] = -\flow\bigr\vert_\phi S'[\phi']
	&=
	-\int_x  \fder{S'[\phi']}{\phi'(x,\Lambda)} \flow\bigr\vert_\phi \phi'(x,\Lambda)
	-\flow\bigr\vert_{\phi'} S'[\phi']
\nonumber
\\
	&=
	\frac{\eta}{2} \int_x \phi'(x,\Lambda)  \fder{S'[\phi']}{\phi'(x,\Lambda)}
	-\flow\bigr\vert_{\phi'} S'[\phi'].
\end{align}
For brevity, we now drop the primes. Indeed, from this point of view it is
more natural to replace~\eq{field-rescale} with the equivalent statement
$\phi(x) \mapsto \phi(x) Z^{1/2}$. The full flow equation reads:
\[
	\left(-\flow + \frac{\eta}{2} \Count_\phi\right) S[\phi]
	=
	\frac{1}{2Z} \classical{S}{\dd}{\Sigma} - \frac{1}{2Z} \quantum{\dd}{\Sigma},
\]
which is almost the same as~\eq{flow-Ball-a}.
However, we find an annoying appearance of $1/Z$s on the \rhs. The solution to this problem
is to exploit the freedom in the blocking transformation and replace $\dd$
with $\dd_{\mathrm{new}} = Z \dd$. (This is, after all, a perfectly good choice satisfying as it
does all the requirements.) With this change, the flow equation is precisely the same as~\eq{flow-Ball-a}
\be
	\left(-\flow + \frac{\eta}{2} \Count_\phi\right) S[\phi]
	=
	\hf \classical{S}{\dd}{\Sigma} - \hf \quantum{\dd}{\Sigma}
\label{eq:flow-Ball}
\ee
where, just to emphasise, the $\dd$s appearing here are still given by~\eq{dd},
the factors of $Z$ having cancelled out.

Were we to set the interaction part of the seed action to zero, then this flow equation
would reduce to the one first written down by Ball et al.~\cite{Ball} (modulo the final
rescalings that we are about to perform). The equation
with general seed action has been considered in~\cite{scalar1,scalar2} where it was
shown that the one-loop and two-loop $\beta$-function coefficients for $\lambda \phi^4$ theory
are independent
of the choice of seed action in four dimensions. We will redo the two-loop calculation, in a rather more
sophisticated way, in \sect{beta-canonical}.

To reduce everything to completely dimensionless form there are two things to do:
scale the canonical dimension out of the field and rewrite everything in terms of
dimensionless coordinates/momenta, as in~\eqs{dimless-fields}{dimless-coords}.
Note, though, that since we have additionally rescaled the field by a factor of $Z$,
$\tilde{\phi} = \tilde{\phi}(\tilde{x},t)$; for the sequel we will indicate all dependencies.
%
Writing $\tilde{S}\bigl[\tilde{\phi}\bigr] = S[\phi]$, and recalling the definition of the RG time $t \equiv \ln \mu/\Lambda$, we have:
\[
	-\Lambda \der{S[\phi]}{\Lambda}\biggr\vert_\phi = \pder{\tilde{S}\bigl[\tilde{\phi}\bigr]}{t}\biggr\vert_{\tilde{\phi}}
	+ \pder{}{t}\biggr\vert_{\phi} \int_{\tilde{x}} \fder{\tilde{S}\bigl[\tilde{\phi}\bigr] }{\tilde{\phi}(\tilde{x},t)} \delta \tilde{\phi}(\tilde{x},t).
\]
We need to take care processing the final term since, in this particular case, we cannot \naively\ take the partial derivative under the integral. Thus we rewrite
\[
\begin{split}
	\pder{}{t} \biggr\vert_\phi & = \der{}{t} - \int_y \pder{\phi(y,\Lambda)}{t} \biggr\vert_{y} \fder{}{\phi(y,\Lambda)}
\\
	& = \der{}{t}  - \Lambda^{(\D-2)/2} \int_y 
	\biggl[
		\pder{\tilde{\phi}(\tilde{y},t)}{t} \biggr\vert_y - \frac{\D-2}{2} \tilde{\phi}(\tilde{y},t)
	\biggr]
	\fder{}{\phi(y,\Lambda)}
\end{split} 
\]
and exploit the fact that the total derivative \emph{can} be taken under the integral. Utilizing
\[
	\pder{\tilde{\phi}(\tilde{y},t)}{t} \biggr\vert_y 
	= 
	\pder{\tilde{\phi}(\tilde{y},t)}{t} \biggr\vert_{\tilde{y}} 
	- \tilde{y}^{\mu} \pder{\tilde{\phi}(\tilde{y},t)}{\tilde{y}^{\mu}}
\]
together with~\eq{derphi-chain} (and remembering that $y \Lambda = \tilde{y}$)
we have:
\[
	-\flow S[\phi] = \partial_t \tilde{S}\bigl[\tilde{\phi}\bigr] 
	+\int_{\tilde{x}} 
	\bigl[
		\bigl( (\D-2)/2 + \tilde{x} \cdot \partial_{\tilde{x}} \bigr) \tilde{\phi}(\tilde{x},t)
	\bigr]
	\fder{\tilde{S}\bigl[\tilde{\phi}\bigr] }{\tilde{\phi}(\tilde{x},t)} 
	.
\]
Notice that $\tilde{x}$ is a dummy variable and so the tilde can be dropped for free. As for the various  other tildes, we will drop them too, mindful that the meaning of $S[\phi]$ must now be interpreted according to context. The \lhs\ of the flow equation now reads:
\[
	\bigl(
		\partial_t -\hat{D}^{-}
	\bigr) S
	= \ldots,
\]
where
\be
	\hat{D}^{\pm} = \int_p \biggl[\frac{\D+2\pm \eta}{2} \phi(p)+ p \cdot \partial_ p \phi(p) \biggr]
	\fder{}{\phi(p)}.
\label{eq:Dhat}
\ee
Notice that, in position space, we have
\be
	\hat{D}^- = -\int_x \bigl[ \bigl(d_\phi + x \cdot \partial_x \bigr) \phi(x) \bigr] \fder{}{\phi(x)},
\label{eq:DilGen}
\ee
where
\be
	d_\phi = (\D-2+\eta)/2
\ee
is seen to be the full scaling dimension of the field. Thus, we can interpret $\hat{D}^-$ as a functional representation of the dilatation generator (see e.g.~\cite{YellowPages}), a point of view which is more
thoroughly explored in~\cite{HO-Remarks}.

On the \rhs\ of the flow equation we take $\ep$ according to~\eq{sensible}, so that
\[
	\dd_\Lambda(p^2) = \frac{2}{\Lambda^2} \cutoff'(\tilde{p}^2),
\]
where, as before, the prime in this context means a derivative \wrt\ the argument of $\cutoff$.
Dropping the tilde, and using the RG time, $t \equiv \ln \mu/\Lambda$, it is now a simple matter to check that the full flow equation reads:
\be
	\bigl(
		\partial_t -\hat{D}^{-}
	\bigr) S
	=
	\classical{S}{\cutoff'}{\Sigma} - \quantum{\cutoff'}{\Sigma}.
\label{eq:alternative}
\ee

It is common to recast the flow equation by taking the last term contributing to $\hat{D}^-$
and integrating by parts. This is sometimes 
finessed further by adding and subtracting $\D$ and defining
\be
	\Count_\partial \equiv \D + \int_{p}  \phi(p) p\cdot \partial_{p} \fder{}{\phi(p)},
\label{eq:Delta_partial}
\ee
so that we arrive at
\be
	\bigr(
		\partial_t + d_\phi \Count_\phi + \Count_\partial -\D
	\bigl) S
	=
	\classical{S}{\cutoff'}{\Sigma} - \quantum{\cutoff'}{\Sigma}.
\label{eq:flow-rescaled}
\ee

The reason that the definition~\eq{Delta_partial} is made is because the operator $\Count_\partial$ has a natural action
on vertices with a single momentum conserving $\delta$-function. For test function $\chi(p)$, this follows on account of
\[
	\int_p \chi(p) p \cdot \partial_p \, \DiracD{\D}{p}
	= -\D \int_p \chi(p)  \DiracD{\D}{p},
	\qquad \Rightarrow \qquad
	p \cdot \partial_p \, \DiracD{\D}{p} = -\D \DiracD{\D}{p}
\]
(where, strictly, the last expression is understood under an integral).
Consequently, hitting a Wilsonian effective action with vertices of this type [of which~\eq{action-exp} is an example, but there is no necessity to expand in powers of the field for the following equation to hold], $\Count_\partial$ can be re-expressed
as
\be
	\Count_\partial = \int_{p}  \phi(p) p\cdot \check{\partial}_{p} \fder{}{\phi(p)},
\label{eq:mom-count}
\ee
where we understand that $\check{\partial}_{p}$ does not strike the momentum conserving
$\delta$-function. This form for $\Count_\partial$ is common in the literature, but it should
be noted that~\eq{Delta_partial} is more primitive, being as it is always true, whereas~\eq{mom-count}
should be understood to act only on vertices out of which one and only one momentum conserving $\delta$-function can and has been extracted. In this case, $\Count_\partial$ can be interpreted as counting the powers of momenta in each vertex.

Loosely, then, the \lhs\ of the flow equation can be interpreted as follows: $\partial_t - \D$ plus a
term which counts the number of fields, weighted by the scaling dimension, plus a term which counts the number of powers of momenta in each vertex. Of course, these counting operators only count in the obvious sense if they hit polynomials, but remembering this structure is an easy way to remember the \lhs\ of the flow equation.

In passing, let us note that we will have cause to consider objects like
\[
	\classical{}{\ep}{}
\]
in rescaled variables. In this case, we find that $\ep_\Lambda(p^2) = \cutoff(p^2/\Lambda^2)/p^2$
is naturally replaced by $ \cutoff(\tilde{p}^2)/\tilde{p}^2$. 
Dropping the tildes, we will denote this latter combination
by $\ep(p^2)$. Indeed, from now on---once again exploiting the
joys of object-orientation---we will usually write the effective propagator as $\ep(p^2)$
(in other words, even in the dimensionful case, we will not generally indicate dependence on $\Lambda$), with the symbol being interpreted according to context (\ie\ whether or not we happen to be working in dimensionless variables).

\subsection{Diagrammatics}
\label{sec:Diagrammatics}

It is often useful, both from the point of view of doing certain calculations and for getting a better feeling for
the flow equation, to introduce a diagrammatic representation. The starting point for this
is to expand both the seed action and Wilsonian effective action in powers of the field, as in~\eq{action-exp}. 
Stripping off the integrals, symmetry factors,
fields and momentum conserving $\delta$-function, we are left with just the
\emph{vertex coefficient functions}---\ie\ the $\hSvert{n}$ or $\Svert{n}$---which are the
objects which we represent diagrammatically, with all momenta flowing in:
\be
	\ensuremath{\begin{array}{c}\input{pstex/Seed-npt.pstex_t} \end{array}} \equiv \hSvert{n}(p_1,\ldots,p_n).
\ee
The string of small dots represents the legs which have not been explicitly drawn.
If, instead, we wanted to consider vertices of the interaction part of the
Wilsonian effective action then we
would simply replace the $\hS$ sitting inside the circle by an $\Sint$. Similarly, we
could place a $\Sigma$ inside the circle. If we preferred, we could shrink the circle to
a point, with $n$ legs emanating from it; but then we would no longer be able to
conveniently specify whether the vertex belongs to $\hS$, $\hSint$ $S$, $\Sint$, $\Sigma$ or $\Sigint$.

The fact that the vertices are `fattened up' also serves to remind us that the Wilsonian
effective action vertices follow (in principle) from the \emph{full, nonperturbative solution}
of the infinite tower of coupled diagrammatic equations. Thus, the diagrammatics
contains nonperturbative information. Given a small parameter, one
can of course expand the tower of coupled equations in a perturbation series, and
solve it order by order. But, by definition, this will provide only the perturbative pieces of
the solution.

The idea now is to substitute the expansion~\eq{action-exp}, together with its analogue for
$\hS$, into the flow equation. To illustrate this, we will take the generalized Polchinski equation, \eq{ProtoFlow}.
Identifying terms with the same number of fields
will give an infinite tower of coupled equations for the $\Siv{n}$, which we represent diagrammatically. 
As an example, let us see how this works for the flow of the $n$-point vertex. On the \lhs\ of the flow equation we have (with fields stripped off but symmetry factor retained, for the time being):
\be
	-\flow \frac{1}{n!} S(p_1,\ldots,p_n) 
	=
	-\frac{1}{n!}\flow \ensuremath{\begin{array}{c}\input{pstex/WEA-npt.pstex_t} \end{array}}
	=
	\ldots
\label{eq:flow-lhs-ex}
\ee

On the \rhs\ of the flow equation, let us start by 
considering how the quantum term, $\delta/\delta  \phi  \cdot \dd   \cdot   \delta \Sigma /\delta \phi$ contributes to this flow. Since the quantum term involves
two functional derivatives hitting the same vertex, this vertex must have $n+2$ fields in order to contribute to the $n$-point flow. In detail we have:
\begin{multline}
	\quantum{\dd}{} 
	 \int_{p_1,\ldots,p_{n+2}}
	\frac{1}{(n+2)!}
	\Sigma^{(n+2)} (p_1,\ldots,p_{n+2}) \phi(p_1)\cdots \phi(p_{n+2})\deltahat{p_1+\cdots+p_{n+2}}
\\
=
	\int_{p_1,\ldots,p_{n};q}
	\frac{1}{n!}
	\Sigma^{(n+2)} (p_1,\ldots,p_{n};q,-q) \phi(p_1)\cdots \phi(p_n)\deltahat{p_1+\cdots+p_n} \dd(q^2),
\label{eq:quantum-explicit}
\end{multline}
where we have exploited the permutation symmetry of the vertex to arrive at the net factor of $1/n!$.
Stripping off the integrals, fields, and momentum conserving $\delta$-function, this has the diagrammatic
representation
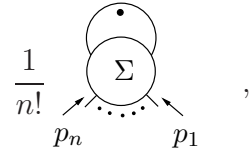
\[
	\frac{1}{n!}\ensuremath{\begin{array}{c}\input{pstex/Padlock.pstex_t} \end{array}},
\]
where the notation $\DummyKernel$ (which in the diagram has been bent round in a loop) stands for $\dd$. 
Since this object attaches to the vertex in two-places, the $\Sigma$ vertex in this example does indeed have $n+2$ legs. 
Again, modulo inconveniences of labelling, we could shrink the inner circle to a point, with $n+2$ legs emanating from it, two of which are tied together. This serves to emphasise that the places where
$\DummyKernel$ attaches to the circle are absolutely not to be considered as three-point vertices, as is evident from~\eq{quantum-explicit}. As we might have anticipated, the factor of $1/n!$ will cancel with the identical factor
in~\eq{flow-lhs-ex}, when we put everything together. The final point to make is that this diagram has a loop, which is why the corresponding term in the flow equation is often called the quantum term.

The last term to analyse is the classical term, $\delta S /\delta \phi \cdot \dd \cdot \delta \Sigma /\delta \phi$. In this case, the functional derivatives hit different vertices. If these vertices have $m+1$ and $m'+1$ legs, then we must sum over all $m,m'$ for which $m+m'=n$.  Now, after the functional derivatives have
acted, the overall symmetry factor of the diagram is $1/(m! m'!)$:
\begin{multline*}
	\sum_{m+m'=n}
	\frac{1}{m!} 
	\int_k \dd(k^2)
	\int_{p_1,\ldots,p_m}
	S^{(m+1)}(p_1,\ldots,p_m,k)
	\phi(p_1)\cdots \phi(p_m)
	\deltahat{p_1+\cdots+ p_m +k}
\\
	\times
	\frac{1}{m'!} 
	\int_{q_1,\ldots,q_{m'}}
	\Sigma^{(m'+1)}(q_1,\ldots,q_{m'},k)
	\phi(q_1)\cdots \phi(q_{m'})
	\deltahat{q_1+ \cdots + q_{m'}-k}
\end{multline*}
Of course, we would like to somehow cancel this symmetry factor against the $1/n!$ common to
the other two terms. To do this, consider the effect of permuting the $p$s and the $q$s 
in the above expression, not counting permutations of the $p$s amongst themselves or
the $q$s amongst themselves. Since there are a total of $m+m'=n$ fields, the effect of
what we are doing is equivalent to asking how many ways there are of partitioning $n$ fields
into two sets of $m$ and $m'$ fields. The answer is, of course, just $n!/(m!m'!)$. So, if we want to
replace the above expression by a sum over such permutations, we had better divide by $1/n!$
in order that the final combinatoric factor reduces to $1/(m!m'!)$, as above. Diagrammatically,
this amounts to considering all independent permutations of the external legs between two vertices,
where by independent we mean that we do not count permuting the legs of either one of the vertices amongst themselves. 
Relabelling $q_i = p_{m+i}$, diagrammatically we have:
\[
	\frac{1}{n!} \sum_{m} \left[ \ensuremath{\begin{array}{c}\input{pstex/Tree.pstex_t} \end{array}} \hspace{-2em}+ \mathrm{permutations} \right].
\]

Before writing the full diagrammatic flow equation, we will refine the diagrammatics~\cite{Primer,NonRenorm,mgiuc}.
Rather than explicitly decorating the various terms in the flow equation with the $n$-legs,
we will imagine pulling the legs off, with the prescription that they are to be reattached in
all independent ways. This allows us to get rid of both the sum and the `$+$ permutations'
above. To be specific, let us denote by $(p_1,\ldots,p_n)$ a set of $n$ legs, each carrying the 
indicated momentum into some vertex. 
Taking account of the factors of $1/2$ on the \rhs\ of the flow equation, \eq{ProtoFlow}, together with the signs of the quantum and classical terms we write the diagrammatic flow equation
as:
\be
	-\flow
	\dec{
		\ensuremath{\begin{array}{c}\begin{picture}(0,0)%
\epsfig{file=pstex/S.pstex}%
\end{picture}%
\setlength{\unitlength}{3947sp}%
\begingroup\makeatletter\ifx\SetFigFont\undefined%
\gdef\SetFigFont#1#2#3#4#5{%
  \reset@font\fontsize{#1}{#2pt}%
  \fontfamily{#3}\fontseries{#4}\fontshape{#5}%
  \selectfont}%
\fi\endgroup%
\begin{picture}(446,446)(2347,-1169)
\put(2508,-1005){\makebox(0,0)[lb]{\smash{{\SetFigFont{11}{13.2}{\rmdefault}{\mddefault}{\updefault}{\color[rgb]{0,0,0}$S$}%
}}}}
\end{picture}%
 \end{array}}
	}{(p_1, \ldots, p_n)}
	=
	\frac{1}{2}
	\dec{
		\ensuremath{\begin{array}{c}\input{pstex/Tree-S-Sig.pstex_t} \end{array}} - \ensuremath{\begin{array}{c}\begin{picture}(0,0)%
\epsfig{file=pstex/Padlock-Sig.pstex}%
\end{picture}%
\setlength{\unitlength}{3947sp}%
\begingroup\makeatletter\ifx\SetFigFont\undefined%
\gdef\SetFigFont#1#2#3#4#5{%
  \reset@font\fontsize{#1}{#2pt}%
  \fontfamily{#3}\fontseries{#4}\fontshape{#5}%
  \selectfont}%
\fi\endgroup%
\begin{picture}(455,661)(2343,-1169)
\put(2515,-1005){\makebox(0,0)[lb]{\smash{{\SetFigFont{11}{13.2}{\rmdefault}{\mddefault}{\updefault}{\color[rgb]{0,0,0}$\Sigma$}%
}}}}
\end{picture}%
 \end{array}}
	}{(p_1, \ldots, p_n)}.
\label{eq:diagrammatic-flow}
\ee

On the \lhs, decoration with the $n$-legs is trivial: they must all decorate the same vertex and
there is only one way to do this. Similarly with the quantum term (although in gauge theory, the
kernel $\dd$ can be decorated, giving a richer diagrammatics~\cite{Primer,mgierg1,qed,qcd}). It is the classical
term where things get interesting: we must distribute the $n$ legs in all independent ways
between the two vertices. 

\subsection{Other ERGs for Scalar Field Theory}
\label{sec:otherflow}

Flow equations with a structure like Polchinski's are not the only one on the market.
Wilson's version is rather similar following, as it does, from the general approach to ERGs that
we have taken. In dimensionless variables, Wilson's equation reads
\be
	\left(\partial_t + \frac{\D}{2} \Count_\phi + \Count_\partial -\D \right) \Swil
	=
	\int_p h(p)
	\left[
		\dfder{\Swil}{\phi(-p)}{\phi(p)} - \fder{\Swil}{\phi(-p)} \fder{\Swil}{\phi(p)} +
		\phi(p) \fder{\Swil}{\phi(p)}
	\right].
\label{eq:flow-Wilson}
\ee
In~\cite{Wilson}, Wilson \& Kogut made the choice $h(p) = a(t) + 2p^2$, where $a(t) = 1 - \eta(t)/2$. Wegner~\cite{Wegner_CS,WegnerInv} derived Wilson's ERG from the generalized approach we have been following by taking (in dimensionless variables)
\be
	\Psi_{\mathrm{W}}(p) = h(p)
	\left[
		\fder{\Swil}{\phi(-p)} - \phi(p)
	\right].
\ee
Notice, though, that to reproduce~\eq{flow-Wilson} requires that the field is taken to
have dimension $\D/2$ (since this gives, upon transferring to dimensionless variables,
the $\D/2 \, \Count_\phi$ term on the \lhs). This is consistent with taking, in \emph{dimensionful}
variables, 
\[
	\Psi_{\mathrm{W}}(p) =
	\left[a(t) + 2\frac{p^2}{\Lambda^2}\right]
	\left[
		\fder{\Swil}{\phi(-p)} - \phi(p)
	\right].
\]
By choosing things in this way, it is apparent that 
$\phi(p)$ and $\delta /\delta \phi(p)$ share the same dimensionality.
But since
\[
	\fder{\phi(p)}{\phi(q)} = \deltahat{p-q},
\]
with the \rhs\ having mass-dimension $-\D$, we conclude that $\left[\phi(p)\right] = -\D/2$
and, therefore, that $\left[\phi(x)\right] = +\D/2$.

Alternatively, Wilson's equation can be derived using fields with canonical scaling
dimension. This approach highlights the relationship between this equation and
Polchinski's---see~\cite{HO-Remarks} (and also~\cite{Golner-GreenFunctions}). 
To this end, let us 
recall~\eq{blocked} and take
\be
	\Psi(p) = \frac{1}{\Lambda^2} L'(p^2/\Lambda^2) \fder{S}{\phi(-p)} - \psi(p),
\ee
where $L'$ is dimensionless (ensuring that the field carries canonical dimension)
but, this restriction aside, remains to be chosen. The object $\psi(p)$ carries the residual
freedom of the blocking transformation.
Now, if we take
$\psi(p) = 2 \ep^{-1}_\Lambda(p^2) L'(p^2/\Lambda^2)\phi(p)$ and identify 
$L$ with $\cutoff$,
then we arrive at Polchinski's equation. On the other hand, if we take 
$\psi(p) = \bigl[L'(p^2/\Lambda^2) + 1\bigr] \phi(p)$ and identify 
$L'(p^2/\Lambda^2)  = -h(p)$
then, after rescaling to dimensionless variables, we arrive at Wilson's equation.


Contemporaneous with Wilson's ERG is an ERG equation with a sharp cutoff, written down by Wegner and Houghton~\cite{WH}. In fact, the term ERG was coined essentially simultaneously in these two works. However, a sharp cutoff introduces its own difficulties---not least non-analyticity in momenta~\cite{TRM-MomScale}.

\section{The Exact Renormalization Group as a Heat Equation}
\label{sec:heat}

ERG equations, as mentioned in \sect{General_ERG}, are non-linear in the Wilsonian effective action. However, as we will discuss in this section, they can be recast as linear equations via a change of variables.%
\footnote{This linearization is an exact operation and is completely different from linearizing a flow equation in the vicinity of a fixed-point [\cf~\eq{perturb}].} 
We begin, in \sect{Linear-Pol}, by showing that the Polchinski equation can be readily cast in the form of a heat equation. Whilst this observation is nothing new~\cite{Salmhofer}, it seems not to have been much exploited. Part of the reason for this is that although solving the linearized equations is trivial, picking out physically viable solutions is not---as we will discuss further, below. Nevertheless, carrying on from~\cite{Trivial,Susy-Chiral}, we will continue to develop an understanding of linearized flow equations and will find (particularly in \sects{solve}{CorrFns}) that we gain some deep insights. In \sect{Linear-Gen} we present the linear form of some of the Polchinski equation's cousins, as part of which we derive an equation which will play an important role in later sections. In \sect{Linear-Diags} a diagrammatic approach is explored and we finish in \sect{Linear-Choose} with a brief discussion of some aspects of the physical interpretation of $\Lambda$.

\subsection{The Linear Form of Polchinski's Equation}
\label{sec:Linear-Pol}

To cast the Polchinski equation as a heat equation, let us start by defining the operator, $\op$, according to
\be
	\op \equiv \hf \quantum{\ep}{}.
\label{eq:op}
\ee
As is our wont, in most circumstances we will deduce whether the variables are dimensionful or dimensionless from the context. However, for much of this section it will pay to make the $\Lambda$-dependence explicit in the former case and so we understand
\be
	\op_\Lambda = \hf \int_p \fder{}{\phi(-p)} \frac{\cutoff(p^2/\Lambda^2)}{p^2} \fder{}{\phi(p)}.
\label{eq:op-dimful}
\ee
Taking $\dot{\op} \equiv - \flow \op$, the Polchinski equation~\eq{Pol} can be recast in linear form:
\be
	-\flow e^{-\Sint[\phi]} = -\dot{\op} e^{-\Sint[\phi]}.
\label{eq:Pol-linear}
\ee
This has the structure of a heat equation (with $\Lambda$-dependent coefficient on the \rhs).%
\footnote{%
A similar-looking equation can be found in the book of Salmhofer~\cite{SalmhoferBook}, but there are some important differences: $e^{-\Sint}$ is replaced by the partition function, regularized at the \emph{IR} scale, $\Lambda$. An overall UV cutoff, $\Lambda_0$ is present, and the analogue of $\op$ is
$\hf \delta/\delta \phi \cdot \left(\ep_{\Lambda_0} - \ep_{\Lambda}\right) \cdot \delta/\delta \phi$.} As pointed out in~\cite{HO-Remarks}, this structure implies that in order for evolution with decreasing $\Lambda$ to correspond, in general, to a well-posed problem we must take $\cutoff'(p^2/\Lambda^2) < 0$, for $p^2/\Lambda^2 < \infty$. In particular, note that we must take $\cutoff'(0) <0$ (which is not always done in the literature), a condition that we will see reappear several times.

Temporarily ignoring the potentially troublesome issue of IR divergences, let us introduce the `dual action',
\be
	-\dual[\phi] \equiv
	\ln \left( e^{\op} e^{-\Sint[\phi]} \right).
\label{eq:dual}
\ee
It is apparent from~\eq{Pol-linear} that this is an invariant under the flow:
\be
	-\flow \dual[\phi] = 0.
\label{eq:dual-flow}
\ee
However, we must take care due to the fact that $\ep_\Lambda(p^2) \sim 1/p^2$ for $p^2/\Lambda^2 \ll 1$. Indeed, in $\D=2$ the Fourier transform of $\ep_\Lambda(p^2)$ blows up and, moreover, even in higher dimensions the
action of $e^{\op}$ might generate IR divergences (as is clear from a diagrammatic perspective, which will be introduced shortly). With this in mind, let us introduce a new scale $\Lambda' \leq \Lambda$ and define
\be
	-\dual_{\Lambda'}[\phi] \equiv
	\ln 
	\Bigl(
		e^{\op_{\Lambda} - \op_{\Lambda'}} e^{-\Sint_\Lambda[\phi]}
	\Bigr).
\label{eq:dual'}
\ee
Just like the dual action, this satisfies
\be
	-\flow \dual_{\Lambda'}[\phi]  = 0
\label{eq:flow-dual'}
\ee
but, in contrast, it is IR finite. The reason for this that, for $p^2/\Lambda^2 \ll 1$, we have $\ep_\Lambda(p^2) - \ep_{\Lambda'}(p^2) = \order{p^0}$, which follows from the fact that $\cutoff(0) = 1$.

We would now like to consider whether or not we can take the limit $\Lambda' \rightarrow 0$.%
\footnote{
I would like to thank Tim Morris for providing the essential elements of the following argument.
}
 First of all,
we note from~\eq{flow-dual'} that we can evaluate $\dual_{\Lambda'}[\phi] $ at any convenient value of $\Lambda$. With this in mind, it follows from taking the limit $\Lambda \rightarrow \Lambda'$ in~\eq{dual'}
that $\dual_{\Lambda'}[\phi] =  \Sint_{\Lambda'}[\phi]$. Therefore, the question as to whether $\lim_{\Lambda' \rightarrow 0} \dual_{\Lambda'}[\phi] $ exists amounts to determining whether $ \Sint_{\Lambda=0}[\phi]$
exists (there is no need to retain the prime on the $\Lambda$). Now, as demonstrated in~\cite{TRM-ApproxSolns} and as we will discuss in much greater detail in \sect{CorrFns}, the low energy limit of the Wilsonian effective action is very closely related to the correlation functions.%
\footnote{%
There is potential for confusion here. Consider a theory in the critical surface of
the Gaussian fixed-point. It is tempting to say that since the theory flows into
the Gaussian fixed-point, $\Sint_{\Lambda=0}[\phi]  = 0$, the correlation functions
are therefore trivial. But this does not make sense: for momenta near the bare scale, the correlation
functions are distinctly non-trivial. The resolution to this apparent paradox is that,
after rescaling to dimensionless variables, it is $S_t$ which sinks into the fixed-point
as $t\rightarrow \infty$, with dimensionless field held constant. Reinstating the
appropriate powers of $\Lambda$ to make things dimensionful, we do not find
a trivial limit of $S_\Lambda$ when we take 
$\Lambda \rightarrow 0$ with dimensionful field held constant. 
This can be illustrated with
the following simple example (for which we will take $\D=4$). Consider an action in the vicinity of the Gaussian fixed-point which possesses a term
$e^{-2t} \FourInt{\tilde{x}} \tilde{\phi}^6(\tilde{x})$, in dimensionless variables. Clearly, the $t\rightarrow \infty $ limit  (with $\tilde{\phi}$ held constant)  vanishes. However, in dimensionful variables this term becomes 
$\frac{1}{\mu^2}\FourInt{x} \phi^6(x)$, which does not vanish as $\Lambda \rightarrow 0$ (with $\phi$ held constant).
Let us also note that it is quite permissible for one limit to yield something quasi-local,
whereas the other does not, an example of which will be encountered at the end
of \sect{comp-corr-fns}.
\label{foot:limit}
}
Indeed, in the case that the action flows according to the Polchinski equation, the precise relationship is:
\begin{subequations}
\begin{align}
	\eval{\phi(p_1)\cdots \phi(p_n)}_{\conn}
	&= -\hat{\delta}(p_1+\cdots+p_n) \Siv{n}_{\Lambda=0} (p_1,\ldots,p_n) \prod_{i=1}^n \ep_b(p_i^2),
	\qquad n>2,
\label{eq:CCF^n}
\\
	\eval{\phi(p) \phi(q)}_{\conn}
	& =  \hat{\delta}(p+q) \ep_b(p^2) \left[ 1 - \Siv{2}_{\Lambda=0} \ep_b(p^2) \right],
\label{eq:CCF^2}
\end{align}
\end{subequations}
where $\conn$ stand for connected and $\ep_b(p^2) = \cutoff(p^2/\Lambda_0^2)/p^2$ with the understanding that, if we sit at a fixed-point or on a renormalized trajectory, $\Lambda_0$ is sent to infinity, in which case $\ep_b(p^2) \rightarrow 1/p^2$. Thus we see that, compared to the correlation functions, the IR behaviour of $ \Siv{n}_{\Lambda=0}[\phi]$  is \emph{improved} by a factor of momentum squared on each leg. Therefore, it is quite permissible for $ \Sint_{\Lambda=0}[\phi]$ to exist in a theory for which the correlation functions are IR divergent (we will see an example of this in \sect{Gen-TP}).
Indeed, it is not even necessary for the vertices of $ \Sint_{\Lambda=0}[\phi]$ to be IR finite; rather, since all momenta are integrated over---as in~\eq{action-exp}---it need only be true that there are no IR divergences in  $ \Sint_{\Lambda=0}[\phi]$ as strong or stronger than $1/\mathrm{mom}^d$.
 With this in mind, we will henceforth assume that $ \Sint_{\Lambda=0}[\phi]$ does indeed exist, implying that so too does the dual action as defined by~\eq{dual}.

With the relationship between the dual action and the correlation functions in our minds, the presence of the logarithm in~\eq{dual} becomes clear: it ensures that the vertices of $\dual$ are related to the
 \emph{connected} correlation functions. Indeed, supposing that we can expand the dual action in powers of the field, the vertices are defined according to
\be
	\dual[\phi] = \sum_n \frac{1}{n!}
	\int_{p_1,\ldots,p_n} \dualv{n}(p_1,\ldots,p_n) \phi(p_1)\cdots \phi(p_n)
	\deltahat{p_1+\cdots+p_n}.
\label{eq:dualv}
\ee
As an aside, let us note that, since the dual action is related to the Wilsonian effective 
action at $\Lambda=0$, there is no reason to expect the vertices of the former to be quasi-local.

Given the relationship between the vertices of the dual action and the connected correlation functions, 
one might wonder why
a name for $\mathcal{D}$ reflecting this property has not been chosen. The point is that this interpretation of the dual action is only exact when we are working with the Polchinski equation. 
We have already commented that, even if we choose the simplest seed action, when we perform rescalings it is desirable to take a flow equation slightly different from the Polchinski equation. For this flow equation, the relationship between the two-point correlation function and the two-point dual action vertex that one finds in the Polchinski case breaks down for large momenta. Taking a non-trivial seed action makes matters much more complex, as we now go on to discuss.

\subsection{The Linear form of some Generalized Flow Equations}
\label{sec:Linear-Gen}

Let us now consider the flow of the dual action when we take the modified flow equation~\eq{flow-Ball}, written out here with the splitting~\eq{split} performed:
\be
	\left(-\flow + \frac{\eta}{2} \Count_\phi\right) \Sint
	=\hf \classical{\Sint}{\dd}{\Sigint}
	-\hf \quantum{\dd}{\Sigint}
	- \phi \cdot \ep^{-1} \dd \cdot \fder{\hSint}{\phi}
	-\frac{\eta}{2} \phi \cdot \ep^{-1} \cdot \phi.
\label{eq:reduced-flow}
\ee
(To recall: in this flow equation the anomalous dimension of the field has been scaled out, but no further rescalings have been performed.) As before, let us start our analysis by blithely ignoring any possible IR subtleties. In this case then, as stated in~\cite{Trivial}, and as we explicitly show in \app{Dual-flow}, 
the flow of the dual action is now given, up to a discarded vacuum energy term, by
\be
	\left(
		\flow + \frac{\eta}{2} \Count_\phi
	\right) \dual[\phi] 
	= \frac{\eta}{2} \phi \cdot \ep^{-1} \cdot \phi
	+ 
	e^{\dual} \phi \cdot \ep^{-1} \dd \cdot
	e^{\op} 
	\,
	\fder{\hSint}{\phi} e^{-\Sint}.
\label{eq:dualflow-Seed}
\ee

There are several comments worth making. First of all, notice that, on the \lhs, the two terms come with the same sign, in contrast to~\eq{reduced-flow}. 
Secondly, if the seed action is set to zero, we are left with an linear equation for $\dual$, of first order in derivatives. However, for non-zero seed action the equation is linear when written in terms of $e^{-\dual}$
and contains higher order derivatives. Finally,
we can already see why taking a flow equation different from Polchinski's spoils the relationship
between the dual action and the correlation functions. If the \rhs\ were zero, then we would have
that the $\dualv{n}$ are scale independent, up to factors of $\sqrt{Z}$ on each leg, as we would
expect for correlation functions of the rescaled field. 
However, the \rhs\ is not zero. For $\hSint=0$,
the \rhs\ only possesses a two-point term which, since $\ep^{-1}(p^2,\Lambda) \sim p^2 + \order{p^4}$,
vanishes for small momenta. This justifies the earlier comment that, for the flow equation of Ball et al.~\cite{Ball}, the dual action exhibits the Polchinski-like relation to the correlation functions
automatically for $n>2$ but only for small momenta when $n=2$.

If the seed action is non-trivial, it is tempting to conclude that, although the \rhs\ now contributes beyond
the two-point level, the \rhs\ still vanishes in the small momentum limit, since both terms involve a $\ep^{-1}$. However, the second term also depends on positive powers of $\ep(p^2) \sim 1/p^2$ (through $e^\op$), so this conclusion is too hasty. This issue deserves further investigation.

Let us now return to the issue of IR divergences, this time in the context of the dual action defined by a Wilsonian effective action which is a solution of~\eq{reduced-flow}. To begin with, let us set the seed action to zero and focus on the IR regularized dual action~\eq{dual'}. Modifying the analysis of \app{Dual-flow}, it is straightforward to show that
\be
	\biggl[
		-\flow - \frac{\eta}{2} \Count_\phi - \frac{\eta}{2} \phi \cdot \ep^{-1}_\Lambda \cdot \phi
		+ \eta \,\phi \cdot \ep^{-1}_\Lambda \ep_{\Lambda'} \cdot \fder{}{\phi}
		+ \frac{\eta}{2} \fder{}{\phi} \cdot \bigl(\ep_\Lambda - \ep_{\Lambda'} \bigr) \ep_{\Lambda'} \ep^
		{-1}_\Lambda \cdot \fder{}{\phi}
	\biggr]
	e^{-\dual_{\Lambda'}[\phi]} = 0.
\label{eq:dual'-modifiedflow}
\ee
Now consider taking the limit $\Lambda' \rightarrow 0$. As before, we will simply assume that $\lim_{\Lambda' \rightarrow 0} \dual_{\Lambda'}[\phi]$ exists. However, there is a subtlety not encountered when we dealt with the Polchinski equation: the existence of the limit is not, by itself, sufficient to guarantee that $ \dual_{\Lambda'=0}[\phi]$ is a solution of~\eq{dualflow-Seed} with $\hSint = 0$.
Indeed, consider~\eq{dual'-modifiedflow} for small $\Lambda'$. We might suppose that the final two terms in the big square brackets are sub-leading in this regime. This is certainly true of the first of these terms, which is IR finite [the $1/p^2$ of $\ep_{\Lambda'}(p^2)$ is cancelled by the $p^2$ of $\ep^{-1}_\Lambda(p^2)$]. However, in the second term, this limit potentially generates an IR divergence, meaning that the term might have non-zero support at vanishing momentum.%
\footnote{%
This can be illustrated by considering the following one-dimensional integral designed to mimic the problematic term:
$
I(a,\epsilon) = \int_\epsilon^\infty dz e^{-z/a+z} \bigl(e^{-z} - e^{-z/a} \bigr) /z ,
$
where we have chosen an exponential UV cutoff and have introduced an IR cutoff, $a$. The reason that we have set the lower limit to be $\epsilon$ is so that we can evaluate the integral in terms of $E_1(y) = \int_y^\infty e^{-z} /z \, dz = -\gamma -\ln y + \order{y}$, where $\gamma$ is the Euler-Mascheroni constant. Combining terms, it is easy to show that $I(a,0) = \ln (2-a)$, which manifestly does not vanish in the limit $a\rightarrow 0$.} 
Henceforth, we will assume that this is not the case. The picture that will be built up in this paper based on this assumption is both consistent and compelling. Nevertheless, it is clear that this issue requires further investigation. In the case that $\hSint \neq 0$, we assume that the dual action exists and that its flow is given by~\eq{dualflow-Seed}.

For later use, let us note that if we set $\hSint=0$ in~\eq{dualflow-Seed}, and perform the usual rescalings
$\phi(p) \mapsto \phi(p) \Lambda^{-(\D+2)/2}$, $p\mapsto p\Lambda$, then we find that
\be
	\left(\partial_t + \frac{\D-2-\eta}{2} \Count_\phi + \Count_\partial - \D \right) 
	\dual[\phi] 
	= -\frac{\eta}{2} \phi \cdot \ep^{-1} \cdot \phi.
\label{eq:RescaledDualFlow}
\ee
This equation will play an important role, especially in \sect{solve}.

Returning to~\eqs{dual}{dual-flow}, it might seem that we have solved the Polchinski equation.
Unsurprisingly, matters are rather more complicated than this! To understand what is going on,
let us utilize the fact that the solutions to~\eq{dual}---scale-independent functionals of the field---are essentially the connected correlation functions. Thus, by trying to find solutions of the Polchinski equation by first solving~\eq{dual-flow}, we are trying to solve an `inverse problem': given the correlation functions (which we choose) and a flow equation, we wish to reconstruct the Wilsonian effective action.
\emph{Formally}, this can be done by inverting~\eq{dual}:
\be
	-\Sint[\phi] = \ln \left( e^{-\op} e^{-\dual[\phi]} \right).
\label{eq:invert}
\ee
(Note that the pair of relationships~\eqs{dual}{invert} essentially provides a realization of the Dominicis-Englert theorem~\cite{Dominicis}. For a recent and interesting application of this theorem in the context of perturbatively renormalizable theories, see~\cite{Pivovarov-Naturalness}.)
So, if everything we have done is well defined then we can choose the correlation functions to be whatever we like and, from these, can reconstruct the corresponding Wilsonian effective action (the scale dependence of the Wilsonian effective action is generated by the scale dependence of $\op$).
Of course, this reconstruction is precisely what we do not expect to be well defined, in general. We require that a good Wilsonian effective action both  exists and is quasi-local (at least away from $\Lambda=0$). 
For any old choice of correlation functions, we expect to run foul of one or other of these requirements. Indeed, we will see a specific example of this in \sect{Gen-TP}. Thus, although we have in some sense solved the Polchinski equation, we have an embarrassment of riches: the useful solutions are part of an infinite set including an uncountable infinity of useless ones. 

One might imagine that it is possible to try to pick out the useful solutions for $\dual$ by some sort of fine-tuning procedure. However, inverse problems of this type for heat equations are ill-posed, in
the sense that $\Sint$ is expected to have excruciating sensitivity on $\dual$. This does not present a difficulty in the case where we can find exact solutions, as we will see in \sect{Gen-TP} for a simple example and at the end of~\sect{Redux} for a much more complicated case. Usually, however, some form of approximation is necessary and here the method would presumably run into severe practical problems. 
Whether any inspiration can be found in the techniques developed for inverse problems, see \eg~\cite{Inverse-MomentTheory,Inverse-Glasko}, remains to be seen.

This should not, however, leave one with the impression that this approach provides nothing useful.
In \sect{FP-dual}, the dual action will play a central role in proving that the spectrum of $\eta_\star$
is quantized at critical fixed-points. One might worry that the caveats discussed around~\eq{dual'-modifiedflow} limit the scope of this proof. However, once the dual action has been used to elucidate the
general structure, it becomes obvious how to proceed without using the dual action at all. Time and again in this paper, we will find that the dual action provides a useful scaffolding for obtaining results, which can ultimately be removed. Indeed, the exact two-point, fixed-point solutions of the rescaled flow equation (with trivial seed action) are most easily found using the dual action formalism and will use the dual action to find a simple expression for a particular redundant operator which plays an important role at critical fixed-points. In \sect{WF}, the formalism will be employed to uncover a novel way of finding the Wilson-Fisher fixed-point whilst in \sect{CorrFns} we will flesh out the relationship between the dual action and the correlation functions.
Moreover, in~\cite{Trivial}, certain consistency conditions on the vertices $\dualv{n}$ are used to argue that there are no physically acceptable, non-trivial fixed-point in scalar field theory for $\D\geq4$; the analysis is extended to the supersymmetric case in~\cite{Susy-Chiral}. 

\subsection{Diagrammatics}
\label{sec:Linear-Diags}

Some additional insights into the dual action can be provided by looking at its diagrammatic representation.
To this end, we expand $e^{\op} = \sum_i \op^i/i!$. Next, we allow the derivatives in $\op$ to strike $e^{-\Sint}$ \emph{before} summing over $i$. Although this procedure is used, for example, in \app{Dual-flow} to quite correctly show that \eg\ $\left[ e^{\op}, {\hf} \Count_\phi \right] =e^{\op} \op$,
we anticipate problems in the current context due to the infinite series generated, the (re)summability of which is not obvious. (Though note that if $\Sint[\phi]$ is at most quadratic in the field, then it is easy
to sum the series, as we will see later in this section and again in \sect{CorrFns}.)
We will make some further comments regarding this interchange in a moment. 

Now, if we suppose that the Wilsonian effective action can be expanded in powers of the field, then the $\dualv{n}$ just consist of all connected diagrams that can be constructed from $\Siv{n}$ and $\ep$. Conversely, from the relationship~\eq{invert}, the Wilsonian effective action can be formally reconstructed from all connected diagrams built from $\dualv{n}$ and $-\ep$. This is illustrated in \fig{dual} for the two-point case where, in both equations, the first ellipsis represents all
remaining 1PI diagrams, whereas the second ellipsis denotes the remaining one-particle reducible (1PR) diagrams. The 1PI diagrams have been ordered according to the number of explicit loops (implicit loops are carried by the vertices which incorporate quantum corrections).
\bcf[h]
	\begin{align*}
		\dualv{2} & = \ensuremath{\begin{array}{c}\begin{picture}(0,0)%
\epsfig{file=pstex/ReducedWEA-2.pstex}%
\end{picture}%
\setlength{\unitlength}{3947sp}%
\begingroup\makeatletter\ifx\SetFigFont\undefined%
\gdef\SetFigFont#1#2#3#4#5{%
  \reset@font\fontsize{#1}{#2pt}%
  \fontfamily{#3}\fontseries{#4}\fontshape{#5}%
  \selectfont}%
\fi\endgroup%
\begin{picture}(358,579)(1629,-672)
\put(1730,-448){\makebox(0,0)[lb]{\smash{{\SetFigFont{11}{13.2}{\rmdefault}{\mddefault}{\updefault}{\color[rgb]{0,0,0}$\Sint$}%
}}}}
\end{picture}%
 \end{array}} + \frac{1}{2} \ensuremath{\begin{array}{c}\begin{picture}(0,0)%
\epsfig{file=pstex/Padlock-2.pstex}%
\end{picture}%
\setlength{\unitlength}{3947sp}%
\begingroup\makeatletter\ifx\SetFigFont\undefined%
\gdef\SetFigFont#1#2#3#4#5{%
  \reset@font\fontsize{#1}{#2pt}%
  \fontfamily{#3}\fontseries{#4}\fontshape{#5}%
  \selectfont}%
\fi\endgroup%
\begin{picture}(418,565)(1606,-593)
\put(1727,-457){\makebox(0,0)[lb]{\smash{{\SetFigFont{11}{13.2}{\rmdefault}{\mddefault}{\updefault}{\color[rgb]{0,0,0}$\Sint$}%
}}}}
\end{picture}%
 \end{array}}  
		-\frac{1}{6} \ensuremath{\begin{array}{c}\input{pstex/TP-TL.pstex_t} \end{array}} + \frac{1}{8} \ensuremath{\begin{array}{c}\begin{picture}(0,0)%
\epsfig{file=pstex/Padlockx2-2.pstex}%
\end{picture}%
\setlength{\unitlength}{3947sp}%
\begingroup\makeatletter\ifx\SetFigFont\undefined%
\gdef\SetFigFont#1#2#3#4#5{%
  \reset@font\fontsize{#1}{#2pt}%
  \fontfamily{#3}\fontseries{#4}\fontshape{#5}%
  \selectfont}%
\fi\endgroup%
\begin{picture}(576,716)(1523,-744)
\put(1734,-457){\makebox(0,0)[lb]{\smash{{\SetFigFont{11}{13.2}{\rmdefault}{\mddefault}{\updefault}{\color[rgb]{0,0,0}$\Sint$}%
}}}}
\end{picture}%
 \end{array}}  + \cdots
		- \ensuremath{\begin{array}{c}\input{pstex/Dumbbell-2.pstex_t} \end{array}} + \cdots
	\\
		\Siv{2} & =  \ensuremath{\begin{array}{c}\begin{picture}(0,0)%
\epsfig{file=pstex/Dual-2.pstex}%
\end{picture}%
\setlength{\unitlength}{3947sp}%
\begingroup\makeatletter\ifx\SetFigFont\undefined%
\gdef\SetFigFont#1#2#3#4#5{%
  \reset@font\fontsize{#1}{#2pt}%
  \fontfamily{#3}\fontseries{#4}\fontshape{#5}%
  \selectfont}%
\fi\endgroup%
\begin{picture}(358,579)(1629,-672)
\put(1730,-448){\makebox(0,0)[lb]{\smash{{\SetFigFont{11}{13.2}{\rmdefault}{\mddefault}{\updefault}{\color[rgb]{0,0,0}$\dual$}%
}}}}
\end{picture}%
 \end{array}} 
		- \frac{1}{2} \ensuremath{\begin{array}{c}\begin{picture}(0,0)%
\epsfig{file=pstex/Dual-Padlock-2.pstex}%
\end{picture}%
\setlength{\unitlength}{3947sp}%
\begingroup\makeatletter\ifx\SetFigFont\undefined%
\gdef\SetFigFont#1#2#3#4#5{%
  \reset@font\fontsize{#1}{#2pt}%
  \fontfamily{#3}\fontseries{#4}\fontshape{#5}%
  \selectfont}%
\fi\endgroup%
\begin{picture}(418,565)(1606,-593)
\put(1747,-457){\makebox(0,0)[lb]{\smash{{\SetFigFont{11}{13.2}{\rmdefault}{\mddefault}{\updefault}{\color[rgb]{0,0,0}$\dual$}%
}}}}
\end{picture}%
 \end{array}}
		+\frac{1}{6} \ensuremath{\begin{array}{c}\input{pstex/Dual-TP-TL.pstex_t} \end{array}} + \frac{1}{8}\ensuremath{\begin{array}{c}\begin{picture}(0,0)%
\epsfig{file=pstex/Dual-Padlockx2-2.pstex}%
\end{picture}%
\setlength{\unitlength}{3947sp}%
\begingroup\makeatletter\ifx\SetFigFont\undefined%
\gdef\SetFigFont#1#2#3#4#5{%
  \reset@font\fontsize{#1}{#2pt}%
  \fontfamily{#3}\fontseries{#4}\fontshape{#5}%
  \selectfont}%
\fi\endgroup%
\begin{picture}(576,716)(1523,-744)
\put(1747,-457){\makebox(0,0)[lb]{\smash{{\SetFigFont{11}{13.2}{\rmdefault}{\mddefault}{\updefault}{\color[rgb]{0,0,0}$\dual$}%
}}}}
\end{picture}%
 \end{array}} + \cdots
		+ \ensuremath{\begin{array}{c}\input{pstex/Dual-Dumbbell-2.pstex_t} \end{array}} + \cdots
	\end{align*}
\caption{The diagrammatic expression for the two-point dual action vertex in terms of Wilsonian effective action vertices and vice-versa. Momentum arguments are suppressed.}
\label{fig:dual}
\ecf

The combinatorics for the diagrams is as follows. Let us write the diagrammatic
expansion for the $\dualv{n}$ in the compact form:
\be
	\dualv{n}(k_1, \ldots, k_n) \equiv \sum_{s=0}^{\infty} \sum_{j=1}^{s+1} \diagnorm_{s,j}
		\dec{\dec{\ensuremath{\begin{array}{c}\begin{picture}(0,0)%
\epsfig{file=pstex/ReducedWEA.pstex}%
\end{picture}%
\setlength{\unitlength}{3947sp}%
\begingroup\makeatletter\ifx\SetFigFont\undefined%
\gdef\SetFigFont#1#2#3#4#5{%
  \reset@font\fontsize{#1}{#2pt}%
  \fontfamily{#3}\fontseries{#4}\fontshape{#5}%
  \selectfont}%
\fi\endgroup%
\begin{picture}(358,358)(1629,-562)
\put(1730,-448){\makebox(0,0)[lb]{\smash{{\SetFigFont{11}{13.2}{\rmdefault}{\mddefault}{\updefault}{\color[rgb]{0,0,0}$\Sint$}%
}}}}
\end{picture}%
 \end{array}}}{j}}{\ep^s \ (k_1, \ldots,k_n)}
\label{eq:dual^n-neat}
\ee
with, for non-negative integers $a$ and $b$, the definition
\be
\label{eq:norm}
	\diagnorm_{a,b} \equiv \frac{(-1)^{b+1}}{a!b!} \left(\frac{1}{2}\right)^{a}.
\ee
We understand the notation of~\eq{dual^n-neat} as follows. The \rhs\ stands for all
independent, connected $n$-point diagrams which can be created from $j$ vertices
belonging to $\Sint$, $s$ internal lines (\ie\ effective propagators)
and $n$ external fields carrying momenta $k_1,\ldots,k_n$.
(It is the constraint of connectedness which restricts the sum over $j$.)
The combinatorics for generating fully fleshed out diagrams is simple
and intuitive. As an example of how it works, consider the diagram shown in
\fig{Decorate}.
\bcf[h]
	\ensuremath{\begin{array}{c}\input{pstex/Example.pstex_t} \end{array}}
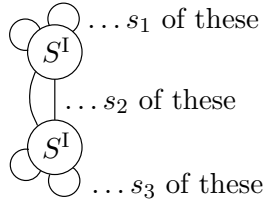
\caption{An example of a diagram represented by the \rhs\ of~\eq{dual^n-neat}, prior to decoration with the external fields.}
\label{fig:Decorate}
\ecf

The number of ways of generating this diagram can be worked out in two
parts. First, consider the effective propagators. To create the diagram,
we need to divide the $s$ effective propagators into sets 
containing $s_1$, $s_2$ and $s_3$ effective propagators.
The rule is that the number of ways of doing this is
\[
	\nCr{s}{s_1} \nCr{s-s_1}{s_2} \nCr{s-s_1-s_2}{s_3} = \frac{s!}{s_1! s_2! s_3!}.
\]
Next, we note that every effective propagator whose ends attach to a different
vertex comes with a factor of two, representing the fact that each of
these lines can attach either way round. This yields a factor
of $2^{s_2}$. The rule for the vertices is that they come with a
factor $j!/\sym$, where $\sym$ is the symmetry factor of the diagram.
Thus, including the numerical factors buried in $\diagnorm$, the
overall factor of our example diagram is
\[
\frac{1}{s_1!s_2!s_3!} \left(\frac{1}{2}\right)^{s_1+s_3} \frac{1}{\sym}.
\]

The diagrammatic expression for the dual action should make it obvious that
we can re-express the dual action vertices in terms of 
1PI components. Let us denote 1PI contributions by a bar so that, for example,
the 1PI contribution to $\dualv{2}$ is denoted by $\dopiv{2}$. From the diagrammatics
it is apparent that (as usual)
\be
	\dualv{2}(p) =
	\frac{
		\dopiv{2}(p)
	}
		{
		1 + \ep(p^2) \dopiv{2}(p)
	}.
\label{eq:D2-1PI}
\ee
However, this relationship holds independently of any diagrammatic representation.
Indeed, we will take the inverted version of this equation as the \emph{definition}
for $\dopiv{2}$. Note that the more
standard notation for $\dopiv{2}$ is $\Pi(p)$ (see \eg~\cite{WeinbergI}), which we
use from now on:
\be
	\Pi(p) \equiv \frac{
		\dualv{2}(p)
	}
		{
		1 - \ep(p^2) \dualv{2}(p)
	}.
\ee
Similarly, at the four-point level, the 1PI piece is defined via
\begin{subequations}
\begin{align}
	\dopiv{4}(p_1,p_2,p_3,p_4) &\equiv 
	\dualv{4}(p_1,p_2,p_3,p_4)
	\prod_{i=1}^4\left[1 + \ep(p_i^2) \Pi(p_i)\right],
\\
	\Rightarrow
	\dualv{4}(p_1,p_2,p_3,p_4) & = 
	\frac{\dopiv{4}(p_1,p_2,p_3,p_4)}{
		\prod_{i=1}^4\left[1 + \ep(p_i^2) \Pi(p_i)\right]
	}.
\label{eq:D4-1PI}
\end{align}
\end{subequations}
At this point, it is natural to introduce the dressed effective propagator,
\be
	\dep(p^2) \equiv \frac{1}{\ep^{-1}(p^2) + \Pi(p)}.
\label{eq:dep}
\ee

Note that resummations such as~\eqs{D2-1PI}{D4-1PI} cure a troubling problem with
the diagrammatic expansions of the dual action vertices. Since these expansions
contain arbitrarily reducible contributions, and since $\ep(p) \sim 1/p^2$, it
looks like the dual action vertices are arbitrarily divergent for vanishing external momenta.
However, the resummation of these reducible pieces ameliorates this problem.

It is worth taking a few moments to assess what the diagrammatic expressions for the dual action vertices in fact represent, since the resummability of the corresponding infinite series is far from
obvious. The first comment to make is that the vertices which appear in the series are (in principle) full, nonperturbative solutions to the flow equation. Consequently, we expect in general (an exception will be given in a moment) for the diagrammatic series to contain more than just perturbation theory; perturbation theory can be recovered by \emph{additionally} performing a perturbative expansion of the vertices (as will be illustrated in \sect{beta}) but this approximation is not made in the initial diagrammatic expressions. From this point of view, we might wonder if the diagrammatic expression is something like~\eq{proper-exp}, and so could, in principle, be resummed. 

With this in mind, let us consider a $\lambda \phi^4$-type theory in $\D<4$.
There are two cases to look at, depending on whether or not we sit on a renormalized trajectory.
Let us suppose, first of all, that we are on an interacting renormalized trajectory. Furthermore,
we will choose to evaluate the dual action at a high scale. As discussed at great length earlier,
that we are on a renormalized trajectory means that we must replace the usual notion of
the bare action with the perfect action in the vicinity of the appropriate fixed-point. This perfect action, whilst well approximated by perturbation theory for the case under discussion, nevertheless contains nonperturbative pieces. In this case, the diagrammatic expression always contains nonperturbative pieces.

Next let us suppose that we are not on a renormalized trajectory and, moreover, let
us chose to take the interaction part of the bare action to have just a $\lambda \phi^4$ term.
If we evaluate the diagrammatic expression for any $\Lambda < \Lambda_0$ then, again,
the diagrammatic expression will contain nonperturbative pieces. If, however, we take
$\Lambda=\Lambda_0$ then it is apparent
that we are doing perturbation theory in the bare coupling, $\lambda_0$, with a UV regularized propagator. Note, though, that we should not understand the cutoff function as merely providing regularization, since we cannot send the bare scale to infinity. Indeed, for such non-renormalizable theories, the cutoff function partly defines the theory, with different cutoff functions giving different theories. The diagrammatic expression will therefore, in this case, contain irremovable
dependence on the bare scale.

For each of these cases---sitting on a renormalized trajectory and considering a non-renormalizable trajectory with both $\Lambda < \Lambda_0$ and $\Lambda =\Lambda_0$---it would be desirable to
understand how much of the full nonperturbative expression $-\ln e^{\op} e^{-\Sint}$ is contained
by the diagrammatics. 

Let us conclude by noting that a partial resummation of the diagrammatic expressions can always
be performed in which classes of diagram are summed up such that all internal lines become dressed as in~\eq{dep}. Since the dressed internal lines are expressed in terms of the exact $\Pi(p)$, we can
expect that partially resummed diagrammatic expressions of this type have better behaviour than the
original ones; this is the basis of the approach taken in~\cite{Trivial,Susy-Chiral}.

\subsection{The Physical Interpretation of $\Lambda$}
\label{sec:Linear-Choose}

The dual action enables us to clarify certain issues regarding the physical interpretation of $\Lambda$. To this end, we return to the plain Polchinski equation as this makes the following analysis particularly simple. Thus, let us reconsider the pair of equations~\eqs{dual}{dual-flow} and recall that, in the current context, the vertices of the dual action essentially correspond to the connected correlation functions.

Independence of $\dual$ on $\Lambda$ confirms a statement made in \sect{General_ERG} that universal quantities know nothing about $\Lambda$. However, the definition of the dual action~\eq{dual} also confirms the flip side of this that the Wilsonian effective action can be used to evaluate universal quantities. Indeed~\eqs{dual}{dual-flow} tell us that, in principle, we can evaluate the correlation functions by using the Wilsonian effective action at any scale of our choosing. Now, if we had the luxury of knowing the full solution for $\Sint_\Lambda$ then we might as well simply set $\Lambda = 0$, thereby recovering the exact universal physics; we have no need to do anything else. Of course, except in very special circumstances, we do not have access to exact solutions.

With this in mind, let us take the opposite extreme. Suppose that we deal with a theory with bare action, $S_{\Lambda_0}$. Rather than using this boundary condition to compute the Wilsonian effective action, let us try to evaluate the dual action simply by setting $\Lambda = \Lambda_0$ in~\eq{dual}.  In order to make headway, we will use a diagrammatic approach and so, as mentioned in the last section, what we are doing amounts to bare perturbation theory. For the sake of argument, we are presuming that we have a small coupling at hand to control the loop expansion. Nevertheless, as is very well known, this is not necessarily a good way to do perturbation theory, particularly in the presence of large logarithms (see \eg~\cite{WeinbergII}). In this circumstance, we would be much better off choosing a value of $\Lambda$ whereby these logarithms are rendered harmless%
\footnote{In this approach we would have to compute $\Sint_\Lambda$ up to an appropriate number of loops, with the perturbative order of $\dual$ coming partly from explicit loops in the diagrams and partly from the loop order of the various vertices \cf~\fig{dual}. See \sect{beta} for some examples of perturbative computations within the ERG.}
and thus would expect this choice to coincide with the characteristic energy of the problem at hand.
To emphasise: if we were able to do things exactly then it would not matter what we choose for $\Lambda$; but in the absence of this we can hope to improve our approximate calculations by making a sensible choice. In this way, there are circumstances where $\Lambda$ can have a quasi-physical interpretation.

All of this is on the firmest footing when we have the luxury of a small parameter. Addressing issues such as those above in strongly coupled problems is much harder; we will discuss this further in \sect{Truncations} when we describe some nonperturbative approximation schemes supported by the ERG.

\section{Properties of Exact Solutions}
\label{sec:solve}

In this section, we will discuss some of the properties exhibited by exact solutions
of the flow equation. It will be useful to write the flow equation in the form
\be
	\partial_t \Sint = \fpop\left(\eta,\Sint\right).
\ee
To simplify things, we will use the flow equation of Ball et al.~\cite{Ball},
which can be obtained from~\eq{flow-rescaled} by setting $\hSint =0$
[and rewriting using~\eq{split}]:
\be
	\left(
		\partial_t + d_\phi \Count_\phi + \Count_\partial -\D
	\right) \Sint
	=
	\classical{\Sint}{\cutoff'}{\Sint} - \quantum{\cutoff'}{\Sint} - \frac{\eta}{2} \phi \cdot \ep^{-1} \cdot \phi.
\label{eq:Ball}
\ee
It is obvious that, in this case,
\be
	\fpop\left(\eta,\Sint\right)
	=
	\classical{\Sint}{\cutoff'}{\Sint} - \quantum{\cutoff'}{\Sint} 
	-\left( d_\phi \Count_\phi + \Count_\partial -\D
	\right) \Sint
	- \frac{\eta}{2} \phi \cdot \ep^{-1} \cdot \phi.
\label{eq:H-Ball}
\ee

As we have intimated already, it is presumably impossible to solve the flow
equation in complete generality. Nevertheless, there are some precise statements
that we can make about putative solutions and there are some suitably simple (but instructive)
cases where exact solutions can be found.

Rather than working with the full flow equation from the start, we will begin
by considering the somewhat simpler (but still complex)
task of finding fixed-points. After some general considerations in \sect{GenCon}, we
will make a first pass at the Gaussian fixed-point in \sect{GFP} to illustrate some of the basic ideas. Armed with the lessons learnt from this, we will refine our analysis using the dual action in \sect{FP-dual}. This will allow us to arrive
at a fuller understanding of fixed-point solutions; in particular, we will demonstrate that the
spectrum of the anomalous dimension at critical fixed-points is quantized. As part of this, we will explicitly construct an infinite family of redundant operators which exist at every fixed-point. 
From a practical point of view, we are also able to 
quickly and efficiently uncover all two-point fixed-point solutions, as we will see in \sect{Gen-TP}.

We will move on to discuss scale-dependent solutions in \sect{renorm}. Our focus here will be
on renormalized trajectories where we will refine the analysis of~\cite{TRM-Elements}
pertaining to nonperturbative renormalizability. Finally, we will deal with the issue first mentioned
in footnote~\ref{foo:self-similar} as to why we are justified in picking out, from the general solution to the linearized flow equation, those eigenperturbations for which the $t$-dependence separates.

\subsection{Fixed-Points}
\label{sec:FP}

\subsubsection{General Considerations}
\label{sec:GenCon}

As we have already discussed at great length, the
fixed-point criterion in dimensionless variables is simply
\be
	\partial_t \Sint_\star = 0
	\qquad
	\Rightarrow
	\qquad
	\fpop(\eta_\star,\Sint_\star) = 0.
\label{eq:H-fp}
\ee
For the flow equation~\eq{Ball} it is apparent that
\be
	\fpop(\eta_\star,\Sint_\star) =
	\classical{\Sint_\star}{\cutoff'}{\Sint_\star} 
	- \quantum{\cutoff'}{\Sint_\star} -\frac{\eta_\star}{2} \phi \cdot \ep^{-1} \cdot \phi
	-\left(
		d_{\star} \Count_\phi + \Count_\partial -\D
	\right) \Sint_\star = 0.
\label{eq:FP-eq}
\ee
Recall that $d_\phi$ depends on $\eta$ and so we take $d_\star \equiv (\D-2+\eta_\star)/2$.
The first thing to notice is that $\eta_\star$ seems like a free parameter, suggesting that
there are exists a continuous infinity of fixed-points. As the analysis proceeds, we will
build up an understanding of why this is not the case.

As discussed in  
\sect{renormalizability}, it greatly aids in understanding the nature of fixed-points to consider linearizing the flow equation around a fixed-point solution. For what follows, we will suppose that
the dependence on $t$ separates so that, just
as in~\eq{perturb}, we write
\[
	S_t[\phi] = S_\star[\phi]  + \sum_i \alpha_i e^{\lambda_i t} \eop_i[\phi].
\]
We will return to the issue of the general solution to the linearized flow equation in \sect{self-similar}.

Having reached the quantitative phase of the discussion we can now be explicit about the operator in the eigenvalue equation~\eq{eigen}:
\be
	 \classifier = 2\classical{\Sint_\star}{\cutoff'}{} - \quantum{\cutoff'}{} - d_\star \Count_\phi - \Count_
	 \partial + d,
\label{eq:classifier}
\ee
so that
\be
	\left(
		2\classical{\Sint_\star}{\cutoff'}{} - \quantum{\cutoff'}{} - d_\star \Count_\phi - \Count_\partial + d
	\right)
	\eop_i[\phi] =  \lambda_i \eop_i[\phi].
\label{eq:full-eigen}
\ee
Note that it does not matter how we normalize the eigenoperators, since such normalizations
are scale-independent. Shifting such constants between the $\eop_i$ and the $\alpha_i$
amounts to redefining the associated coupling constant by a scale-independent factor. However, in the flow equation approach to noncommutative theories,
things are much more subtle~\cite{RG+OJR}.

As before, the RG eigenvalues, $\lambda_i$, are divided up into those which are relevant, irrelevant or marginal\footnote{Wegner introduces one further classification~\cite{Wegner_CS}: the constant eigenoperator is referred to as `special' since, although it has positive eigenvalue $+\D$, it does not affect the critical behaviour and is therefore distinct from the rest of the relevant operators.}; the latter may, upon analysis beyond leading order, either turn out to remain exactly marginal, or to become marginally relevant/irrelevant. 

The corresponding eigenoperators are additionally classified according to whether or not they are \emph{redundant}. Redundant operators correspond to infinitesimal, quasi-local field redefinitions%
\footnote{We will use $\varepsilon$ for generic small quantities, reserving $\epsilon$ for use in the context of the $\epsilon$-expansion, in which deviations from some given dimensionality of Euclidean space are considered.},
\be
	\phi'(p) = \phi(p) + \varepsilon  \Theta(p).
\ee
Recalling~\eq{G_tra}, any operator---defined at the fixed-point $S_\star$---that, for \emph{quasi-local} $\Theta(x)$, can be written
in the form
\be
\begin{split}
	\eop^{\mathrm{R}}[\phi;\Theta] 
	&= 
	\int_p
	\left\{
	 	\Theta(p) \fder{S_\star[\phi]}{\phi(p)}
		- \fder{\Theta(p)}{\phi(p)}
	\right\}
\\
	& =
	\int_p
	\left\{
		\Theta(p) \ep^{-1}(p^2) \phi(-p)
	 	+\Theta(p) \fder{\Sint_\star[\phi]}{\phi(p)}
		- \fder{\Theta(p)}{\phi(p)}
	\right\}
\end{split}
\label{eq:redundant}
\ee
corresponds to a quasi-local change of variables and therefore has no effect on physics. Such
operators are redundant. To put things another way, an infinitesimal perturbation of a fixed-point action in a redundant direction can be undone by a quasi-local change of variables. Wegner noted that,
for very general ERGs,  the redundant operators form a closed subspace under the flow in the vicinity of a fixed-point~\cite{Wegner_CS,WegnerInv}. For the case of the flow equation we are considering, O'Dwyer and Osborn confirmed this by demonstrating that~\cite{JOD+HO}
\be
	\classifier \eop^{\mathrm{R}}[\phi;\Theta] 
	=
	\eop^{\mathrm{R}}
	\Bigl[\phi;\bigl(\classifier -\D + d_\star- p\cdot \partial_p \bigr)\Theta(p)\Bigr]
	+  \eop^{\mathrm{R}}\Bigl[\phi;2\Theta \ep^{-1} \cutoff'\Bigr],
\label{eq:closed}
\ee
which can be checked by direct substitution (see~\app{Menagerie} for some similar, albeit simpler, calculations).

So let us now consider perturbing fixed-point actions in various ways. If the change is in either a relevant or irrelevant direction---discounting for the moment those which are only marginally so---then a flow is induced. Contrariwise, suppose that we perform an infinitesimal perturbation of a fixed-point in a marginal direction: $S_\star \rightarrow S_\star + \varepsilon \marginal$. Whatever happens beyond leading order
in $\varepsilon$, \emph{at leading order}  we have a new fixed-point. 

The strategy for going beyond leading order (of which we will see an explicit example in \sect{SFM}) is to write $S_t[\phi] = S_\star[\phi]+\mathscr{P}_t[\phi]$, where $\mathscr{P}$ satisfies the flow equation up to $\order{\varepsilon^2}$ and reduces to $\varepsilon \marginal$ at $\order{\varepsilon}$. Assuming that the eigenoperators of the putative fixed-point form a complete basis in theory space, then $\mathscr{P}_t$ will have the structure
\be
	\mathscr{P}_t[\phi] = \chi(t) \marginal[\phi] + \sum_i \mu_i(t) \eop_i[\phi],
\label{eq:T_t}
\ee
where the sum runs over all operators besides the marginal one that has been singled out and the $\mu_i(t)$ are understood to be quadratically small in $\varepsilon$.%
\footnote{%
The $\mu_i(t)$ (which are nothing to do with the arbitrary scale, $\mu$, buried inside $t$)
are sometimes called `scaling fields'. These fields are not fields in the sense of $\phi$. This terminology is much less confusing in the original context of critical phenomena, where the action is a functional of `spins'. It is only in the context of QFT, where these `spins' are more naturally referred to as fields, that overuse of the word `field' occurs in this way.
}
 If the projection of $\mathscr{P}_t[\phi]$ on to the $\marginal$ direction depends on $t$ then our operator is either marginally relevant or marginally irrelevant (which of these it is must be computed). 

However, it might be that the flow in the $\marginal$ direction still vanishes
\ie\ $d\chi / dt = \order{\varepsilon^3}$.
Supposing that this is the case, we would like to know whether it is possible to tune things such that, to $\order{\varepsilon^2}$, we also have $d \mu_i /d t = \order{\varepsilon^3}, \forall i$. Substituting~\eq{T_t}
into the flow equation~\eq{Ball}, we find that
\be
	\der{\mu_i}{t} = \lambda_i \mu_i + b_i \chi^2,
\ee
where the $b_i$ must be computed. The last term represents the feedback of $\marginal$ into the flows of the other operators. In the case that this feedback is zero, we can kill the flows simply by setting $\mu_i = 0$. For the cases where this does not happen, we can kill the flows by choosing
$\mu_i = -b_i\chi^2 /\lambda_i$, so long as $\lambda_i \neq0$.
The result of this analysis is that if the marginal direction remains marginal at $\order{\varepsilon^2}$ then, provided that this operator does not contribute to the flow of some other marginal operator (at the same order) then we can arrange for $\partial_t \mathscr{P} = \order{\varepsilon^3}$. We can imagine that
marginal operators might exist for which this picture holds true to every order in $\varepsilon$ 
(and also for any contributions to $\mathscr{P}$ which are nonperturbative in $\varepsilon$, should they exist).
In this case, our operator is said to be \emph{exactly marginal}, and there exists a line of fixed-points, since we can
go a finite distance away from the original fixed-point without generating a flow.

It is sometimes said that an exactly marginal operator generates a line of fixed-points. There is nothing wrong with this statement, but it can be a bit confusing. Let us emphasise that, due to the feedback of $\marginal$ into the flow of other operators, perturbing a fixed-point action in \emph{just} an exactly marginal direction yields another fixed-point only up to $\order{\varepsilon}$. To generate the line of fixed-points associated with an exactly marginal operator requires figuring out which other operators must become non-zero, as we go along the line, in order for the flow of the action to remain zero. 

Note also that, once a line of fixed-points has been found, we can linearize about any action along the line. Generically, in each case, the exactly marginal operator will be of a different form since \emph{by itself} the exactly marginal operator generates only an infinitesimal perturbation along the line. This situation is illustrated in \fig{marginal}.
\bcf[h]
	\[
	\ensuremath{\begin{array}{c}\input{pstex/Marginal.pstex_t} \end{array}}
	\]
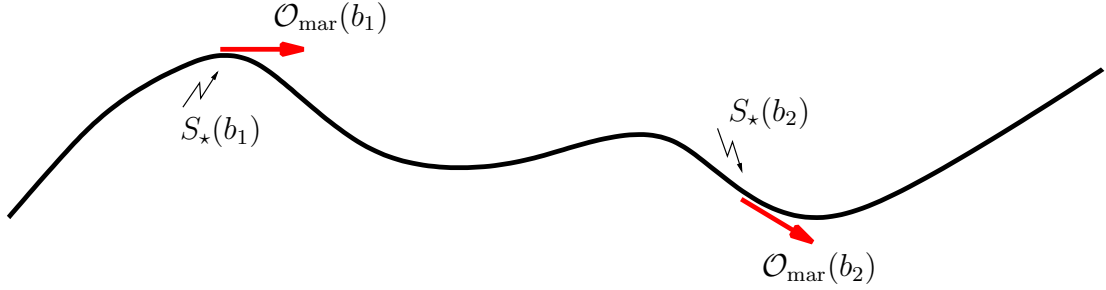
\caption{A portion of a line of fixed-points, parametrized by $b$. The flow equation can be linearized using any action along
the line, in each case yielding a different expression for the exactly marginal operator. In each case,
a perturbation in the exactly marginal direction takes the action an infinitesimal distance along the line, as indicated for two values of $b$, $b=b_1$ and $b=b_2$.}
\label{fig:marginal}
\ecf

We have not yet specified whether our putative exactly marginal direction is redundant or not. If the operator is redundant, then the fixed-points along the line are all equivalent, being as they are
related to each other by a quasi-local change of variables. Consequently, they all encode the same physics. However, if the operator is not redundant, then each of the fixed-points along the line are physically distinct. An example of the latter case is $\mathcal{N}=4$ super Yang--Mills, in $\D=4$, which is (thought to be) conformal for any value of the coupling.

Now, all critical fixed-points turn out to possess a marginal, redundant operator~\cite{Wilson+Bell,Wilson+Bell-FiniteLattice,RGN,JOD+HO,WegnerInv,Wegner_CS} associated with the normalization of the field. As argued in~\cite{RGN}, this operator is exactly marginal.
\emph{Consequently, every critical fixed-point exists as a line of equivalent fixed-points.}%

For the flow equation we are working with (\ie\ with $\hSint=0$), this operator has been explicitly constructed by O'Dwyer and Osborn~\cite{JOD+HO}. First of all, let us define
\be
	\varrho(p^2) \equiv - p^{2(\eta_\star/2)} \cutoff (p^2) \int_0^{p^2} dq^2
	\left[
		\frac{1}{\cutoff (q^2)}
	\right]'
	q^{-2(\eta_\star/2)},
\label{eq:HO-b}
\ee
(where the prime denotes a derivative \wrt\ momentum squared). Notice that for the integral to be well defined at its lower limit, we must take $\eta_\star <2$. This has a physical origin: as we will argue in \sect{Redux}, only those fixed-point for which $\eta_\star <2$ are critical. Next construct
\be
	\Theta(p) = [\varrho(p^2) + 1] \phi(p) + 
	\ep(p^2) \varrho(p^2) \fder{\Sint_\star}{\phi(-p)}
\label{eq:HO-Theta}
\ee
and  substitute this into~\eq{redundant} to yield an operator we will call $\eop'^{\mathrm{R}}_\mathrm{mar}$:
\be
	\eop'^{\mathrm{R}}_\mathrm{mar} = \phi \cdot \ep^{-1} \bigl(\varrho+1 \bigr) \cdot \phi
	+ \phi \cdot \bigl(2\varrho+1 \bigr)\cdot \fder{\Sint_\star}{\phi}
	+
	\classical{\Sint_\star}{\ep\varrho}{\Sint_\star}
	-
	\quantum{\ep\varrho}{\Sint_\star}.
\label{eq:HO-mro}
\ee
That this operator is indeed marginal can be checked by using~\eq{full-eigen}, as we show in \app{Menagerie-mro}.
Now, since $\varrho$ is quasi-local, starting at $\order{p^2}$, this means that
the combination $\ep \varrho$ is quasi-local. Therefore, given  the assumed quasi-locality of the 
action, $\Theta(p)$ is quasi-local and so $\eop'^{\mathrm{R}}_\mathrm{mar} $ is redundant.

However, for many purposes, there
is a neater way of writing the marginal, redundant operator at least for $\eta_\star <2, \ \neq 0$.
Defining the `cutoff function counting operator', 
$\Count_{\cutoff} \equiv \cutoff \cdot \delta/\delta \cutoff$ observe that
\be
	\marginal [\phi] \equiv
	\left(\hf \Count_\phi + \Count_{\cutoff}\right) \Sint_\star[\phi]
	\equiv
	\hat{\Count} \Sint_\star[\phi]
\label{eq:emro}
\ee
is marginal. We see this by substituting this expression into~\eq{full-eigen}
and recognizing that
\be
	\left[ \hat{\Count}, \classical{}{\cutoff'}{}\right] = 0,
\label{eq:Deltahat-com}
\ee
upon which we are left with
\begin{align}
\nonumber
	\classifier \hat{\Count} \Sint_\star
	&=
	\hat{\Count}
	\left[
		\classical{\Sint_\star}{\cutoff'}{\Sint_\star} - \quantum{\cutoff'}{\Sint_\star} 
		- \left(d_\phi \Count_\phi + \Count_\partial - \D\right) \Sint_\star
	\right]
\\
	& =
	\hat{\Count} 
	\left(\frac{\eta_\star}{2} \phi \cdot \ep^{-1} \cdot \phi\right) = 0,
\label{eq:zeromode}
\end{align}
where we have used the fixed-point equation~\eq{FP-eq}.\footnote{We are assuming that
$\Count_{\cutoff} \eta_\star = 0$. By this we mean that the values of the quantized $\eta_\star$ corresponding to quasi-local fixed-points are independent of the cutoff function. This is to be expected on physical grounds, though I am unaware of a general proof. To be safe, we could understand $\Count_{\cutoff}$ to act at constant $\eta_\star$.}
As we will show in \app{redundant},
\be
	\eop'^{\mathrm{R}}_\mathrm{mar}[\phi] = 
			-2\marginal[\phi],
			\qquad \eta_\star <2, \ \neq 0.
\label{eq:simple}
\ee

The situation for $\eta_\star = 0$ is as follows. There is certainly one fixed-point with this
$\eta_\star$: the Gaussian one. In this case, we show in  \app{redundant} that
\be
	\eop'^{\mathrm{R}}_\mathrm{mar}[\phi] = 
	\frac{2(1-\intconst)}{\intconst} \marginal[\phi], \qquad \mbox{Gaussian fixed-point},
\label{eq:simple-GFP}
\ee
where $\intconst$ parametrizes the line of equivalent Gaussian fixed-points (see the next section).
This leaves the obvious question as to whether there are other fixed-points with
$\eta_\star=0$ and, if so, what role $\hat{\Count} \Sint_\star$ plays in this case. For integer dimension
$\geq 2$, there is a theorem due to Pohlmeyer~\cite{Pohlmeyer} which implies that the only critical
fixed-point with $\eta_\star = 0$ is the Gaussian one. Deferring until \sect{LPA} what we mean by solutions to the flow equation in non-integer dimensions, it is claimed in~\cite{Trivial} that the same is true for any $\D \geq 4$, though the level of rigour is certainly not that of a theorem. 
It is tempting to speculate that it is generally true that the only critical fixed-point with $\eta_\star=0$ is the Gaussian one.

With this in mind, let us note that writing the marginal, redundant operator in the form~\eq{emro} has a distinct advantage: it is possible to derive a very simple (new) expression for the associated line of fixed-points. Indeed, given some fixed-point $\Sint_\star$, there exists a family of fixed-points given by
\be
	e^{b \hat{\Count}}  \Sint_\star( b_0 ) = \Sint_\star(f(b_0,b)),
\label{eq:line}
\ee
where $b_0$ and $b$ are real parameters and $f$ is some function. This function can be determined by operating on the left with $e^{b' \hat{\Count}}$, from which it is apparent that consistency demands:
$
	f(b_0,b+b') = f(f(b_0,b),b')
	\	
	\Rightarrow
	\
	f(b,b') = b+b'.
$
Therefore,
\be
	e^{b \hat{\Count}} \,  \Sint_\star(b_0) = \Sint_\star(b_0+b),
\label{eq:line-explicit}
\ee
so long as no singularities are encountered between $b_0$ and $b_0+b$. Henceforth, we will not bother to indicate dependence on $b_0$.
Now, the above suggests a way to prove that the only critical fixed-point with $\eta_\star = 0$ is the Gaussian one: 
suppose that we can show that all such fixed-points are connected with the Gaussian one in the limit 
$b \rightarrow -\infty$, as is certainly plausible from the form of~\eq{line-explicit}. Then, equivalence of these fixed-points follows because we know from the form of~\eq{line-explicit} that the Gaussian fixed-point is approached along its marginal, redundant direction. This scenario has been confirmed in~\cite{OJR-Pohl}.

To prove~\eq{line},  let us recall~\eqs{FP-eq}{full-eigen}, upon which it is apparent that
\be
	\fpop(\eta_\star, e^{b\hat{\Count}} \Sint_\star) 
	=
	\classifier 
	\bigl(e^{b\hat{\Count}} -1 \bigr) \Sint_\star
	+
	\classical{\bigl(e^{b\hat{\Count}} -1 \bigr)\Sint_\star}{\cutoff'}{\bigl(e^{b\hat{\Count}} -1 \bigr)\Sint_\star},
\label{eq:zero}
\ee
where we have written $e^{b\hat{\Count}} \Sint_\star = \Sint_\star + (e^{b\hat{\Count}}-1) \Sint_\star$
and used the fact that $\fpop(\eta_\star, \Sint_\star) =0$. To show that~\eq{zero} vanishes, let us differentiate \wrt\ $b$:
\be
	\der{}{b}\, \fpop(\eta_\star, e^{b\hat{\Count}} \Sint_\star)	=
	\classifier \, e^{b\hat{\Count}} \hat{\Count} \Sint_\star
	+
	2
	\classical{\bigl(e^{b\hat{\Count}} -1 \bigr)\Sint_\star}{\cutoff'}{\, e^{b\hat{\Count}} \hat{\Count} \Sint_\star}.
\label{eq:differentiated}
\ee
Remembering that $\hat{\Count} \Sint_\star$ is marginal, we know that
$\classifier \, \hat{\Count} \Sint_\star = 0$ and so
\be
	\classifier \, e^{b\hat{\Count}} \hat{\Count} \Sint_\star
	=
	\bigl[\classifier , e^{b\hat{\Count}}\bigr] \hat{\Count} \Sint_\star.
\label{eq:classify-line}
\ee
The commutator can be processed using standard tricks:
\begin{align}
	\bigl[
		\classifier, e^{b \hat{\Count}}
	\bigr]	
	& =
	\int_0^1 e^{sb  \hat{\Count}}
	\bigl[
		\classifier, b \hat{\Count}
	\bigr]	
	e^{-sb \hat{\Count}}
	e^{b \hat{\Count}}
\nonumber
\\
	& =
	-\Bigl(
	\bigl[
		 b\hat{\Count}, \classifier
	\bigr]	
	+
	\frac{1}{2!} 
	\bigl[
	b\hat{\Count},
	\bigl[
		 b\hat{\Count}, \classifier
	\bigr]
	\bigr]	
	+
	\frac{1}{3!}
	\bigl[ 
	b\hat{\Count},
	\bigl[
	b\hat{\Count},
	\bigl[
		 b\hat{\Count}, \classifier
	\bigr]
	\bigr]	
	\bigr]
	+\ldots
	\Bigr) e^{b \hat{\Count}}
\nonumber
\\
	& =
	-2\classical{\bigl(e^{b\hat{\Count}} -1\bigr)\Sint_\star}{\cutoff'}{} e^{b\hat{\Count}},
\label{eq:tricks}
\end{align}
where the last line is obtained using
\be
	\bigl[\hat{\Count}, \classifier\bigr]
	=
	2\classical{\hat{\Count} \Sint_\star}{\cutoff'}{}.
\label{eq:commutator}
\ee
Substituting~\eq{tricks} into~\eq{classify-line}
it is immediately apparent that the \rhs\ of~\eq{differentiated} vanishes. Integrating up, the integration constant can be seen to be zero by noting that the \rhs\ of~\eq{zero} vanishes for $b=0$.
Therefore,
\be
	\fpop(\eta_\star, e^{b\hat{\Count}} \Sint_\star) 
	=0,
\label{eq:demonstrated}
\ee
from which~\eq{line} follows directly.

Let us summarize what we have learnt so far. The eigenperturbations at a fixed-point
can be divided into those which are redundant and those which are not. The former
correspond to quasi-local field redefinitions and carry no physics. Every critical fixed-point
possesses an exactly marginal, redundant operator [which, for the flow equation we are using, is given by~\eq{HO-mro}] meaning that every such fixed-point appears as a line of equivalent fixed-points in theory space. For the case of $\eta_\star<2\neq0$, this operator is related to $\hat{\Count} \Sint_\star$ via~\eq{simple}. At the Gaussian fixed-point, the two operators are related by~\eq{simple-GFP}. Given the result of~\cite{OJR-Pohl} that the Gaussian fixed-point is the only critical fixed-point with $\eta_\star = 0$ (subject to positivity of the connected two-point correlation function), there are no other cases with $\eta_\star = 0$ to treat (we will say a little more about non-critical theories in \sects{GFP}{Gen-TP}).

However, there is more. In~\cite{Wegner_CS}, Wegner 
demonstrated that, if the spectrum of $\eta_\star$ is quantized, 
then
there necessarily exists a marginal, redundant operator. Wegner's proof was formulated
for completely general flow equations; here we will reproduce it for the special case of the
flow equation we are focusing on in this section, \eq{Ball}. Recalling~\eq{FP-eq},
let us consider
\be
	\fpop
	\biggl(
		\eta_\star + \varepsilon,\Sint_\star+ \sum_i \alpha_i \eop_i
	\biggr) =
	\sum_i \alpha_i \lambda_i  \eop_i
	+\varepsilon
	\pder{\fpop(\eta_\star,\Sint_\star) }{\eta_\star}
	+ \order{\varepsilon^2},
\label{eq:quantized-proof}
\ee
where $\critexp_i$ are the critical exponents [recall~\eq{full-eigen}] and we take $\alpha_i \sim \order{\varepsilon}$. Next observe that
\be
	\pder{\fpop(\eta_\star,\Sint_\star) }{\eta_\star}
	=
	\hf
	\left(
		\Count_\phi \Sint_\star +  \phi \cdot \ep^{-1} \cdot \phi
	\right)
\ee
is a redundant operator, as can be seen by taking $\Theta(p) = \phi(p)$ in~\eq{redundant} [neglecting the (divergent) constant piece, as usual].
Since redundant operators form a closed subspace [recall~\eq{closed}], we can therefore
write
\be
	\varepsilon \pder{\fpop(\eta_\star,\Sint_\star) }{\eta_\star}
	=
	\sum_{j=\{\mathrm{R}\}} \tilde{\alpha}_j \eop^{\mathrm{R}}_j,
\label{eq:H-deriv-decomp}
\ee
where $j$ runs only over the redundant operators, and the $\tilde{\alpha}_j$
are some set of numbers distinct from the $\alpha_i$, but again of $\order{\varepsilon}$. Consequently, we
can cast~\eq{quantized-proof} in the form
\be
	\fpop
	\biggr(
		\eta_\star + \varepsilon,\Sint_\star + \sum_i \alpha_i  \eop_i
	\biggl) =
	\sum_{i\neq \{\mathrm{R}\}} \alpha_i \lambda_i \eop_i
	+
	\sum_{j=\{\mathrm{R}\}} 
	\bigl(
		\tilde{\alpha}_j + \alpha_j \lambda_j
	\bigr)
	\eop^{\mathrm{R}}_j
	+ \order{\varepsilon^2}.
\label{eq:quantized-fp}
\ee

Now for the point: since, by assumption, $\eta_\star$ is quantized, \eq{H-fp} only has a
discrete spectrum of solutions and so 
the \lhs\ of~\eq{quantized-fp} cannot
vanish for infinitesimal $\varepsilon$; in other words, there must always be a non-vanishing term at order $\varepsilon$.
With this in mind,
notice that the first term on the \rhs\ can always be made to vanish by choosing those $\alpha_i$ corresponding to scaling operators to vanish.
Moreover, if none of the $\lambda_j$ vanish then the $\alpha_j$ can always to chosen to make the
second term on the \rhs\ vanish. Therefore we conclude 
that there must be at least one value of $j$ for which $\lambda_j =0$
and $\tilde{\alpha}_j \neq 0$. As a result, quantization of $\eta_\star$ implies the existence
of a marginal, redundant operator.

Note that this argument can be turned around. Suppose that a marginal, redundant direction exists for every critical fixed-point. Furthermore, suppose that
 the corresponding quantities $\partial \fpop(\eta_\star , \Sint_\star) / \partial \eta_\star$ all have a component in the appropriate marginal, redundant direction. Then it follows
the spectrum of $\eta_\star$ is quantized.\footnote{I would like to thank Hugh Osborn for pointing
this out to me.} Now, we already know that the first of these criteria is true for critical
fixed-points; in \sect{FP-dual} we will prove the second. 
Before embarking on this proof, we will illustrate some of the considerations of this
section with a simple example.

\subsubsection{The Gaussian Fixed-Point}
\label{sec:GFP}

By inspection of~\eq{FP-eq}, there is a very simple solution: $\Sint_\star=0$ together with $\eta_\star=0$
(this is encouraging, since the solution $\Sint_\star=0$ occurs only for a special value of $\eta_\star$).
This solution corresponds, of course, to the Gaussian fixed-point. Recalling the splitting~\eq{split},
we see that
\[
	S^{\mathrm{Gaussian}}_\star[\phi] = \hf \phi \cdot \ep^{-1} \cdot \phi
	= \hf \int_p \phi(-p) p^2 \cutoff^{-1}(p^2) \phi(p),
\]
where we remember that we are now working with dimensionless momenta.
There are several points worthy of comment. First of all, as we will see later, there are many physically inequivalent two-point solutions to the flow equation; we will reserve the term `Gaussian' for this one. The second, rather more disturbing point is that, due to the presence of the cutoff function in the action, the fixed-point action is not dilatation invariant!%
\footnote{%
This is easy to see. Recall that a representation of the dilatation generator is given by~\eq{DilGen}. At the Gaussian fixed-point, $\eta_\star=0$ and it is easy to check that $\hat{D}^- \int_p \phi(p) \phi(-p) p^2  = 0$. This invariance is obviously spoilt by a cutoff function. But see the conclusion of~\cite{HO-Remarks} for some indications of a grander picture.
}
We will discuss this in much greater detail in \sect{FP-scaling}. For the time being we note that 
whilst it is a general feature of fixed-points within the ERG formalism that the implementation of a cutoff spoils dilatation invariance of the \emph{action}, the correlation functions---which are more directly related to physics---are automatically dilatation covariant at a fixed-point. The final point to make is that this Gaussian fixed-point is in fact only a representative of a line of \emph{equivalent} fixed-points~\cite{WegnerInv,TRM-Elements}.
Let us recall that, by equivalent, we mean that they all describe exactly the same universal physics; as we will see (and as we expect), this is because they are related to one another by a quasi-local field redefinition.

To see this line of equivalent fixed-points, we note that there is a more general solution to~\eq{FP-eq}, \emph{for which $\eta_\star$ is still zero}, given by:
\be
	\Sint_\star[\phi] = \hf \int_p \phi(-p) \frac{\intconst p^2}{1-\intconst \cutoff (p^2)} \phi(p)
	\qquad \Rightarrow \qquad
	S^{\mathrm{Gaussian}}_\star[\phi]
	= \hf \int_p \phi(-p) \frac{\ep^{-1}(p^2)}{1-\intconst \cutoff (p^2)}  \phi(p),
\label{eq:Gen-Gauss}
\ee
where $\intconst$ is an integration constant. Recalling from \sect{Linear-Pol} that we must take $\cutoff (p^2)$ to be monotonically decreasing---and that $\cutoff (0)=1$---it is apparent that we must restrict to $\intconst<1$. For $B=1$, the denominator starts at $\order{p^2}$ and the theory has a mass term, meaning that it is non-critical; we will say more about this in \sect{Gen-TP}. For $B>1$, the kinetic term is of the wrong sign (which, as we will see in \sect{CorrFns-Ball}, leads to a violation of positivity of the two-point correlation function). Consequently, for the remainder of this section we will focus on the case $B<1$. 
This general Gaussian solution can be checked by direct substitution but we will give a more sophisticated derivation later.

Let us now classify the eigenperturbations of the Gaussian fixed-point. We will do this for the general Gaussian solution~\eq{Gen-Gauss} in \sect{Gen-TP}, using more sophisticated machinery. For the time being we will focus on the simplest representative, $\intconst=0$. 

Given some fixed-point, the eigenperturbations are found by linearizing the flow equation around the fixed-point solution whilst separating the variables $t$ and $\phi$. Anticipating this, and
anticipating that physically acceptable perturbations of the Gaussian fixed-point will be labelled by two integers, $n$ and $r/2$
 (essentially counting fields and powers of momenta), we introduce the integration constants, $\alpha_{n,r}$, and the scaling exponents, $\lambda_{n,r}$, and write
\be
	\Sint_t[\phi] = \Sint_\star[\phi] + \sum_{n,r} \alpha_{n,r} e^{\lambda_{n,r} t} \mathcal{G}'_{n,r}[\phi],
\ee
where the $\mathcal{G}'_{n,r}$ are the eigenperturbations at the simplest representative of the
Gaussian fixed-point.
At linear order, these eigenperturbations satisfy the equation
\be
	\lambda_{n,r} \mathcal{G}'_{n,r}
	=
	-\left(\quantum{\cutoff'}{}
	+\frac{\D-2}{2} \Count_\phi + \Count_\partial - \D
	\right)\mathcal{G}'_{n,r}.
\ee
To solve this equation, we follow Wegner~\cite{Wegner_CS}---who analysed the analogous equation derived from Wilson's version of the ERG equation. Recalling the definition~\eq{op},
\[
	\op \equiv 
	\hf
	\left(
		\classical{}{\ep}{}
	\right),
\]
we observe that since
\be
	\left[\op, \frac{\D-2}{2}\Count_\phi + \Count_\partial -\D \right] = 
	-\quantum{\cutoff'}{}
\ee
(as can be easily checked) it follows that
\be
	 \lambda_{n,r}
	e^\op \mathcal{G}'_{n,r}
	=
	-
	\left(
		\frac{\D-2}{2} \Count_\phi + \Count_\partial -\D
	\right) e^\op \mathcal{G}'_{n,r}.
\label{eq:Gaussian-eigen}
\ee
Deferring the issue of general solutions to the linearized flow to \sect{self-similar}, for the time being we notice that  one set of solutions is given by:
\begin{subequations}
\begin{align}
	\mathcal{G}'_{n,r}[\phi]
	& =
	e^{-\op} 
	\int_{q_1,\ldots,q_n}
	\frac{1}{n!} 
	v_r(q_1,\ldots,q_n) \phi(q_1)\cdots \phi(q_n) \hat{\delta}(q_1+\cdots+q_n),
\label{eq:SimpleGaussian-eop}
\\
	\lambda_{n,r} & = \D - r - \frac{n(\D-2)}{2},
\label{eq:scaling_exp}
\\
	v_r(a q_1,\ldots,a q_n) & = a^r v_r(q_1,\ldots,q_n).
\end{align}
\end{subequations}
So long as we take the eigenperturbations to be quasi-local, 
 $v_r(q_1,\ldots,q_n)$ is a homogeneous polynomial with $r/2$  a non-negative integer: the RG eigenvalues are quantized. 

Let us now analyse what we have found. The eigenperturbations look like a generalization of Hermite polynomials: $\mathcal{G}'_{n,r}$ has a term with $n$ fields, $n-2$ fields\ldots, all the way down to a (divergent) constant piece. 
In the standard  lingo, the presence of $e^{-\op}$ in~\eq{SimpleGaussian-eop} amounts to normal ordering.%
\footnote{In $\D=2$, the presence of normal ordering generates IR divergences which causes various operators to cease to be well defined. Consequently, in this case one should work at finite volume. For a careful derivation of the flow equation at finite volume see~\cite{HO-Remarks}.
} 
The even, non-negative integer, $r$, carries the order in momenta of the vertex coefficient function. It is perhaps easiest to see what is going on in pseudo-diagrammatic form, as illustrated
in \fig{eigen}.
\bcf[h]
	\[
	\mathcal{G}'_{4,r}[\phi] =
	\frac{1}{4!}
	\ensuremath{\begin{array}{c}\input{pstex/Eigen_4-r-0.pstex_t} \end{array}}
	\
	-
	\frac{1}{4}
	\ensuremath{\begin{array}{c}\input{pstex/Eigen_4-r-1.pstex_t} \end{array}}
	\
	+\frac{1}{8}
	\ensuremath{\begin{array}{c}\begin{picture}(0,0)%
\epsfig{file=pstex/Eigen_4-r-2.pstex}%
\end{picture}%
\setlength{\unitlength}{3947sp}%
\begingroup\makeatletter\ifx\SetFigFont\undefined%
\gdef\SetFigFont#1#2#3#4#5{%
  \reset@font\fontsize{#1}{#2pt}%
  \fontfamily{#3}\fontseries{#4}\fontshape{#5}%
  \selectfont}%
\fi\endgroup%
\begin{picture}(485,716)(1631,-737)
\put(1747,-435){\makebox(0,0)[lb]{\smash{{\SetFigFont{11}{13.2}{\rmdefault}{\mddefault}{\updefault}{\color[rgb]{0,0,0}$v_r$}%
}}}}
\end{picture}%
 \end{array}}
	\]
\caption{The diagrammatic expression for $\mathcal{G}'_{4,r}$. Since each term has a different
number of fields, neither the fields nor the symmetry factors can be stripped off. Thus, integrals
over the momenta carried by the fields are implied.}
\label{fig:eigen}
\ecf

The link with Hermite polynomials is clearest when we focus on the case where $r=0$ (\ie\ the vertices do not have any momentum dependence)~\cite{Wegner_CS}. If we could simply forget about
the fact that the $\phi$ carry momenta---and, along these lines, just ignore the associated momentum
integrals---then, defining
\[
	I_0 \equiv \int_p \ep(p^2),
\]
we could write the $\mathcal{G}'_{n,0}$ as
\[
	\frac{I_0^{n/2}}{n!} H_n (\phi/I_0^{1/2}),
\]
where $H_n$ is a Hermite polynomial of degree $n$. Actually, neglecting the momentum dependence in this way essentially amounts to the lowest order of the derivative expansion. But, of course, there is no need to do this here; our purpose has simply been to elucidate the relationship of the eigenperturbations to Hermite polynomials.

The RG eigenvalues can be extracted from~\eq{scaling_exp}. Recall that if $\lambda>0$ then the associated operator is relevant, since it increases with $t$, whereas those with $\lambda<0$ are irrelevant. If $\lambda=0$, the corresponding operator is marginal and we must go beyond leading order to determine whether it is marginally relevant, marginally irrelevant or exactly marginal.

Let us assume a $\phi\leftrightarrow -\phi$ symmetry and take $\D=4$.
The term $n=r=0$ is a vacuum term, and does not interest us in this treatment. The marginal and relevant operators in $\D=4$ are:
\[
	\begin{array}{cccl} 
		n \hspace{1em} & r \ & \lambda_{n,r} &
	\\ \hline
		2 \hspace{1em} & 0 \ & 2 & \mathrm{relevant}
	\\
		2  \hspace{1em} & 2 \ & 0 &\mathrm{marginal}
	\\
		4 \hspace{1em} & 0 \ & 0 &\mathrm{marginal}	
	\end{array}
\]

This is telling us that there is a two-point, momentum-independent term which is relevant: this is the mass term. There is a four-point, momentum-independent term which is marginal: this corresponds to the scalar coupling, $\lambda$. (Actually, we need to be careful with this identification, since our eigenperturbations have a structure similar to Hermite polynomials, rather than monomials. We will
deal with this in \sect{renorm}.) This classification is very familiar from standard treatments of scalar field theory in four dimensions. Indeed, from this we know that, at next to leading order in perturbations about the Gaussian fixed-point, the four-point coupling turns out to be marginally irrelevant.\footnote{%
This is so long as the coupling is taken to be positive. As recognized by Symanzik~\cite{Symanzik-SmallDistance,Symanzik-Computable}, if the coupling is negative then the theory is asymptotically free. Unfortunately, it is also thought to be sick~\cite{Gross+Wilczek-UV,Politzer-AF,Gross-RG-Applications} but see~\cite{Kleefeld}.}
Finally, there is a two-point term, at order $p^2$, which is also marginal. We will return to this in a moment.

First, though, we note that~\eq{scaling_exp} reproduces the expected classification of operators (in the vicinity of the Gaussian fixed-point) in all dimensions. There is always a relevant mass operator present, with scaling exponent $+2$. Below four dimensions the four-point coupling becomes relevant and therefore allows for the construction of interacting renormalized trajectories out of the Gaussian fixed-point. When we hit three dimensions, the six-point, momentum-independent coupling becomes marginal and when we hit $\D=2$, there are an infinite number of marginal couplings.

Let us now return to the operator with $n=r=2$, noticing that it is marginal in \emph{any} dimension. Ignoring the associated constant, this operator takes the form
\[
	\hf \int_p \phi(-p) p^2 \phi(p),
\]
and so simply changes the normalization of the kinetic term. Clearly, the effect of this operator can be undone by a local field definition: it is redundant. Now, suppose that we perturb the Gaussian solution in this redundant direction:
\be
	S_\star^{\mathrm{Gaussian}} = \hf  \int_p \phi(-p) p^2 \cutoff^{-1}(p^2) \phi(p)
	+ \frac{\varepsilon}{2}  \int_p \phi(-p) p^2 \phi(p).
\ee
Immediately, we see that we would get the same result by taking $\intconst = \varepsilon$ in~\eq{Gen-Gauss}
and expanding to leading order. Thus we see that the marginal, redundant direction
of the simplest representative of the Gaussian fixed-point takes us an infinitesimal step along the line of equivalent fixed-points, precisely as anticipated. Comparing with~\eq{line-explicit}, it is apparent that the simplest representative of the Gaussian fixed-point must correspond to $b=-\infty$. We will make this more explicit,  in \sect{Gen-TP}, where general representatives of the Gaussian fixed-point are treated.

Let us close this section by tying up a loose end. We have ascribed physical meaning to the momentum-independent two-point and four-point eigenoperators, but we have not actually checked that they are scaling operators. In fact, the proof is trivial and automatically applies to all eigenoperators with $r=0$. 
The game is, using~\eq{redundant} (with $\Sint_\star =0$), to try to find a $\Theta$ which generates the $\mathcal{G}'_{n,0}$. The point is that the first term in
the last line of~\eq{redundant} is at least $\order{p^2}$. Since $\Theta$ is quasi-local and since we are considering momentum-independent eigenoperators, we must try to cancel this term against the
last in~\eq{redundant} (remember that the second term vanishes since we are taking $\Sint_\star=0$). But this will never work, since if the highest-point contribution to $\Theta$ has $n$ fields, then the first term in~\eq{redundant} has a contribution with $n$ fields but the final term
does not.

\subsubsection{The Dual Action at Fixed-Points}
\label{sec:FP-dual}

We can gain deep insights into fixed-point solutions---as well as simplifying the above analysis---by
using the dual action. Since we have rescaled to
dimensionless variables, $\cutoff (p^2/\Lambda^2) \mapsto \cutoff (p^2)$, and so the operator
$\op$ appearing in the definition of the dual action [see~\eqs{op}{dual}] satisfies $\partial_t \op = 0$.
Recalling the discussion around~\eq{dual'-modifiedflow}, we assume both that the dual action exists and that
its flow is given by~\eq{RescaledDualFlow}. Thus we have that 
\be
	\partial_t \Sint_\star[\phi] = 0
	\qquad 
	\Rightarrow
	\qquad
	\partial_t
	\dual_\star[\phi] = 0
\ee
and so, at a fixed-point, the dual action satisfies
\be
	\left( \frac{\D-2-\eta_\star}{2} \Count_\phi + \Count_\partial - \D \right) 
	\dual_\star[\phi] 
	= -\frac{\eta_\star}{2} \phi \cdot \ep^{-1} \cdot \phi.
\label{eq:dual-FP}
\ee

To solve this equation, let us introduce a function $\h(p^2)$ and define
\be
	\homog[\phi] = -\hf \phi \cdot \h \cdot \phi + \dual_\star[\phi].
\label{eq:homog}
\ee
If we choose $h$ such that it satisfies
\be
	-\frac{2+\eta_\star}{2} \h(p^2) + p^2 \h'(p^2) 
	= -\frac{\eta_\star}{2} \ep^{-1}(p^2),
\label{eq:h-condition}
\ee
(where the prime denotes a derivative \wrt\ momentum squared)
then we find that
\be
	\left( \frac{\D-2-\eta_\star}{2} \Count_\phi + \Count_\partial - \D \right) 
	\homog[\phi] 
	= 0.
\ee
We will look for solutions to this equation in which the dual action has an expansion in powers of the field, as in~\eq{dualv}:
\be
	\homog^{(n)}(a p_1,\ldots, a p_n) = a^r \homog^{(n)}
	(p_1,\ldots,p_n),
	\qquad
	r = \D -  n \frac{\D-2-\eta_\star}{2}.
\label{eq:homog-solution}
\ee
To complete the solution for $\dual_\star$, we must solve~\eq{h-condition}:
\be
	\h(p^2) =
	-\tilde{\const}_{\eta_\star} p^{2(1+\eta_\star/2)}
	- \frac{\eta_\star}{2} p^{2(1+\eta_\star/2)}
	 \int^{p^2} dq^2 \frac{\cutoff^{-1}(q^2)}{q^{2(1+\eta_\star/2)}},
\label{eq:h-solution}
\ee
where $\tilde{\const}_{\eta_\star}$ is an integration constant, one for each fixed-point. The integration constant is chosen as follows. First let us note that, for some other constants $\intconst_{\eta_\star}$,
\be
	\homog^{(2)}(p) \equiv \homog^{(2)}(p,-p) = -\intconst_{\eta_\star}  p^{2(1+\eta_\star/2)}
\label{eq:constants}
\ee
(the reason for the choice of sign will become apparent in \sect{CorrFns}).
We choose the $\tilde{\const}_{\eta_\star}$ by demanding that $h(p^2)$ has no pieces exhibiting this momentum dependence. Thus, for example, we choose $\tilde{\const}_0 = 0$. Note that, at the two-point level, it is trivially the case that
\be
	\dualv{2}_\star(p) = -\intconst_{\eta_\star}  p^{2(1+\eta_\star/2)} + h(p^2).
\label{eq:D2-solution}
\ee

It will prove useful to recast~\eq{h-solution} by integrating by parts:
\be
	\h(p^2) =
	-\tilde{\const}_{\eta_\star} p^{2(1+\eta_\star/2)}
	+
	\ep^{-1}(p^2)
	-
	p^{2(1+\eta_\star/2)} 
	\int^{p^2} dq^2
	\left[
		\frac{1}{\cutoff (q^2)}
	\right]'
	q^{-2(\eta_\star/2)}.
\label{eq:h}
\ee
Notice the similarity of
the second term to the object, $\varrho$, 
appearing in the marginal, redundant operator of O'Dwyer and Osborn [recall~\eq{HO-b}]. Indeed, with this in mind, let us 
recast~\eq{h}, for $\eta_\star<2$:
\be
	\h(p^2) = -\const_{\eta_\star} p^{2(1+\eta_\star/2)}
	+ \ep^{-1}(p^2) \bigl[1+ \varrho(p^2)\bigr],
\label{eq:h-lessthan}
\ee
where the $\const_{\eta_\star}$ are constants are related to the $\tilde{\const}_{\eta_\star}$.%
\footnote{To see an example of where $\const_{\eta_\star} \neq \tilde{\const}_{\eta_\star}$,
consider the case $K(q^2) = e^{-q^2}$, $\eta_\star = -2$. Then 
$\int^{p^2}_0 dq^2 e^{q^2} q^2 = \int^{p^2} dq^2 e^{q^2} q^2 + 1$: by putting in a lower limit
on the integral, we are effectively supplementing the integration constant, in this particular case.}
As we now discuss, it is easy to see that
\be
	\const_{\eta_\star} = 
	\left\{
		\begin{array}{ll}
			1, \ & \eta_\star = 0
		\\
			0, \ & \eta_\star <2, \ \neq 0
		\end{array}
	\right.
\label{eq:c}
\ee
The first case is simple to check: for $\eta_\star = 0$, $h(p^2) = (1-\const_{\eta_\star}) p^2$. But since we have defined $h(p^2)$ such that it does not have any contributions which transform in the same way as $\homog^{(2)}$, it must be that $\const_{\eta_\star} =1$. The second case follows upon exploiting quasi-locality of the cutoff function:
\be
	\ep^{-1}(p^2) \bigl[1+ \varrho(p^2)\bigr] = p^2 + \order{p^4},
\label{eq:rho-Taylor}
\ee
making it immediately apparent that, for $\eta_\star <2, \ \neq 0$, this term cannot supplement the $\const_{\eta_\star}$ piece.

Before moving on, it will be useful to consider the action of 
$\Count_\cutoff \equiv \cutoff \cdot \delta/ \delta \cutoff$ on the dual action. For $\eta_\star=0$,
all vertices of $\dual_\star[\phi]$ transform homogeneously with momenta and
therefore cannot depend on the cutoff function, which does not transform in this way. It thus follows that
\be
	\Count_\cutoff \dual_\star[\phi] = 0, \qquad \eta_\star = 0.
\label{eq:Count_K-=0}
\ee

For $\eta_\star \neq 0$, the two-point dual action vertex does not transform homogeneously with momentum and so we must work a little harder. To proceed, we observe
that~\eq{dual-FP} implies
\be
	\left( \frac{\D-2-\eta_\star}{2} \Count_\phi + \Count_\partial - \D \right) 
	\Count_\cutoff \dual_\star[\phi] 
	= +\frac{\eta_\star}{2} \phi \cdot \ep^{-1} \cdot \phi,
\ee
from which we deduce that
\be
	\left(
		-2 -\eta_\star + \Count_\partial
	\right)
	\left(1+ \Count_\cutoff\right)
	\hf
	\phi \cdot \dualv{2}_\star \cdot \phi = 0.
\ee
Therefore, the vertex belonging to $\left(1+ \Count_\cutoff\right) \phi \cdot \dualv{2}_\star \cdot \phi$
transforms homogeneously with momentum---precisely as the two-point contribution to $\homog$ does.
Recalling that, for $\eta_\star <2 \neq 0$, $\const_{\eta_\star} = 0$, it is thus apparent that 
\be
	\Count_\cutoff \dual_\star[\phi] =
	-\hf
	\phi \cdot  \ep^{-1} \bigl(1+ \varrho \bigr) \cdot \phi , \qquad \eta_\star <2,\ \neq 0.
\label{eq:Count_K-neq0}
\ee
It is the difference between \eqs{Count_K-=0}{Count_K-neq0} that accounts for the 
difference between~\eqs{simple}{simple-GFP}, as can be seen in \app{redundant}.

Let us now return to the solutions~\eqs{homog-solution}{h-solution} and attempt to understand what they are telling
us. At first sight, each of the $\homog^{(n)}$ is largely arbitrary. Although each must 
behave with the correct net powers of momenta, there are many ways of achieving this. Moreover, at the two-point level,
the constant $\intconst_{\eta_\star}$ in~\eq{constants} is undetermined and $\eta_\star$ appears to be a free parameter.
This seems to be a problem: since the Wilsonian effective action can apparently be reconstructed from the dual action according to~\eq{invert}, our solutions for the dual action appear to imply a continuous infinity of fixed-points. 

However, two things can go potentially go wrong with this reconstruction.
First, it could be that particular $\dual_\star$s with particular $\eta_\star$s give rise
to an ill-defined Wilsonian effective action. To see one way in which this might occur, recall the
diagrammatic expression of \fig{dual}  for the Wilsonian effective action in terms of the dual action.
Looking at~\eq{homog-solution}, it is apparent that the dual action vertices can have
large, negative powers of momenta. Consequently, it might be that the expression for the Wilsonian effective action is ill-defined, as a consequence of IR divergent integrals.
Even if we do end up with a Wilsonian effective action which is finite, it may be that  it is  not quasi-local.

Indeed, an explicit example
of the latter will be given in \sect{Gen-TP} where we will find that, at the two-point level, it is the
requirement of a quasi-local Wilsonian effective action which quantizes $\eta_\star$.
Let us emphasise that, in this case, everything can be solved exactly. Furthermore, the dual action
can be thought of as a crutch to be discarded after the intermediate steps have been carried out:
the Wilsonian effective
action can be reconstructed from the dual action, at which stage it can be checked that
the former is actually a solution of the flow equation, without ever referring back to the dual action.

Whilst it is nice to be able to see that it is a restriction to quasi-locality which quantizes the
spectrum of two-point fixed-point solutions, it is natural to ask whether there is any underlying
reason why this had to occur. The answer is yes: as promised earlier, we can use the dual action
to help show that the spectrum of quasi-local, critical fixed-points is quantized. In order to do this, it is necessary
to understand first how the dual action formalism can also be used to analyse the eigenperturbations of a fixed-point.
Notice that shifting a fixed-point action according to~\eq{perturb} induces a change in the dual action,
$\dual_t = \dual_\star + \delta \dual_t$, with
\be
	\delta \dual_t[\phi] = \sum_i \alpha_i e^{\lambda_i t}
	e^{\dual_\star[\phi] } e^{\op} e^{-\Sint_\star[\phi]}
	\eop_i[\phi].
\label{eq:dual-shift}
\ee
Directly from~\eq{RescaledDualFlow}, which is linear in $\dual$, we find that
\be
	\left(\lambda_i + \frac{\D-2-\eta_\star}{2} \Count_\phi + \Count_\partial -\D\right) 
	e^{\dual_\star[\phi] } e^{\op} e^{-\Sint_\star[\phi]}
	\eop_i[\phi] = 0.
\label{eq:dual-eigen}
\ee
Let us tentatively write the solution to this equation as
\[
\begin{split}
	e^{\dual_\star[\phi] } e^{\op} e^{-\Sint_\star[\phi]}
	\eop_{i}[\phi] 
	& \stackrel{?}{=}
	\frac{1}{n!}
	\int_{q_1,\ldots,q_n}
	P_r(q_1,\ldots,q_n) \phi(q_1)\cdots \phi(q_n) \hat{\delta}(q_1+\cdots+q_n)
\\
	 \lambda_{i} & = \D - r - \frac{n(\D-2-\eta_\star)}{2},
\end{split}
\]
were $P_r(q_1,\ldots,q_n)$ satisfies
\[
	P_r(a q_1,\ldots,a q_n) = a^r P_r(q_1,\ldots,q_n).
\]

It will become clear, in a moment, why we have not identified $i$ with 
the pair of non-negative integers
$(n,r/2)$, as in the Gaussian case.
First let us note that it looks like we have solved the problem of the spectrum of eigenperturbations
at a generic fixed-point. As should by now be unsurprising, this is illusory. The point is that we need to
constrain $r$. If it were the case that  $e^{\dual_\star[\phi] } e^{\op} e^{-\Sint_\star[\phi]} \eop_i[\phi] $
were quasi-local, then we would be done:  $r/2$ would be a non-negative integer, as before. But this does not occur, in general.

What is true is that the $\eop_{i}$ should always to taken to be quasi-local.
But only in special circumstances does this imply that
$e^{\dual_\star[\phi] } e^{\op} e^{-\Sint_\star[\phi]} \eop_{i}[\phi]$
is quasi-local. Underlying this is, of course, precisely the same mechanism that
generates correlation functions which are not quasi-local from a quasi-local action.
Now we can see why we have not identified $i$ with $n,r$. 

Since $r/2$ is not generally expected be a non-negative integer, it is quite possible that there are several different values of $n,r$ which
yield the same $\lambda_i$ and so we should write:
\be
\begin{split}
	e^{\dual_\star[\phi] } e^{\op} e^{-\Sint_\star[\phi]}
	\eop_{i}[\phi] 
	& =
	\sum_{n_i,r_i} 
	\frac{1}{n_i!}
	\int_{q_1,\ldots,q_{n_i}}
	P^{(i)}_{r_i}(q_1,\ldots,q_{n_i}) \phi(q_1)\cdots \phi(q_{n_i}) \hat{\delta}(q_1+\cdots+q_{n_i})
\\
	 \lambda_{i} & = \D - r_i - \frac{n_i(\D-2-\eta_\star)}{2}.
\end{split}
\label{eq:eigen-soln}
\ee
To be clear: the sum over $n_i$ and $r_i$ is over all values required for the quasi-locality of $\eop_i$, with  all of these pairs giving the same $\lambda_i$. We additionally label the $P$s with a subscript `$(i)$'
to remove any degeneracy in notation in the case that there is more than one $\eop_i$ with the same
value of $\lambda_i$, sharing some pair of values of $(n_i,r_i)$.

It is in instructive to see an example. To this end, let us recall that every fixed-point possesses a 
marginal operator which, we recall from~\eq{emro}, is given by
\[
	\marginal [\phi] =
	\left(\hf \Count_\phi + \Count_{\cutoff}\right) \Sint_\star[\phi]
	\equiv
	\hat{\Count} \Sint_\star[\phi].
\]
Observe that
\be
	e^{\dual_\star[\phi] } e^{\op} e^{-\Sint_\star[\phi]} \marginal [\phi] 
	=
	-e^{\dual_\star[\phi] } e^{\op} \hat{\Count}  e^{-\Sint_\star[\phi]}
	=
	\hat{\Count}\dual_\star[\phi],
\label{eq:O_mar-proj}
\ee
where we have used the fact that $e^{\dual_\star[\phi] } e^{\op} e^{-\Sint_\star[\phi]}  =1$
(so long as there is nothing which follows on which the operator, $\op$, can act)
together with $[\hat{\Count},\op] = 0$. Consequently, for $\marginal [\phi]$, the corresponding $P^{(i)}_{r_i}$s can be read of from the
vertices of $\hat{\Count} \dual_\star[\phi]$. Let us check the consistency of this: 
operating on both sides of~\eq{dual-FP} with $\hat{\Count}$,
it is apparent that
\be
	\left( \frac{\D-2-\eta_\star}{2} \Count_\phi + \Count_\partial - \D \right) 
	\hat{\Count} \dual_\star[\phi] 
	= 0.
\ee
Therefore, the vertices of $\hat{\Count} \dual_\star[\phi]$ correspond to $P^{(i)}_{r_i}$s with 
$\lambda_i=0$---precisely as they must for a marginal operator.

Although we have emphasised that, in general, there is no reason for $r_i/2$ to
satisfy any obvious constraint there is one set of operators---which exists at every
fixed-point---for which $r_i/2$ turns out to be a positive integer (greater than 1). These operators
satisfy
\be
\begin{split}
	e^{\dual_\star[\phi] } e^{\op} e^{-\Sint_\star[\phi]} \eop_{2,r}[\phi]
	&=
	\hf \int_q P_{r}(q) \phi(q) \phi(-q), \qquad r=4,6,8\ldots,
\\
	\lambda_{2,r} &= 2+ \eta_\star -r.
\end{split}
\ee
With this in mind, we note that~\eq{useful-commutator-b} gives
\be
	\Bigl[
		e^{-\op} , \phi \cdot P_r \cdot \phi
	\Bigr]
	=
	-\deltahat{0} \ep \cdot P_r
	e^{-\op}
	-
	2\phi \cdot \ep^2 P_r \cdot \fder{}{\phi}
	e^{-\op}
	+ \classical{}{\ep^2 P_r}{} e^{-\op},
\ee
from which it is straightforward to show that
\begin{multline}
	\eop_{2,r}[\phi] = 
	e^{\Sint_\star[\phi]} e^{-\op}e^{- \dual_\star[\phi] } 
	\hf \phi\cdot P_r \cdot \phi
\\
	=
	\hf \phi\cdot P_r \cdot \phi
	+ \phi \cdot \ep P_r \cdot \fder{\Sint_\star}{\phi}
	+
	 \hf \classical{\Sint_\star}{\ep^2 P_r}{\Sint_\star}
	 -\hf \quantum{\ep^2 P_r}{\Sint_\star},
\label{eq:redundant-family}
\end{multline}
with $\lambda_{2,r} = 2+\eta_\star -r$.
Note that any worries about
either the existence of the dual action or the validity of
 inverting the operator $e^{\op}$ should, in this case, be allayed:
the dual action has been used as a crutch to obtain the answer~\eq{redundant-family}, the
veracity of which can be checked by direct substitution into~\eq{full-eigen}---see \app{Menagerie-D^2}.
A sufficient condition
for this operator to be quasi-local is that $\ep^2 P_r$ is quasi-local, which requires that
$r\geq 4$.%
\footnote{At least for theories with a local potential (by which we mean that the potential contains at least some contributions which do not have any derivatives). The Gaussian fixed-point, for example, does not have a local potential and so the effect of the constraint of quasi-locality is weakened, resulting in the condition $r\geq0$. Note, though, that in this context we nevertheless require $r\geq 2$ if we want the operator to be expressible as a quasi-local field redefinition. The upshot of this is that the $r=0$ term must be a scaling operator and, indeed, it is obvious that it corresponds to the mass.
}
 Let us note that these operators are redundant, since they can be constructed from~\eq{redundant} by making the choice
\be
	\Theta_{2,r}(p) = 
	\phi(p) \ep(p^2) P_r(p) + \ep^2(p^2) P_r(p) \fder{\Sint_\star}{\phi(-p)}.
\ee

We are now in a position to prove that the spectrum of the anomalous dimension at quasi-local, critical fixed-points is quantized.
Let us recall from the discussion at the end of \sect{GenCon} that a necessary condition for this
to occur is that
\be
	\pder{\fpop(\eta_\star,\Sint_\star) }{\eta_\star}
	=
	\hf
	\bigl(
		\Count_\phi \Sint_\star +  \phi \cdot \ep^{-1} \cdot \phi
	\bigr)
\label{eq:eta-deriv}
\ee
has a component in the marginal, redundant direction. At first sight it is not obvious how to go about proving this, since it is perhaps not clear how to project the \rhs\ onto some particular axis in theory space. 
We will search for inspiration by using the dual action.
The trick is to start by using the result~\eq{useful-manip} which, up to an uninteresting vacuum term, implies that
\be
	e^{\dual_\star[\phi] } e^{\op} e^{-\Sint_\star[\phi]} \hf
	\left(
		\Count_\phi \Sint_\star +  \phi \cdot \ep^{-1} \cdot \phi
	\right)
	=
	\hf  \phi \cdot \ep^{-1} \cdot \phi - \hf \Count_\phi \dual_\star[\phi].
\ee

Let us now add and subtract $\Count_K \dual_\star[\phi]$ on the \rhs. The subtracted term
will be combined with the final term to yield $-\hat{\Count} \dual_\star[\phi]$ whereas
we substitute for the added term using~\eqs{Count_K-=0}{Count_K-neq0}, yielding:
\be
e^{\dual_\star[\phi] } e^{\op} e^{-\Sint_\star[\phi]} 
	\pder{\fpop(\eta_\star,\Sint_\star) }{\eta_\star}
	=
	\left\{
		\begin{array}{ll}
		\ds
			-\hat{\Count}\dual_\star[\phi] + \hf \phi \cdot \ep^{-1} \cdot \phi, & 
			\eta_\star = 0,
		\\[2ex]
		\ds
			-\hat{\Count}\dual_\star[\phi] - \hf \phi \cdot \ep^{-1} \varrho \cdot \phi, 
			& \eta_\star < 2, \ \neq 0.
		\end{array}
	\right.
\ee
In the second case, observe that
\[
	-\ep^{-1}(p^2) \varrho(p^2) = p^{2(1+\eta_\star/2)}
	\int_0^{p^2} dq^2 
	\left[
		\frac{1}{\cutoff(q^2)}
	\right]'
	q^{-2\eta_\star/2}.
\]
Taylor expanding the cutoff function, we see that this term starts at $\order{p^4}$
and can thus be written as a linear combination of the redundant operators denoted
by $\eop_{2,r}$, above. For the case $\eta_\star<2, \ \neq 0$, at any rate, we have therefore
shown how to decompose $\partial \fpop(\eta_\star,\Sint_\star) / \partial \eta_\star$ into
a sum of eigenperturbations. The presence of $- \hat{\Count} \dual_\star[\phi]$ means that
there is a component in  the marginal, redundant direction.  Therefore, we have shown (for critical fixed-points) 
that the spectrum of those $\eta_\star$ satisfying $\eta_\star<2, \ \neq 0$,  is quantized. Indeed,
this implies that we can drop the last condition, thereby encompassing all those critical fixed-points with $\eta_\star <2$.
Note that none of this says anything as to whether or not there
is more than one fixed-point with a particular value of $\eta_\star$.

Having used the dual action to elucidate the basic structure of the argument, we
can now rephrase the proof without mentioning it. This will have the added benefit
of treating the $\eta_\star =0$ case along with all the others.
To this end, return to \eq{eta-deriv}.
With this equation in our minds, let us now operate on~\eq{FP-eq} with $\hf \Count_\phi$
to yield:
\be
	\hf\classifier  \Count_\phi \Sint_\star
	-\classical{\Sint_\star}{\cutoff'}{\Sint_\star} 
	+\quantum{\cutoff'}{\Sint_\star} 
	-\frac{\eta_\star}{2} \phi \cdot \ep^{-1} \cdot \phi = 0,
\label{eq:Count_phi-Sint-eq}
\ee
where we recall that $\classifier$ is given by~\eq{classifier}.
We will solve this equation by making the guess (inspired from what we have learnt using the dual action)
\be
	\hf \bigl( \Count_\phi \Sint_\star +  \phi \cdot \ep^{-1} \cdot \phi \bigr)
	= \sum_{r=4}^{\infty} a_r \eop_{2,r} \ + \ \sum_{i = \{\mathrm{marginal}\}} b_i \eop_{i},
\label{eq:guess}
\ee
where the final sum is over all marginal operators (which are, of course, killed by $\classifier$) and the $a_r$ and $b_i$ are to be determined. 
Substituting this equation into~\eq{Count_phi-Sint-eq} 
(and recalling that the $\eop_{2,r}$ have RG eigenvalues $2+\eta_\star -r$)
gives a condition for the $a_r$:
\be
	\sum_{r=4}^{\infty} (2+\eta_\star -r)  a_r \eop_{2,r}
	- \phi \cdot \ep^{-2} \cutoff' \cdot \phi  - 2\phi \cdot \ep^{-1} \cutoff' \cdot \fder{\Sint_\star}{\phi}
	- \classical{\Sint_\star}{\cutoff'}{\Sint_\star} + \quantum{\cutoff'}{\Sint_\star} = 0.
\ee
Comparing with~\eq{redundant-family} we see that
\be
	\hf \sum_{r=4}^{\infty} (2+\eta_\star -r)  a_r p^{2r/2} = p^4 \cutoff^{-2}(p^2)\cutoff'(p^2)
\ee
Dividing through by $p^{2(2+\eta_\star/2)}$ and
integrating up, it is easy to check that
\be
	\sum_{r=4}^{\infty} a_r p^{2r/2} = -\ep^{-1}(p^2) \varrho(p^2),
\ee
where we have used~\eq{HO-b}. Substituting this back into~\eq{guess} and using~\eq{redundant-family}, which gives the explicit form for the $\eop_{2,r}$, we see that
\be
	 \sum_{i = \{\mathrm{marginal}\}} b_i \eop_{i}
	 =
	 \phi \cdot \ep^{-1} \bigl(\varrho+1 \bigr) \cdot \phi
	+ \phi \cdot \bigl(2\varrho+1 \bigr)\cdot \fder{\Sint_\star}{\phi}
	+
	\classical{\Sint_\star}{\ep\varrho}{\Sint_\star}
	-
	\quantum{\ep\varrho}{\Sint_\star}.
\ee
Comparing with~\eq{HO-mro}, we identify the \rhs\ as $\eop'^{\mathrm{R}}_\mathrm{mar}$, making it clear that, on the \lhs, the $b_i$ should be chosen so as to pick out just this term.
Putting everything together, we find that
\be
	\pder{\fpop(\eta_\star,\Sint_\star) }{\eta_\star}
	=
	\hf
	\bigl(
		\Count_\phi \Sint_\star +  \phi \cdot \ep^{-1} \cdot \phi
	\bigr)
	=
	\eop'^{\mathrm{R}}_\mathrm{mar}
	+
	\sum_{r=4}^\infty a_r \eop_{2,r}.
\label{eq:eta-deriv-expr}
\ee
Thus, for $\eta_\star <2$, $\partial \fpop(\eta_\star,\Sint_\star) / \partial \eta_\star$  has a component in the exactly marginal, redundant direction and so the spectrum of $\eta_\star$ corresponding to critical
fixed-points is quantized. 

With this result in mind, let us now return to the issue of reconstructing a valid Wilsonian
effective action from solutions for the dual action, via~\eq{invert}. As mentioned
already, and as we will see explicitly in the next section, the dual action can be used to
readily uncover a \emph{continuum} of two-point fixed-point solutions, parametrized by $\eta_\star$,
only a discrete
subset of which are quasi-local (for $\eta_\star<2$). Consequently, it must be true that the quantization of the spectrum of $\eta_\star<2$ only holds for quasi-local fixed-points.
It is worth understanding, then, where quasi-locality was used in our proof
of the quantization of the spectrum of $\eta_\star <2$. Indeed,
we should phrase the condition for quantization of $\eta_\star$ as follows:
\begin{quote}
	If a marginal, redundant operator exists at some quasi-local fixed-point,
	if $\partial \fpop(\eta_\star , \Sint_\star) / \partial \eta_\star$ has a component in this direction, 
	and if we allow only quasi-local deformations of the fixed-point, then the value
	of $\eta_\star$ at this fixed-point is isolated.
\end{quote}

But now we seem to arrive at a paradox. Consider the eigenoperators at the Gaussian fixed-point, given by~\eq{SimpleGaussian-eop}. We can relax the constraint of quasi-locality simply by allowing $r/2$ to take values other than $0,1,2,\ldots$. If we incorporate this modification into the sum over $i$ appearing in~\eq{quantized-fp}, then it seems that the above argument goes through as before, and we again conclude that the spectrum of $\eta_\star <2$ is quantized; but we know that it is not if we allow non-local fixed-points.

To see the resolution to this problem, let us do things carefully. Sticking with
 the Gaussian fixed-point, if we allow non-local eigenperturbations, then the sum over $i$ above decomposes not into a sum over $n$ and a sum over $r$ but into a sum over $n$ and an integral over $r$. In fact, we can take $n=2$ since this is all we need to go along the line of two-point solutions.
In this case~\eq{quantized-proof} becomes:
\be
	\fpop
	\biggr(
		\eta_\star + \varepsilon,\Sint_\star + 
		\integral{r} \alpha_2(r)  \eop_2(r)
	\biggl) =
	\integral{r} \alpha_2(r)(2-r) \eop_2(r)
	+
	\sum_{r=\{\mathrm{R}\}} 
	\tilde{\alpha}_2(r)
	\eop^{\mathrm{R}}_2(r)
	+ \order{\varepsilon^2},
\label{eq:quantized-fp-nonlocal}
\ee
where we have used~\eq{scaling_exp} to set $\lambda_2(r) =2-r$ and (up to an unimportant constant)
\be
	\eop^{\mathrm{R}}_2(r) = \hf \int_p \phi(p) \phi(-p) p^{2r/2},
	\qquad r = 4,6,8,\ldots
\ee
 Note that the final term is a sum over the discrete values of $r$ corresponding to the (two-point) redundant operators of the Gaussian fixed-point. This term is exactly the same as in the previous analysis, since it is the perturbations of the fixed-point, and not the fixed-point itself, which have become non-local in this particular case. In other words, \eq{H-deriv-decomp} is unchanged, as is the conclusion that  $\tilde{\alpha}_2(2) \neq 0$.

There are two possible resolutions to the paradox. First, the assumption that the non-local 
extensions of the eigenperturbations~\eq{SimpleGaussian-eop} 
span the non-local theory space could be incorrect [note that eigenperturbations of the
form~\eq{SimpleGaussian-eop} do not exhaust the solutions of the linearized flow equation, as we discuss in \sect{self-similar}]. If this is true then it is not possible to go along the line of inequivalent fixed-points using the eigenperturbations we are considering, and there is no paradox. Alternatively, it might be that the $\alpha_2(r)$ can be adjusted such that the \rhs\ vanishes.

Let us consider the latter option. Denoting the values of $r$ which pick out the redundant operator(s) 
for which $\tilde{\alpha}_2(r) \neq 0$, it would seem that we can take
\[
	\alpha_2(r) = - \frac{1}{2-r} \tilde{\alpha}_2(r) \sum_i \delta(r-r_i).
\]
However, this clearly does not work. First of all, it amounts to discarding all the non-local operators, taking us back to the case of quasi-local deformations. Moreover, one of the operators that it must pick out is the marginal, redundant direction corresponding to $r=2$: we know from the above analysis that $\partial \fpop/\partial \eta_\star$ has a
component in this direction, meaning that $\tilde{\alpha}_2(2) \neq 0$. Consequently, our choice of $\alpha_2(r)$ would cause the perturbation of the fixed-point, $\integral{r} \alpha_2(r)  \eop_2(r)$, to blow up and so it seems that
the $\alpha_2(r)$ cannot be chosen to make the \rhs\ of~\eq{quantized-fp-nonlocal} vanish. 
Therefore, we conclude that the resolution to the paradox is the alternative possibility:  the non-local extensions of the eigenperturbations~\eq{SimpleGaussian-eop}  do not span the non-local theory space.

\subsubsection{General Two-Point Solutions}
\label{sec:Gen-TP}

In this section, we use the dual action formalism to very quickly uncover the complete
set of two-point fixed-point solutions, at least for $\hSint=0$. 
The first point to make is that
if the Wilsonian effective action does not have higher than two-point vertices, then the
dual action only has a two-point contribution. In this case (and only in this case)
we can write
\be
	\Siv{2}_\star(p) = 
	\frac{
		\dualv{2}_\star(p)
	}
		{
		1 - \ep(p^2) \dualv{2}_\star(p)
	}.
\label{eq:Siv2-invert}
\ee 
It is easiest to derive this expression diagrammatically. Just as the dual action is composed of all connected diagrams composed from vertices of $\Sint$ and $\ep$s, so is the Wilsonian effective action composed of all connected diagrams composed from vertices of $\dual$ and $-\ep$s (recall \fig{dual}). If the Wilsonian effective
action has only two-point contributions, then the same is true of the dual action. Therefore, the possible diagrams are very simple, as shown in \fig{Sint-TP}, and they can be summed to give~\eq{Siv2-invert}. The game now is to substitute~\eq{D2-solution} into~\eq{Siv2-invert} and to analyse what we
find. 
\bcf[h]
	\[
	\ensuremath{\begin{array}{c}\begin{picture}(0,0)%
\epsfig{file=pstex/ReducedWEA-2.pstex}%
\end{picture}%
\setlength{\unitlength}{3947sp}%
\begingroup\makeatletter\ifx\SetFigFont\undefined%
\gdef\SetFigFont#1#2#3#4#5{%
  \reset@font\fontsize{#1}{#2pt}%
  \fontfamily{#3}\fontseries{#4}\fontshape{#5}%
  \selectfont}%
\fi\endgroup%
\begin{picture}(358,579)(1629,-672)
\put(1730,-448){\makebox(0,0)[lb]{\smash{{\SetFigFont{11}{13.2}{\rmdefault}{\mddefault}{\updefault}{\color[rgb]{0,0,0}$\Sint$}%
}}}}
\end{picture}%
 \end{array}} = \ensuremath{\begin{array}{c}\begin{picture}(0,0)%
\epsfig{file=pstex/Dual-2.pstex}%
\end{picture}%
\setlength{\unitlength}{3947sp}%
\begingroup\makeatletter\ifx\SetFigFont\undefined%
\gdef\SetFigFont#1#2#3#4#5{%
  \reset@font\fontsize{#1}{#2pt}%
  \fontfamily{#3}\fontseries{#4}\fontshape{#5}%
  \selectfont}%
\fi\endgroup%
\begin{picture}(358,579)(1629,-672)
\put(1730,-448){\makebox(0,0)[lb]{\smash{{\SetFigFont{11}{13.2}{\rmdefault}{\mddefault}{\updefault}{\color[rgb]{0,0,0}$\dual$}%
}}}}
\end{picture}%
 \end{array}} + \ensuremath{\begin{array}{c}\input{pstex/Dual-Dumbbell-2.pstex_t} \end{array}} 
	+ \ensuremath{\begin{array}{c}\input{pstex/Dual-Dumbbell-3.pstex_t} \end{array}} 
	+\cdots
	\]
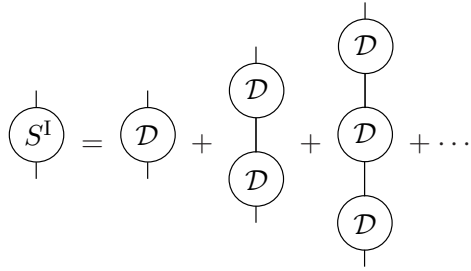
\caption{Diagrammatic expression for the Wilsonian effective action in terms of the dual action, in the case where the Wilsonian effective action (and hence the dual action) has only two-point pieces.}
\label{fig:Sint-TP}
\ecf

\paragraph{Critical Fixed-Points}

Let us start by looking at critical fixed-points, for which we should take $\eta_\star <2$ (as we have mentioned already). With this restriction, we can substitute~\eqs{h-lessthan}{c} into the expression~\eq{D2-solution} for $\dualv{2}_\star$ to yield:
\be
	\Siv{2}_\star(p) = 
	\frac{
		\bigl(\intconst_{\eta_\star} + \const_{\eta_\star} \bigr) p^{2(1+\eta_\star/2)}
		- p^2 \cutoff^{-1}(p^2) [1+\varrho(p^2)]
	}
		{
		\varrho(p^2) - 
		\bigl(\intconst_{\eta_\star} + \const_{\eta_\star} \bigr) \cutoff(p^2) p^{2\eta_\star/2}
	}.
\label{eq:Gen-TPFP}
\ee
Focusing on quasi-local fixed-points (and recalling that we are taking $\eta_\star <2$), we immediately conclude that this restriction forces us to take $\eta_\star/2 = \mbox{non-positive integer}$. To see this, simply multiply through by $p^{-2\eta_\star/2}$ and recall that $\varrho$ is quasi-local (an exception is if $\intconst_{\eta_\star} + \const_{\eta_\star} =0$, in which case we are dealing with a non-critical fixed-point). Thus it is apparent that, at the two-point level, the countable quasi-local fixed-points are embedded in an uncountable number of non-local fixed-points. On this basis, it is tempting to speculate that there is, in general, a vastly bigger spectrum of non-local fixed-points as compared to quasi-local ones. Moreover, it would not be surprising if it turns out that the spectrum of  $\eta_\star<2$ is only quantized, in complete generality (rather than just at the two-point level), when the fixed-points are restricted to being quasi-local. It would be interesting to explore this further. Anyhow, returning to the two-point case, there are two classes of critical solution.

\subparagraph{The Gaussian Solution: $\eta_\star = 0$}

Noting that, for $\eta_\star=0$ we have $\varrho(p^2) = \cutoff(p^2)-1$, yields
\be
	\dualv{2}_\star(p) = \intconst p^2
	\qquad
	\Rightarrow
	\qquad
	\Siv{2}_\star(p) = \frac{\intconst p^2}{1-\intconst \cutoff (p^2)},
\label{eq:Gen-Gauss-Sint}
\ee
where we have identified $\intconst_0 = -\intconst$, thereby recovering~\eq{Gen-Gauss}. Notice that the dual action is IR safe, even if $\D\leq 2$. Recalling the discussion around~\eqs{CCF^n}{CCF^2}, this gives an example where the dual action is perfectly well defined, even though the correlation functions are not.

We now employ the dual action formalism to classify the eigenperturbations,
for which we need to use~\eq{dual-eigen}, with $\eta_\star =0$. Anticipating
the result, we will identify the index $i$ with two non-negative integers, $n$ and $r/2$.
Immediately, for the simplest representative of the Gaussian fixed-point, 
$\Sint_\star = 0 \ \Rightarrow \dual_\star = 0$ we see  that we obtain
\be
	\left(\lambda_{n,r} + \frac{\D-2}{2} \Count_\phi + \Count_\partial -\D\right) 
	e^{\op} \mathcal{G}'_{n,r}[\phi] = 0,
\ee
recovering the previous result~\eq{Gaussian-eigen}.

In the more general case we have:
\be
	\left(\lambda_{n,r} + \frac{\D-2}{2} \Count_\phi + \Count_\partial -\D\right) 
	e^{\dual_\star[\phi] } e^{\op} e^{-\Sint_\star[\phi]}
	\mathcal{G}_{n,r}[\phi] = 0,
\label{eq:Gaussian-eigen-gen}
\ee
with $\Sint_\star$ given by~\eq{Gen-Gauss-Sint} and $\dual_\star = \intconst p^2$.
[Note that we have dropped the prime on $\mathcal{G}_{n,r}$ since we are now dealing
with the eigenperturbations of a generic representative of the Gaussian fixed-point,
rather than the special (primed case) corresponding to $\Sint_\star=0$.]
It is tempting---and in this case correct---to say that the \emph{entire} object to the right
of the big brackets is quasi-local, and so the $\lambda_{n,r}$ are the same as
before. Therefore, as expected, the RG eigenvalues are the same for all representatives
of the Gaussian fixed-point.

Note that by writing out the explicit solution for the $\mathcal{G}_{n,r}[\phi]$, we can say something
about the speed with which the cutoff function must decay. Specifically:
\be
	\mathcal{G}_{n,r}[\phi] = e^{\Sint_\star[\phi]} e^{-\op} e^{-\dual_\star[\phi] }
	\int_{q_1,\ldots,q_n}
	\frac{1}{n!} 
	v_r(q_1,\ldots,q_n) \phi(q_1)\cdots \phi(q_n) \hat{\delta}(q_1+\cdots+q_n)
\label{eq:GenGaussPerts}
\ee
where, as before, $r/2$ is a non-negative integer. Allowing the $e^{-\op}$ to act will generate loop integrals.
These are very similar to those in \fig{eigen}, with the difference that the internal lines should be
replaced with $\ep/(1+\dualv{2}_\star \ep)$.
Since $r$ can be arbitrarily large, for all these integrals to converge it must be that the cutoff function
falls off faster than any power.

The exactly marginal, redundant direction is easy to find using~\eq{emro}:
\be
	\mathcal{G}_\mathrm{mar}^{\mathrm{R}}[\phi] = 
	\hat{\Count}
	\left[
	\hf
	\int_p
	\phi(-p)
		\frac{\intconst p^2}{1-\intconst \cutoff (p^2)}
	\phi(p)
	\right]
	=
	\frac{\intconst}{2}
	\int_p 
	\phi(-p)
		\frac{p^2}{[1-\intconst \cutoff (p^2)]^2}
	\phi(p).
\label{eq:mro-Gauss}
\ee
Note, though, that we need to take care at the simplest representative, $\intconst = 0$, since then the above expression vanishes. In this case, we would be better off using~\eq{simple-GFP}, instead. Now, repeatedly applying $\hat{\Count}$ to this expression, it is straightforward to check that
\be
	e^{b \hat{\Count}} \Sint_\star(\intconst) = \Sint_\star(\intconst e^b),
\ee
consistent with~\eq{line-explicit} (once we identify $B=e^{b_0}$).

Demonstrating that the eigenoperators with $r=0$ are scaling operators is only slightly more involved for
the general Gaussian solution than for the simplest representative. In this case~\eq{redundant} becomes
\begin{align*}
	\eop^{\mathrm{R}}[\phi;\Theta] 
	& =
	\int_p
	\left\{
		\Theta(p) \ep^{-1}(p^2)
		\left[
			1 + \Siv{2}_\star(p) \ep(p^2)
		\right]
		\phi(-p)
		-
		\fder{\Theta(p)}{\phi(p)}
	\right\}
\\
	& =
	\int_p
	\left\{
		\Theta(p) 
		\frac{\ep^{-1}(p^2)}{
			1 - \dualv{2}_\star(p) \ep(p^2)
		}
		\phi(-p)
		-
		\fder{\Theta(p)}{\phi(p)}
	\right\}.
\end{align*}
Since, at the Gaussian fixed-point, the combination $\dualv{2}_\star(p)\ep(p^2)$ is quasi-local, the
proof proceeds as before.

\subparagraph{The Non-Unitary Fixed-Points: $\eta_\star = -2, -4,\dots$} 

The leading behaviour of the dual action and Wilsonian effective action two-point vertices
are given by:
\be
	\dualv{2}_\star(p) = -\intconst_{\eta_\star} p^{2(1+\eta_\star/2)} + p^2 + \ldots,
	\qquad
	\Rightarrow
	\qquad
	\Siv{2}_\star(p) = -p^2 + \order{p^4}.
\ee
The crucial point to observe is that when we compute the \emph{full} Wilsonian effective
action the order $p^2$ piece of $\hf \phi \cdot \ep^{-1} \cdot \phi$ is exactly removed [\cf~\eq{split}].
Consequently, upon continuation to Minkowski space, the theory is presumably non-unitary.

Let us now compute the spectrum of eigenoperators at these non-unitary
fixed-points. To do this, we return to~\eq{eigen-soln}, and employ the condition
that $\eop_{i}$ is quasi local. 
Now let us see if we can deduce anything
about the momentum dependence of 
$e^{\dual_\star[\phi] } e^{\op} e^{-\Sint_\star[\phi]}
	\eop_{i}[\phi]$.
This object is derived from the dual action which, we recall, consists only of
connected pieces. This feature is thus inherited by the object under
consideration. From a diagrammatic point of view, one subset of these
connected diagrams can be resummed into a decoration of each external
leg. This is illustrated in \fig{decorate}.
\bcf[h]
	\[
		\ensuremath{\begin{array}{c}\begin{picture}(0,0)%
\epsfig{file=pstex/Decorate-primitive.pstex}%
\end{picture}%
\setlength{\unitlength}{3947sp}%
\begingroup\makeatletter\ifx\SetFigFont\undefined%
\gdef\SetFigFont#1#2#3#4#5{%
  \reset@font\fontsize{#1}{#2pt}%
  \fontfamily{#3}\fontseries{#4}\fontshape{#5}%
  \selectfont}%
\fi\endgroup%
\begin{picture}(924,273)(2089,-547)
\end{picture}%
 \end{array}}
		-
		\ensuremath{\begin{array}{c}\begin{picture}(0,0)%
\epsfig{file=pstex/Decorate-1.pstex}%
\end{picture}%
\setlength{\unitlength}{3947sp}%
\begingroup\makeatletter\ifx\SetFigFont\undefined%
\gdef\SetFigFont#1#2#3#4#5{%
  \reset@font\fontsize{#1}{#2pt}%
  \fontfamily{#3}\fontseries{#4}\fontshape{#5}%
  \selectfont}%
\fi\endgroup%
\begin{picture}(924,998)(2089,-1272)
\put(2475,-1048){\makebox(0,0)[lb]{\smash{{\SetFigFont{11}{13.2}{\rmdefault}{\mddefault}{\updefault}{\color[rgb]{0,0,0}$\Sint_\star$}%
}}}}
\end{picture}%
 \end{array}}
		+
		\ensuremath{\begin{array}{c}\input{pstex/Decorate-2.pstex_t} \end{array}}
		-\cdots
	\]
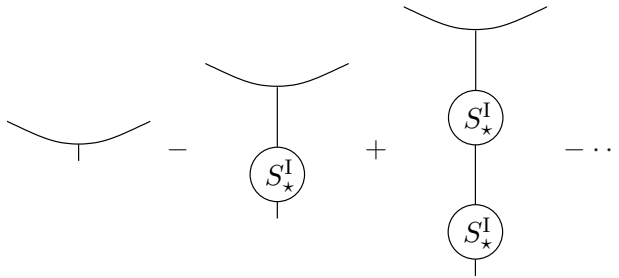
\caption{Decoration of an external leg belonging to some object in the case that the
fixed-point action only has a two-point piece.}
\label{fig:decorate}
\ecf

Therefore, every leg is decorated with a factor
\be
	\frac{1}{1+\ep(p^2) \Siv{2}_\star(p)} = 1 - \ep(p^2) \dual_\star(p)
	= p^{2\eta_\star/2} \times \mbox{quasi-local},
\ee
where we have used~\eq{Siv2-invert}.
Consequently, each leg possesses a non-quasi-local piece going like $p^{2\eta_\star/2}$
(remember that $\eta_\star/2$ is a negative integer). Totting up the contributions from $n$ legs,
we find that
\be
	r = n \eta_\star + r', \qquad r'/2 =0,1,2,\ldots.
\ee
and so, just as in the Gaussian case, $i$ can be identified with two integers.
Thus we find that
\be
	\lambda_{n,r'} = \D -r' - n\frac{\D-2+\eta_\star}{2}
\ee
and, as observed by Wegner, something rather interesting occurs. If $\D - 2 + \eta_\star \leq 0$
then there are an infinite number of relevant directions (again, it is easy to show that those with $r'=0$, at any rate, are scaling directions~\cite{Wegner_CS}). We have already stated that such theories are non-unitary, and are therefore of no interest to particle physics. Could they be of interest in statistical mechanics? Well, if there are an infinite number of relevant directions, then there are an infinite number of `knobs that must be dialled' to approach the critical point, and so presumably physical samples of such systems cannot be experimentally induced to undergo a second order phase transition. So let us try to avoid this scenario. Since the least negative value of $\eta_\star$ is $-2$, we must therefore take $\D\geq 4$. Of course, this is not very useful for statistical systems of practical interest! 

\paragraph{Non-Critical Fixed-Points}

As we will argue in \sect{Redux}, a sufficient condition for a theory to be non-critical is $\eta_\star \geq 2$. However, this is not a necessary condition and, indeed, we will illustrate some key properties of non-critical theory with one for which $\eta_\star=0$. As mentioned earlier, this can be obtained from our general Gaussian solution~\eq{Gen-Gauss} by setting $B=1$, giving
\be
	S^{\mathrm{non-crit}}_\star[\phi]
	= \hf \int_p \phi(-p) \frac{\cutoff^{-1}(p^2) p^2}{1- \cutoff (p^2)}  \phi(p).
\ee
Taylor expanding the two-point vertex in momentum, we see that the leading contribution is $\order{p^0}$ and not $\order{p^2}$.%
\footnote{Notice that were it legal to take $\cutoff'(0) = 0$, then this fixed-point would disappear. But as discussed under~\eq{Pol-linear}, this is not an option.} 
Being non-critical, the theory does not possess the usual marginal, redundant direction. This can be seen from~\eq{mro-Gauss} where it is apparent that, for $B=1$, the candidate operator is non-local. Moreover, as pointed out by Wegner~\cite{Wegner_CS}, it is straightforward to check that all eigenperturbations are redundant (\cf\ the discussion at the end of \sect{GFP}). Since this fixed-point does not possess any relevant, scaling directions, it cannot be used to construct renormalized trajectories. However, it can act as a sink for RG trajectories (recall that operators which are redundant local to this fixed-point need not be redundant \wrt\ some other fixed-point, from which we imagine an RG trajectory is initiated). Consequently, this fixed-point can unambiguously be referred to
as IR fixed-points---a property which is expected to be a hallmark of non-critical fixed-points in general.

That our illustrative non-critical fixed-point does not possess a marginal, redundant direction suggests that this is generally true of non-critical fixed-points. With this in mind, let us return to~\eq{Gen-TPFP} and recall that quasi-local solutions exist for any $\eta_\star$ if we take $\intconst_{\eta_\star} + \const_{\eta_\star} =0$. All of the resulting fixed-points are non-critical---the two-point vertex starts at $\order{p^0}$ in every case---and there is no quantization of $\eta_\star$.

\subsection{Scale-Dependent Solutions}
\label{sec:renorm}

Ignoring exotic RG behaviour such as limit cycles (as mentioned in \sect{renormalizability}, we will say a little bit more about this in \sect{cfn}) there are two types of scale-dependent solution to the flow equation. The first are those corresponding to renormalized trajectories, which we recall arise from perturbing a fixed-point solution in one or more of its relevant (scaling) directions. The second class of solutions are those which follow from specifying some bare action as a boundary condition and then evolving the flow into the IR. We will confine our interest to the former case in this review, recapping and improving Morris' argument~\cite{TRM-Elements} (see also~\cite{TRM-3D}) as to why renormalized trajectories really are renormalizable nonperturbatively.

For the sake of simplicity, we will continue to work with a single scalar field and will
consider a fixed-point with $j$ relevant directions, none of which are marginal. 
Now, a renormalized trajectory is one for which, as we reverse the flow and climb into the UV as $t\rightarrow -\infty$, the action sinks back into the UV fixed-point action. Therefore, the boundary condition of the flow is
\be
	S_t[\phi] \sim  S_\star[\phi] + 
	\sum_{i=1}^j \alpha_i e^{\lambda_i t}
	\eop_i[\phi]
	\qquad \mbox{for $t \sim -\infty$}.
\label{eq:flow-bc}
\ee
Had we taken some marginally relevant directions, then there would be terms which sink into the fixed-point only like $1/t$ \ie\ logarithmically slowly. Clearly, irrelevant directions cannot be included in the sum since terms with a negative $\lambda_i$ blow up, rather than vanish, in the UV limit.

Now, at any point along the flow, it is apparent that the boundary condition~\eq{flow-bc},
together with the flow equation~\eq{flow-rescaled} (which explicitly depends on the anomalous dimension of the field, via $d_\phi$), implies that we can write
\be
	S_t[\phi] = S_t[\phi](\alpha_1,\ldots,\alpha_j;\eta(t)),
\label{eq:flow-alpha}
\ee
Let us recall an important point made in \sect{renormalizability}. Although the
boundary condition involves perturbing the fixed-point action in just the relevant directions,
if these directions are non-trivial (\ie\ interacting) then \emph{all possible interactions will be generated
along the flow}. However the couplings of the nascent irrelevant operators will not be new, independent couplings but will depend on the $\alpha_i$. Of course, computing this dependence is non-trivial! 

Morris' next step is to define the renormalized couplings, $g_i(t)$, and the running anomalous dimension, $\eta(t)$. Actually, this step is perhaps done a little too hurriedly in~\cite{TRM-Elements}.
The basic idea is that the natural (but not only---see below) definition of the renormalized couplings is to identify $g_i(t)$ as the coefficient in front of $\eop_i[\phi]$
in the action. But there is a subtlety here, which can be easily seen by returning to scalar field theory and recalling the Gaussian solution~\eq{scaling_exp} (there is no reason to complicate matters by taking a generic representative of the Gaussian fixed-point),
\[
	\mathcal{G}'_{n,r} = 
	e^{-\op}
	\int_{q_1,\ldots,q_n}
	\frac{1}{n!} 
	v_r(q_1,\ldots,q_n) \phi(q_1)\cdots \phi(q_n) \hat{\delta}(q_1+\cdots+q_n).
\]
As a consequence of the $e^{-\op}$, $\mathcal{G}'_{4,0}$ (for example)
has both a four-point piece and a two-point piece. With this in mind, imagine perturbing the Gaussian fixed-point in $\D=4-\epsilon$ in both the $n=2,r=0$ (mass) and $n=4,r=0$ directions.
The momentum-independent part of the two-point contribution to the action---which is a natural definition of the mass---clearly receives contributions from more than one eigenoperator!

To see the resolution to this problem (at least in principle), we will remain in scalar 
field theory, but consider an arbitrary fixed-point. We assume that the eigenperturbations, 
$\eop_{i}[\phi]$, span theory space. Therefore, all the way along the
flow we can write:
\be
	S_t[\phi] = S_\star[\phi] + \sum_{i} f_{i}(t) \eop_{i}[\phi],
\label{eq:action-decomp}
\ee
where the $f_i (t)$ would have to be determined by computation. From~\eq{eigen-soln}
we see that
\begin{multline}
	e^{\dual_\star[\phi] } e^{\op} e^{-\Sint_\star[\phi]} 
	\left(S_t[\phi] - S_\star[\phi]\right)
\\
	=
	\sum_i f_{i}(t) 
	\sum_{n_i,r_i}
	\int_{q_1,\ldots,q_{n_i}}
	P^{(i)}_{r_i}(q_1,\ldots,q_{n_i}) \phi(q_1)\cdots \phi(q_{n_i}) \hat{\delta}(q_1+\cdots+q_{n_i}).
\label{eq:extract}
\end{multline}
To proceed, we suppose that we have already computed the $P^{(i)}_{r_i}(q_1,\ldots,q_{n_i})$ and that we can evaluate the \lhs\
(perhaps needless to say,  it is this supposition which limits this procedure to being a solution in principle, at least for non-trivial fixed-points!). Now we can pick out any of the $f_i$s. For some value of $i$, we focus on the largest $n_i$. This determines a value of $r_i$ via~\eq{eigen-soln}. If this pair of labels $(n_i,r_i)$ is unique, then we are done: the coupling $f_i$ is easy to pick out. If this pair of labels are not unique then we proceed to the next largest value of $n_i$  (presuming it exists). Suppose that we go down the complete tower of pairs $(n_i,r_i)$ for a given eigenoperator and find that none of them are unique. 
Then we should broaden our view and consider together all eigenoperators that are sharing various pairs in this tower. Clearly, all of these eigenoperators have the same RG eigenvalue (though note that operators sharing the same RG eigenvalue do not necessarily share pairs of labels). If we assume that the members of this set are linearly independent and finite in number,  then we should be able to pick out the corresponding couplings.
Since the $g_i(t)$ are just the $f_{i}(t)$ belonging to the
relevant couplings (\wrt\ our UV fixed-point of choice, of course)  
we recover Morris' condition that
\be
	 g_i(t) \sim  \alpha_i e^{\lambda_i t}, \qquad \mbox{for $t \sim -\infty$}.
\label{eq:coupling-bc}
\ee

This still leaves the determination of the anomalous dimension. Let us recall that the fixed-point anomalous dimension is associated with a marginal, redundant direction, and that this yields a line of equivalent fixed-points. Now suppose that
we look at one particular representative and choose this one about which to linearize the flow equation. Clearly, since we are at this representative and not some other, we have not nudged this representative along its exactly marginal, redundant direction, whose value of $i$ we denote
by $i_{\mathrm{R}}$. Consequently, it must be
that
\be
	f_{i_{\mathrm{R}}}(t) = 0, \qquad \mbox{for $t \sim -\infty$}.
\ee

Now, the anomalous dimension at the fixed-point is a universal quantity. However, along the flow, $\eta(t)$ is subject to how we choose it to be defined. One apparently natural choice is to define it
such that $f_{i_{\mathrm{R}}}(t) = 0, \forall t$, presuming that this definition is globally well defined. This means that there is a term which exists in the action, coming from the fixed-point action, which is never corrected along the flow and so its 
 coefficient does not change. To look at it another way, this term is telling us that the field has had its anomalous scaling removed at all scales and so this procedure would seem to be a sensible way to define $\eta(t)$. Alternatively, suppose it is known that the flow is between two fixed-points. Then at both ends of the flow a universal value is obtained by $\eta$. Along the flow we should be free (within reason) to define $\eta(t)$ to be any function we like so long as it has the correct limits as $t \rightarrow \pm \infty$.

To see an example of how the first scheme works, let us return to the simplest representative of the Gaussian fixed point,
\[
	S^{\mathrm{Gaussian}}_\star[\phi] = 
	\hf \int_p \phi(-p) p^2 \cutoff^{-1}(p^2) \phi(p),
\]
with exactly marginal, redundant direction
\[
	\hf \int_p \phi(-p) p^2 \phi(p).
\]
By taking the coefficient of this redundant operator to be zero at all scales, we enforce that the \emph{total action}---see~\eq{action-decomp}---has unit coefficient in front of $\hf \int_p \phi(-p) p^2 \phi(p)$,
so long as we choose $\cutoff (0)=1$. Thus, this choice corresponds to canonical normalization of the kinetic term.

It is well worth pointing out that we can define the couplings 
in other ways, that might be slightly more convenient from the point of view of performing actual
calculations. It was pointed out before, in the case of the interacting renormalized trajectory in $\D=4-\epsilon$, that the momentum-independent  pieces of the two-point vertex receives contributions from more than one operator (in fact, it will generically receive contributions from an infinite number of operators, at a generic point along the flow). Nevertheless, we can still use this contribution to
the action to define the mass. This definition will differ from the previous one but is still perfectly good.
The point is that, if we have $j$ relevant couplings, then we need $j$ independent conditions on the action---\emph{which are compatible with the boundary conditions}---to serve as definitions. 

What we mean by this is best illustrated by example. Suppose that we need definitions for the mass and the four-point coupling. Then taking them to be given by the momentum-independent contributions to the
four-point and two-point vertices is fine, since both of these contributions to the action are present in the $t\sim-\infty$ boundary condition. Obviously, trying to define the four-point coupling through the six-point vertex is a silly thing to attempt, even though the eigenoperator whose highest-point vertex is six-point does indeed have a four-point contribution. This is because this operator does not contribute to the boundary condition.

Given two different definitions of the couplings it is, of course, in principle possible to relate them.
Universal quantities will be independent of this definition. We will discuss universality of the $\beta$-function in four-dimensional scalar field theory in \sect{beta}.

After this brief detour, we can continue with Morris' proof of renormalizability. To this end, we stick with the definition of the couplings which involves identifying them as the coefficients in front of the associated eigenoperators. Having read off the couplings directly from~\eq{extract}, we can invert the
$g_i(t)$ to obtain $t$ as a function of the couplings. Moreover, 
the $\alpha_i$ can be extracted from~\eq{coupling-bc} by
observing that, for $t \sim -\infty$, $e^{-\lambda_i t} g_i(t) \sim \alpha_i$. (If we were to take a different definition of the couplings, then this limit would give $j$ independent coupled equations for the $\alpha$s.) Consequently, we can trade the $\alpha$ and $t$ dependence of~\eq{flow-alpha} for dependence on the couplings:
\be
	S_t[\phi](\alpha_1,\ldots,\alpha_j;\eta(t)) = S[\phi](g_1(t),\ldots,g_j(t);\eta(t)).
\ee
Thus, as repeatedly emphasised by Morris, the action along a renormalized trajectory can be cast in self-similar form, which is no less than a nonperturbative statement of renormalizability.

\subsection{The Full Linearized Flow}
\label{sec:self-similar}

In this section we will return to an important and subtle issue that has, until now, been glossed over: the general solution to the linearized flow equation.%
\footnote{%
This analysis has grown out of a highly illuminating discussion with Hugh Osborn.
}
It is simplest to approach this using the dual action formalism. Given some fixed-point, we consider a perturbation, $\delta \Sint_t [\phi]$, which induces a perturbation in the dual action:
 $\dual_t[\phi] = \dual_\star[\phi] + \delta \dual_t[\phi]$. The precise relationship is:
 \be
 	\delta \dual_t[\phi] = 
	e^{\dual_\star[\phi] } e^{\op} e^{-\Sint_\star[\phi]}
	\delta \Sint_t [\phi].
\label{eq:induce}
 \ee
 However, we will not assume that $\delta \Sint_t$ can be written as in~\eq{perturb}, meaning that $\delta \dual_t$ does not necessarily reduce to~\eq{dual-shift}. Rather, working in position space we have that
\be
 	\biggl[
 	\partial_t
	+
	\int_x
	 \phi(x)
	\biggl(
		\frac{d-2-\eta_\star}{2} + x \cdot \partial_x
	\biggr)
	\fder{}{\phi(x)}
	\biggr]
	 \delta \dual_t [\phi] = 0,
 \ee
 where we take any deviation of $\eta$ from $\eta_\star$ to be of second order in the perturbation from the fixed-point under consideration.
 It is straightforward to check that the solution to this equation is
 \be
 	\delta \dual_t[\phi]
	=
	\sum_n
 	e^{ndt}
 	\int_{x_1,\ldots,x_n} 
	\mathcal{F}_n
	\bigl(
	\phi(x_1) e^{-(d-2-\eta_\star)t/2}, \ldots, \phi(x_n) e^{-(d-2-\eta_\star)t/2}; x_1 e^t,\ldots,x_n e^t
	\bigr),
\label{eq:Perturbations-GenSol}
 \ee
 where the $\mathcal{F}_n$ are arbitrary functions of their arguments.
 
 We now restrict the form of the $\mathcal{F}_n$ by applying the following conditions. First of all, we assume that there exist various $\delta \dual_t[\phi]$s for which the limits $t \rightarrow \pm \infty$ exist.
Specifically, we suppose that there exist trajectories which sink into the fixed-point as $t \rightarrow +\infty$ (\ie\ trajectories in the critical surface) and others which sink back into the fixed-point as  $t \rightarrow -\infty$  (\ie\ renormalized trajectories).
Moreover, we insist that the number of directions both into and out of the fixed-point are countable.
 
This suggests that we demand that the $\mathcal{F}_n$ are homogeneous functions of their arguments,
\begin{multline}
	e^{ndt}
 	\int_{x_1,\ldots,x_n} 
	\mathcal{F}_n
	\bigl(
	\phi(x_1) e^{-(d-2-\eta_\star)t/2}, \ldots, \phi(x_n) e^{-(d-2-\eta_\star)t/2}; x_1 e^t,\ldots,x_n e^t
	\bigr)
\\
	=
	e^{[nd - m(d-2-\eta_\star)/2 + s-ud]t}
	\int_{x_1,\ldots,x_n}  P_m\bigl(\phi(x_1),\ldots,\phi(x_n)\bigr) P_{s,u}\bigl(x_1,\ldots,x_n \bigr).
\end{multline}
Note that $P_{s,u}$ contains pieces that look like \eg\ $(x_1 -x_2)^{2s/2}$ together with a total of $u$ $\delta$-functions. To achieve a countable number of operators, and using the fact that since the dual action is intimately related to the correlation functions it had better have an expansion in powers of $\phi$, we further demand that $P_m$
is a polynomial. Indeed, because $P_{s,u}$ can legitimately contain $\delta$-functions of the coordinates, then since we are ultimately summing over $n$ we can demand, without loss of generality, that
\[
P_m\bigl(\phi(x_1),\ldots,\phi(x_n)\bigr) = \phi(x_1)\cdots\phi(x_n).
\]
We now assume that the allowed values of $s$ and $i$ are quantized by the requirement of quasi-locality of $\delta \Sint_t[\phi]$. Notice that if we transfer to momentum space then we have
\[
	e^{[d - r -(d-2-\eta_\star)n/2] t}
	\int_{p_1,\ldots,p_n}  P_r(p_1,\ldots,p_n) \phi(p_1)\cdots \phi(p_n) \hat{\delta}(p_1+\cdots+p_n),
\]
where we have used dimensional considerations to recognize that $r-d = -nd -s + ud$.
Comparing this with~\eq{eigen-soln}, it is clear we have recovered the eigenperturbations we found before. 

Let us now return to~\eq{Perturbations-GenSol} and consider what will happen if $\mathcal{F}_n$ is not a homogeneous function of $\phi$. In this case, if we cook up an $\mathcal{F}_n$ such that $\delta \dual_t[\phi]$ exhibits one of the limits $t \rightarrow \pm \infty$, then there will not be a quantization condition on such perturbations. This is reminiscent of the work of Halpern and Huang~\cite{Halpern+Huang} who (within the local potential approximation---see \sect{Trunc-DE}) constructed non-standard eigenperturbations for the Gaussian fixed-point. However, as vigorously pointed out by Morris~\cite{TRM-Comment,TRM-3D,TRM-Elements} these perturbations lack a quantization condition and so are inappropriate from a physical standpoint.

\section{The $\beta$-Function}
\label{sec:beta}

For scalar field theory formulated in dimensions near to four, a special
role is played by the coupling, $\lambda$, which essentially corresponds 
to the momentum-independent part of the four-point vertex (we will discuss
various precise definitions of this coupling in a moment). Considering
perturbations about the Gaussian fixed-point, all
scaling operators besides the mass are irrelevant
at linear order for $\D>4$. In exactly four dimensions, $\lambda$ becomes
marginal at linear order. Although, beyond leading order, $\lambda$
turns out to be marginally irrelevant, it dominates in the IR over the other
couplings for flows within the critical surface of the Gaussian fixed-point.
Moreover, for $\D<4$, $\lambda$ becomes relevant; not only does this
allow for the construction of interacting renormalized trajectories emanating
from the Gaussian fixed-point, but it is also intimately tied up with the
famous `Wilson-Fisher fixed-point'~\cite{W+F-3.99}, which we will rediscover
in \sect{WF}.

With these points in mind, this section will be
primarily devoted to studying the $\beta$-function:
\be
	\beta(\lambda) \equiv \Lambda \der{\lambda}{\Lambda}.
\ee
To actually compute the $\beta$-function requires that we define
what we mean by $\lambda$.  One part of the
definition comes from saying how we pick $\lambda$ out
of the action, and there are two ways we will do this.
The first is what one might call the canonical
definition: $\lambda$ is simply taken to be the momentum
independent part of the four-point vertex. In the second
definition, $\lambda$ is identified as the coupling of
the eigenperturbation $\mathcal{G}_{4,0}$, for which we recall~\eq{GenGaussPerts}.%
\footnote{Note that by using this explicit form we are implicitly making a choice of
flow equation. If we deform the flow equation, say by taking a non-trivial seed action,
we expect that $\mathcal{G}_{4,0}$ will still exist but will be different.}
This is the eigenperturbation whose highest-point
contribution is four-point and momentum independent.
But as we discussed in \sects{GFP}{renorm}, not only does this
eigenperturbation also come with lower-point contributions,
but there are other eigenperturbations which supply contributions
to the \emph{total} four-point, momentum-independent piece
of the action.

The second part of the definition of $\lambda$ is implicit
in the choice of flow equation: for two different flow equations,
the various couplings will flow in different ways and thus
can be expected, in general, to have different $\beta$-functions.
Nevertheless, given certain restrictions to be discussed in \sect{beta-canonical},
we expect the $\beta$-function coefficients at one and two loops to
agree between different definitions of the coupling, and this is
precisely what we will find. (The perturbative calculations presented 
here represent a huge refinement of those done in~\cite{scalar1,scalar2}. For other
computations of the $\beta$-function in scalar field theory see~\cite{P+W,Kopietz-2loop,Hughes+Liu-Beta}.)

Actually, we will show much more than this. Using the flow equation~\eq{flow-Ball} and 
taking the canonical
way of picking $\lambda$ out of the action, 
we will find that all \emph{explicit} dependence on the seed action
cancels out nonperturbatively! Given what we have said above,
this cancellation is expected to happen up to two
loops, but there is no obvious reason why it should happen beyond.
That this occurs seems to be a generic feature of generalized flow
equations, since the same thing has been found in QED~\cite{resum},
QCD~\cite{mgiuc,qcd} and the Wess--Zumino model~\cite{Susy-Chiral}. As for \emph{implicit}
dependence on the seed action and explicit dependence on the cutoff function, 
this will be shown to cancel out at one and two loops by direct calculation 
(using a different method to~\cite{scalar2}, where this has been done in the past). 
It was speculated in~\cite{RG2005} that these latter cancellations might
also persist beyond two loops, and it might be interesting to revisit this issue.

\subsection{The Canonical Definition of the Coupling}
\label{sec:Canonical}

\subsubsection{General Considerations}

To set up the machinery for computing the $\beta$-function, 
there is no particular advantage in scaling the
canonical dimensions out of the field and momenta, and so we will use the flow equation~\eq{flow-Ball},
for which only the anomalous dimension of the field has been taken into account. Actually, it is convenient to perform an additional rescaling:
\be
	\phi \mapsto \phi/\sqrt{\lambda}.
\label{eq:rescale-lambda}
\ee
The reason for executing this standard operation is that a factor of $1/\lambda$ now appears in front of the action. Consequently, the expansion in terms of $\lambda$ coincides with the expansion in $\hbar$,
meaning that our formalism is naturally adapted to doing perturbation theory. Of course, until such time
as we actually perform a perturbative expansion, everything we say is exact.
The flow equation that we will be using reads:
\be
	\left(-\flow + \frac{\gamma}{2} \Count_\phi\right) S_\lambda[\phi]
	=
	\frac{\lambda}{2} \classical{S_\lambda}{\dd}{\Sigma_\lambda} 
	- \frac{\lambda}{2} \quantum{\dd}{\Sigma_\lambda},
\label{eq:flow-lambda}
\ee
where 
\begin{subequations}
\begin{gather}
	\beta(\lambda) 
	\equiv \Lambda \der{\lambda}{\Lambda},
	\qquad
	\gamma(\lambda) \equiv \eta - \frac{\beta}{\lambda},
\label{eq:gamma}
\\
	\Sigma_\lambda 
	\equiv S_\lambda - 2\hS_\lambda,
\end{gather}
\end{subequations}
with $S_\lambda$ and $\hS_\lambda$ being appropriate to the rescaled field. In other words,
had we written $\phi_\lambda(p) \equiv \phi(p) \sqrt{\lambda}$, then we would have
 $S_\lambda[\phi_\lambda] = S[\phi]$. Note that, for $S_\lambda$ and $\hS_\lambda$,
 the splittings~\eq{split} become:
 \be
 	S_\lambda[\phi] =  \frac{1}{2\lambda} \phi \cdot \ep^{-1} \cdot \phi + \Sint_\lambda[\phi],
	\qquad
	\hS_\lambda[\phi] = \frac{1}{2\lambda} \phi \cdot \ep^{-1} \cdot \phi + \hSint_\lambda[\phi],
\label{eq:split-lambda}
 \ee
 so that we can rewrite the flow equation~\eq{flow-lambda}:
 \begin{multline}
	\left(-\flow + \frac{\gamma}{2} \Count_\phi\right) \Sint_\lambda
	=\frac{\lambda}{2} \classical{\Sint_\lambda}{\dd}{\Sigint_\lambda}
	-\frac{\lambda}{2} \quantum{\dd}{\Sigint_\lambda}
\\
	- \phi \cdot \ep^{-1} \dd \cdot \fder{\hSint_\lambda}{\phi}
	-\frac{1}{2\lambda} \left(\gamma + \frac{\beta}{\lambda} \right)\phi \cdot \ep^{-1} \cdot \phi.
\label{eq:reduced-flow-lambda}
\end{multline}
Given the rescaling~\eq{rescale-lambda}, we also redefine the dual action:
\be
	-\dual_\lambda[\phi] \equiv
	\ln
	\left\{
		\exp \left(\frac{\lambda}{2} \classical{}{\ep}{} \right) e^{-\Sint_\lambda[\phi]}
	\right\}
\ee

Repeating the calculation that lead to~\eq{dualflow-Seed} but remembering  
(whenever appropriate) to differentiate
$\lambda$ \wrt\ $\Lambda$, we arrive at:
\be
	\left[
		\flow + \left(\frac{\gamma}{2} + \frac{\beta}{\lambda} \right) \Count_\phi
	\right] \dual_\lambda
	=
	\left(\frac{\gamma}{\lambda} + \frac{\beta}{\lambda^2} \right) \hf \phi \cdot \ep^{-1} \cdot \phi
	+
	e^{\dual_\lambda} \phi \cdot \ep^{-1} \dd \cdot
	e^{\op} 
	\,
	\fder{\hSint_\lambda}{\phi} e^{-\Sint_\lambda}.
\label{eq:dual_lambda-flow}
\ee
The job now is to extract, from this expression, a formula for the $\beta$-function. We choose to
do this in the massless theory, since it is here that we expect to find agreement with the `universal' results at one and two loops. For the time
being, let us ignore the fact that we have not specified the boundary condition for the flow
(nor even fixed the dimensionality). However, whatever we end up doing, we will certainly
need to define what we mean by $\lambda$ and $\gamma$, and must ensure that the mass is zero.

Bearing in mind our rescaling~\eq{rescale-lambda}, in this section we will define
$1/\lambda$ as the coupling in front of the momentum-independent piece of the four-point vertex and $\gamma$ by demanding canonical normalization of the kinetic term. Writing out these conditions yields:
\be
	S_\lambda^{(4)}(0,0,0,0) = \frac{1}{\lambda},
	\qquad
	S_\lambda^{(2)}(p) = \frac{1}{\lambda}
	\left[
		\sigma(\lambda) \Lambda^2 + p^2 + \order{p^4}
	\right].
\label{eq:rc}
\ee
The mass is set to zero by tuning $\sigma$ such that $\Pi_\lambda(0) = 0$ where, taking account
of the rescaling~\eq{rescale-lambda}, we recall from~\eq{dep} that $\Pi_\lambda$ enters the dressed effective propagator according to
\be
	\dep(p^2) \equiv \frac{1}{\ep^{-1}(p^2) + \lambda \Pi_\lambda (p)}.
\label{eq:dep-lambda}
\ee
Note that the renormalization conditions apply to the Wilsonian effective action and not the seed action.

To derive an expression for the $\beta$-function, we start by using~\eq{dual_lambda-flow}
to find expressions for the flows of the 1PI parts of the two-point and four-point dual action vertices.
For the first pass, we will set $\hSint=0$. This will make the equations simpler and, when we
work with a general seed action, we will actually find that the expression for the $\beta$-function
is unchanged! Due to the rescaling~\eq{rescale-lambda}, \eqns{D2-1PI}{D4-1PI} become:
\begin{subequations}
\begin{align}
	\dualv{2}_\lambda(p) & =
	\frac{
		\Pi_\lambda(p)
	}
		{
		1 + \lambda \ep(p^2) \Pi_\lambda (p)
	},
\label{eq:D2_lambda-1PI}
\\
	\dualv{4}_\lambda(p_1,p_2,p_3,p_4) & = 
	\frac{
		\dopiv{4}_\lambda(p_1,p_2,p_3,p_4)
	}
		{
		\prod_{i=1}^4\left[1 + \lambda \ep(p_i^2) \Pi_\lambda (p_i)\right]
	}.
\label{eq:D4_lambda-1PI}
\end{align}
\end{subequations}
Substituting~\eq{D2_lambda-1PI} into~\eq{dual_lambda-flow}, with $\hSint=0$, gives:
\begin{multline}
	\frac{
		\flow \Pi_\lambda(p) + 
		\lambda \Pi_\lambda(p) \dd(p^2) \Pi_\lambda(p) 
		- \beta \Pi_\lambda(p) \ep(p^2) \Pi_\lambda(p)
	}
		{
		\left[1 + \lambda \ep(p^2) \Pi_\lambda(p)\right]^2
	}
	+
	\left(
		\gamma + \frac{2\beta}{\lambda}
	\right)
	\frac{\Pi_\lambda(p)}{1 + \lambda \ep(p^2) \Pi_\lambda(p)}
\\
	=
	\left(
	\frac{\gamma}{\lambda}
	+\frac{\beta}{\lambda^2}
	\right)
	\ep^{-1}(p^2).
\label{eq:dopi2-flow}
\end{multline}
Separating out $\gamma + 2\beta/\lambda = (\gamma + \beta/\lambda) + \beta/\lambda$
and noting that
\[
	\left(
	\frac{\gamma}{\lambda}
	+\frac{\beta}{\lambda^2}
	\right)
	\left[
		\ep^{-1}(p^2)
		-\frac{\lambda \Pi_\lambda(p)}{1 + \lambda \ep(p^2) \Pi_\lambda(p)}
	\right]
	=
	\left(
	\frac{\gamma}{\lambda}
	+\frac{\beta}{\lambda^2}
	\right)
	\frac{\ep^{-1}(p^2)}{1 + \lambda \ep(p^2) \Pi_\lambda(p)}
\]
we multiply~\eq{dopi2-flow} through by $1 + \lambda \ep(p^2) \Pi_\lambda(p)$ to yield:
\be
	\frac{1}{\lambda^2}
	\bigl(\gamma \lambda + \beta\bigr) \ep^{-1}(p^2)
	=
	\frac{
		\flow \Pi_\lambda(p) + 
		\lambda \Pi_\lambda(p) \dd(p^2) \Pi_\lambda(p) + \beta/\lambda \Pi_\lambda(p)
	}
		{
		1 + \lambda \ep(p^2) \Pi_\lambda(p)
	}.
\label{eq:Pi2-proto}
\ee

Before moving on, we would like to check that our masslessness condition, $\Pi_\lambda(0)=0$,
is a solution to this equation. We must be careful setting $p=0$, due to the $1/p^2$ appearing in
the $\ep(p^2)$ in the denominator. However, we can remove this problem by again multiplying
through by  $1 + \lambda \ep(p^2) \Pi_\lambda(p)$ to give
\[
	\frac{1}{\lambda^2}
	\bigl(\gamma \lambda + \beta\bigr) \ep^{-1}(p^2)
	=
	\flow \Pi_\lambda(p) + 
	\lambda \Pi_\lambda(p) \dd(p^2) \Pi_\lambda(p) 
	-\gamma  \Pi_\lambda(p).
\]
Now we can safely set $p=0$ everywhere: the \lhs, which goes
as $p^2 \cutoff^{-1}(p^2)$, vanishes and so it is apparent that $\Pi_\lambda(0)=0$ is indeed a solution.

The next step is to specialize~\eq{Pi2-proto} to $\order{p^2}$. On the \lhs\ this is easy, since it
yields just $(\gamma \lambda + \beta)/\lambda^2$ (where we understand the $p^2$ to have
been stripped off). On the \rhs, things are a bit more subtle. As we will see below, the $\order{p^2}$
part of $\Pi_\lambda(p)$ contains pieces which are non-polynomial in $p^2$. These come from
the IR end of certain loop integrals and, in a sense which will be made more precise below,
the external momentum can be thought of as playing the role of an IR regulator for such terms.
When we take
into account all terms on the \rhs, these non-polynomial pieces cancel out (as they must, since they are not present on the \lhs). However, at intermediate stages, they most certainly exist. Thus, by
$\Pi_\lambda(p)\bigr\vert_{p^2}$, we mean that we pick out all terms with a $p^2$ dependence
(and, indeed, strip this off) irrespective of whether they have additional non-polynomial dependence on
$p^2$. Therefore, for constants $a$ and $b$ we have, for example:
\[
	a p^2 + bp^2 \times \mbox{non-polynomial} \Bigr\vert_{p^2}
	=  a + b \times \mbox{non-polynomial}.
\]

Since, in the massless case, $\Pi_\lambda(0) =0$, it is apparent that the $\lambda \Pi_\lambda(p) \dd(p^2) \Pi_\lambda(p)$ piece in~\eq{Pi2-proto} cannot contribute at $\order{p^2}$. Note that the fact the $\ep$ is differentiated in this expression is crucial, since this converts a $1/p^2$ to a $1/\Lambda^2$.
The remaining terms in the numerator on the \rhs\ of~\eq{Pi2-proto} are both (up to possible non-polynomial pieces, of course) $\order{p^2}$. Therefore we must take $\order{p^0}$ from the denominator. This means that we are forced to take the $1/p^2$ contribution from the effective propagator, and the $\order{p^2}$ piece of $\Pi_\lambda(p)$. This leads to the simple expression:
\be
	\frac{1}{\lambda^2}
	\bigl(\gamma \lambda + \beta\bigr)
	=
	\frac{
		\flow  \Pi_\lambda(p)\big\vert_{p^2} + \beta/\lambda  \Pi_\lambda(p)\big\vert_{p^2}
	}
		{
		1 + \lambda  \Pi_\lambda(p)\big\vert_{p^2}
	}.
\label{eq:simult-a}
\ee

Now we repeat this procedure at the four-point level. Here, however, we will take the $\order{\mom^0}$
contribution. Again, we generically expect non-polynomial dependence on the external momenta at
intermediate stages of the calculation. With this in mind, we define $\dopiv{4}_\lambda \bigr\vert_{0}$
to be the $\order{\mom^0} \times \mbox{non-polynomial}$ pieces. These non-polynomial pieces
could depend on any of the external momenta $p_1,\ldots,p_4$ and blow up as these momenta go to zero. As in the two-point case (and as we will see below) this non-polynomial dependence comes from the IR end of loop integrals, and the
external momenta can be thought of as providing IR regularization. Since these non-polynomial pieces
exactly cancel out, we can treat all of them (in whatever combinations they occur) as if the IR regularization is provided by a single momentum, $p$. Equivalently, as we will see later, we can work
in $\D=4+\epsilon$ dimensions, whereupon we really can set the external momenta to zero everywhere
since the increased dimensionality serves to regularize any IR divergences.
With this in mind, substituting~\eq{D4_lambda-1PI} into~\eq{dual_lambda-flow} gives:
\be
	\left(
		\flow
		+ 2\gamma + \frac{4\beta}{\lambda}
	\right)
	\frac{
		\dopiv{4}_\lambda\bigr\vert_0
	}
		{
		\left[1 + \lambda \Pi_\lambda(p)\bigr\vert_{p^2}\right]^4
	}
	=0.
\label{eq:Dopi^4}
\ee
Cranking the handle once more yields:
\be
	-2 \gamma - \frac{4\beta}{\lambda}
	= \frac{\flow \dopiv{4}_\lambda\bigr\vert_0}{\dopiv{4}_\lambda\bigr\vert_0}
	- \frac{4}{1+ \lambda \Pi_\lambda(p)\bigr\vert_{p^2}}
	\left[
 		\beta \Pi_\lambda(p)\bigr\vert_{p^2}
		+ \lambda \flow \Pi_\lambda(p)\bigr\vert_{p^2}
	\right].
\label{eq:simult-b}
\ee
Finally, then, we can solve~\eqs{simult-a}{simult-b} for $\beta$:
\be
	\frac{\beta}{\lambda^2}
	=
	\flow \Pi_\lambda(p)\bigr\vert_{p^2}
	-\frac{1}{2\lambda}
	\left[
		1+ \lambda \Pi_\lambda(p)\bigr\vert_{p^2}
	\right]
	\frac{
		\flow \dopiv{4}_\lambda\bigr\vert_0 
	}
		{
		\dopiv{4}_\lambda\bigr\vert_0 
	}.
\label{eq:beta-direct}
\ee

There are two noteworthy ways of rewriting this equation. In the first,
we write it in as compact a form as possible, whereas in the second
we note that there are additional incidences of the $\beta$-function
which can be extracted from the \rhs\ by writing $\flow =
\flowlam  + \beta \partial_\lambda$:
\begin{subequations}
\begin{align}
	\frac{\beta}{\lambda}
	& =
	\flow
	\ln
	\left[
		\frac{
			1+ \lambda \Pi_\lambda(p)\bigr\vert_{p^2}
		}
			{
			\sqrt{\dopiv{4}_\lambda\bigr\vert_0 }
		}
	\right]
\label{eq:beta-compact}
\\[2ex]
	\frac{\beta}{\lambda^2}
	&
	=
	\frac{
		\flowlam  \Pi_\lambda(p)\bigr\vert_{p^2}
		-1/2\lambda
		\left[
			1+ \lambda \Pi_\lambda(p)\bigr\vert_{p^2}
		\right]
		\flowlam  \ln \dopiv{4}_\lambda\bigr\vert_0
	}
		{
		1 + \lambda/2
		\left[
			1+ \lambda \Pi_\lambda(p)\bigr\vert_{p^2}
		\right]
		\partial_\lambda \ln \dopiv{4}_\lambda\bigr\vert_0
		-\lambda^2
		\partial_\lambda  \Pi_\lambda(p)\bigr\vert_{p^2}
	}.
\label{eq:beta-convenient}
\end{align}
\end{subequations}
It is the latter equation, though apparently more complicated, from which 
the $\beta$-function can be most easily evaluated in perturbation theory.

Before moving on, we will demonstrate that the  expression for
the $\beta$-function remains the same in the presence of a non-trivial
seed action. From~\eq{dual_lambda-flow}, it follows that 
the \lhs\ of~\eq{dopi2-flow} picks up 
a term
\be
	2\ensuremath{\begin{array}{c}\input{pstex/InverseEP-dd-hS-1PI-dressed.pstex_t} \end{array}}
	=
	\frac{2}{1+ \lambda \ep(p^2) \Pi_\lambda(p)} \ensuremath{\begin{array}{c}\input{pstex/InverseEP-dd-hS-1PI.pstex_t} \end{array}}.
\ee
The vertex $\hDopi_\lambda$ is understood to be a version of $\Pi_\lambda$ in which one
vertex has been replaced by a seed action vertex (leading to a change in the
combinatorics). \emph{Note also that we understand that it is this vertex which is
attached to the $\dd$.} The thickened-up external leg in the first diagram
is dressed and can re-expressed as indicated, where the 
factor of $1/[1+ \lambda \Pi_\lambda(p)]$ can be
expanded out to give a 1PI diagram plus the usual tower of one-particle \emph{reducible} (1PR)
diagrams. The overall factor of two arises because either of the external
fields can be used to decorate the bottom vertex. 

Working at $\order{p^0}$, the presence of the $\ep^{-1}(p^2)$ ensures that
the masslessness condition $\Pi(0)=0$ is still a solution for non-trivial seed action.
At
$\order{p^2}$
(with the usual proviso about non-polynomial dependence),
all polynomial dependence comes from the $\ep^{-1}$ and so~\eq{simult-a}
which, we recall, involves combining terms in~\eq{dopi2-flow} and, crucially, multiplying
through by $1 + \lambda \ep(p^2) \Pi_\lambda(p)$ becomes:
\be
	\frac{1}{\lambda^2}
	\bigl(\gamma \lambda + \beta\bigr)
	=
	\frac{
		\flow  \Pi_\lambda(p)\big\vert_{p^2} + \beta/\lambda  \Pi_\lambda(p)\big\vert_{p^2}
	}
		{
		1 + \lambda  \Pi_\lambda(p)\big\vert_{p^2}
	}
	+
	2 \left. \ensuremath{\begin{array}{c}\begin{picture}(0,0)%
\epsfig{file=pstex/dd-hS-1PI.pstex}%
\end{picture}%
\setlength{\unitlength}{3947sp}%
\begingroup\makeatletter\ifx\SetFigFont\undefined%
\gdef\SetFigFont#1#2#3#4#5{%
  \reset@font\fontsize{#1}{#2pt}%
  \fontfamily{#3}\fontseries{#4}\fontshape{#5}%
  \selectfont}%
\fi\endgroup%
\begin{picture}(446,836)(2350,-545)
\put(2465,-98){\makebox(0,0)[lb]{\smash{{\SetFigFont{11}{13.2}{\rmdefault}{\mddefault}{\updefault}{\color[rgb]{0,0,0}$\hDopi_\lambda$}%
}}}}
\end{picture}%
 \end{array}} \right\vert_{p^0}.
\label{eq:simult-a-seed}
\ee

Next let us move on to the modification of~\eq{Dopi^4}, which can again
be read off from~\eq{dual_lambda-flow}. Since we work to
$\order{\mom^0}$, the only seed action terms which will survive are those
for which the $\mom^2$ coming from the $\ep^{-1}$ is ameliorated. Thus
we must take diagrams which are 1PR. All such contributions can be
summed up to give a new term on the \rhs:
\be
	4 \left.\ensuremath{\begin{array}{c}\input{pstex/InverseEP-dd-hS-1PI-dep-d4.pstex_t} \end{array}} \right\vert_{\mom^0},
\label{eq:tall-diagram}
\ee
where the external lines are dressed, as before, and
the thick internal line stands for a dressed effective propagator.

Now,
to go from~\eq{Dopi^4} to~\eq{simult-b} involves multiplying through
by $\left[1 + \lambda \Pi_\lambda(p)\bigr\vert_{p^2}\right]^4$
and dividing through by $\dopiv{4}_\lambda\bigr\vert_0$.  The effect
of the former operation on the seed action term is to remove the
aforementioned dressings (up to higher order terms in momenta), leaving
behind three undressed external legs and an internal $\ep$. 
The effect of the latter operation is to
remove the $\dopiv{4}_\lambda$ vertex. The final step is to observe
that the now undressed $\ep$ combines with the $\ep^{-1}$
at the bottom of the diagram in~\eq{tall-diagram}
to yield unity. Therefore~\eq{simult-b} becomes:
\be
	-2 \gamma - \frac{4\beta}{\lambda}
	= \frac{\flow \dopiv{4}_\lambda\bigr\vert_0}{\dopiv{4}_\lambda\bigr\vert_0}
	- \frac{4}{1+ \lambda \Pi_\lambda(p)\bigr\vert_{p^2}}
	\left[
 		\beta \Pi_\lambda(p)\bigr\vert_{p^2}
		+ \lambda \flow \Pi_\lambda(p)\bigr\vert_{p^2}
	\right]
	-
	4
	\left. \ensuremath{\begin{array}{c}\begin{picture}(0,0)%
\epsfig{file=pstex/dd-hS-1PI.pstex}%
\end{picture}%
\setlength{\unitlength}{3947sp}%
\begingroup\makeatletter\ifx\SetFigFont\undefined%
\gdef\SetFigFont#1#2#3#4#5{%
  \reset@font\fontsize{#1}{#2pt}%
  \fontfamily{#3}\fontseries{#4}\fontshape{#5}%
  \selectfont}%
\fi\endgroup%
\begin{picture}(446,836)(2350,-545)
\put(2465,-98){\makebox(0,0)[lb]{\smash{{\SetFigFont{11}{13.2}{\rmdefault}{\mddefault}{\updefault}{\color[rgb]{0,0,0}$\hDopi_\lambda$}%
}}}}
\end{picture}%
 \end{array}} \right\vert_{p^0}.
\label{eq:simult-b-seed}
\ee
Combining~\eqs{simult-a-seed}{simult-b-seed}, we see that the seed
action terms exactly cancel, reproducing~\eq{beta-direct}. Let us emphasise
that this result is nonperturbative and, as indicated earlier, in some sense 
quite surprising.

\subsubsection{Perturbation Theory}
\label{sec:beta-canonical}

In this section, we will perform a perturbative analysis to evaluate the one 
and two-loop $\beta$-function coefficients for the massless theory in  $\D=4$. 
It should be emphasised that 
the ERG is not being advocated as the best overarching framework in which to do 
perturbation theory. However, perturbation theory is a good way of getting a feeling for how the ERG works. Moreover, it will hopefully become apparent that \emph{given equal levels of familiarity}, the
illustrative calculations that we will do are of
comparable difficulty to the analogous calculations performed using more conventional approaches. That this is the case is a new development,
arising as a consequence of~\eq{beta-convenient}, which appears nowhere else in the literature (though similar expressions have been derived in QED~\cite{resum} and the Wess--Zumino model~\cite{Susy-Chiral}).

\Eqn{beta-convenient} allows us to immediately write down the set of diagrams from which
the perturbative $\beta$-function coefficients can be readily extracted; this is our starting point.
Previously~\cite{scalar1,scalar2}, the flow equation was the starting point, with the set
of diagrams encoded in~\eq{beta-convenient} being laboriously derived, loop order by loop order,
using elaborate diagrammatic techniques. It is well worth comparing the approach of~\cite{scalar2}
to the current one, since the level of simplification is prodigious.

To generate the perturbation series, we introduce the expansions of the actions which follow from~\eq{rescale-lambda}:
\be
	S_\lambda \sim \sum_{i=0}^\infty \lambda^{i-1} S_i,
	\qquad
	\hS_\lambda \sim \sum_{i=0}^\infty \lambda^{i-1} \hS_i,
	\qquad
	\dual_\lambda \sim \sum_{i=0}^\infty \lambda^{i-1} \dual_i.
\label{eq:action-pert}
\ee
Thus we understand $S_0$ to be the classical (\aka\ tree-level) action, $S_1$ to be the one-loop
correction and so forth.
Anticipating the results of our perturbative analysis, we can introduce similar expansions
for $\beta$ and $\gamma$:
\be
	\beta \sim \sum_{i=1}^\infty \lambda^{i+1} \beta_i, 
	\qquad
	\gamma \sim \sum_{i=1}^\infty \lambda^{i} \gamma_i.
\label{eq:beta-exp}
\ee

Following~\cite{scalar1,scalar2}, we will use a trick in order to simplify the perturbative treatment:
we will exploit the fact that, as discussed in \sect{AS}, $\lambda \phi^4$ theory in $\D=4$ is self-similar, \emph{within perturbation theory}. Of course, as has been described in great detail, this catastrophically breaks down beyond perturbation theory. But, if we are happy to shut our eyes and ignore this, then the perturbative analysis---which is all that interests us here---can be formulated in a very pleasing manner.
Recalling that we are working in the massless case, and given perturbative self-similarity, it follows---supposing for the moment that we scale out all canonical dimensions---that $S_{\lambda,\Lambda} = S_\lambda(\lambda(\Lambda),\gamma(\Lambda))$. The
presence of $\gamma(\Lambda)$---which is itself just a function of $\lambda$---is to remind us that the actual solution for $\gamma$ requires a renormalization condition separate for the one for $\lambda$.

The benefit of exploiting `self-similarity' in this way is that the $\beta$-function can now be computed simply by specifying renormalization conditions for $\beta$ and $\gamma$, seeing what these conditions imply, and cranking the handle. There is never any mention of the bare action, nor the notion of taking $\Lambda_0 \rightarrow \infty$ at the end of the calculation. In the case at hand, it cannot be overemphasised that this amounts to a sleight of hand, since perturbation theory cannot be unambiguously resummed without including $\Lambda/\Lambda_0$ terms which manifestly violate self-similarity.%
\footnote{In massless QED~\cite{resum} and the massless Wess--Zumino model~\cite{Susy-Chiral}, it has been argued that the $\beta$-function as computed in the ERG is in fact free of nonperturbative power corrections. This implies that in these cases the $\beta$-function can be resummed. One the one hand, this suggests that the Landau pole exists beyond perturbation theory since triviality means that $\lambda$ should be aware of the bare scale; on the other hand, there is no
reason why the perturbative series for any of the other couplings in these theories can be resummed.}
If we were to go beyond perturbation theory then, because of this
lack of self-similarity, we would have to specify a boundary condition for the flow at the bare scale. This would amount to providing a definition for all possible couplings in the theory, rather than just $\lambda$ and $\gamma$. Note, though, that the perturbative calculation we will do 
provides a template for doing computations directly in terms of renormalized parameters for field theories which exhibit bona-fide self-similarity, such as $\SU(N)$ Yang--Mills~\cite{mgierg2} and QCD~\cite{qcd}.

As a final point, let us recall the argument as to why the one and two-loop coefficients
of the $\beta$-function agree for certain classes of renormalization schemes in $\D=4$~\cite{WeinbergII,aprop}.
Suppose
that we have two definitions of $\lambda$ which are equivalent at the 
classical level. Then we can write
\be
	\frac{1}{\lambda} = \frac{1}{\tilde{\lambda}} + \kappa + \order{\lambda},
\ee
where $\lambda$ and $\tilde{\lambda}$ correspond to our
two different definitions, and $\kappa$ is a dimensionless, one-loop matching
coefficient. Hitting both sides with $\flow$ yields:
\be
	\tilde{\beta}_1 + \tilde{\beta_2}	\lambda
	=
	\beta_1 + \beta_2 \lambda - \flow \kappa + \order{\lambda}.
\ee
In four dimensions, the canonical dimension of $\lambda$ is zero,
and so $\kappa$ is dimensionless. But, if we have self-similarity and masslessness then 
we can write the scale dependence of
all dimensionless quantities---such as $\kappa$---in terms of $\lambda$, upon
which $\kappa$ does not to depend, by construction. Consequently, for the massless
theory, $\flow \kappa=0$. Therefore, the $\beta$-function
coefficients for these two definitions of the couplings agree at one and two loops.
Of course, this agreement can be spoilt if there are any additional
scales in the game. In four dimensions, this is the case beyond
perturbation theory. Also, taking a non-zero mass would spoil things.%
\footnote{There are
more elaborate reasons why the one and two-loop $\beta$-function coefficients
might not agree between different schemes. This is particularly pertinent to gauge theory and is
discussed further in~\cite{aprop,mgierg2}.}

To compute the $\beta$-function, we must use the renormalization conditions~\eq{rc}.
A vital point to make is that the condition on the four-point vertex is saturated
at tree-level. This is immediately apparent upon comparing this renormalization condition
with~\eq{action-pert}. Consequently, the momentum-independent part of the four-point
vertex does not receive quantum corrections. Precisely the same is true for the $\order{p^2}$
part of the two-point vertex. Indeed, we can go further: since we have taken $\cutoff(0)=1$, the splitting~\eq{split-lambda} tells us that, the $\order{p^2}$ part of $\Siv{2}(p)$ is zero, even at tree-level.
Finally, we note that $\sigma$ vanishes at tree level and can be self-consistently determined
(should one so desire) from one loop onwards.

The final ingredients that we need are the expressions for
$\Pi_\lambda$ and $\dopiv{4}_\lambda$, up to whatever loop order necessary. Contenting ourselves
with two loops and focusing first on the former we have:
\begin{align}
	\Pi_\lambda(p) \bigr\vert_{p^2}=
	\left.
	\hf
	\ensuremath{\begin{array}{c}\begin{picture}(0,0)%
\epsfig{file=pstex/Padlock2_0.pstex}%
\end{picture}%
\setlength{\unitlength}{3947sp}%
\begingroup\makeatletter\ifx\SetFigFont\undefined%
\gdef\SetFigFont#1#2#3#4#5{%
  \reset@font\fontsize{#1}{#2pt}%
  \fontfamily{#3}\fontseries{#4}\fontshape{#5}%
  \selectfont}%
\fi\endgroup%
\begin{picture}(418,580)(1606,-593)
\put(1768,-446){\makebox(0,0)[lb]{\smash{{\SetFigFont{11}{13.2}{\rmdefault}{\mddefault}{\updefault}{\color[rgb]{0,0,0}$0$}%
}}}}
\end{picture}%
 \end{array}}
	-\frac{\lambda}{6} \ensuremath{\begin{array}{c}\input{pstex/S4_0-ep3.pstex_t} \end{array}}
	+\frac{\lambda}{8} \ensuremath{\begin{array}{c}\begin{picture}(0,0)%
\epsfig{file=pstex/Padlockx2-2_0.pstex}%
\end{picture}%
\setlength{\unitlength}{3947sp}%
\begingroup\makeatletter\ifx\SetFigFont\undefined%
\gdef\SetFigFont#1#2#3#4#5{%
  \reset@font\fontsize{#1}{#2pt}%
  \fontfamily{#3}\fontseries{#4}\fontshape{#5}%
  \selectfont}%
\fi\endgroup%
\begin{picture}(576,746)(1523,-759)
\put(1765,-442){\makebox(0,0)[lb]{\smash{{\SetFigFont{11}{13.2}{\rmdefault}{\mddefault}{\updefault}{\color[rgb]{0,0,0}$0$}%
}}}}
\end{picture}%
 \end{array}}
	+\frac{\lambda}{2} \ensuremath{\begin{array}{c}\begin{picture}(0,0)%
\epsfig{file=pstex/Padlock2_1.pstex}%
\end{picture}%
\setlength{\unitlength}{3947sp}%
\begingroup\makeatletter\ifx\SetFigFont\undefined%
\gdef\SetFigFont#1#2#3#4#5{%
  \reset@font\fontsize{#1}{#2pt}%
  \fontfamily{#3}\fontseries{#4}\fontshape{#5}%
  \selectfont}%
\fi\endgroup%
\begin{picture}(418,580)(1606,-593)
\put(1764,-448){\makebox(0,0)[lb]{\smash{{\SetFigFont{11}{13.2}{\rmdefault}{\mddefault}{\updefault}{\color[rgb]{0,0,0}$1$}%
}}}}
\end{picture}%
 \end{array}}
	+\order{\lambda^2}
	\right\vert_{p^2}.
\label{eq:Pi-diag}
\end{align}
There are several points to make. 
The number inside each vertex
refers to the order in perturbation theory of said vertex, \cf~\eq{action-pert}.
All vertices belong to $\Sint$ but since it is only at the two-point, classical level 
that there is a difference between $S$ and $\Sint$
there is no need to tag any of the vertices in the above expression with an `I'.
It is taken as understood that the external momenta flowing into each diagram
are $p$ and $-p$.
Had we not restricted ourselves to looking at $\order{p^2}$ (up to non-polynomial pieces),
the diagrams
\[
	\frac{1}{\lambda} \ensuremath{\begin{array}{c}\begin{picture}(0,0)%
\epsfig{file=pstex/Sint_0.pstex}%
\end{picture}%
\setlength{\unitlength}{3947sp}%
\begingroup\makeatletter\ifx\SetFigFont\undefined%
\gdef\SetFigFont#1#2#3#4#5{%
  \reset@font\fontsize{#1}{#2pt}%
  \fontfamily{#3}\fontseries{#4}\fontshape{#5}%
  \selectfont}%
\fi\endgroup%
\begin{picture}(358,579)(1629,-672)
\put(1744,-450){\makebox(0,0)[lb]{\smash{{\SetFigFont{11}{13.2}{\rmdefault}{\mddefault}{\updefault}{\color[rgb]{0,0,0}$0^{\mathrm{I}}$}%
}}}}
\end{picture}%
 \end{array}} + \ensuremath{\begin{array}{c}\begin{picture}(0,0)%
\epsfig{file=pstex/Sint_1.pstex}%
\end{picture}%
\setlength{\unitlength}{3947sp}%
\begingroup\makeatletter\ifx\SetFigFont\undefined%
\gdef\SetFigFont#1#2#3#4#5{%
  \reset@font\fontsize{#1}{#2pt}%
  \fontfamily{#3}\fontseries{#4}\fontshape{#5}%
  \selectfont}%
\fi\endgroup%
\begin{picture}(371,579)(1629,-672)
\put(1770,-445){\makebox(0,0)[lb]{\smash{{\SetFigFont{11}{13.2}{\rmdefault}{\mddefault}{\updefault}{\color[rgb]{0,0,0}$1$}%
}}}}
\end{picture}%
 \end{array}} + \order{\lambda}
\]
would be included in~\eq{Pi-diag}.
However, as mentioned above,
these terms do not contribute at $\order{p^2}$. Finally,
we have dressed all internal lines, as indicated by their thickening,
so that they represent dressed effective propagators~\eq{dep-lambda}.

On account of this latter step, every diagram thus contributes both
at the \naive\ order of perturbation theory indicated by the 
power of $\lambda$ in front of every diagram and at every subsequent
order. For some of the terms (but not all---this is the point of dressing the effective propagators)
it will be necessary to expand the dressed effective propagators as a perturbation series.
We obtain, from~\eq{dep-lambda}:
\be
	\dep_\lambda(p^2) = \frac{\ep(p^2)}{1 + \ep(p^2) \Pi_{0}(p)} + \order{\lambda},
\label{eq:dep-expansion}
\ee
where $\Pi_{0}(p)$ is the classical contribution to $\Pi_{\lambda}(p)$, comprising the
vertex $\Siv{2}_0(p)$. Now, due to 
the masslessness of the theory and the renormalization condition for the $\order{p^2}$ part
of $\Siv{2}_0(p)$, $\Pi_{0}(p)$ first contributes at $\order{p^4}$. Therefore we find a result
which will prove to be very useful:
\be
	\dep_\lambda(p^2) = \frac{1}{p^2} + \order{p^0,\lambda}.
\label{eq:dep-leading}
\ee

Let us now move on to $\dopiv{4}\bigr\vert_0$:
\begin{multline}
	\dopiv{4}\bigr\vert_0 =
		\frac{1}{\lambda} \ensuremath{\begin{array}{c}\begin{picture}(0,0)%
\epsfig{file=pstex/Sint4_0.pstex}%
\end{picture}%
\setlength{\unitlength}{3947sp}%
\begingroup\makeatletter\ifx\SetFigFont\undefined%
\gdef\SetFigFont#1#2#3#4#5{%
  \reset@font\fontsize{#1}{#2pt}%
  \fontfamily{#3}\fontseries{#4}\fontshape{#5}%
  \selectfont}%
\fi\endgroup%
\begin{picture}(424,420)(1597,-592)
\put(1770,-444){\makebox(0,0)[lb]{\smash{{\SetFigFont{11}{13.2}{\rmdefault}{\mddefault}{\updefault}{\color[rgb]{0,0,0}$0$}%
}}}}
\end{picture}%
 \end{array}}
		-\frac{3}{2} \ensuremath{\begin{array}{c}\input{pstex/S_04-depx2.pstex_t} \end{array}}
		+\hf \ensuremath{\begin{array}{c}\begin{picture}(0,0)%
\epsfig{file=pstex/Padlock4_0.pstex}%
\end{picture}%
\setlength{\unitlength}{3947sp}%
\begingroup\makeatletter\ifx\SetFigFont\undefined%
\gdef\SetFigFont#1#2#3#4#5{%
  \reset@font\fontsize{#1}{#2pt}%
  \fontfamily{#3}\fontseries{#4}\fontshape{#5}%
  \selectfont}%
\fi\endgroup%
\begin{picture}(433,632)(1594,-645)
\put(1768,-446){\makebox(0,0)[lb]{\smash{{\SetFigFont{11}{13.2}{\rmdefault}{\mddefault}{\updefault}{\color[rgb]{0,0,0}$0$}%
}}}}
\end{picture}%
 \end{array}}
		+\frac{3\lambda}{4}
		\ensuremath{\begin{array}{c}\input{pstex/S_04x3-depx4.pstex_t} \end{array}}
		+3\lambda
		\ensuremath{\begin{array}{c}\input{pstex/S_04x3-depx4-b.pstex_t} \end{array}}
		-\frac{2\lambda}{3}
		\ensuremath{\begin{array}{c}\input{pstex/S_04x2-dep3.pstex_t} \end{array}}
\\
		-3\lambda \ensuremath{\begin{array}{c}\input{pstex/S_04-S_14-depx2.pstex_t} \end{array}}
	\left.
		-\frac{3\lambda}{2}
		\ensuremath{\begin{array}{c}\input{pstex/S_04-depx2-Padlock4_0.pstex_t} \end{array}}
		+\frac{\lambda}{2}
		\ensuremath{\begin{array}{c}\begin{picture}(0,0)%
\epsfig{file=pstex/Padlock4_1.pstex}%
\end{picture}%
\setlength{\unitlength}{3947sp}%
\begingroup\makeatletter\ifx\SetFigFont\undefined%
\gdef\SetFigFont#1#2#3#4#5{%
  \reset@font\fontsize{#1}{#2pt}%
  \fontfamily{#3}\fontseries{#4}\fontshape{#5}%
  \selectfont}%
\fi\endgroup%
\begin{picture}(433,632)(1594,-645)
\put(1771,-452){\makebox(0,0)[lb]{\smash{{\SetFigFont{11}{13.2}{\rmdefault}{\mddefault}{\updefault}{\color[rgb]{0,0,0}$1$}%
}}}}
\end{picture}%
 \end{array}}
		+\frac{\lambda}{8} \ensuremath{\begin{array}{c}\begin{picture}(0,0)%
\epsfig{file=pstex/Padlockx24_0.pstex}%
\end{picture}%
\setlength{\unitlength}{3947sp}%
\begingroup\makeatletter\ifx\SetFigFont\undefined%
\gdef\SetFigFont#1#2#3#4#5{%
  \reset@font\fontsize{#1}{#2pt}%
  \fontfamily{#3}\fontseries{#4}\fontshape{#5}%
  \selectfont}%
\fi\endgroup%
\begin{picture}(685,539)(1468,-645)
\put(1768,-446){\makebox(0,0)[lb]{\smash{{\SetFigFont{11}{13.2}{\rmdefault}{\mddefault}{\updefault}{\color[rgb]{0,0,0}$0$}%
}}}}
\end{picture}%
 \end{array}}
		+\order{\lambda^2}
	\right\vert_0
\label{eq:D^4bar-diags}
\end{multline}
Note that higher order analogues of the first diagram do not appear, as a consequence of
the renormalization condition~\eq{rc}. Compared to conventional approaches, where there
is no need to consider vertices with more than four legs, 
the above expression looks rather unwieldy, particularly at two loops. However,
we will find that
most contributions actually drop out of the two-loop $\beta$-function. 

Our calculations of the $\beta$-function will use~\eq{beta-convenient}. This equation (though defined nonperturbatively) can be decomposed, loop order by
loop order. Noting that $\Pi_\lambda(p)\bigr\vert_{p^2}$ starts at one loop and that
the tree-level contribution to $\dopiv{4}_\lambda\bigr\vert_0$ is just $1/\lambda$ we have:
\begin{subequations}
\begin{align}
	\Pi_\lambda(p)\bigr\vert_{p^2} 
	& =  \Pi_1(p)\bigr\vert_{p^2} + \lambda \Pi_2(p)\bigr\vert_{p^2} + \order{\lambda^2},
\\
	\dopiv{4}_\lambda\bigr\vert_0 
	&= \frac{1}{\lambda}
	+ \dopiv{4}_0 \bigr\vert_0 + \lambda \dopiv{4}_1\bigr\vert_0 + \order{\lambda^2}.
\end{align}
\end{subequations}
Substituting these expressions into~\eq{beta-convenient} we find that,
as expected, the $\beta$-function receives no contribution at tree-level. The one and
two-loop expressions are:
\begin{subequations}
\begin{align}
	\beta_1 
	& = 
	\flowlam
	\left[
		2 \Pi_1(p)\bigr\vert_{p^2} - \dopiv{4}_1 \bigr\vert_0 
	\right],
\label{eq:beta_1}
\\
	\beta_2
	& =
	\flowlam
	\left\{
		2 \Pi_2(p)\bigr\vert_{p^2} - \dopiv{4}_2 \bigr\vert_0 
		+
		\left[\Pi_1(p)\bigr\vert_{p^2} - \dopiv{4}_1 \bigr\vert_0\right]^2
	\right\}.
\label{eq:beta_2}
\end{align}
\end{subequations}

Focusing first on $\beta_1$, we write out the \rhs\ of~\eq{beta_1} diagrammatically:
\be
	\beta_1 = 
	 -
	 \left[
	\left.\ensuremath{\begin{array}{c}\begin{picture}(0,0)%
\epsfig{file=pstex/Padlock2_0.pstex}%
\end{picture}%
\setlength{\unitlength}{3947sp}%
\begingroup\makeatletter\ifx\SetFigFont\undefined%
\gdef\SetFigFont#1#2#3#4#5{%
  \reset@font\fontsize{#1}{#2pt}%
  \fontfamily{#3}\fontseries{#4}\fontshape{#5}%
  \selectfont}%
\fi\endgroup%
\begin{picture}(418,580)(1606,-593)
\put(1768,-446){\makebox(0,0)[lb]{\smash{{\SetFigFont{11}{13.2}{\rmdefault}{\mddefault}{\updefault}{\color[rgb]{0,0,0}$0$}%
}}}}
\end{picture}%
 \end{array}}\right\vert_{p^2}
	+
	\left.
	\frac{3}{2}
	\ensuremath{\begin{array}{c}\input{pstex/S_04-depx2.pstex_t} \end{array}}
	-\hf \ensuremath{\begin{array}{c}\begin{picture}(0,0)%
\epsfig{file=pstex/Padlock4_0.pstex}%
\end{picture}%
\setlength{\unitlength}{3947sp}%
\begingroup\makeatletter\ifx\SetFigFont\undefined%
\gdef\SetFigFont#1#2#3#4#5{%
  \reset@font\fontsize{#1}{#2pt}%
  \fontfamily{#3}\fontseries{#4}\fontshape{#5}%
  \selectfont}%
\fi\endgroup%
\begin{picture}(433,632)(1594,-645)
\put(1768,-446){\makebox(0,0)[lb]{\smash{{\SetFigFont{11}{13.2}{\rmdefault}{\mddefault}{\updefault}{\color[rgb]{0,0,0}$0$}%
}}}}
\end{picture}%
 \end{array}}
	\right\vert_{0}
	\right]^{\bullet}
	+\order{\lambda},
\label{eq:beta_1-diagrams}
\ee
where $[\cdots]^\bullet \equiv -\flowlam [\cdots]$ and
we have retained the dressings of the effective propagators for
reasons that will become apparent [this is why the $+\order{\lambda}$
appears
on the \rhs: the dressed effective propagators contribute to all orders in
perturbation theory].
We start the evaluation of these terms by looking at the first one. If we expand the
dressed effective propagator to zeroth order in perturbation theory then we have, 
recalling~\eq{dep-expansion}:
\be
	\flowlam 
	\left\{
		\MomInt{4}{k} \Svert{4}_0(p,-p,k,-k;\Lambda) \frac{\cutoff(k^2/\Lambda^2)}{k^2} 
		\frac{1}{1 + \ep(k^2) \Pi_{0}(k)}
	\right\}_{p^2}.
\ee

The $\order{p^2}$ part of this expression is dimensionless, as must be
true (and as can be readily checked) since it contributes to the dimensionless object $\beta_1$. Stripping off the $p^2$ (which must come from Taylor expanding the vertex to this order, since there is no $p$-dependence anywhere else), we therefore have something of the form:
\[
	\flowlam \  \left[\mbox{dimensionless quantity}\right] .
\]
Now for the point: within perturbation theory we have self-similarity, meaning that the only objects on which the action depends are $\lambda$ and $\Lambda$. All $\lambda$-dependence has been 
factored out in our perturbative treatment. Furthermore, there are no available scales with which to combine $\Lambda$ to form a dimensionless quantity. Consequently,  we conclude that the contribution of the diagram under analysis is zero, this property remaining true if we take the internal line to
be fully dressed. (This observation will simplify the two-loop calculation.)
Beyond perturbation theory, it is a different matter, since we know that the scale $\Lambda_0$ is floating around. Strictly, then, we have that
\be
	\flowlam 
	\left.\ensuremath{\begin{array}{c}\begin{picture}(0,0)%
\epsfig{file=pstex/Padlock2_0.pstex}%
\end{picture}%
\setlength{\unitlength}{3947sp}%
\begingroup\makeatletter\ifx\SetFigFont\undefined%
\gdef\SetFigFont#1#2#3#4#5{%
  \reset@font\fontsize{#1}{#2pt}%
  \fontfamily{#3}\fontseries{#4}\fontshape{#5}%
  \selectfont}%
\fi\endgroup%
\begin{picture}(418,580)(1606,-593)
\put(1768,-446){\makebox(0,0)[lb]{\smash{{\SetFigFont{11}{13.2}{\rmdefault}{\mddefault}{\updefault}{\color[rgb]{0,0,0}$0$}%
}}}}
\end{picture}%
 \end{array}}\right\vert_{p^2}
	\sim
	\Or \left(\frac{\Lambda}{\Lambda_0}\right).
\ee
The same result obtains for the final term in~\eq{beta_1-diagrams}.

Given this, one might wonder how a non-zero contribution to the $\beta$-function can ever arise within perturbation theory. The answer becomes apparent upon analysis of the second diagram 
in~\eq{beta_1-diagrams}. To analyse this diagram, we will replace the dressed effective
propagators with just $\ep$. Note that this is not quite the same as expanding the dressed
effective propagators to zeroth order in perturbation theory, since~\eq{dep-expansion} tells us that
the dressed effective propagators pick up contributions as tree-level.
However, as we will see, these extra terms contribute nothing. Thus, we consider:
\begin{multline*}
	\frac{3}{2}
	\flowlam
	\Biggl\{
	\MomInt{4}{k}
	\biggl[
		\Svert{4}_{0}(p_1,p_2,k-p_1-p_2,-k;\Lambda) \Svert{4}_{0}(k,-k+p_1+p_2,p_3,p_4;\Lambda) 
\\		
		\frac{\cutoff((k-p_1-p_2)^2/\Lambda^2)}{(k-p_1-p_2)^2}
		\frac{\cutoff(k^2/\Lambda^2)}{k^2}
	\biggr]
	\Biggr\}_{\mom^0},
\end{multline*}
where we have taken the external momenta flowing into the diagram to be $p_1,\ldots,p_4$
(with $p_1+p_2 = -p_3-p_4$). Here, we need to be very careful setting the external momenta
to zero: for if we do so immediately, then  the integral over $k$ would diverge in the IR,
as a result of making the replacement $(k-p_1-p_2)^2 k^2 \rightarrow 1/k^4$. Note, though, that
we are quite at liberty to set the external momenta to zero in all quasi-local terms---\ie\ in the vertices and the cutoff functions. Therefore our expression reduces to
\[
	\frac{3}{2}
	\flowlam
	\Biggl[
	\MomInt{4}{k}
		\left[\Svert{4}_{0}(0,0,k,-k;\Lambda)\right]^2 
		\frac{\cutoff^2(k^2/\Lambda^2)}{(k-p_1-p_2)^2 k^2}
	\Biggr]_{\mom^0}.
\]
Once again, we arrive at the $\Lambda$-derivative of a dimensionless quantity. But there is a major difference compared to the last case: we can form a dimensionless quantity involving $\Lambda$ by using the $p_1+p_2$ which must be kept in order to prevent the loop integral from diverging in the IR. Thus we expect to find a contribution at $\order{\mom^0}$ coming from:
\be
	\flow \ln (p_1+p_2)^2/\Lambda^2 = -2.
\ee
This structure is only present whenever $p_1+p_2$ must be kept non-zero at intermediate
stages of a calculation to provide IR regularization.%
\footnote{Note that because of this, and because at the end of
the calculation of the $\beta$-function all such non-polynomial terms cancel out, we could replace
all combinations of momenta which act as IR regulators simply by $p$. This strategy has been
explicitly employed in the denominator of~\eq{Dopi^4}.}
Consequently, we can set $k=0$ in the vertex coefficient functions, 
which
then reduce to unity as a consequence of the renormalization condition. 
(Taking powers of $k$ from the vertices---which must be positive as a consequence of quasi-locality---obviates the need to keep $p_1+p_2 \neq 0$ and so such contributions are killed after differentiation \wrt\ $\Lambda$.)

At this stage it should be clear
why we were able to neglect the tree-level contributions to the dressed effective propagator:
as~\eq{dep-leading} informs us, these contributions do not affect the $1/\mom^2$ behaviour of the
effective propagator, which is what governs the part of the term which survives differentiation \wrt\ $\Lambda$.
We need to be careful
doing likewise with the cutoff function, since non-trivial $k$-dependence is required for UV
regularization. So, we have reduced our problem to that of evaluating
\be
	\frac{3}{2}
	\Biggl[
	\totalflow
	\MomInt{4}{k}
		\frac{\cutoff^2(k^2/\Lambda^2)}{(k-p)^2 k^2}
	\Biggr]_{\mom^0}.
\ee

There are several different ways to evaluate this expression. One of them involves taking the
derivative inside the integral and explicitly
differentiating the cutoff functions~\cite{bo,scalar2}. This is a simple way to do things in the case
at hand since, for this particular example, we can replace (under the integral) $\flow \rightarrow -2
d/dk^2$. However, there is a different way to proceed which 
is more sympathetic to the fact that any contributions from the integral that survive differentiation
\wrt\ $\Lambda$ must come from the IR end of the integrand. Moreover, this method is technically easier
for higher-loop diagrams or in gauge theories~\cite{aprop,mgierg2,qed,qcd}.

With this in mind, let us use a trick~\cite{aprop}: we can evaluate the differentiated integral by
temporarily working in $\D=4+\epsilon$ since, for \emph{positive} $\epsilon$, the integral is IR finite even if we set $p=0$. Consequently, we must evaluate
\be
	\frac{3}{2}
	\lim_{\epsilon \rightarrow 0^+}
	\totalflow
	\MomInt{\D}{k} \frac{\cutoff^2(k^2/\Lambda^2)}{k^4}
	=
	\frac{3}{2}
	\lim_{\epsilon \rightarrow 0^+}
	\flow
	\left[
	\Lambda^{\epsilon}
	\Omegasl{\D}
	\int_0^\infty du \frac{\cutoff(u^2)}{u^{1-\epsilon}},
	\right]
\ee
where we have defined $u^2 \equiv k^2/\Lambda^2$ and, taking $\Omega_{\D}$ to be the
 area of the $\D$-dimensional unit sphere,
\be
	\Omegasl{\D} \equiv \frac{\Omega_{\D}}{(2\pi)^{\D}}
	=
	\frac{2}{\Gamma(2+\epsilon/2)} \frac{1}{(4\pi)^{\D/2}}
	=
	\frac{2}{(4\pi)^{2}}
	+ \order{\epsilon}.
\label{eq:Omega}
\ee
 Notice that the $\Lambda$-derivative
pulls down a power of $\epsilon$; therefore the only term that will survive the limit $\epsilon \rightarrow 0^+$ is the one for which the integral generates a power of $1/\epsilon$. With this in mind, we
can perform the final step. Let us suppose that the cutoff function starts cutting off modes at a scale,
$\alpha$. (In previous works~\cite{aprop,mgierg2,qed,qcd}, this scale has assumed to be unity corresponding, in dimensionful units, to $\Lambda$. Whilst this seems natural, there is actually no good reason why the cutoff function cannot cutoff modes at some related scale. For example, $e^{-4k^2/\Lambda^2}$ is a perfectly good choice of cutoff function).  Now rescale $u \mapsto u/a$,
so that our expression becomes
\[
	\frac{3}{(4\pi)^2}
	\lim_{\epsilon \rightarrow 0^+}
	\totalflow
	\left[
	\Lambda^{\epsilon} 
	a^{\epsilon}
	\int_0^\infty du \frac{\cutoff(u^2/a^2)}{u^{1-\epsilon}}
	\right].
\]
The cutoff function, $\cutoff(u^2/a^2)$, cuts off modes above $u=1$. Therefore, we can pick out the
$1/\epsilon$ pole of the integral by Taylor expanding the cutoff function, discarding all
terms beyond leading order, so long as we replace the upper limit of the integral with
unity. (In other words, we can think of the cutoff as a sharp cutoff, plus corrections.)
Putting everything together reproduces the standard answer:
\be
	\beta_1 = 
	\lim_{\epsilon \rightarrow 0^+}
	\frac{3}{(4\pi)^2} \epsilon \left[\frac{u^{\epsilon}}{\epsilon}\right]_0^1
	= \frac{3}{(4\pi)^2}.
\label{eq:beta_1-value}
\ee
Thus, all dependence on the non-universal details (seed action and cutoff function) has
cancelled out. Note that we can substitute this expression for $\beta_1$ back into~\eq{simult-a}
or~\eq{simult-b} to find $\gamma_1$. Considering the case where the interaction part of the seed action is set to zero, we find that $\gamma_1 = -\beta_1$, and so $\eta_1 = 0$ [see~\eq{gamma}].
But this result is not universal and so is changed by taking a non-zero seed action.

This might have seemed like a rather long calculation. But what have we really done? 
We wrote out~\eq{beta-convenient} as the one-loop diagrammatic expression~\eq{beta_1-diagrams}. We then noticed that (within perturbation
theory) the only term which survives differentiation \wrt\ $\Lambda$ is the one with a
non-trivial structure in the IR. Given familiarity with the advocated method for evaluating this
term, this is actually an easy calculation.

In preparation for the two-loop calculation let us recall that, even with the dressings of
the effective propagators, the first and last terms in~\eq{beta_1-diagrams} vanish after differentiation
\wrt\ $\Lambda$. 
Consequently, we can throw away contributions of these diagrams to $\flowlam \Pi_2(p)$
(though, as we will see, it will be necessary to retain them elsewhere in the calculation).
However,
for the second term in~\eq{beta_1-diagrams}, we must remember to include the $\order{\lambda}$ piece of the dressed effective propagators as a contribution to $\flowlam \Pi_2(p)$.
Rather than immediately converting~\eq{beta_2} into a diagrammatic
expression for $\beta_2$, we can simplify things by taking account of these points.

Let us  begin by focusing on $\flowlam \Pi_2(p)\bigr\vert_{p^2}$.
Referring to~\eq{Pi-diag}, the contributions at two loops coming from the first, third and fourth
diagrams are killed by the $\Lambda$-derivative. Next let us move on to $\flowlam  \dopiv{4}_2 \bigr\vert_0$, for which we refer to~\eq{D^4bar-diags}. Clearly contributions from the third, penultimate and last diagrams can be thrown away.
So too can contributions from the sixth and seventh diagrams, since the IR structure is trivial
in the sense that the external momenta can be safely set to zero, even before differentiation \wrt\ $\Lambda$. Notice that in the latter
case this is guaranteed by the renormalization condition: the four-point one-loop vertex
must start at $\order{\mom^2}$. Since the external momenta are set to zero, these two
powers of momenta must be loop momenta. This kills any hope of the diagram having an
interesting IR structure.

With these simplifications made, we have:
\begin{multline}
	\beta_2
	=
	\left[
		\left.
		\frac{1}{3}
		\ensuremath{\begin{array}{c}\input{pstex/S4_0-dep3.pstex_t} \end{array}}
		\right\vert_{p^2}
		\left.
		+\frac{3}{4} \ensuremath{\begin{array}{c}\input{pstex/S_04x3-depx4.pstex_t} \end{array}}
		+3 \ensuremath{\begin{array}{c}\input{pstex/S_04x3-depx4-b.pstex_t} \end{array}}
		-\frac{3}{2} 
		\ensuremath{\begin{array}{c}\input{pstex/S_04-depx2-Padlock4_0.pstex_t} \end{array}}
		-\frac{3}{2}
		\left(
			\ensuremath{\begin{array}{c}\input{pstex/S_04-depx2.pstex_t} \end{array}} - \ensuremath{\begin{array}{c}\input{pstex/S_04-depx2-pdress.pstex_t} \end{array}}
		\right)
		\right\vert_0
	\right]^\bullet
\\
	-
	\hf
	\left[
		\left.\ensuremath{\begin{array}{c}\begin{picture}(0,0)%
\epsfig{file=pstex/Padlock2_0.pstex}%
\end{picture}%
\setlength{\unitlength}{3947sp}%
\begingroup\makeatletter\ifx\SetFigFont\undefined%
\gdef\SetFigFont#1#2#3#4#5{%
  \reset@font\fontsize{#1}{#2pt}%
  \fontfamily{#3}\fontseries{#4}\fontshape{#5}%
  \selectfont}%
\fi\endgroup%
\begin{picture}(418,580)(1606,-593)
\put(1768,-446){\makebox(0,0)[lb]{\smash{{\SetFigFont{11}{13.2}{\rmdefault}{\mddefault}{\updefault}{\color[rgb]{0,0,0}$0$}%
}}}}
\end{picture}%
 \end{array}}\right\vert_{p^2}
		+
		\left.
		3
		\ensuremath{\begin{array}{c}\input{pstex/S_04-depx2.pstex_t} \end{array}}
		- \ensuremath{\begin{array}{c}\begin{picture}(0,0)%
\epsfig{file=pstex/Padlock4_0.pstex}%
\end{picture}%
\setlength{\unitlength}{3947sp}%
\begingroup\makeatletter\ifx\SetFigFont\undefined%
\gdef\SetFigFont#1#2#3#4#5{%
  \reset@font\fontsize{#1}{#2pt}%
  \fontfamily{#3}\fontseries{#4}\fontshape{#5}%
  \selectfont}%
\fi\endgroup%
\begin{picture}(433,632)(1594,-645)
\put(1768,-446){\makebox(0,0)[lb]{\smash{{\SetFigFont{11}{13.2}{\rmdefault}{\mddefault}{\updefault}{\color[rgb]{0,0,0}$0$}%
}}}}
\end{picture}%
 \end{array}}
		\right\vert_{0}
	\right]
	\left[
		\left.\ensuremath{\begin{array}{c}\begin{picture}(0,0)%
\epsfig{file=pstex/Padlock2_0.pstex}%
\end{picture}%
\setlength{\unitlength}{3947sp}%
\begingroup\makeatletter\ifx\SetFigFont\undefined%
\gdef\SetFigFont#1#2#3#4#5{%
  \reset@font\fontsize{#1}{#2pt}%
  \fontfamily{#3}\fontseries{#4}\fontshape{#5}%
  \selectfont}%
\fi\endgroup%
\begin{picture}(418,580)(1606,-593)
\put(1768,-446){\makebox(0,0)[lb]{\smash{{\SetFigFont{11}{13.2}{\rmdefault}{\mddefault}{\updefault}{\color[rgb]{0,0,0}$0$}%
}}}}
\end{picture}%
 \end{array}}\right\vert_{p^2}
		+
		\left.
		3
		\ensuremath{\begin{array}{c}\input{pstex/S_04-depx2.pstex_t} \end{array}}
		- \ensuremath{\begin{array}{c}\begin{picture}(0,0)%
\epsfig{file=pstex/Padlock4_0.pstex}%
\end{picture}%
\setlength{\unitlength}{3947sp}%
\begingroup\makeatletter\ifx\SetFigFont\undefined%
\gdef\SetFigFont#1#2#3#4#5{%
  \reset@font\fontsize{#1}{#2pt}%
  \fontfamily{#3}\fontseries{#4}\fontshape{#5}%
  \selectfont}%
\fi\endgroup%
\begin{picture}(433,632)(1594,-645)
\put(1768,-446){\makebox(0,0)[lb]{\smash{{\SetFigFont{11}{13.2}{\rmdefault}{\mddefault}{\updefault}{\color[rgb]{0,0,0}$0$}%
}}}}
\end{picture}%
 \end{array}}
		\right\vert_{0}
	\right]^{\bullet}
	+\order{\lambda},
\label{eq:beta_2-diagrams}
\end{multline}
where the dotted internal lines stand for effective propagators with tree-level
dressing,  \cf~\eq{dep-expansion}; the diagram to which these objects belong is 
designed to subtract off the one-loop contributions from its sister diagram.

In the second line,
it looks like we have kept some terms which vanish after differentiation \wrt\ $\Lambda$.
For example, we expect the $\Lambda$-derivative to kill the first term and third term
in the final square brackets. However, we must be careful, since this bracket is
multiplied by undifferentiated terms. Let us suppose that we work in $\D=4+\epsilon$,
as before.%
\footnote{As in the one-loop case, it is possible to perform the calculation directly in $\D=4$,
whereupon it is found that the $\beta$-function can be expressed as the integral of a total
momentum derivative~\cite{scalar2}. 
This structure is precisely what we would expect from universality, since
the cutoff function is only universal at zero and infinite momentum. Let us note, in passing, that a similar structure has recently been observed in a two-loop calculation in $\mathcal{N}=1$ super Yang--Mills, regularized by covariant higher derivatives~\cite{Stepanyantz-N=1}.} 
Then the $\Lambda$-derivative of the first term and last term in the
final square brackets $\sim \epsilon$. However, the second term in the preceding
brackets goes like $1/\epsilon$, yielding a finite contribution, overall!
[Note, though, that the combination of first (or third) term in the first square brackets
and the first (or third) term in the second square brackets does indeed vanish
in the limit $\epsilon \rightarrow 0$.]

Let us focus on a pair of terms that survives the $\epsilon \rightarrow 0$ limit:
\[
	\frac{3}{2}
	\left\{
	\ensuremath{\begin{array}{c}\begin{picture}(0,0)%
\epsfig{file=pstex/Padlock4_0.pstex}%
\end{picture}%
\setlength{\unitlength}{3947sp}%
\begingroup\makeatletter\ifx\SetFigFont\undefined%
\gdef\SetFigFont#1#2#3#4#5{%
  \reset@font\fontsize{#1}{#2pt}%
  \fontfamily{#3}\fontseries{#4}\fontshape{#5}%
  \selectfont}%
\fi\endgroup%
\begin{picture}(433,632)(1594,-645)
\put(1768,-446){\makebox(0,0)[lb]{\smash{{\SetFigFont{11}{13.2}{\rmdefault}{\mddefault}{\updefault}{\color[rgb]{0,0,0}$0$}%
}}}}
\end{picture}%
 \end{array}}
	\left[
		\ensuremath{\begin{array}{c}\input{pstex/S_04-depx2.pstex_t} \end{array}}
	\right]^\bullet
	+
	\ensuremath{\begin{array}{c}\input{pstex/S_04-depx2.pstex_t} \end{array}}
	\left[
		\ensuremath{\begin{array}{c}\begin{picture}(0,0)%
\epsfig{file=pstex/Padlock4_0.pstex}%
\end{picture}%
\setlength{\unitlength}{3947sp}%
\begingroup\makeatletter\ifx\SetFigFont\undefined%
\gdef\SetFigFont#1#2#3#4#5{%
  \reset@font\fontsize{#1}{#2pt}%
  \fontfamily{#3}\fontseries{#4}\fontshape{#5}%
  \selectfont}%
\fi\endgroup%
\begin{picture}(433,632)(1594,-645)
\put(1768,-446){\makebox(0,0)[lb]{\smash{{\SetFigFont{11}{13.2}{\rmdefault}{\mddefault}{\updefault}{\color[rgb]{0,0,0}$0$}%
}}}}
\end{picture}%
 \end{array}}
	\right]^\bullet
	\right\}_0.
\]
Since the vertices are quasi-local, we are always at liberty to
Taylor expand them in momenta, irrespective of whether
or not we are allowed to set the external momenta to zero along the internal lines. 
From the four-point vertices,
we must take the $\order{\mom^0}$ part: on the one hand,
we are instructed to set all external momenta to zero whereas, on
the other, if we take
any powers of internal momenta, we lose the $1/\epsilon$
keeping these terms alive. In the six-point vertex, we must set
all four external momenta to zero. Recalling that
 the momentum-independent
part of the four-point vertex is just unity, on account of the renormalization
condition, we can thus re-express this set of diagrams as:
\be
	\frac{3}{2}
	\left[
		\ensuremath{\begin{array}{c}\input{pstex/S_04-depx2-Padlock4_0.pstex_t} \end{array}}
	\right]^{\bullet}_0
	+ \order{\epsilon}.
\ee
Notice, then, that this diagram cancels the fourth diagram in~\eq{beta_2-diagrams}
when we take the $\epsilon \rightarrow 0$ limit.

Next let us consider the combination
\be
	\frac{3}{2}
	\left\{
		\left.\ensuremath{\begin{array}{c}\begin{picture}(0,0)%
\epsfig{file=pstex/Padlock2_0-pdress.pstex}%
\end{picture}%
\setlength{\unitlength}{3947sp}%
\begingroup\makeatletter\ifx\SetFigFont\undefined%
\gdef\SetFigFont#1#2#3#4#5{%
  \reset@font\fontsize{#1}{#2pt}%
  \fontfamily{#3}\fontseries{#4}\fontshape{#5}%
  \selectfont}%
\fi\endgroup%
\begin{picture}(418,565)(1603,-586)
\put(1768,-446){\makebox(0,0)[lb]{\smash{{\SetFigFont{11}{13.2}{\rmdefault}{\mddefault}{\updefault}{\color[rgb]{0,0,0}$0$}%
}}}}
\end{picture}%
 \end{array}}\right\vert_{p^2}
		\flowlam
		\left[
			\ensuremath{\begin{array}{c}\input{pstex/S_04-depx2-pdress.pstex_t} \end{array}}
		\right]_0
		+
		\left.
			\ensuremath{\begin{array}{c}\input{pstex/S_04-depx2-pdress.pstex_t} \end{array}}
		\right\vert_0
		\flowlam
		\left[
			\left.\ensuremath{\begin{array}{c}\begin{picture}(0,0)%
\epsfig{file=pstex/Padlock2_0-pdress.pstex}%
\end{picture}%
\setlength{\unitlength}{3947sp}%
\begingroup\makeatletter\ifx\SetFigFont\undefined%
\gdef\SetFigFont#1#2#3#4#5{%
  \reset@font\fontsize{#1}{#2pt}%
  \fontfamily{#3}\fontseries{#4}\fontshape{#5}%
  \selectfont}%
\fi\endgroup%
\begin{picture}(418,565)(1603,-586)
\put(1768,-446){\makebox(0,0)[lb]{\smash{{\SetFigFont{11}{13.2}{\rmdefault}{\mddefault}{\updefault}{\color[rgb]{0,0,0}$0$}%
}}}}
\end{picture}%
 \end{array}}\right\vert_{p^2}
		\right]
	\right\},
\label{eq:combo}
\ee
where we have made a concession to the order in perturbation
theory to which we are working by taking only the tree-level
dressing of the effective propagators. 
The fact that we take the $\order{\mom^2}$ part of
the indicated diagram means that we can re-express this set of
terms as follows:
\be
	3
	\flowlam
	\left[
		\ensuremath{\begin{array}{c}\input{pstex/S_04-pdepx3-S_01.pstex_t} \end{array}}
		+\hf
		\ensuremath{\begin{array}{c}\input{pstex/S_04-pdepx3-padlock2_0-pdress.pstex_t} \end{array}}
	\right]_0 + \order{\epsilon}.
\ee
The reason for the appearance of the first term is as follows. Let us take the loop momentum 
shared by the three internal lines forming a triangle to be $k$.  Now,
\be
	\Pi_1(k) = \ensuremath{\begin{array}{c}\begin{picture}(0,0)%
\epsfig{file=pstex/Sint_1.pstex}%
\end{picture}%
\setlength{\unitlength}{3947sp}%
\begingroup\makeatletter\ifx\SetFigFont\undefined%
\gdef\SetFigFont#1#2#3#4#5{%
  \reset@font\fontsize{#1}{#2pt}%
  \fontfamily{#3}\fontseries{#4}\fontshape{#5}%
  \selectfont}%
\fi\endgroup%
\begin{picture}(371,579)(1629,-672)
\put(1770,-445){\makebox(0,0)[lb]{\smash{{\SetFigFont{11}{13.2}{\rmdefault}{\mddefault}{\updefault}{\color[rgb]{0,0,0}$1$}%
}}}}
\end{picture}%
 \end{array}} + \hf \ensuremath{\begin{array}{c}\begin{picture}(0,0)%
\epsfig{file=pstex/Padlock2_0-pdress.pstex}%
\end{picture}%
\setlength{\unitlength}{3947sp}%
\begingroup\makeatletter\ifx\SetFigFont\undefined%
\gdef\SetFigFont#1#2#3#4#5{%
  \reset@font\fontsize{#1}{#2pt}%
  \fontfamily{#3}\fontseries{#4}\fontshape{#5}%
  \selectfont}%
\fi\endgroup%
\begin{picture}(418,565)(1603,-586)
\put(1768,-446){\makebox(0,0)[lb]{\smash{{\SetFigFont{11}{13.2}{\rmdefault}{\mddefault}{\updefault}{\color[rgb]{0,0,0}$0$}%
}}}}
\end{picture}%
 \end{array}}.
\ee
Since we are working in the massless theory, for which $\Pi_\lambda(0)=0$, 
the zero-momentum contribution of this pair of diagrams must vanish. So,
the first non-trivial contributions come at $\order{k^2}$. There is no such piece from the
first diagram, on account of the renormalization condition. The $\order{k^2}$ contribution
of the second term recovers the original expression~\eq{combo}. Higher order contributions in
momentum vanish in the $\epsilon \rightarrow 0$ limit. Consequently, the combination of
diagrams in~\eq{combo} cancels the
pair of
diagrams in the round brackets in~\eq{beta_2-diagrams},  up to
$\order{\epsilon}$ terms.

As a result of these diagrammatic cancellations, we can write a simple
expression for the $\beta$-function,
\be
	\beta_2 = 
	\left[
		\frac{1}{3}
		\left.\ensuremath{\begin{array}{c}\input{pstex/S4_0-pdep3.pstex_t} \end{array}}\right\vert_{p^2}
		\left.
			+\frac{3}{4} \ensuremath{\begin{array}{c}\input{pstex/S_04x3-pdepx4.pstex_t} \end{array}}
			+3 \ensuremath{\begin{array}{c}\input{pstex/S_04x3-pdepx4-b.pstex_t} \end{array}}
			-\frac{9}{4}
			\ensuremath{\begin{array}{c}\input{pstex/S_04-depx2-pdress.pstex_t} \end{array}}
			\ensuremath{\begin{array}{c}\input{pstex/S_04-depx2-pdress.pstex_t} \end{array}}
		\right\vert_0
	\right]^{\bullet} + \order{\epsilon},
\label{eq:beta_2-diags-final}
\ee
where we have now explicitly discarded all pieces
which are too high order in $\lambda$.
This coincides with the expression obtained in~\cite{scalar2}. But let us emphasise
once again that whilst this expression took many pages to obtain in~\cite{scalar2},
here we were able to start the analysis with~\eq{beta_2-diagrams}, eliminating
almost all of the hard work!

To evaluate the first term, which we will denote by $\beta_2^{(1)}$,
let us route momenta such that the three internal
lines carry $k$, $l+k$ and $l+p$:
\[
	\beta_2^{(1)} =
	\frac{1}{3}
	\left[
	\MomInt{\D}{k} \MomInt{\D}{l}
	\frac{\cutoff(k^2/\Lambda^2) \cutoff(l^2/\Lambda^2) }{k^2 (l+k)^2 (l+p)^2}
	\right]^\bullet_{p^2} + \order{\epsilon},
\]
where the $\order{\epsilon}$ term arises from cutoff functions we have thrown away
and the tree-level dressing of the effective propagator (note that we have anticipated that this diagram will turn out to be IR finite after differentiation \wrt\ $\Lambda$). Since we are working in
$\D=4+\epsilon$, the $\order{p^2}$ contribution can be picked out by Taylor expanding,
since the resulting IR divergence---which is ultimately killed when we take 
the $\Lambda$-derivative---is regularized at intermediate steps. It is well worth noting
that an IR divergence of this type is really a \emph{pseudo} divergence, appearing as it
does only as a result of the way we choose to do the calculation.
Thus we are left with:
\begin{multline*}
	\beta_2^{(1)}=
	\frac{1}{3}
	\left[
	\MomInt{\D}{k} \MomInt{\D}{l}
	\frac{\cutoff(k^2/\Lambda^2) \cutoff(l^2/\Lambda^2) }{k^2 (l+k)^2 l^2}
	\left(\frac{4 (l.p)^2 }{l^4} -\frac{p^2}{l^2}\right)
	\right]^\bullet_{p^2} + \order{\epsilon}
\\
	=
	\frac{1}{3}
	\left(\frac{4}{\D} -1\right)
	\left[
	\MomInt{\D}{k} \MomInt{\D}{l}
	\frac{\cutoff(k^2/\Lambda^2) \cutoff(l^2/\Lambda^2) }{k^2 (l+k)^2 l^4}
	\right]^\bullet + \order{\epsilon},
\end{multline*}
where we have exploited Euclidean invariance to replace $l_\mu l_\nu \rightarrow l^2/\D\, \delta_{\mu\nu}$, under the $l$ integral.

To proceed, we use another trick~\cite{mgierg2}. By inspection, the $l$-integral is UV finite even
in the absence of the cutoff function but has an IR divergence which turns out to be dimensionally regularized. (The latter statement is most obvious if we do the $k$-integral first.)
Suppose that we are interested only in the contribution to the term as a whole
coming from this IR divergence (it turns out that this contribution is the only one which survives the $\epsilon \rightarrow 0^+$ limit).  Then when we throw away the cutoff function we can leave the range of the $l$-integration \emph{unrestricted}. Remember: the $l$-integral is, by lucky hap, regularized whether or not the cutoff function is there. The point of this is that the $l$-integral is much easier to evaluate taking
this course of action. Differences between this approach and restricting the range of integration are sub-leading.

Focusing just on the $l$-integral, we combine denominators using the Feynman
parameter, $\alpha$, then we shift $l \mapsto l -\alpha k$ and finally
perform the resulting integral using dimensional regularization (see \eg~\cite{P+S}):
\be
\label{eq:two-loop-inner}
\begin{split}
	\MomInt{\D}{l}
	\frac{1}{(l+k)^2 l^4}
	&=
	2\int_0^1 d\alpha (1-\alpha) \MomInt{\D}{l}
	\frac{1}{\left[l^2 + k^2 x(1-x) \right]}
\\
	&=
	\frac{\Gamma(1-\epsilon/2)}{(4\pi)^{\D/2}} 
	\int_0^1 (1-x)^{\epsilon/2} x^{-1+\epsilon/2} \frac{1}{k^{2(1-\epsilon/2)}}
\\
	&=
	\frac{1}{(4\pi)^{\D/2}} \frac{1}{k^{2(1-\epsilon/2)}} 
	\frac{
		\Gamma(\epsilon/2)\Gamma(1+\epsilon/2)\Gamma(1-\epsilon/2)
	}{\Gamma(1+\epsilon)}.
\end{split}
\ee
Finally, we perform the integral over $k$, which we do just as in the one loop case (though we will not bother to go through the procedure of rescaling to ensure that the cutoff function cuts off modes at the scale $\Lambda$: having seen how this works already, here we will just assume that the cutoff function is already of this type).
First we change to the dimensionless variable, $u^2 \equiv k^2/\Lambda^2$, and
then we drop the cutoff function whilst restricting the range of the radial integral
to unity:
\be
	\beta_2^{(1)}
	\equiv
	\frac{1}{3}
	\left[\ensuremath{\begin{array}{c}\input{pstex/S4_0-pdep3.pstex_t} \end{array}}\right]^\bullet_{p^2}
	=
	\frac{\epsilon}{12} \frac{\Omegasl{4}}{(4\pi)^2} \frac{2}{\epsilon}
	\totalflow \Lambda^{2\epsilon} \int_0^1 \frac{du}{u^{1-2\epsilon}}
	+\order{\epsilon}
	=
	\frac{1}{3}
	\frac{1}{(4\pi)^4}
	+\order{\epsilon}.
\label{eq:beta_2^(1)}
\ee
As anticipated, $\beta_2^{(1)}$ is IR finite, justifying that the terms we threw away various stages
do indeed vanish in the limit $\epsilon \rightarrow 0$.

The remaining three terms in~\eq{beta_2-diags-final}, which we will collectively denote by $\beta_2^{(2)}$, must be evaluated together. Notice that
each of these diagrams, including the second, has at least one copy of the same 
one-loop, four-point sub-diagram. Indeed, we can write the second term of this set as
\be
	3 \left[\ensuremath{\begin{array}{c}\input{pstex/S_04x3-pdepx4-b.pstex_t} \end{array}}\right]^{\bullet}_0
	=
	3\left[ \MomInt{\D}{l} \frac{\cutoff(l^2)}{l^4}
	\hspace{1em}
	\ensuremath{\begin{array}{c}\input{pstex/S_04-depx2-pdress-route.pstex_t} \end{array}} 
	\right]^{\bullet}+ \order{\epsilon},
\label{eq:beta_2-a}
\ee
where the little zeros indicate that the vertices are to be Taylor expanded
to zeroth order in their external momenta. Two of these lines are external
to the diagram as a whole, whereas two are internal to the diagram as a whole.
These latter two carry $\pm l$. If we take non-zero powers of $l$ from these
vertices, then the diagram as a whole loses all interesting IR structure and
vanishes after differentiation \wrt\ $\Lambda$.
However, we can take any number of powers of the
momentum, $k$, which is internal to the sub-diagram. Suppose that we do take
such contributions. Although this means that we do not take the most
IR divergent possible contribution to the diagram as a whole, such
terms do survive even after differentiation \wrt\ $\Lambda$: the divergence
carried by the integral over $l$ is enough to ensure this. Note, though, that
if we were ever to kill the divergence in the $l$-integral, then the diagram as a whole
only contributes at $\order{\epsilon}$. [Such has been the fate of the tree-level
dressings of the $\ep(l^2/\Lambda^2)$.]

We can sum up the contributions coming from the last three terms
in~\eq{beta_2-diags-final} in which there is a divergence in one sub-diagram
(such that the diagram as a whole survives the $\epsilon \rightarrow 0$ limit)
but the divergence in the other is killed by taking too many powers of momentum.
There are two ways of doing this in the first and third diagrams and one in the second.
The sum of these contributions is zero:
\[
	\frac{3}{4} \times 2 + 3 - \frac{9}{4} \times 2 = 0.
\]
Consequently, the only surviving terms from the sum of these three
diagrams arise when we take no extra powers of momentum from
any of the vertices, nor any from the internal lines.

Temporarily retaining those cutoff functions necessary to ensure UV regularization
we have:
\be
	\beta_2^{(2)}
	=
	\frac{3}{2}
	\MomInt{\D}{k} \MomInt{\D}{l}
	\left[
		2 \frac{\cutoff(k^2/\Lambda^2)}{k^2 (l-k)^2 l^4}
		-
		\frac{\cutoff(k^2/\Lambda^2)}{k^4} \frac{\cutoff(l^2/\Lambda^2)}{l^4} 
	\right]^\bullet
	+\order{\epsilon}.
\label{eq:beta_2^(2)-start}
\ee
We have computed both of these terms already, the first in
the two-loop calculation leading to~\eq{beta_2^(1)}, and the second from the 
one-loop calculation leading to~\eq{beta_1-value}.
This time, we need to keep
the sub-leading terms in $\epsilon$. 

It is worth pausing on this point. In the earlier two-loop calculation of
$\beta_2^{(1)}$, the term came with an overall factor of $4/\D -1 \sim \epsilon$.
Here, this is not the case, and so even after differentiation \wrt\ $\Lambda$,
there will be a $1/\epsilon$ left over. Of course, this will cancel against
a $1/\epsilon$ coming from the other term. Nevertheless, 
we might worry that we can no longer play the trick of leaving the range of
the $l$-integral unrestricted in the second term of~\eq{beta_2^(2)-start}. However,
corrections from doing so are of the type which we have already argued cancel
between the three diagrams contributing to $\beta_2^{(2)}$ (see also~\cite{mgierg2,scalar2}).
Keeping track of the sub-leading terms which do not cancel by this mechanism gives:
\[
	\beta_2^{(2)}
	=
	-\frac{3\Omegasl{\D}}{2\epsilon}
	\totalflow \Lambda^{2\epsilon}
	\left[
		\frac{1}{(4\pi)^{\D/2}}
		\frac{
			\Gamma(\epsilon/2)\Gamma(1+\epsilon/2)\Gamma(1-\epsilon/2)
		}{\Gamma(1+\epsilon)}
		-
		\frac{\Omegasl{\D}}{\epsilon}
	\right] + \order{\epsilon}.
\]
We can evaluate $\beta_2^{(2)}$ by utilizing
the following expressions for the $\Gamma$ function:
\be
	\Gamma(\epsilon/2) = \frac{2}{\epsilon} - \emc + \order{\epsilon},
	\qquad
	\Gamma(1+\epsilon) =  1-\emc \epsilon,
	\qquad
	\Gamma(2+\epsilon/2) = 1 -\emc\epsilon/2 + \epsilon/2,
\ee
where $\emc$ is the Euler-Mascheroni constant. Noticing that the second of these
expressions implies that $\Gamma(1+\epsilon/2)\Gamma(1-\epsilon/2) = 1 + \order{\epsilon^2}$,
we have:
\begin{multline}
	\beta_2^{(2)}
	=
	 -3 \frac{\Omegasl{\D}}{(4\pi)^{\D/2}}
	 \left[
	 	\frac{\Gamma(\epsilon/2)}{\Gamma(1+\epsilon)} -
		\frac{2}{\epsilon} \frac{1}{\Gamma(2+\epsilon/2)}
	 \right] + \order{\epsilon}
\\
	 =
	-\frac{6}{(4\pi)^4}
	\left[
		\left(\frac{2}{\epsilon} - \emc\right) (1 + \emc \epsilon)
		- \frac{2}{\epsilon} \left(1 + \emc \frac{\epsilon}{2} - \frac{\epsilon}{2} \right)
	\right] + \order{\epsilon}
	=
	-\frac{6}{(4\pi)^4} + \order{\epsilon}.
\label{eq:beta_2^(2)}
\end{multline}

Adding together~\eqs{beta_2^(1)}{beta_2^(2)}, and taking the limit $\epsilon \rightarrow 0^+$,
we recover the standard result:
\be
	\beta_2 = -\frac{17}{3} \frac{1}{(4\pi)^4}.
\label{eq:beta_2-ans}
\ee

In the context of more standard ways of computing the $\beta$-function, where dimensional
regularization might be used to pick out \emph{UV} divergent contributions, from which
the $\beta$-function is determined, our approach has a perverse appeal: for we have
arranged our calculation such that
dimensional regularization is used to pick out \emph{IR} divergences, and it is these which
determine the $\beta$-function!

Let us conclude this section by commenting on a possible source of confusion. It follows from
the analysis of \sect{heat} that we expect the $\Lambda \rightarrow 0$ limit of the dual action to kill all diagrams
possessing an internal line. In this section, however, we have seen that loop integrals
generate contributions
to the dual action which (in $\D=4$) go  like $p^2 \ln p^2/\Lambda^2$ and
which thus seem to \emph{diverge} in this limit. The point is that the $\Lambda\rightarrow 0$
behaviour of the order $p^2 \times \mbox{non-polynomial}$ pieces of a function
are not necessarily diagnostic of the behaviour of the function as a whole. This
is amply illustrated by considering \eg\ $1/(1+p^2 \ln p^2/\Lambda^2)$.

\subsection{The Scaling Field Method}
\label{sec:SFM}

In this section, we will take a rather different approach to computing the
$\beta$-function. Having classified the eigenoperators in the vicinity of the Gaussian
fixed-point by linearizing the flow equation as in \sects{GFP}{Gen-TP}, we will now
identify $\lambda$ as the coupling in front of $\mathcal{G}_{4,0}$. Actually, because this is a different
definition of the coupling from the one used in the last section, we will call it $\tilde{\lambda}$.
As discussed
in \sect{renorm}, this is a perhaps a rather natural definition in the context of the ERG, 
if somewhat more awkward to work with than the  definition used in the previous section.
As the flow develops, the $\beta$-function is computed by considering how the non-linear
term in the flow equation generates contributions to  $\mathcal{G}_{4,0}$. This is the `Scaling Field Method' of Golner \& Riedel~\cite{Golner+Riedel-75,Golner+Riedel-76,RGN}
(see also~\cite{Wegner_CS} and~\cite{H+H}).

Compared to the previous section, our technology is rather less sophisticated.
We will take $\hSint=0$, since in this case we know the form of the  
$\mathcal{G}_{n,r}$. In fact, we will take the simplest representative of
the Gaussian fixed-point, $\Sint_\star =0$, so that the $\mathcal{G}_{n,r}$
reduce to the simpler $\mathcal{G}'_{n,r}$ of~\eq{SimpleGaussian-eop}.
A special role will be played by $\mathcal{G}'_{4,0}[\phi]$ which we will write
as just $H[\phi]$, for brevity.
Moreover, we will work to just one loop, since this is sufficient to
get the idea. Also, we will return to our completely rescaled flow equation~\eq{Ball},
\be
	\left(
		\partial_t + d_\phi \Count_\phi + \Count_\partial -\D
	\right) \Sint
	=
	\classical{\Sint}{\cutoff'}{\Sint} - \quantum{\cutoff'}{\Sint} - \frac{\eta}{2} \phi \cdot \ep^{-1} \cdot \phi.
\label{eq:Ball-redux^2}
\ee
(We will \emph{not} additionally rescale the field by $\sqrt{\tilde{\lambda}}$ in this section.)

The game now is to consider a perturbation of the Gaussian fixed-point in the
$ H[\phi]$ direction. This operator is, of course, marginal and so satisfies
\be
	 \left(d_\phi \Count_\phi + \Count_\partial -\D
	 +
	 \quantum{K'}{}\right) H[\phi]= 0.
\label{eq:G'_(4,0)}
\ee
This is the result of linearizing the flow equation about a fixed-point. Beyond linear order, we
go along the lines of~\eq{T_t} and write
\be
	\Sint_t[\phi] = \Sint_\star[\phi] + \mathscr{P}_t[\phi],
	\qquad
	\mathscr{P}_t[\phi] = \tilde{\lambda}(t)H[\phi]  + \sum_i \mu_i(t) \eop_i[\phi],
\label{eq:quadratic-flow}
\ee
where the sum runs over all operators besides the one that has been singled out. The coupling
$\tilde{\lambda}$ is considered to be linear in the perturbation about the fixed-point, whereas the other 
couplings---and $\eta(t)$---are quadratic in the perturbation. As mentioned above, we
will take $\Sint_\star = 0$.

Now we substitute~\eq{quadratic-flow} into~\eq{Ball-redux^2}, using~\eq{G'_(4,0)}.
Focusing just on
the contributions to $H[\phi]$, and discarding terms which only contribute
beyond quadratic order yields:
\be
	H[\phi]\partial_t \tilde{\lambda}
	=
	\tilde{\lambda}^2
	\left.\classical{H}{K'}{H}\right\vert_{H}.
\label{eq:quadratic-flow-H}
\ee
To extract the contributions to $H$ coming from the \rhs, we operate on this equation
with $e^{\op}$. Recalling the diagrammatic notation of \fig{eigen} note that
\be
	e^{\op} H[\phi]
	=
	\frac{1}{4!} \ensuremath{\begin{array}{c}\input{pstex/Eigen_4-v_0.pstex_t} \end{array}},
\ee
where $v_0$ has no momentum dependence \ie\ is just a constant.

To process the \rhs\ of~\eq{quadratic-flow-H}, we notice that
\be
	\classical{H}{K'}{H}
	=
	\frac{1}{3!3!}
	\
	\ensuremath{\begin{array}{c}\input{pstex/v_0x2-Kpr.pstex_t} \end{array}}
	-\frac{1}{3!}
	\
	\ensuremath{\begin{array}{c}\input{pstex/v_0-K-padlock.pstex_t} \end{array}}
	\hspace{1em}
	+\frac{1}{4}
	\ensuremath{\begin{array}{c}\input{pstex/padlockx2-K.pstex_t} \end{array}},
\ee
where the dashed lines denote instances of $K'$. Operating on this with $e^{\op}$ yields:
\begin{multline}
	e^{\op}
	\classical{H}{K'}{H}
	=
\\
	\frac{1}{3!3!}
	\
	\ensuremath{\begin{array}{c}\input{pstex/v_0x2-Kpr.pstex_t} \end{array}}
	+
	\hspace{1em}
	\frac{1}{4}
	\ensuremath{\begin{array}{c}\input{pstex/v_0x2-ep-K.pstex_t} \end{array}}
	+
	\left(
		\frac{1}{3!} - \frac{1}{3!}
	\right)
	\ensuremath{\begin{array}{c}\input{pstex/v_0-K-padlock.pstex_t} \end{array}}
	\hspace{1em}
	+\mbox{two-point terms}.
\end{multline}

Due to the cancellation of the second four-point term we find that the only contribution
to $e^{\op} H$ comes from the second---rather familiar looking---term.
Indeed, \eqn{quadratic-flow-H} becomes:
\be
	-\partial_t \tilde{\lambda}^{-1}
	=6v_0^2
	\MomInt{4}{k} \frac{\cutoff(k^2)}{k^2} \der{\cutoff(k^2)}{k^2}
	=
	6v_0^2 \Omegasl{4}
	\int_0^\infty \frac{dk^2}{2}
	\hf
	\der{\cutoff^2(k^2)}{k^2}
	=
	-\frac{3 v_0^2}{(4\pi)^2}.
\ee
(Remember that our momenta are dimensionless in this section, so the cutoff function
just depends on $k^2$.)
As discussed earlier, we are free to normalize the eigenoperators however we
choose, and we will take $v_0 = 1$, ensuring that $\lambda$ and $\tilde{\lambda}$
agree at the classical level. Noting that $\partial_t \tilde{\lambda} = -\tilde{\beta}$,
we get agreement with our earlier calculation:
\be
	\tilde{\beta}_1 = \frac{3}{(4\pi)^2}.
\ee

\subsection{The Wilson-Fisher Fixed-Point}
\label{sec:WF}

It is irresistible, particularly given some of the work that we have already done, 
to briefly discuss the $\epsilon$-expansion and use it to
find the celebrated Wilson-Fisher fixed-point~\cite{W+F-3.99}. 
(For a historical perspective on the birth of the $\epsilon$-expansion
and further references, see section~XI of~\cite{Fisher-Rev}.)
To provide some
novelty, we will make use of the dual action formalism.
The basic idea is to consider a $\phi^4$-type theory where 
both the four-point coupling and $\epsilon=4-\D>0$ to be small (this is a
slightly
different definition of $\epsilon$ compared to the one used in
\sect{beta-canonical}, where we took $\epsilon= \D-4>0$). 
With this in
mind, we will analyse the two-point and four-point contributions to the dual
action which, at a fixed-point, follow as solutions to~\eq{dual-FP}.

We recall from~\eq{D2-solution} together with~\eqs{h-lessthan}{c} that, for a critical fixed-point with $\eta_\star <2, \ \neq 0$,
\begin{align*}
	\dualv{2}_\star(p) & = -\intconst_{\eta_\star} p^{2(1+\eta_\star/2)}
	+ \ep^{-1}(p^2) [1+\varrho(p^2)],
\\
	\varrho(p^2) & \equiv - p^{2(\eta_\star/2)} \cutoff (p^2) \int_0^{p^2} dq^2
	\left[
		\frac{1}{\cutoff (q^2)}
	\right]'
	q^{-2(\eta_\star/2)},	
\end{align*}
where, for a given fixed-point, $\intconst_{\eta_\star}$ is an integration constant
labelling the line of equivalent fixed-points.
From these equations, we deduce that
\be
	\Pi_\star(p)
	\equiv
	\frac{\dualv{2}_\star(p)}{
		1 - \ep(p^2) \dualv{2}_\star(p)
	}
	=
	\frac{1}{\intconst_{\eta_\star}} p^{2(1-\eta_\star/2)} - p^2 + \cdots.
\label{eq:Pi_star-exp}
\ee

Now let us move on to the four-point level, where
\eq{dual-FP} tells us that $\dualv{4}_\star$ satisfies:
\be
	\left(-\epsilon -2\eta_\star + \sum_{i=1}^4 p_i \cdot \pder{}{p_i}\right)
	\dualv{4}_\star(p_1,p_2,p_3,p_4) = 0.
\label{eq:dual_4-eqn}
\ee
We would now like to see what~\eqs{Pi_star-exp}{dual_4-eqn}
tell us about the 1PI vertex $\dopiv{4}_\star$. To this end, let us recall
that
\be
	\dualv{4}_\star(p_1,p_2,p_3,p_4)
	=
	\frac{\dopiv{4}_\star(p_1,p_2,p_3,p_4)}{
	\prod_{i=1}^4
	\left[
		1+ \ep(p_i) \Pi_\star(p_i)
	\right]
	}.
\ee
For small momenta, the denominator contains leading contributions of the
form $p_i^{2(\eta_\star/2)}$. When these are hit by the momentum derivatives
in~\eq{dual_4-eqn}, factors of $\eta_\star$ will be pulled down. Since this
is meant only to be an illustrative calculation, let us make life easy for ourselves
by utilizing the fact that we expect $\eta_\star = \order{\epsilon^2}$. This allows
us to deduce from~\eq{dual_4-eqn} that:
\be
	\lim_{p_i\rightarrow 0, \epsilon \rightarrow 0} 
	\left(-\epsilon + \sum_{i=1}^4 p_i \cdot \pder{}{p_i}\right)
	\dopiv{4}_\star(p_1,p_2,p_3,p_4) = 0.
\label{eq:dopi_4-leading}
\ee
Of course, there is no need to throw away the $\eta_\star$ terms at this stage;
if we kept them in we would simply end up determining that $\eta_\star = \order{\epsilon^2}$.
From~\eq{dopi_4-leading}, we see that $\dopiv{4}_\star(p_1,p_2,p_3,p_4)$ must have
non-polynomial dependence on its momenta.

With this in mind, the next step in our strategy is to examine the diagrammatic expression for
$\dopiv{4}_\star$. We have essentially done this already in our first computation
of the $\beta$-function, but this time we would like to keep the external momenta
non-zero:
\be
	\dopiv{4}_\star(p_1,p_2,p_3,p_4)
	=
	\ensuremath{\begin{array}{c}\input{pstex/Sint4_star.pstex_t} \end{array}}
	-\frac{1}{4}
	\left(
		\ensuremath{\begin{array}{c}\input{pstex/Sint2_starx2-epx2.pstex_t} \end{array}}
		+ \mathrm{permutations}
	\right)
	+\cdots
\label{eq:dopi^4_star-diagrams}
\ee
where there are a total of $\nCr{4}{2} = 6$ independent diagrams included inside
the brackets (following from all independent ways of arranging the external momenta).
The important point about the second diagram is that it is (within our approximation scheme
of taking the four-point coupling to be small) the first term in the expansion of
$\dopiv{4}_\star$ which can generate non-polynomial dependence on the external
momenta. Indeed, we could immediately deduce what this dependence must be,
from our calculation of the $\beta$-function. But let us do an independent calculation,
to show explicitly how everything hangs together. 

Since we are interested in the leading behaviour for small external momenta, we
can Taylor expand the vertices to zeroth order in their momenta; we will denote
this component of the vertices by $w_\star$ (and not by $\lambda_\star$, as we might have expected). 
Unlike the calculation of the one and two-loop $\beta$-function performed earlier,
it is important that we do not throw away the tree-level dressings of the internal
lines. This is because, in the current case, we have not canonically normalized
our kinetic term. It is rather instructive to leave the kinetic term alone and
so we will do so. In actual fact, the easiest way to proceed is to substitute directly
for the \emph{completely} dressed internal lines, seeing as we have a formula
for them in terms  of $\eta_\star$:
\be
	\dep(p^2)
	=
	\frac{\ep(p^2)}{
		1 +\ep(p^2) \Pi_\star(p)
	},
\ee
where $\Pi_\star$ is given by~\eq{Pi_star-exp}. Now, since we are supposing that
$\eta_\star \sim \order{\epsilon^2}$, we have that
\be
	\dep(p^2)
	= \intconst_{\eta_\star} \frac{\cutoff(p^2)}{p^2} + \order{\epsilon^2}.
\ee
Thus, up to terms which are sub-leading in $\epsilon$, we are led to evaluate
\[
	-\frac{\intconst_{\eta_\star}^2 w_\star^2}{4} \MomInt{\D}{k} \frac{1}{k^2 (k+p_1+p_2)^2},
\]
and its five friends involving different combinations of the momenta $p_1,\ldots,p_4$. (We have discarded all cutoff functions since, as we are in $\D=4-\epsilon$, the integral is UV regularized without them.)
Rewriting the denominator using the Feynman parameter, $\alpha$, we have:
\[
	-\frac{\intconst_{\eta_\star}^2 w_\star^2}{4}
	\int_0^1 d\alpha \MomInt{\D}{k}  \frac{1}{\left[k^2 + \alpha(1-\alpha) (p_1+p_2)^2\right]^2}
	=
	-\frac{\intconst_{\eta_\star}^2 w_\star^2}{32\pi^2 \epsilon} 
	\left[
		(p_1+p_2)^{-2\epsilon/2} -1 +\order{\epsilon}
	\right].
\]
Substituting this expression into~\eq{dopi^4_star-diagrams} yields:
\be
	\dopiv{4}_\star(p_1,p_2,p_3,p_4)
	= w_\star -
	\frac{\intconst_{\eta_\star}^2 w_\star^2}{32\pi^2 \epsilon} 
	\biggl\{
	\left[(p_1+p_2)^{-2\epsilon/2}-1 \right] + \mathrm{permutations}  + \cdots
	\biggr\}
	+ \cdots,
\label{eq:dopi^4_star-epsilon}
\ee
where the first ellipsis includes terms higher order in $\epsilon$ coming from the associated
terms, and the second ellipsis includes additional terms higher order in momenta and/or $\epsilon$.
As we will find, $w_\star \sim \epsilon$, so the terms represented by both ellipses---including the one in the curly bracket
which is \naively\ multiplied by $1/\epsilon$---are sub-leading.
Substituting~\eq{dopi^4_star-epsilon} into~\eq{dopi_4-leading} yields:
\be
	-\epsilon
	\left\{
		w_\star -
		\frac{\intconst_{\eta_\star}^2 w_\star^2}{32\pi^2 \epsilon} 	
			\left[2(p_1+p_2)^{-2\epsilon/2} - 1+ \mathrm{permutations} \right] 
	\right\}
	 + \cdots
	=0
\ee
where, again, the ellipsis denotes terms higher order in momenta and/or $\epsilon$.
Expanding $(p_1+p_2)^{-\epsilon} = 1 + \order{\epsilon}$. The non-trivial solution to
this equation is:
\be
	w_\star= \frac{(4\pi)^2\epsilon}{3 \intconst_{\eta_\star}^2}  + \order{\epsilon^2}.
\label{eq:w_star}
\ee
Let us note, at this stage, that it seems rather natural to make the following definition:
$\lambda_\star = w_\star \intconst_{\eta_\star}^2$, but this seems to be more a matter
of labelling than anything profound.

Now we move to the two-point level, where we have the familiar diagrammatic
expansion
\be
	\Pi_\star(p)  =
	\ensuremath{\begin{array}{c}\begin{picture}(0,0)%
\epsfig{file=pstex/ReducedWEA-2.pstex}%
\end{picture}%
\setlength{\unitlength}{3947sp}%
\begingroup\makeatletter\ifx\SetFigFont\undefined%
\gdef\SetFigFont#1#2#3#4#5{%
  \reset@font\fontsize{#1}{#2pt}%
  \fontfamily{#3}\fontseries{#4}\fontshape{#5}%
  \selectfont}%
\fi\endgroup%
\begin{picture}(358,579)(1629,-672)
\put(1730,-448){\makebox(0,0)[lb]{\smash{{\SetFigFont{11}{13.2}{\rmdefault}{\mddefault}{\updefault}{\color[rgb]{0,0,0}$\Sint$}%
}}}}
\end{picture}%
 \end{array}} + \frac{1}{2} \ensuremath{\begin{array}{c}\begin{picture}(0,0)%
\epsfig{file=pstex/Padlock-2-pdress.pstex}%
\end{picture}%
\setlength{\unitlength}{3947sp}%
\begingroup\makeatletter\ifx\SetFigFont\undefined%
\gdef\SetFigFont#1#2#3#4#5{%
  \reset@font\fontsize{#1}{#2pt}%
  \fontfamily{#3}\fontseries{#4}\fontshape{#5}%
  \selectfont}%
\fi\endgroup%
\begin{picture}(418,580)(1606,-593)
\put(1727,-457){\makebox(0,0)[lb]{\smash{{\SetFigFont{11}{13.2}{\rmdefault}{\mddefault}{\updefault}{\color[rgb]{0,0,0}$\Sint$}%
}}}}
\end{picture}%
 \end{array}} 
	-\frac{1}{6} \ensuremath{\begin{array}{c}\input{pstex/TP-TL-pdress.pstex_t} \end{array}} + \cdots
	=
	\frac{1}{\intconst_{\eta_\star}} p^{2(1-\eta_\star/2)}  + \cdots
	=
	-\frac{\eta_\star}{\intconst_{\eta_\star}} \frac{p^2 \ln p^2}{2}
	+ \cdots,
\label{eq:Pi_star-almost}
\ee
and we have used the result that $p^{-2\eta_\star/2} = 1- \eta_\star/2\, \ln p^2 + \cdots$.
It should come as no surprise that we look to the third diagram to
generate (at the current order of approximation) the non-polynomial
term:
\begin{multline}
	-\frac{\intconst_{\eta_\star}^3}{6}
	\MomInt{\D}{k}\MomInt{\D}{l} \frac{\cutoff(l^2) \cutoff(k^2)}{k^2 (l+k)^2 (l+p)^2}	
	=
	\frac{\intconst^3_{\eta_\star} w_\star^2}{6(4\pi)^4} \frac{p^2\left(p^{-2\epsilon/2} -1\right)}{\epsilon}
	+\cdots
\\
	=
	-\frac{\epsilon^2}{54 \intconst_{\eta_\star}} \frac{p^2 \ln p^2}{2} + \cdots,
\label{eq:non-poly-epsilon}
\end{multline}
where we have substituted for $w_\star$ using~\eq{w_star}.
Comparing~\eqs{Pi_star-almost}{non-poly-epsilon}, it is immediately
apparent that
\be
	\eta_\star = \frac{\epsilon^2}{54},
\label{eq:WFFP}
\ee
which is the standard result~\cite{Wilson-D<4}. 
Notice how $\intconst_{\eta_\star}$ cancelled out, as it had to.
It is interesting to point out that by taking the internal lines
to be fully dressed, rather than dressed at just tree-level,
we are in some sense working beyond $\order{\epsilon^2}$.
We cannot see this in the final answer~\eq{WFFP} because
we assumed that $\eta_\star \sim \order{\epsilon^2}$ from
the start and threw away instances of $\eta_\star$ whenever
they were sub-leading. Had we kept them in then we would
presumably
find that~\eq{WFFP} would receive corrections to all orders
in $\epsilon$. Beyond $\order{\epsilon^2}$ these would not,
of course, be the complete contributions, since we terminated
the diagrammatic expansion for $\Pi_\star$ at the third term.
Nevertheless, this suggests a way of improving the $\epsilon$
expansion which merits further investigation.

\section{Nonperturbative Truncations}
\label{sec:Truncations}

In this section, we will describe some intrinsically nonperturbative truncation schemes supported by the ERG. After an overview in \sect{Trunc-Over}, we give further details of the famous derivative expansion in \sect{Trunc-DE}. As part of this we recall in \sect{cfn} how, within the lowest order of the derivative expansion, it is possible to construct a function which decreases monotonically along the flow; for flows between fixed-points, at any rate, this functions shares important properties with Zamolodchikov's $c$-function~\cite{Zam-cfn}. Finally, in \sect{Trunc-Opt} we discuss some of the issues associated with optimizing truncation schemes.

\subsection{Overview}
\label{sec:Trunc-Over}

If any of the menagerie of flow equations could be solved, in generality, this would amount to a complete
solution of the QFT in question. Actually, this is an even stronger statement
than it may first appear (and even at a first glance it is rather strong!). Solving the flow equation would mean more than
solving the theory corresponding to one particular type of bare action. A general
solution of the flow equation would yield all trajectories in theory space and so would
amount to a solution of all possible theories with the given field content! Surely,
then, it is not possible to exactly solve the flow equation.
(Modulo the interesting twist to this argument  discussed in \sect{heat}.)

An obvious question to ask is whether the simpler, fixed-point equation can be exactly solved.
This would yield the complete set of fixed-points (critical or otherwise) of the system in question; 
unsurprisingly, it is only known how to find the simplest fixed-points, analytically. The intractability of
the flow equation might seem rather problematic since, in general, there is no small parameter present in the ERG equation with which to perform some type of perturbation theory. Of course, there
are exceptions: notably perturbation theory in the case where a $\lambda \phi^4$ theory is
considered with small $\lambda$, the $\epsilon$-expansion and (for $N$-component theories) the $1/N$ expansion. The first two have been discussed \sect{beta}; a review of the $1/N$
expansion in QFT can be found in~\cite{Zinn-LargeN}. All of these method are discussed in the
context of QFT and critical phenomena in Zinn-Justin's book of the same name~\cite{Z-J}.
For a particularly clear analysis of the how various flow equations simplify in the
large-$N$ limit, see~\cite{TRM-LargeN}.

In this section we describe one of the particular strengths of the ERG approach: specifically, that it is amenable to various approximations which are intrinsically nonperturbative (whether or not at a fixed-point). The basic idea behind all of these schemes is to \emph{truncate} the
space of allowed interactions, so that $S_\Lambda$ is constrained to some hypersurface in the space
of all possible $S_\Lambda$s. All terms generated by the flow equation which are outside of the truncation scheme are simply discarded. It is, perhaps needless to say, very difficult to assess the errors in such a procedure. One can certainly hope that extending a truncation by allowing new terms will improve it, but the convergence of such a procedure is by no means guaranteed. We will discuss some of these issues further in \sect{Trunc-Opt}.

Nevertheless, such truncations have allowed computations to be performed in situations---such as the strong-coupling domain of QCD (see \sect{Other-Overview} references)---where any results are of interest. Moreover, in certain theories, particular truncations are known to work very well, in practice. The most celebrated example of this is the derivative expansion in scalar field theory, whereby interactions are classified according to the number of derivatives which hit the fields; in momentum space, this amounts to expanding in powers of momenta. We will discuss the derivative expansion further in \sect{Trunc-DE}; excellent reviews can be found in~\cite{TRM-Elements,B+B}.

It is probably fair to say that the derivative expansion is on the safest ground as far as truncations of the ERG go. Unfortunately, it is not always practical (or appropriate) to use it. In gauge theories, each order of the derivative expansion involves a set of coupled equations for each of the gauge invariant objects that can be constructed. This is prohibitively complicated in cases of interest: for example, in four dimensional $\SU(N)$ Yang--Mills, the lowest order in the derivative expansion would involve 34 invariants~\cite{TRM-U1}!

Consequently (and also in cases where one expects the momentum dependence of vertices to be particularly important) other truncations have been used. One such is to expand the action in powers
of the field and to truncate at some point.%
\footnote{In a similar vein, one can write the action as a linear combination of
the eigenoperators as defined at some fixed-point. This is the scaling field method, discussed earlier in \sect{SFM}, which, perhaps needless to say, has only ever been practically applied using the eigenoperators of the Gaussian fixed-point.} 
In other words, starting from~\eq{action-exp}, all $\Siv{n>n'}$ are---for some choice of $n'$---artificially set to zero. Consequently, the flow equation reduces to a finite number of coupled equations for the surviving vertices. It is precisely this truncation in which spurious fixed-point solutions can occur~\cite{TRM-Truncations}, though it seems to be an empirical fact that the order at which
the truncation starts to diverge can be substantially increased by expanding about the minimum of
the effective potential~\cite{Aoki:1998um,Tetradis+Wetterich-Exponents,Alford}.

Sitting somewhere between the derivative expansion and the vertex expansion is the `BMW' scheme~\cite{BMW,BMW-npt-I,BMW-npt-II,BMW-FirstNumerical,GMW-CorrFns,BMW-O(N),BMW-full}. In this approach, the entire tower of equations for the vertices is kept, and some---but crucially not all---of the momentum dependence is discarded.

\subsection{The Derivative Expansion}
\label{sec:Trunc-DE}

\subsubsection{The LPA and Beyond}
\label{sec:LPA}

The leading order of the derivative expansion is the so-called Local Potential Approximation (LPA) which, whilst first written down by Nicoll, Chang \& Stanley~\cite{NCS}
has since been rediscovered---apparently independently---several times~\cite{Tokar,H+H,TRM-Deriv,Wetterich-1PI}. In each case, the authors have there own pet way of obtaining the
truncated form of the flow equation, but the method used by Hasenfratz \& Hasenfratz is particularly elegant. In position space, the Wilsonian effective action (or effective average action, if one prefers this formalism) is written as
\be
	\Sint_t[\phi] \sim
	\Int{x}
	\Bigl[
		V_t(\phi) + W_t(\phi) \partial_\mu \phi \partial_\mu \phi + \order{\partial^4}
	\Bigr],
\label{eq:action-DE}
\ee
where $V$ and $W$ possess no derivatives. [Notice the minor change in notation compared with~\eq{FullAction-DE}.]
For the rest of this section, we will work in momentum space, and so henceforth understand $\phi = \phi(p)$.
Hasenfratz \& Hasenfratz picked out the first term above by applying the projector, $\proj$,
which acts on some arbitrary functional of the fields, $X$,
according to
\be
	\proj(\zeta) X[\phi]
	=
	\left. \exp \left( \zeta \pder{}{\phi(0)}\right) X[\phi] \right\vert_{\phi=0}.
\label{eq:proj}
\ee
To see how this works, let us return to the field expansion of the action~\eq{action-exp} (but this time for $\Sint$)
\be
	\Sint_t[\phi] = \sum_n \int_{p_1,\ldots,p_n}
	\frac{1}{n!}
	\Siv{n}_t(p_1,\ldots,p_n) \phi(p_1)\cdots \phi(p_n)\deltahat{p_1+\cdots+p_n}
\label{eq:Sint-exp}
\ee
and write
\be
	\Siv{n}_t(p_1,\ldots,p_n) = V^{(n)}_t + \frac{1}{n(n-1)} \bigl(p_1^2 + \cdots + p_n^2\bigr) W^{(n-2)}_t
	+ \order{p^4}.
\ee
It is thus apparent that
\be
	\proj (\zeta) \Sint_t[\phi] = \deltahat{0} V_t[\zeta].
\ee
Hasenfratz \& Hasenfratz removed the $\delta$-function by working in a finite volume, so their projector is actually slightly different from~\eq{proj}, but this is of no real consequence (see also~\cite{Comellas+Travesset}). Note that the projector
replaces the field, $\phi(p)$, with the variable $\zeta$, and so the flow equation reduces, in the LPA, to
a partial differential equation. Specifically, if we define
\be
	\tilde{I}_0 \equiv - \int_p \cutoff'(p^2)	,
\ee
then the flow equation~\eq{Ball} projects down to
\be
	\partial_t V_t(\zeta) = \tilde{I}_0 V'' + \cutoff'(0) V'^2 - d_\zeta \zeta V' + \D V,
\ee
where here we use primes to denote derivatives \wrt\ $\zeta$.
At the level of the LPA, the anomalous dimension is undetermined and so is usually set to zero, meaning that we take
\be
	d_\zeta \equiv \frac{\D-2}{2}.
\ee
Performing the rescalings%
\footnote{The reason for the minus sign is that, as we recall from \sect{Linear-Pol}, the cutoff function to be monotonically decreasing. Note also that this forbids the singular case $\cutoff'(0) = 0$.} 
$V \mapsto -\tilde{I}_0  V / \cutoff'_0, \ \zeta \mapsto \sqrt{\tilde{I}_0} \zeta$,
gives an equation which is manifestly independent of the cutoff function:
\be
	\partial_t V_t(\zeta) = V'' - V'^2 -d_\zeta \zeta V' + \D V.
\label{eq:LPA}
\ee

Before moving on, let us note that a common feature of the various approximation schemes mentioned in \sect{Trunc-Over} is that, in each case, the functional flow equation decomposes into a tower of coupled partial differential equations. These towers depend on $\D$ in such a way that gives meaning to the notion of solving the flow equation in non-integer dimensions. In particular, we need never define precisely what is meant by expressions such as~\eq{action-DE}---which involve integrals over the fields---for non-integer dimension. Rather, we have only to deal with equations such as~\eq{LPA}, for which there is no difficulty taking $\D$ to be arbitrary.

As the name suggests, the LPA involves keeping only those interactions which contribute to the local potential,  $V_t(\zeta)$, throwing away all interactions with derivatives. This sounds like a rather severe thing to do. But it should be emphasised that there are no restrictions placed on the local potential, itself. Indeed, this serves to highlight what has been a recurring theme throughout this review: the Wilsonian effective action [or, in this case, its truncation to $V_t(\zeta)$] follows as a \emph{solution} of the flow equation (given boundary conditions). We do not put in any prior restrictions (beyond those involved in any truncation scheme), such as a stipulation that the potential must have \eg\ a $\phi^4$-type behaviour. 

At a fixed-point, a truncation to the LPA still results in an equation which is too hard to solve analytically. But it can be solved numerically and doing so amounts to scanning the \emph{complete space} of local potentials (within the limits of the numerics) for fixed-point solutions. This is a powerful approach! In three dimensions, for example, the LPA can be used to find the Wilson-Fisher fixed-point, to show that no further non-trivial fixed-points exist at this level of approximation, and to compute the critical exponents to reasonable accuracy---see~\cite{B+B,TRM-Elements,Wetterich-Rev} for detailed discussions and further references. Again, it is worth remembering that there is no small parameter available.

Moreover, the use of the LPA is by no means limited to fixed-points. Of the various applications
that can be found in the aforementioned reviews let us mention, in particular, that this \emph{nonperturbative} technique has been applied to the interesting and topical problem of
the upper bound of the Higgs mass~\cite{H+N}.

Before leaving the LPA behind us, there are a few comments to make. First, we note that just as~\eq{LPA} was derived, so too can one derive the corresponding equation within the effective average action formalism~\cite{NCS,Wetterich-1PI,TRM-Deriv} or from the Wegner-Houghton equation~\cite{H+H}. In each case, the equation takes a different form and, in the case of the former, depends on the cutoff. See~\cite{TRM-Rebuttal} for some comments pertaining to relationships between certain realizations of the LPA. Let us also comment that there
have been some recent developments in computational techniques~\cite{BJL,Bervillier-AnalyticI,Bervillier-AnalyticII}.

To go beyond the LPA, one must project onto the higher order terms in~\eq{action-DE}. In fact,
in Hasenfratz \& Hasenfratz's paper, they work with the Wegner-Houghton equation, which
has a sharp cutoff. Consequently, should one wish to go beyond the LPA in this approach, the `momentum scale expansion'~\cite{TRM-MomScale}---in which one expands in $\sqrt{p_\mu p_\mu}$---must be used instead of the derivative expansion. Anyhow, sticking to the latter, one can use the projector (which, to the best of my knowledge has never been explicitly written down), $\proj_2$, which is defined via
\be
	\proj_2 X[\phi]
	=
	\left. \exp \left( \zeta \pder{}{\phi(0)}\right) 
	\hf
	\pder{}{p^2} \dfder{}{\phi(p)}{\phi(-p)}
	X[\phi] \right\vert_{\phi=0}.
\ee
Alternatively, of course, one can use the other methods of obtaining the derivative expansion on the market~\cite{TRM-Deriv,Wetterich-1PI}. Either way, one obtains a tower of coupled partial differential equations.

There are several papers in which calculations have been done to $\order{\partial^2}$ in 
the derivative expansion for theories of a single scalar field, using a Wilson/Polchinski-like equation~\cite{Ball,Comellas,Golner-NonPert,Filippov+Radievsky,Bervillier-d^2} and the effective average
action approach~\cite{TRM-Deriv,TRM-2D, Aoki:1998um,Seide+Wetterich,Opt-Canet,Ballhausen:2003gx,Bervillier-AnalyticII}. 
In addition to an incomplete treatment at 
$\order{\partial^4}$~\cite{Ballhausen:2003bu}, there  even exists one treatment of
the full $\order{\partial^4}$ equations~\cite{Canet-d^4}. O$(N)$ scalar field theory
has been treated to $\order{\partial^2}$ in only a handful of papers~\cite{TRM-O(N),Gersdorff+Wetterich,Bervillier-d^2}. In the noteworthy contribution of Tetradis and Wetterich~\cite{Tetradis+Wetterich-Exponents},
the computations are not fully $\order{\partial^2}$, since the running of the wavefunction
renormalization is neglected.

Of all of these papers, perhaps~\cite{TRM-2D}
provides the most compelling evidence that the derivative expansion really can perform well in intrinsically nonperturbative situations (though this is not to say that the other papers are not convincing!). The purpose of this beautiful paper by Morris
 was to compare the output of the flow equation to known results from conformal field theory. Working in two dimensions,%
\footnote{%
In the context of $\D=2$, it is worth mentioning a series of works in which the ERG
has been applied to the sine--Gordon model, initiated in~\cite{SG-Genesis}. The majority
of subsequent studies~\cite{SG-Layered,MassiveSG-EffPtl,SG-RenormParam,SG-Universality,Nandori-Schwinger,SG-SymBreaking,SG-Censorship} are performed within the LPA to the Wegner-Houghton
equation. A comparison between this approach and a perturbative one is given in~\cite{SG-PertComparison}.
The analysis of~\cite{SG-FRG} sits between the LPA and
a fully fledged $\order{\partial^2}$ approximation within the effective average action approach, whilst a treatment of scheme dependence within the LPA for a variety of flow equations can be found in~\cite{SG-SchemeDependence}.}
 and to $\order{\partial^2}$ in the derivative expansion,  twenty multicritical fixed-points were uncovered and roughly 100 associated quantities computed, all of which turned out to be reasonably accurate, at worst, and highly accurate in many cases. There can be little doubt, then, that the ERG can be an effective, \emph{practical} nonperturbative tool.

\subsubsection{$c$-Functions and the Like}
\label{sec:cfn}

A very interesting feature of the LPA equation~\eq{LPA} is that a function of the couplings can be constructed which (for real Euclidean action) decreases monotonically along the flow~\cite{Zumbach-Almost,Zumbach-LPA,Zumbach-O(N),TRM-cfn}.%
\footnote{An analysis of comparatively limited scope, in the
context of the LPA to the Wegner-Houghton equation, can be found in~\cite{Latorre-Gradient}.}
Consequently, limit cycles and so forth are forbidden, at least to this level of approximation. To see this, we begin by rewriting~\eq{LPA} in terms of $\gibbs_t(\zeta) = e^{-V_t(\zeta)}$:
\be
	\partial_t \gibbs = \gibbs'' - d_\zeta \zeta \gibbs' + \D \gibbs \ln \gibbs.
\label{eq:LPA-equiv}
\ee
The next step is to introduce the operator
\be
	\adjop \equiv \frac{\partial^2}{\partial \zeta^2} - d_\zeta \zeta \pder{}{\zeta},
\label{eq:adjop}
\ee
so that~\eq{LPA-equiv} can be written as
\be
	\partial_t \gibbs = \adjop \gibbs + \D \gibbs \ln \gibbs.
\label{eq:LPA-simple}
\ee

Inspired by Zumbach~\cite{Zumbach-Almost,Zumbach-LPA,Zumbach-O(N)}, we now introduce an inner product
\be
	\inner{X}{Y} \equiv \frac{1}{\normalization}
	\integral{\zeta} \weight (\zeta) \, XY,
	\qquad
	\normalization \equiv \integral{\zeta} \weight (\zeta),
\label{eq:inner-product}
\ee
where $X$ and $Y$ are square-integrable functions of $\zeta$. The weight function, $\weight (\zeta)$, is determined
by demanding that $\adjop$ is Hermitean \wrt\ to this inner product:
\be
	\inner{X}{\adjop Y} = \inner{\adjop X}{Y}.
\label{eq:adj}
\ee
By substituting~\eq{adjop} into~\eq{adj} and using~\eq{inner-product}, it is easy to check that
\be	
	\weight (\zeta) = e^{-\frac{d_\zeta}{2} \zeta^2}.
\ee

With this in mind, we now construct the following functional of $\gibbs$~\cite{Zumbach-Almost,Zumbach-LPA,Zumbach-O(N),TRM-cfn}:
\be
	F_t[\gibbs] = 
	-\frac{\normconst}{\normalization}
	\integral{\zeta} \weight (\zeta)
	\left[
		\hf \gibbs \adjop \gibbs
		-
		\frac{\D}{4} \gibbs^2
		\left(
			1 - 2 \ln \gibbs
		\right)
	\right],
\label{eq:F}
\ee
where $\normconst$ is a positive constant, which will be determined below.
The point of all this becomes apparent when we take the total derivative \wrt\ $t$. Differentiating under
the integral on the \rhs\ yields
\be
	\der{F_t[\gibbs]}{t}
	=
	-\frac{\normconst}{\normalization}
	\integral{\zeta} \weight (\zeta) \partial_t \gibbs
	\left[
		\adjop \gibbs + \D \gibbs \ln \gibbs
	\right]
	=
	-\frac{\normconst}{\normalization}
	\integral{\zeta} \weight (\zeta) \left(\partial_t \gibbs\right)^2,
\label{eq:F-flow}
\ee
where we have used~\eq{LPA-simple} in the last step. (The adjoint nature of $\adjop$ has
been exploited by noting that $\inner{\gibbs}{\adjop \gibbs} = \inner{\adjop \gibbs}{\gibbs}$, so that
$\partial_t\inner{\gibbs}{\adjop \gibbs}  = 2 \inner{\partial_t \gibbs}{\adjop \gibbs} $.) Since $\weight$, $\normconst$ and $\normalization$ are positive definite, it therefore follows that if $\gibbs$ is real then
$F_t[\gibbs]$ decreases monotonically along the flow.

It is natural to try to compare $F_t$ with Zamolodchikov's $c$-function~\cite{Zam-cfn}, the properties of which we now recall. Working in $\D=2$, and assuming Euclidean invariance, positivity and renormalizability (in the full nonperturbative, Wilsonian sense---of course!), Zamolodchikov constructed a function of the couplings, $c (g_i) \geq 0$ which satisfies the following criteria:
\begin{enumerate}
	\item The $c$-function decreases monotonically along the RG flow,
	\be	
		\der{c}{t} = \beta_i \pder{c}{g_i} \leq 0,
	\ee
	(summation is  implied by the repeated index), with the inequality being saturated only
	at fixed-points.

	\item The $c$-function is stationary at fixed-points%
	\footnote{%
		Zamolodchikov considered critical fixed-points but our analysis  deals
		with non-critical fixed-points, also.
	}:	
	\be
		\left.\pder{c}{g_i}\right\vert_{g_i=g_{i\star}} = 0.
	\ee

	\item The value of $c(g_i)$ at a fixed-point is the same as the corresponding Virasoro
		algebra central charge~\cite{BPZ}.
\end{enumerate}

Although the last property only makes sense in $\D=2$, it nevertheless tells us that
$F_t$ is not of the right form to compare, directly, with the $c$-function. The point is that
the Virasoro central charge essentially counts massless degrees of freedom and so
is \emph{extensive}. Suppose that we have $N$ scalar fields which do not interact with each
other (though we do not prohibit any of the scalar fields exhibiting self-interactions) and that each of the scalar fields is at a fixed-point. Then Zamolodchikov's $c$-function
will simply sum up the $c$s for each of the individual scalar field theories. With this in mind,
let us consider $F_t$ at a fixed-point. Substituting~\eq{LPA-simple} into~\eq{F},
it is apparent that
\be
	F_t[\gibbs] = 
	-\frac{\normconst}{\normalization}
	\integral{\zeta} \weight (\zeta)
	\left[
		\hf \gibbs \partial_t \gibbs - \frac{\D}{4} \gibbs^2
	\right]
\ee
and, therefore,
\be
	F_\star[\gibbs] = 
	\frac{\D \normconst}{4 \normalization}
	\integral{\zeta} \weight (\zeta) \gibbs^2_\star.
\label{eq:Fstar}
\ee
We can generalize this to $N$ scalar fields very easily [recall the discussion around~\eq{flow-N}]:
\be
	F^{(N)}_\star[\gibbs] =
	\frac{\D \normconst^N}{4 \normalization^N}
	\Nintegral{\zeta} \weight(\zeta_1)\cdots \weight(\zeta_n) \gibbs^2_\star(\zeta_1,\ldots,\zeta_N)
\label{eq:F^N}
\ee
Now, the point is that, for mutually non-interacting fields, $\gibbs(\zeta_1,\ldots,\zeta_N)
= \gibbs(\zeta_1)\cdots \gibbs(\zeta_N)$. Consequently, for mutually non-interacting fields,
$F^{(N)}_\star[\gibbs]$ factorizes. To arrive at something extensive Generowicz, Harvey-Fros and Morris therefore took the logarithm~\cite{TRM-cfn}. To be precise, they defined their $c$-function, which we will denote by
$\tilde{c}$, according to
\be
	F^{(N)}_t[\gibbs] = \frac{\D \normconst^{\tilde{c}}}{4}.
\label{eq:TRM-cfn}
\ee
Notice that if the $N$ scalars are not interacting with each other and, moreover, each of them is at its Gaussian fixed-point ($\gibbs = 1$) then, by comparing with~\eq{F^N}, it is apparent that $\tilde{c}=N$: the normalization is such that $\tilde{c}$ counts one
for each Gaussian scalar. The constant, $\normconst$, was fixed by demanding that $\tilde{c}$ counts zero
at the high-temperature (infinite-mass) fixed-point, with the result~\cite{TRM-cfn}
\be
	\normconst = e^{-2/\D} \left(\frac{\D+2}{\D-2}\right)^{1/2}.
\ee
Notice that $b>1$, at least for $\D\geq2$, though it becomes infinite for $\D=2$.

From the definition~\eq{TRM-cfn} and~\eqn{F-flow}, it is easy to check 
that
\be
\der{\tilde{c}}{t} = -\frac{1}{{F}^{(N)}_t \ln b} 
	\frac{b^N}{\mathcal{N}^N}  \Nintegral{\zeta} 
	\weight (\zeta_1)\cdots \weight(\zeta_N) \left(\partial_t \gibbs\right)^2.
\ee
In~\cite{TRM-cfn}, it was now asserted that, since $b>1$, $\tilde{c}$ is monotonically decreasing along the flow. But this seems to miss something: for this to be true, it must also be that $F^{(N)}_t$
is positive everywhere along the flow. The conditions under which this holds have not been established. Certainly,
given that Zamolodchikov required Euclidean invariance, positivity and renormalizability to prove his theorem, it is reasonable to expect that one or more of these plays a role. Indeed, for flows between two
fixed-points, $F_t$ must be positive at both ends of the flow [see~\eq{Fstar}] and, due to its monotonically decreasing character, must therefore be positive everywhere along the flow. 
Consequently, having a flow which starts at one fixed-point and ends at another is a sufficient condition for positivity of $F_t$; but what the
necessary and sufficient conditions are does not appear to be known.

Although this issue has not been properly addressed, let us continue to follow~\cite{TRM-cfn}, 
and to this end define the metric
\be
	\mathcal{G}_{ij} \equiv
	\frac{1}{{F}^{(N)}_t \ln b} 
	\frac{b^N}{\mathcal{N}^N}  \Nintegral{\zeta} \weight(\zeta_1)\cdots \weight(\zeta_N) \partial_i \gibbs  \, \partial_j \gibbs,
\ee
where $\partial_i \equiv \partial/\partial g_i$. Since we have that
$\partial_t \gibbs_t[\zeta] = \beta_i \partial_i \gibbs$ and $d\tilde{c}(g_i) / dt = \beta_j \partial_j \tilde{c}$, it
is clear that
\be
	\partial_i \tilde{c} = - \mathcal{G}_{ij} \beta_j.
\ee 
If the metric is indeed positive definite (the conditions for which, we emphasise, have not been determined), then $\tilde{c}$
exhibits a so-called `gradient flow'~\cite{Wallace+Zia-GradientFlow,Wallace+Zia-GradientProperties} and manifestly satisfies the first two of Zamolodchikov's criteria. The question remains whether, in $\D=2$, $\tilde{c}_\star$ coincides with the Virasoro central charge (Zamolodchikov's third criterion). The normalization, $\normconst$, has been chosen with this in mind, but to prove that it does its job presumably requires that an explicit link with Zamolodchikov's $c$-function is found. Note, however, that entirely independently of these considerations (and in particular those pertaining to the positivity of $F_t$), limit cycles and other exotic RG flows are forbidden, within the LPA, by the fact that $F_t$ is monotonically decreasing along the flow. The subtleties creep in when we try to construct an extensive function which does likewise.

Finally, let us observe an interesting point which, to the best of my knowledge, has not been made before.
Suppose that we linearize the LPA equation~\eq{LPA} about a fixed-point, $V_t = V_\star + v_t$ (we will work with $N=1$, for brevity, but the generalization to arbitrary $N$ is trivial):
\be
	\partial_t v_t =   \bigl(\adjop + \D \bigr) v_t -2V'_\star v'_t \equiv \adjopb v_t.
\ee
Obviously, $\adjopb$ is just the LPA version
of the operator which classifies the RG eigenvalues, $\classifier$ [(see~\eq{classifier})].
Writing
\be
	v_t(\zeta) = \sum_i \alpha_i e^{\lambda_i t} u_i(\zeta)
\ee
we obtain
\be
	\adjopb u_i = \lambda_i u_i.
\ee

With this in mind, let us construct a second inner product,
\be
	\inner{X}{Y}' \equiv \frac{1}{\normalization}
	\integral{\zeta} \weight'_\star (\zeta) \, XY,
\ee
(with $\normalization$ as before) where $\weight'_\star$ is chosen such that $\adjopb$ is Hermitean \wrt\ this inner
product:
\be
	\inner{X}{\adjopb Y}' = \inner{\adjopb X}{Y}'.
\label{eq:adjb}
\ee
Proceeding as before, it is easy to check that
\be
	\weight'_\star(\zeta) = \weight(\zeta) \gibbs_\star^2.
\ee
Looking at~\eq{Fstar}, which gives the expression for $F_t$ at a fixed-point, we observe that
\be
	F_\star[\gibbs] 
	= \frac{\D\normconst}{4 \normalization}  \integral{\zeta} \weight' (\zeta)
	= \frac{\D\normconst}{4} \inner{1}{1}'.
\ee
This has a very interesting consequence. For let us suppose that we perturb the fixed-point action
in the direction of one of the eigenoperators:
\be
	V_\star \mapsto V_\star + \varepsilon e^{\lambda_i t} u_i,
	\qquad
	\Rightarrow
	\qquad
	\weight'(\zeta) \mapsto \weight'(\zeta)
	\left(
		1 - 2 \varepsilon e^{\lambda_i t} u_i
	\right) + \order{\varepsilon^2}.
\ee
Therefore, under this perturbation,
\be
	\delta_\varepsilon \inner{1}{1}'
	=
	\inner{1}{1}'
	- 2\varepsilon e^{\lambda_i t} \inner{1}{u_i}' + \order{\varepsilon^2}.
\ee
Now for the point: $u_i$ is an eigenfunction of $\adjopb$ with eigenvalue $\lambda_i$,
whereas unity is an eigenfunction of $\adjopb$ with eigenvalue $\D$. So, if
$\lambda_i \neq \D$, then $\inner{1}{u_i}' =0$. This follows simply because $u_i$ and unity are
both eigenfunctions (presumed to have different eigenvalues) of the operator \wrt\ which
the inner product is Hermitean. Assuming that the special operator is the only one with
RG eigenvalue $+\D$, we have therefore shown that the directional derivative of $F_\star$ in any direction besides the constant one is zero.

Let us wrap up our discussion of the $c$-function by making the obvious point that it would be wonderful if this analysis could be extended beyond the LPA or, better still, could be realized at the level of the exact flow equation, without any recourse to a derivative expansion.

\subsection{Reparametrization Invariance \& Optimization}
\label{sec:Trunc-Opt}

To conclude our discussion of truncations, it is important to mention that they generically spoil certain features of
exact flow equations. Most obviously, independence of universal quantities on the cutoff function (or, more generally, the complete set of non-universal inputs of whatever flow equation is used)
is lost. (An exception is the LPA of the Polchinski equation which, as we have seen, can be written in a form which is manifestly independent of the cutoff function). This naturally raises the question as to whether the cutoff function can be `optimized', in order to yield answers that are expected to be closest to the physical ones.%
\footnote{It would be interesting to explore, within the framework of generalized ERGs, whether it is worthwhile
trying to optimize the seed action within various truncation schemes.
} 
This important issue has been discussed by Litim~\cite{DFL-Opt1,DFL-Opt2,DFL-DerivExp,DFL-CriticalExps,DFL-Gap}, by Canet and collaborators~\cite{Opt-Canet,Canet-Opt-II}, by Andersen et al.~\cite{Andersen+Strickland,Andersen-Review} and by Liao et al.~\cite{Liao-Opt}. The most ambitious approach is due to Pawlowski, we which will describe shortly~\cite{JMP-Review}.

More subtly, truncations generically spoil the reparametrization invariance of the flow equation discussed in \sect{solve}. At a critical fixed-point, this means that the expected line of equivalent fixed-points fragments into a line of \emph{inequivalent} fixed-points. Consequently, predictions become ambiguous since it matters which of these fixed-points is chosen. This issue has received attention since the early days of the ERG, with a particularly noteworthy contribution being provided by Bell and Wilson~\cite{Wilson+Bell-FiniteLattice}. More recently, attention has focused on the derivative expansion beyond leading order.

Using the Polchinski-like flow equation of Ball et al.~\eq{Ball}, the derivative expansion breaks
reparametrization invariance at any finite order. In this setting,  Comellas advocates a scheme,
based on the `principle of minimum sensitivity'~\cite{Stevenson-Opt}, in which one strives to realize the reparametrization invariance as well as possible~\cite{Comellas}. However, a word of caution
should be made, since the principle of minimum sensitivity is known to fail badly in certain circumstances~\cite{BLM}.\footnote{I would like to thank Stan Brodsky for pointing
this out to me.}

If one is to take reparametrization invariance as  seriously as possible then, within
the effective average action approach, a cutoff function can be chosen which preserves reparametrization invariance~\cite{TRM-Deriv}, but at a considerable price: with such a choice, the derivative expansion ceases to make sense beyond a certain order~\cite{TRM-Convergence}.\footnote{Also within the effective average action formalism, a sharp cutoff preserves reparametrization invariance, but then one is forced to use the momentum scale expansion.}
An alternative point of view, advocated particularly by Litim, is to regard reparametrization invariance as something of a red-herring and to focus instead
on stability properties of the flow, taking this as the guiding principle for optimizing truncations~\cite{Litim-Universal}. However, it turns out that Litim's commonly employed `optimized' cutoff cannot be used beyond
$\order{\partial^2}$ in the derivative expansion~\cite{TRM-Rebuttal}: after this order a momentum scale expansion is required, which is expected to have poor convergence properties~\cite{TRM-MomScale}. (The two papers~\cite{Litim-Universal,TRM-Rebuttal} should be read as a pair, with~\cite{TRM-Rebuttal} providing a strong critique of certain claims of the other.) 

There thus appears to be a recurring theme: cutoff functions chosen according to various sensible criteria turn out not to behave as nicely as one might have hoped. With this in mind, let us mention two interesting ideas.

First of all, we consider Polchinski-like equations. Recently, by making a carefully chosen modification to the first order equations of the derivative expansion (the zeroth order being the LPA), Osborn and Twigg were able to restore reparametrization invariance for
any cutoff~\cite{Osborn+Twigg} function. Subsequent to this initial proposal where the modification was essentially unjustified, it has been put on firm footing~\cite{HO-Remarks}: it was realized that the pertinent equations can be derived by considering not a derivative expansion of the flow equation for $\Sint$, but rather for the `normal ordered' $e^{-\op} \Sint$.

Finally, we describe an ambitious proposal due to Pawlowski, by the name of `Functional Optimization'~\cite{JMP-Review}, which seeks to fully systematize the process of optimization. This is formulated in the context of the effective average action (and more general flows of the same ilk). The basic scheme is as follows.

For simplicity, we will consider a theory sitting on a renormalized trajectory. Whilst the bare scale does not appear, the effective average action depends on $\Lambda$ and also, through dimensional transmutation, on an arbitrary reference scale, $\mu$. Now suppose that we vary the cutoff function, here denoted by $R$ [see the comments below~\eq{Choice-CW}]. Since $\Gamma_{\Lambda=0, \mu}$ is universal, it will be invariant under this procedure. However, if we perform this variation instead for $\Lambda \neq 0$, we will of course find that $\Gamma_{\Lambda,\mu}$ changes.

\bcf[h]
	\ensuremath{\begin{array}{c}\begin{picture}(0,0)%
\epsfig{file=pstex/Opt.pstex}%
\end{picture}%
\setlength{\unitlength}{3947sp}%
\begingroup\makeatletter\ifx\SetFigFont\undefined%
\gdef\SetFigFont#1#2#3#4#5{%
  \reset@font\fontsize{#1}{#2pt}%
  \fontfamily{#3}\fontseries{#4}\fontshape{#5}%
  \selectfont}%
\fi\endgroup%
\begin{picture}(3865,3889)(4033,-5488)
\end{picture}%
 \end{array}}
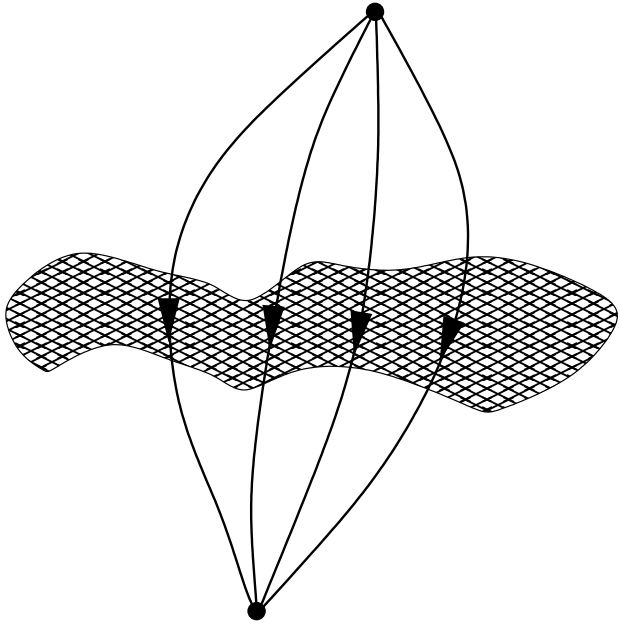
\caption{A family of flows with different cutoff functions for some (renormalizable) theory. 
The hypersurface is defined such that the effective average actions which populate it exhibit
a particular relationship between their private values of the effective scale, $\Lambda$. Loosely
speaking, these effective actions are `all at the same effective scale'.
Note that in contrast to similar pictures elsewhere in this review, this one is in the space of theories written in dimension\emph{ful} variables.}
\label{fig:Flows-opt}
\ecf

The general picture, then, is shown in \fig{Flows-opt}. In dimensionful variables, we consider a flow for a renormalizable theory starting at $\Lambda=\infty$ and running down to $\Lambda=0$. In between these limits, the precise details of the flow depends on the cutoff function. Each of these flows is parametrized by its own private $\Lambda$. With this in mind, let us consider comparing effective average actions on two of these trajectories, say $\Gamma_{\Lambda,\mu}[\Phi,R]$ and $\Gamma_{\Lambda',\mu}[\Phi,R']$. Note that when making this comparison there is no requirement that we set $\Lambda' = \Lambda$---and this is crucial! Indeed, part of the scheme put forward in~\cite{JMP-Review} is as follows.

First of all, a norm is proposed on theory space. This is, perhaps needless to say, a deep issue which certainly requires further attention. Putting aside any reservations we might have, consider an object, $F_\Lambda[\Phi,R]$, derived from $\Gamma_{\Lambda,\mu}[\Phi,R]$ ($F$ might simply be the effective average action, or its second derivative, or something more exotic; ideally, it should be bounded from above and below). Then, given an appropriate space of fields, $\mathcal{S}$, and an appropriate norm, the distance between two theories \wrt\ $F$ is taken to be
\be
	d_F[R_\Lambda,R'_{\Lambda'}] = \sup_{\Phi \in \mathcal{S}} 
	\norm{F_\Lambda[\Phi,R] - F_{\Lambda'}[\Phi,R']}.
\label{eq:distance}
\ee
We now go one step further and define 
\be
	\tilde{d}_F[R, R'](\Lambda) \equiv \min_{\Lambda'} d_F[R(\Lambda), R'(\Lambda')].
\label{eq:distance'}
\ee
This implicitly determines $\Lambda'$ as a function of $\Lambda$ (which we assume to be smooth, though this might require additional constraints~\cite{JMP-Review}). The pair of effective average actions 
$\Gamma_{\Lambda,\mu}[\Phi,R]$ and $\Gamma_{\Lambda'(\Lambda),\mu}[\Phi,R']$ are now said to `live at the same effective scale'. Thus, given a reference cutoff function, $R_\mathrm{ref}$, and a value of $\Lambda$, we can construct a hypersurface populated by all those effective average actions which live at the same effective scale, as indicated in the figure. To move within this surface we can consider performing a variation of the cutoff function \eg\ from $R_\Lambda \rightarrow R'_\Lambda$ (which generally takes us out of the surface), followed by a change $\Lambda \rightarrow \Lambda'(\Lambda)$ (which takes us back in). Alternatively, we note that the change in $\Lambda$ can itself be implemented by a change to the cutoff function, and so we can move within the hypersurface by performing restricted variations of the cutoff function. In~\cite{JMP-Review}, these later restricted variations are denoted by $\delta R_\perp$.

Considering variations of the cutoff function that are restricted in this way, the second part of Pawlowski's scheme is to select a cutoff function for which the variation of $\Gamma_{\Lambda,\mu}$ takes a particular form:
\be
	\delta R_{\perp} \cdot \fder{\Gamma_{\Lambda,\mu}[\Phi,R]}{R}
	\biggl\vert_{R = R_{\mathrm{stab}}}
	=
	\delta (\ln \mu)
	\biggl(
		\mu \pder{}{\mu} +  \eta \, \Phi \cdot \fder{}{\Phi}
	\biggr)
	\Gamma_{\Lambda,\mu}[\Phi,R_{\mathrm{stab}}]
.
\label{eq:Stability}
\ee
 Notice that the operator in big brackets on the \rhs\  annihilates the physical effective action, $\Gamma_{0,\mu}$:
 \[
 	\biggl(
		\mu \pder{}{\mu} +  \eta \, \Phi \cdot \fder{}{\Phi}
	\biggr)
	\Gamma_{0,\mu}[\Phi,R_{\mathrm{stab}}] = 0,
 \]
 this being the form of a textbook RG equation (at least after extracting derivatives \wrt\ the relevant couplings from the scale derivative, whereupon we would find an additional term of the form $\beta_i \partial / \partial g_i$). 
It is asserted in~\cite{JMP-Review} that solutions to~\eq{Stability} correspond to cutoff functions which yield the most stable/unstable flows (obviously, our prime interest is in the former!). 

The justification for this is as follows. First we note that the \rhs\ of~\eq{Stability} contains only implicit dependence on the cutoff function buried in the renormalization scheme (which, moreover, vanishes for universal objects). 
Therefore, we are choosing a cutoff function for which the explicit effects of a (suitably constrained) variation vanish. Now, the most stable/unstable flows are understood as the ones for which the distance between the start and end points is either a minimum or a maximum \wrt\ small variations. With this in mind, consider a pair of effective average actions lying on the path defined by $R_\mathrm{stab}$, at scales $\Lambda$ and $\Lambda - \delta \Lambda$.  If we take only the explicit effects of performing an infinitesimal variation of the cutoff function in the usual hypersurface, then this pair off effective average actions are left invariant. Therefore, the distance between them does not change. Since this is true all the way along the path, the length of this path is stable against infinitesimal variations and so represents, by definition, a flow of either maximal or minimal stability (ignoring the possibility of points of inflexion with vanishing gradient).

It is important to note that, so far, everything is being done at the level of the exact flow equation. Within a given truncation scheme, the game is as follows: differentiate \wrt\ $\Lambda$ and replace $\flow \Gamma_{\Lambda,\mu}$ by the appropriate \emph{approximation} to the \rhs\ of the flow equation. The advocated interpretation is that  now $R_{\mathrm{stab}}$ is optimized \wrt\ the truncation scheme of choice.

So, in essence, the scheme is as follows. Consider a family of trajectories in theory space---each corresponding to a different $R$---parametrized by their own private $\Lambda$, all ending up at the same destination as $\Lambda \rightarrow 0$. Now foliate theory space with hypersurfaces such that all effective average actions on each hypersurface are `at the same effective scale' (this step depends on the choice of norm on theory space). Next consider variations of the cutoff function, such that the resulting effective average actions are constrained to a given hypersurface. Within this (and given a truncation scheme), we choose the cutoff function for which the 
explicit effects of performing an infinitesimal variation vanish. Further details, including comments on the existence of the proposed scheme are given in~\cite{JMP-Review}. Issues which merit further investigation are the norm used on theory space and the freedom to choose the hypersurfaces via different choices of $F$ in~\eqs{distance}{distance'}.

\section{Correlation Functions}
\label{sec:CorrFns}

\subsection{Motivation}

It almost goes without saying that, in any approach to quantum field theory worth its salt, it is understood how to compute correlation functions. However, quite apart from this fundamental motivation, there are some other, very deep reasons why it is worthwhile considering correlation functions within the framework of the ERG, as we will discuss momentarily.
First, though, let us fix the set-up.

The quantitative work of this section will be performed using theories of a single scalar field, $\phi$.
The most primitive correlation functions correspond to the family of expectation values of $n$ fields at different points:
\be
	\eval{\phi(x_1)\cdots \phi(x_n)}
	\sim \frac{1}{\pf}
	\Fint{\phi} \phi(x_1)\cdots \phi(x_n) e^{-S_{\Lambda_0}}.
\label{eq:correlations}
\ee
For a non-renormalizable theory, $S_{\Lambda_0}$ is the boundary condition to the flow \ie\ the bare action. In this case, we can simply replace the $\sim$
with an equality symbol.
For a renormalizable theory, $S_{\Lambda_0}$
is the perfect action in the vicinity of the appropriate UV fixed-point, with the understanding that
we take $\Lambda_0 \rightarrow \infty$ at the end of the calculation.
 In this case, we should keep the $\sim$ until such time as the limit is taken (of course, this limit does not exist in the non-renormalizable case).
Henceforth, in both the renormalizable and non-renormalizable cases, $S_{\Lambda_0}$ 
will be referred to as the bare action, for brevity.

As usual, the expression for the correlation functions~\eq{correlations} can be recast by
adding a source term,  $J\cdot \phi$, to the bare action
\be
	\pf[J] \sim
	\Fint{\phi}
	e^{-S_{\Lambda_0}[\phi] + J\cdot\phi},
\ee
so that we have
\be
	\eval{\phi(x_1)\cdots \phi(x_n)}
	=
	\frac{1}{\pf}
	\left. \fder{}{J(x_1)} \cdots \fder{}{J(x_n)} \pf[J] \right\vert_{J=0}.
\ee
Generally speaking, we will prefer to focus on the connected correlation functions
which (taking $\conn$ to stand for `connected', as before) are written as
\be
	\eval{\phi(x_1)\cdots \phi(x_n)}_{\conn} 
	\sim
	 \left. \fder{}{J(x_1)} \cdots \fder{}{J(x_n)} \ln \pf[J] \right\vert_{J=0}.
\label{eq:ConnCorr}
\ee
In momentum space we write
\be
	G(p_1,\ldots,p_n) \deltahat{p_1+\cdots+p_n} \sim
	 \left. \fder{}{J(p_1)} \cdots \fder{}{J(p_n)} \ln \pf[J] \right\vert_{J=0},
\label{eq:ConnCorr-mom}
\ee
with $G(p_1,p_2)$ traditionally written simply as $G(p_1)$.

For almost all of this section,
we will consider objects of the type shown in~\eqs{ConnCorr}{ConnCorr-mom} and will refer to them simply as \emph{the} connected correlation functions. When we have occasion to distinguish these correlation
functions from ones involving local functions of the field, we will refer to the 
former as the standard correlation functions and the
latter as correlation functions involving composite operators. An example of a composite operator is $\phi^2(x)$.

This should be very familiar from standard approaches to QFT; now we wish to switch 
gear and figure out how to extract the correlation functions using the ERG.
For the Polchinski equation, at any rate, we recall from~\eqs{CCF^n}{CCF^2} the relationship between the correlation functions and the low energy limit of the Wilsonian effective action.
Not only do these equations provide a recipe for computing the correlation functions from the
Wilsonian effective action but also shed light on
an important issue which, up until now, we have glossed over.

So far, our entire discussion of renormalizability has been performed at the level of
the effective action, whereas it is more conventionally phrased in terms of the correlation functions.
In the case of the Polchinski equation, these two notions of renormalizability can be conflated,
for the simple reason that the correlation functions are directly related to the
low energy limit of the Wilsonian effective action, as mentioned above. 
Thus, for the Polchinski equation, we know how to compute the correlation functions
and we understand that their renormalizability is guaranteed if the Wilsonian effective action
is renormalizable. From this perspective, one might wonder
if there is any more to be said about computing correlation functions using the ERG; perhaps needless
to say, there is!

There are two angles that one can take. First, suppose that we do not use the Polchinski
equation but rather some other flow equation. In this case, we would like to know how to compute the correlation functions
and how their renormalizability is related to that of the Wilsonian effective action. Secondly, we would like to understand the \emph{nonperturbative} renormalization properties of
correlation functions of composite operators. 
This is not such an unreasonable request. After all, for the Wilsonian effective
action, we were able to give very simple conditions for nonperturbative renormalizability:
either the action sits at a fixed-point or is on a renormalized trajectory. In particular, we did
not have to employ any of the standard machinery, which is far less intuitive and anyway 
perturbative in nature.
Obviously, it would be very nice to be able to do the same sort of thing for correlation functions involving 
composite operators. 

In this paper, we will make a start at dealing with these issues within a new conceptual framework, to be introduced in the next section.
As an illustration it will be shown in \sect{CorrFns-Ball} how to understand the renormalizability of the standard correlation functions 
when using the flow equation of
Ball et al.~\eq{Ball}. Seeing the
technique in action hopefully opens the door to treating more complicated
flow equations (such as those with a non-trivial seed action) and dealing
with composite operators. Moreover, it sheds light
on  the relationship between the dual action
and the correlation functions, as we will see in \sect{Redux}.

A further motivation for studying correlation functions is that they give a proper understanding of
how dilatation covariance is realized in the ERG at a fixed-point. As mentioned already, fixed-point
actions are manifestly \emph{not} dilatation-invariant as a consequence of the cutoff function. Nevertheless, in \sect{FP-scaling} we will see that these actions are such that dilatation covariance of correlation functions at a fixed-point is automatic, which is rather reassuring!

\subsection{Basic Considerations}
\label{sec:comp-corr-fns}

To compute the connected correlation functions using the ERG, we
follow the defining philosophy and integrate out degrees of freedom between the bare and effective scales (this approach mimics that in~\cite{evalues,univ}; see also~\cite{Polonyi-Rev}). As we do so, both the Wilsonian effective action and the source term will evolve. Compared to
the sourceless case, we can consider the effect of this as inducing a shift of the Wilsonian effective action:
\begin{subequations}
\begin{align}
	\Sint_\Lambda[\phi] &\rightarrow \Tact_\Lambda[\phi,J] = \Sint_\Lambda[\phi] + \jop_{\Lambda}[\phi,J],
\label{eq:shift}
\\
	\lim_{\Lambda \rightarrow \Lambda_0} \jop_{\Lambda}[\phi,J]
	& \sim -J \cdot \phi,
\label{eq:source_bc}
\end{align}
\end{subequations}
where we make the obvious split between the functionals $\Sint$ and $\jop$, so that
all terms which are independent of $J$ reside in the former. Thus we can write
\be
	\eval{\phi(x_1)\cdots \phi(x_n)}_{\conn}
	\sim
	\left. \fder{}{J(x_1)} \cdots \fder{}{J(x_n)} \ln
	 \Fint{\phi}
	e^{-S_\Lambda[\phi] -
	\jop_{\Lambda}[\phi,J]}
	\right\vert_{J=0}.
\ee
Now, integrating all the way down to $\Lambda=0$ (at which point the functional integral has been performed), the $S_{\Lambda=0}$ term does not feature
after differentiation \wrt\ the source. This is just as well since $S_{\Lambda=0}$ is divergent,
due to the inverse cutoff function appearing in the two-point vertex.  Since all modes of the field have been integrated over, the contribution to $\jop_{\Lambda=0}[\phi,J]$ which is independent of the field
must be the one which contains the correlation functions. We project this out by setting the field to zero%
\footnote{\label{foot:wonder} We might wonder if there are other options; we return to this in \sect{CorrFns-Ball}. }
 and so write:
\be
	\eval{\phi(x_1)\cdots \phi(x_n)}_{\conn} 
	\sim
	- \left. \fder{}{J(x_1)} \cdots \fder{}{J(x_n)} \jop_{\Lambda=0}[0,J] \right\vert_{J=0}.
\label{eq:corr-O_0}
\ee
Thus, to evaluate the correlation functions, we need to compute $\jop_{\Lambda}[0,J]$,
which can be done using the flow equation. Indeed, given our flow equation of choice, the flow of $\Tact[\phi,J]$ (from which the flow of $\jop$ can be extracted) follows simply by making the shift~\eq{shift}, as is obvious from~\eq{blocked}.

It is important to point out that almost everything we have done so far goes through
exactly the same whatever operator we happen to couple to the action in the UV.
Of course, the boundary condition~\eq{source_bc}
will change. More subtly, if we are on a renormalized trajectory, whilst we do not expect to encounter any problems taking the $\Lambda_0 \rightarrow \infty$ limit if $J \cdot \phi$ is coupled in the UV, the same
is not true for a generic source term. Indeed, in the general case, it is well known that one expects
additional renormalizations, beyond those necessary for the action, in order that the bare scale can be removed for correlation functions involving composite operators (see \eg~\cite{I+Z,Z-J,WeinbergII}).

How are we to deal with the nonperturbative renormalization of composite operators?
The answer is actually staring us in the face! In the sourceless case, we know that the critical fixed-points
of the Wilsonian effective action form the basis for nonperturbatively renormalizable theories. Perturbations of a fixed-point in either an exactly marginal scaling direction or a relevant direction yield
additional renormalizable theories. In the case where source terms are present, we simply repeat
this statement, but allow both the fixed-points and the perturbations to depend on $J$. It is anticipated that these new, source-dependent fixed-points and the relevant/exactly marginal perturbations thereof
will describe the nonperturbative renormalization properties of correlation functions involving composite operators.

Indeed, we will illustrate some of these considerations using the flow equation of Ball et al., \eq{Ball}.
It will be shown that, given a critical fixed-point, it is always possible to construct a related source-dependent fixed-point such that the source-dependent piece reduces,  in dimensionful variables, to $J \cdot \phi$ in the UV. Since fixed-point theories---be they source-dependent or otherwise---are automatically renormalizable, renormalizability of the standard correlation functions follows directly.
This provides a completely new perspective on why the standard correlation functions are renormalizable if the same is true of the Wilsonian effective action. Furthermore, this approach has the right ingredients to be generalizable to more complicated flow equations and to the renormalization of composite operators. 
These tasks are, however,  left to the future, though see the conclusion for a further discussion of their importance.

Before proceeding any further, let us illustrate some of these ideas in the simplest possible setting. To this end, we use the Polchinski equation, \eq{Pol}---but with $\Sint[\phi]$ replaced by $\Tact[\phi,J]$---to compute the correlation functions at the Gaussian fixed-point. To make life easy, we take the simplest representative of the Gaussian fixed-point, $\Sint_\star=0$, whereupon we find that $\jop_\Lambda[\phi,J]$ itself satisfies the Polchinski equation:
\be
	-\flow \jop_\Lambda[\phi,J] 
	= \hf \classical{\jop_\Lambda[\phi,J]}{\dd}{\jop_\Lambda[\phi,J]} 
	- \hf \quantum{\dd}{\jop_\Lambda[\phi,J]}.
\ee
The boundary condition for the operator is
\be
	\lim_{\Lambda \rightarrow \infty} \jop_\Lambda[\phi,J] 
	= -J \cdot \phi,
\ee
and so we see that
\be
	\jop_\Lambda[\phi,J] = - J\cdot \phi
	+
	\hf \int_p J(-p) \frac{\cutoff (p^2/\Lambda^2) - 1}{p^2} J(p).
\label{eq:O-Gaussian}
\ee
In the limit that $\Lambda \rightarrow \infty$, this correctly reproduces the boundary condition [on account of $\cutoff(0)=1$],
whereas at the other end of the RG trajectory we find%
\footnote{Note that for $\D\leq 2$, the momentum integral blows up in the IR. This well-known problem can be circumvented by considering correlation functions of $\partial_\mu \phi(x)$, rather than $\phi(x)$. We will not comment on this further but refer the reader to~\cite{OJR-Pohl} for details.
}:
\be
	\lim_{\Lambda \rightarrow 0} \jop_\Lambda[0,J] =
	-\hf \int_p J(-p) \frac{1}{p^2} J(p).
\label{eq:lowenergy}
\ee
Therefore, precisely as we should, we obtain
\be
	G(p) 
	= \frac{1}{p^2}.
\label{eq:Gaussian-TPCF}
\ee

Now let us transfer to dimensionless variables: $p \mapsto p\Lambda$, $\phi(p) \mapsto \phi(p) \Lambda^{-(d+2)/2}$, $J(p) \mapsto J(p) \Lambda^{(2-d)/2}$.
As anticipated above, upon doing so it is apparent that
\be
	\partial_t \jop_\star[\phi,J] = 0,
\ee
and $\Tact_\star[\phi,J] = \Sint_\star[\phi] + \jop_\star[\phi,J] $ is indeed a source-dependent fixed-point. Note that we can also think of $\jop_\star[\phi,J]$ as an exactly marginal, source-dependent deformation of the
Gaussian fixed-point. To round off this discussion let us note that~\eq{O-Gaussian} provides a rather nice example of a function which is quasi-local for all $\Lambda >0$ but non-local for $\Lambda=0$. Furthermore, whilst the $\Lambda \rightarrow 0$ limit (with dimensionful $\phi,J$ held constant) is non-local, the $t\rightarrow \infty$ limit, after transferring to dimensionless variables (and holding dimensionless $\phi,J$ constant), is quasi-local, a possibility anticipated in footnote~\ref{foot:limit}.

\subsection{Renormalization}
\label{sec:CorrFns-Ball}

Having seen how the (simplest representative of) the Gaussian fixed-point supports a source-dependent extension which satisfies the boundary condition~\eq{source_bc}, we will now show that the same is true of any critical fixed-point. This analysis follows~\cite{OJR-Pohl}. To begin with, 
 we will take our flow equation to be the generalized flow equation~\eq{flow-Ball} [equivalently~\eq{reduced-flow}].
The flow equation for $\Tact_\Lambda[\phi,J]$ can be obtained simply
by replacing $\Sint$ by $\Tact$ (the subscript $\Lambda$ will henceforth be dropped, for brevity):
\be
	\left(-\flow + \frac{\eta}{2} \Count_\phi\right) \Tact[\phi,J]
	=
	\hf \classical{\Tact}{\dd}{\Sigma_\Tact} - \hf \quantum{\dd}{\Sigma_\Tact}
\label{eq:Ball-source-unscaled}
\ee
where $\Sigma_T \equiv T - 2\hSint$. We now rescale to dimensionless variables. For $\phi$, we know that this entails scaling out the canonical dimension, the anomalous part having been already taken care of. For the source, the most general approach to take is to simply suppose that it has some scaling dimension, which we will denote by $d_J$. This object can then be self-consistently determined in what follows. However, rather than doing this, we will fix it here since it is easy to do so. To do this we note that, starting from the bare field, the full rescaling of $\phi$ is $\phi(x) \mapsto \phi(x) \Lambda^{(\D-2)/2}Z^{1/2}$. To ensure that
the $J\cdot \phi$ term contains no explicit dependence on $\Lambda$
it is clear that (remembering the $\volume{x}$ picks up a factor of $\Lambda^{-\D}$)
we should send $J(x) \mapsto J(x) \Lambda^{(\D+2)/2} Z^{-1/2}$.
This leads us to introduce
\be
	d_J = \D - d_\phi = \frac{\D+2-\eta}{2}.
\label{eq:dim-J}
\ee
Defining $\Count_J \equiv J \cdot \delta/\delta J$ and henceforth setting $\hSint=0$, the flow equation~\eq{Ball-source-unscaled} becomes
\be
	\left(
		\partial_t -\hat{D}^- - \hat{D}^J
	\right) \Tact
	=
	\classical{\Tact}{\cutoff'}{\Tact} - \quantum{\cutoff'}{\Tact} 
	- \frac{\eta}{2} \phi \cdot \ep^{-1} \cdot \phi,
\label{eq:Ball-source-b}
\ee
where we have used~\eq{Dhat} and, along the same lines, define
\be
	\hat{D}^J = 
	 \int_p
	\biggl[
		\biggl( \frac{\D- 2 + \eta}{2} + p \cdot \partial_p \biggr) J(p)
	\biggr]
	\fder{}{J(p)}.
\label{eq:DJ}
\ee

We will now demonstrate the central result of this section: given a critical fixed-point, $\Sint_\star$,
there exists a family of source-dependent fixed-points---so long as we take the dimension of the source to be given by~\eq{dim-J}---given by
\be
	\Tact^a_\star[\phi,J,g]= 
	\Sint_\star[\phi]
	+
	\Bigl[
		e^{-a \bar{J} \cdot (\ep g -1 )  \cdot \delta/\delta \phi}
		-1
	\Bigr]
	\Bigl[
	\Sint_\star[\phi] + \hf \phi \cdot \bigl(\ep g -1 \bigr)^{-1} g \cdot \phi
	\Bigr],
\label{eq:T-general}
\ee
where $a$ is an arbitrary real number, $g=g(p^2)$ is a function to be determined and $\bar{J}(p) \equiv J(p)/p^2$. To prove this, it is useful to interpret the second term on the \rhs\ as an exactly marginal source-dependent deformation of the critical fixed-point $\Sint_\star$. Now, in \sect{GenCon} we saw how the marginal, redundant operator associated with every critical fixed-point could be used to generate an exactly marginal perturbation. By appropriately modifying this analysis, it is straightforward to prove~\eq{T-general}. 

With this in mind, we would now like to consider solutions to the fixed-point equation
\be
	\fpop_J(\Tact^a_\star[\phi,J,g]) = 0,
\label{eq:fp-source}
\ee
where
\be
	\fpop_J(\Tact_\star) =
	\classical{\Tact_\star}{\cutoff'}{\Tact_\star} - \quantum{\cutoff'}{\Tact_\star} 
	+ \Bigl(\hat{D}^-_\star +  \hat{D}^J_\star \Bigr) \Tact_\star
	- \frac{\eta_\star}{2} \phi \cdot \ep^{-1} \cdot \phi.
\ee
Let us now define
\be
	\hat{\mathcal{R}} \equiv \bar{J} \cdot \bigl(\ep g -1 \bigr) \cdot \delta / \delta \phi,
	\qquad
	\Stilde_\star \equiv \Sint_\star + \hf \phi \cdot \bigl(\ep g -1 \bigr)^{-1} g \cdot \phi,
	\qquad
	\coupled_a[\phi,J]
	\equiv
	\bigl(e^{-a \hat{\mathcal{R}}} - 1 \bigr) 
	\Stilde_\star.
\ee
Using the fact that $\fpop_J(\Sint_\star[\phi]) = 0$, we have that
\be
	\fpop_J(\Sint_\star + \coupled_a)
	=
	\classifierJ \coupled_a
	+ \classical{\coupled_a}{\cutoff'}{\coupled_a},
\label{eq:analogue}
\ee
where, recalling~\eq{classifier},
\be
	\classifierJ = \classifier + \hat{D}^J_\star =
	2 \classical{\Sint_\star}{\cutoff'}{} - \quantum{\cutoff'}{} + \hat{D}^{-}_\star +  \hat{D}^J_\star.
\ee
Notice that~\eq{analogue} is the analogue of~\eq{zero}. To build on this analogy, let us observe that
\be
	\bigl[
		\hat{\mathcal{R}}, \hat{D}^{-}_\star +  \hat{D}^J_\star
	\bigr]
	=
	\int_p \frac{J(p)}{p^2}
	\Bigl[
		p \cdot \partial_p \, \ep(p^2) g(p^2) - \eta_\star \ep(p^2) g(p^2) +\eta_\star
	\Bigr]
	\fder{}{\phi(p)}
\label{eq:Rhat-comm}
\ee
from which it is easy to check that, so long as $g(p^2)$ satisfies
\be
	- \frac{2+\eta_\star}{2} g(p^2) + p^2 g'(p^2) +\frac{\eta_\star}{2} \ep^{-1}(p^2) =0,
\label{eq:g-condition}
\ee
we have (up to a neglected constant in the second case)
\be
	\bigl[
		 \hat{\mathcal{R}}, \classifierJ
	\bigr]	
	=
	2\classical{\, \hat{\mathcal{R}}\Stilde_\star}{\cutoff'}{},
	\qquad
	\classifierJ \,
	\hat{\mathcal{R}}
	\Stilde
	= 0.
\ee
These equations are in correspondence with~\eqs{commutator}{zeromode}, respectively. Now we simply repeat the steps leading to~\eq{demonstrated} (though this time differentiating \wrt\ $a$), from which~\eq{T-general} follows. This latter equation can be expanded out to yield
\be
	\jop^a_\star[\phi,J,g] = 
	-a \bar{J} \cdot g \cdot \phi + \frac{a^2}{2} \bar{J} \cdot g \bigl(\ep g-1\bigr) \cdot \bar{J}
	+ 
	\Bigl[
		e^{-a \bar{J} \cdot (\ep g -1) \cdot \delta / \delta \phi}
		- 1
	\Bigr]
	\Sint_\star[\phi].
\label{eq:jop-solution}
\ee

The next step is to determine $g(p^2)$. Remembering that $\bar{J}(p) \equiv J(p)/p^2$, the solution for $g$ must be such that both $g(p^2)/p^2$ and $[\ep(p^2) g(p^2) -1]/p^2$ are quasi-local. Recalling the analysis around~\eq{h-lessthan}, notice that we can achieve this by taking
\be
	g(p^2) = \ep^{-1}(p^2) \bigl[1+ \varrho(p^2)\bigr], \qquad \eta_\star <2.
\label{eq:g-soln}
\ee
(We will treat the case $\eta_\star \geq 2$ shortly.)
Let us now consider the object $g(p^2) /p^2$ in dimensionful variables. Employing~\eq{rho-Taylor} we see that, under the transfer to dimensionful variables,
\be
	\frac{g(p^2)}{p^2} \mapsto 1  + \order{p^2/\Lambda^2}, \qquad \eta_\star <2.
\ee
Next let us render $\phi$ and $J$ dimensionful, by which we mean that we restore not just their canonical dimensions, but their full scaling dimension [which will generate factors of $Z \sim (\Lambda/\mu)^{\eta_\star}$]. Upon doing so, each $J(p)$ and each $\delta /\delta \phi(p)$ pick up a factor of $\Lambda^{d_\star}/\mu^{\eta_\star}$. Remembering to extract a $1/p^2$ from each $\bar{J}(p)$s we find that, in dimensionful variables,
\be
	\lim_{\Lambda \rightarrow \infty} a \bar{J} \cdot  g \cdot \phi
	= a\, J \cdot \phi, \qquad \eta_\star <2,
\ee
making it clear that if we set $a=1$ then we have some hope of satisfying the boundary condition~\eq{source_bc}. Indeed, with $g(p^2) = \ep^{-1}(p^2) \bigl[1+ \varrho(p^2)\bigr] $
we have, under the transfer to dimensionful variables,%
\footnote{Note that the $\order{p^2/\Lambda^2}$ terms would in fact vanish if we took $\cutoff'(0) = 0$, spoiling the following analysis. This is yet another example of the importance of the constraint mentioned under~\eq{Pol-linear} that we must take $\cutoff'(p^2/\Lambda^2) < 0$, for $p^2/\Lambda^2 < \infty$.
}
\be
\begin{split}
	\bar{J} \cdot g \bigl(\ep g -1 \bigr)  \cdot \bar{J}
	& \mapsto
	\frac{\Lambda^{2d_\star-\D}}{\mu^{\eta_\star}} \int_p J(p) J(-p) 
	\, \frac{\Lambda^2}{p^2}  \,
	\order{p^2/\Lambda^2},
\\
	\bar{J} \cdot (\ep g -1 )  \cdot \fder{}{\phi}
	& \mapsto
	\frac{\Lambda^{2d_\star-\D}}{\mu^{\eta_\star}} \int_p J(p) \fder{}{\phi(p)}
	\,  \frac{\Lambda^2}{p^2} \,\order{p^2/\Lambda^2}.
\end{split}
\label{eq:limit}
\ee
Since $2d_\star -\D = \eta_\star -2$ and we are currently considering $\eta_\star <2$, it is apparent that the contributions to
$\jop_\star[\phi,J]$ involving the terms in~\eq{limit} vanish in the $\Lambda \rightarrow \infty$ limit (given that, in dimensionful variables, $\Sint_\star$ has at least some terms which survive the limit). Therefore, we have a source-dependent fixed-point solution 
\be
	\Tact_\star[\phi,J] \equiv \Tact^1_\star[\phi,J,\ep^{-1}(1+\varrho)]  =\Sint_\star[\phi] + 
	\Bigl(
		e^{-\bar{J} \cdot \varrho \cdot \delta / \delta \phi}
		- 1
	\Bigr)
	\Bigl[
		\Sint_\star + \hf \phi \cdot  \ep^{-1} (1+ \varrho^{-1}) \cdot \phi 
	\Bigr],\qquad \eta_\star <2
\label{eq:T-solution}
\ee
which, in dimensionful variables, satisfies the boundary condition~\eq{source_bc}.

What about fixed-points for which $\eta_\star \geq 2$? In this case, for $\eta_\star = 2,4,6,\ldots$ a quasi-local $g(p^2)$ does not exist, due to unavoidable logarithmic corrections. Excluding these cases, we find that $g(p^2) /p^2 = 1 + \order{p^2}$, in dimensionless variables. It therefore follows that, rather than vanishing in the limit $\Lambda \rightarrow \infty$, the terms in~\eq{limit} in fact diverge. So, let us suppose that the operator, $\jop_\star[\phi,J]$,  is the unique exactly marginal, source-dependent perturbation of a fixed-point that, for some range of $\eta_\star$, reduces (in dimensionful variables) to $J \cdot \phi$ in the $\Lambda \rightarrow \infty$ limit. Then the above analysis implies that it is impossible to define the standard correlation functions at fixed-points with $\eta_\star \geq 2$. Can this possibly make sense? The answer is yes, so long as we identify these fixed-points as being non-critical. As mentioned at the end of \sect{renormalizability}, non-critical fixed-points are uniquely IR fixed-points, and are thus reached in the $\Lambda \rightarrow 0$ limit of some flow. 
Since the construct for renormalizable correlation functions requires working in the  $\Lambda \rightarrow \infty$ limit, it thus makes perfect sense that this procedure breaks down for non-critical fixed-points.

So let us now turn things around: \emph{if} the operator, $\jop_\star[\phi,J]$ is unique in the above sense  (which would be nice to prove), then the fact that it only reduces (in dimensionful variables) to $J \cdot \phi$ in the $\Lambda \rightarrow \infty$ limit for $\eta_\star <2$ would be one way of understanding why critical fixed-points necessarily exhibit this restriction on the anomalous dimension. We will make some further comments in the next section.



Before moving on, let us return to the case of $\eta_\star<2$ and emphasise that the existence of the exactly marginal, source dependent operator which satisfies the boundary condition~\eq{source_bc} implies  the renormalizability of the standard correlation 
functions not only at a critical fixed-point, but also along the associated renormalized trajectories. 
This is easy to prove. Consider a fixed-point, $\Sint_\star[\phi]$, with eigenperturbations $\eop_i[\phi]$, and associated eigenvalues $\lambda_i$. The source-dependent extensions of this fixed-point, $\Tact^a_\star[\phi,J,g]$ each possess eigenperturbations
\be
\tilde{\eop}_i[\phi,J] = e^{-a \bar{J} \cdot (\ep g -1) \cdot \delta / \delta \phi} \eop_i[\phi]
=e^{-a\hat{\mathcal{R}}} \eop_i[\phi].
\ee
 with the same eigenvalues. To see this, consider substituting
 \be
	 \Tact^a_t[\phi,J,g] = \Tact^a_\star[\phi,J,g] + \alpha e^{\tilde{\lambda}_i t} e^{-a\hat{\mathcal{R}}}
	 \eop_i[\phi]
 \ee
into the source-dependent flow equation~\eq{Ball-source-b}. Utilizing the expression, \eq{T-general}, we find that
\begin{align*}
	\tilde{\lambda}_i e^{-a\hat{\mathcal{R}}} \eop_i[\phi]
	&=
	\biggl(
		2 \classical{e^{-a\hat{\mathcal{R}}} \Sint_\star}{\cutoff'}{} 
		- 2 a \bar{J} \cdot g \cutoff' \cdot \fder{}{\phi}
		-\quantum{\cutoff'}{} + 
		\hat{D}^-_\star +  \hat{D}^J_\star
	\biggr)
	e^{-a\hat{\mathcal{R}}} \eop_i[\phi]
\\
	& =
	e^{-a\hat{\mathcal{R}}} 
	\biggl(
		2 \classical{\Sint_\star}{\cutoff'}{} 
		-\quantum{\cutoff'}{} + 
		\hat{D}^-_\star
	\biggr)
	\eop_i[\phi]
\\
	&
	\qquad \qquad \qquad
	-e^{-a\hat{\mathcal{R}}} 
	\biggl(
		2 a \bar{J} \cdot g \cutoff' \cdot \fder{}{\phi}
		-a 
		\Bigl[
			\hat{\mathcal{R}}, \hat{D}^-_\star +  \hat{D}^J_\star
		\Bigr]
	\biggr)
	\eop_i[\phi].
\end{align*}
Using~\eqs{Rhat-comm}{g-condition} it is easy to show that the last line is identically zero. We thus find:
\be
	\tilde{\lambda}_i e^{-a\hat{\mathcal{R}}} \eop_i[\phi] = 
	e^{-a\hat{\mathcal{R}}} \classifier  \eop_i[\phi] = \lambda_i  e^{-a\hat{\mathcal{R}}} \eop_i[\phi],
\ee
from which we conclude that $\tilde{\lambda}_i = \lambda_i$, as advertised. 

For the specific case of $\eta_\star <2$, the particular source-dependent fixed-point solution $\Tact_\star[\phi,J]$ generates renormalized correlation functions appropriate to the critical fixed-point $\Sint_\star$. Now, we have just learnt that for every $\Tact_\star$ there exist source-dependent renormalized trajectories that reduce to the usual source-independent ones in the limit $J \rightarrow 0$. We thus conclude that the correlation functions of scale-dependent renormalizable theories [obeying the flow equation~\eq{Ball}] are nonperturbatively renormalizable.

To conclude this section, let us make sure that our solution, \eq{T-solution}, yields the correct answer at the Gaussian fixed-point.
Working in dimensionful variables we have
\[
	\eta_\star = 0,
	\qquad
	\Sint_\star[\phi] = \hf \int_p \phi(-p) \frac{\intconst p^2}{1-\intconst \cutoff (p^2/\Lambda^2)} \phi(p),
	\qquad
	\rho(p^2) = \frac{\cutoff(p^2/\Lambda^2)-1}{p^2/\Lambda^2}
\]
and so we find:
\be
	\jop_\Lambda[\phi,J] = -\int_p J(p)\frac{1-B}{1-B\cutoff(p^2/\Lambda^2)} 
	\biggl[
		\phi(-p)
		-\hf \frac{\cutoff(p^2/\Lambda^2)-1}{p^2} J(-p)
	\biggr].
\ee
In the $\Lambda \rightarrow \infty$ limit this reproduces the boundary condition. At the other end we see that
\be
	\lim_{\Lambda \rightarrow 0} \jop_\Lambda[0,J] = -\frac{(1-\intconst)}{2} \int_p J(p) \frac{1}{p^2} J(-p).
\label{eq:lowenergy-gen}
\ee
For the simplest representative of the Gaussian fixed-point, $\intconst=0$, we recover the earlier solution~\eq{lowenergy}. To understand the solution for the other representatives, let us recall that
\[
	S^{\mathrm{total}}_\star[\phi]
	= \hf \int_p \phi(-p) \frac{\ep^{-1}(p^2)}{[1-\intconst \cutoff (p^2)]}  \phi(p).
\]
Here the $\order{p^2}$ part of the kinetic term is not canonically normalized, going like
$1/[2(1-\intconst)]$, rather than $1/2$. If we so desired, we could remove this $1/(1-\intconst)$ by shifting
$\phi \rightarrow \phi (1-\intconst)^{1/2}$. In order to leave the source term alone, we would also have
to shift $J \rightarrow J (1-\intconst)^{-1/2}$, which would remove the $1-\intconst$ in~\eq{lowenergy-gen}. 

Let us conclude with two comments. First, for the non-critical fixed-point corresponding to $B=1$, we see that the solution disappears, as expected. Secondly, for $B>1$, the sign of $\lim_{\Lambda \rightarrow 0} \jop_\Lambda[0,J] $ flips and so, as anticipated under~\eq{Gen-Gauss}, a wrong sign kinetic term leads to a loss of positivity of the two-point correlation function.

\subsection{The Dual Action, Redux}
\label{sec:Redux}

Further insights follow from understanding the relationship between the correlation functions and the dual action. The idea is to modify the dual action to appropriately take account
of the fact that we have introduced a source term into the action:
\be	
	-\dualshift[\phi,J]
	\equiv
	\ln
	\bigl(
		e^{\op} e^{-\Tact_\Lambda[\phi,J]}
	\bigr),
\label{eq:dual-J}
\ee
where $\Tact_\Lambda[\phi,J]$ is defined in~\eq{shift} and we recall that $\op \equiv \hf \delta/\delta \phi \cdot \ep \cdot \delta/\delta \phi$. (We will assume that there are no IR obstructions to constructing the dual action.) Using the source-dependent flow equation~\eq{Ball-source-unscaled},
the flow of $\dualshift[\phi,J]$ is just the same as 
the flow of $\dual[\phi]$, \eq{dualflow-Seed}, but with $\Sint$ replaced by $\Tact$ and $\dual$ replaced by $\dualshift$ wherever appropriate:
\be
	\left(
		\flow 
		+ \frac{\eta}{2} \Count_\phi
	\right)
	\dualshift[\phi,J] = 
	\frac{\eta}{2} \phi \cdot \ep^{-1} \cdot \phi
	+
	e^{\dualshift} \phi \cdot \ep^{-1} \dd \cdot
	e^\op
	\,
	\fder{\hSint}{\phi} e^{-\Tact}.
\label{eq:D_O-Flow}
\ee
The crucial observation to make is that the \rhs\ vanishes if we set the field to zero (we can always choose the seed action such that it has an expansion about vanishing field).
Assuming that the same is true of
the $\Count_\phi$ term on the left yields:
\be
	\flow \dualshift[0,J] = 0.
\label{eq:dual_J-flow}
\ee
This establishes that the shifted dual action, as a functional of the source at vanishing field, is
independent of scale.%
\footnote{Indeed, independence of scale in dimensionful variables implies that $\dualshift[0,J]$ closely related to physical quantities. Following on from footnote~\ref{foot:wonder} this suggests that, for zero seed action, \emph{any} solution $\phi = \Phi$ to the equation $\phi \cdot \delta \dualshift[\phi,J] /\delta \phi- \phi \cdot \ep^{-1} \cdot \phi = 0$ yields a physical $\dualshift[\Phi,J]$ (the generalization to non-trivial seed action is obvious). This is strongly reminiscent of the quantum equations of motion obtained from the standard effective action and merits further investigation.}
Since we can therefore evaluate $\dualshift[0,J]$ at any scale and get the same answer, 
let us see what
happens as $\Lambda \rightarrow 0$. Consistent with our assumption that there are no IR pathologies, we observe that since $\lim_{\Lambda\rightarrow0} \cutoff (p^2/\Lambda^2) = 0$ we can set $\op$ to zero.
Therefore we have that
\be
	\dualshift[0,J] = \lim_{\Lambda\rightarrow0} \left(\Sint_\Lambda[0] + \jop_{\Lambda}[0,J]\right).
\label{eq:dualshift_0}
\ee
When we take derivatives \wrt\ $J$, the first term is killed and so substituting~\eq{dualshift_0} 
into~\eq{corr-O_0} yields
\begin{align}
	\eval{\phi(x_1)\cdots \phi(x_n)}_{\conn}
	&\sim
	\left.
	- \fder{}{J(x_1)} \cdots \fder{}{J(x_n)} 
	\jop_{\Lambda=0}[0,J]
	\right\vert_{J=0}
\nonumber
\\
	&\sim
	\left.
	- \fder{}{J(x_1)} \cdots \fder{}{J(x_n)} 
	\dualshift[0,J]
	\right\vert_{J=0}.
\label{eq:correlations-dual}
\end{align}
Consequently, the connected correlation functions are just given by the vertices of $\dualshift[0,J]$.
Let us emphasise that this is true for any choice of seed action and, since we are yet to send the bare scale to infinity, for any boundary condition of $\jop$.

This is a good point to pause to see how we can recover our previous result
at the Gaussian fixed-point using the dual action formalism. Since for this fixed-point $\eta_\star=0$, if we set $\hSint=0$ then we are effectively dealing with the Polchinski equation. With this in mind, let us substitute our earlier solution to the Polchinski equation, \eq{O-Gaussian}, into~\eq{dual-J}
so that we have, for the simplest representative of the Gaussian fixed-point:
\be
	-\dualshift[\phi,J]
	=
	-
	\hf \int_p J(-p) \frac{\cutoff (p^2/\Lambda^2) - 1}{p^2} J(p)
	+
	\ln e^{\op} e^{J \cdot \phi}.
\label{eq:dualshift-Gaussian}
\ee
The final term is the sum of all connected diagrams built from
$J \cdot \phi$ vertices connected by $\ep$s. The constraint
of connectedness is highly restrictive, in this case, and all we
have is
\[
	\ln e^{\op} e^{J \cdot \phi} = J \cdot \phi + \hf J \cdot \ep \cdot J.
\]
Substituting this into~\eq{dualshift-Gaussian} yields:
\be
	-\dualshift[\phi,J]
	=
	J \cdot \phi 
	+
	\hf
	\int_p J(-p) \frac{1}{p^2} J(p).
\ee
Finally, then, using this result in~\eq{correlations-dual} recovers the
expected answer~\eq{Gaussian-TPCF}.

Although the result~\eq{correlations-dual} is true for any seed action,
henceforth we will take $\hSint=0$, leaving the general analysis for
the future. In this case, \eq{D_O-Flow} reduces to
\be
	\left(
		\flow 
		+ \frac{\eta}{2} \Count_\phi
	\right)
	\dualshift[\phi,J] = 
	\frac{\eta}{2} \phi \cdot \ep^{-1} \cdot \phi.	
\label{eq:D_O-Flow-simple}
\ee
Transferring to dimensionless variables we obtain
\be
\left(
	\partial_t - \hat{D}^+ - \hat{D}^J -\frac{\eta}{2} \phi \cdot \ep^{-1} \cdot \phi
	\right) 
	e^{-\dualshift_t[\phi,J]} = 0,
\label{eq:dualshift-flow}
\ee
where we recall~\eqs{Dhat}{DJ}. Note that, henceforth, all variables are dimensionless unless explicitly stated otherwise.

Let us now check that this result is consistent with what we found in the previous section.
To do this, we will compute the dual action corresponding to the $\Tact_\star[\phi,J]$ given by~\eq{T-solution}. We begin by noting that
\be
	\exp
	\Bigl(
		-e^{-\bar{J} \cdot \varrho \cdot \delta/\delta \phi} \Sint_\star[\phi]
	\Bigr)
	=
	e^{\bar{J} \cdot \varrho \cdot \delta/\delta \phi} e^{-\Sint_\star[\phi]}.
\label{eq:exp-manip-reverse}
\ee
The proof is simple. First we write
\begin{multline}
	e^{-\bar{J} \cdot \varrho \cdot \delta/\delta \phi } e^{-\Sint_\star[\phi]}
	=
	\sum_{i,j=0}^\infty
	\frac{(-1)^j}{i!j!}
	\biggl(
		-\bar{J} \cdot \varrho \cdot \fder{}{\phi}
	\biggr)^i
	\bigl(
		\Sint_\star[\phi]
	\bigr)^j
\\
	=
	\sum_{i,j=0}^\infty
	\frac{(-1)^j}{i!j!}
	\sum_{i_1=0}^\infty \cdots \sum_{i_j=0}^\infty \delta_{i,i_1+\cdots+i_j}
	\nCr{i}{i_1} \nCr{i-i_1}{i_2} \cdots \nCr{i-i_1 \cdots -i_{j-1}}{i_j}
\\
	\times
	\Biggl\{
	\biggl(
		-\bar{J} \cdot \varrho \cdot \fder{}{\phi}
	\biggr)^{i_1} \Sint_\star[\phi]
	\Biggr\}
	\cdots
	\Biggl\{
	\biggl(
		-\bar{J} \cdot \varrho \cdot \fder{}{\phi}
	\biggr)^{i_j} \Sint_\star[\phi]
	\Biggr\}.
\end{multline}
Expanding out the combinatoric symbols, we are left with a product $1/i_1! \cdots 1/i_j!$, with
all dependence on $i$ cancelling out. Consequently, the sum over $i$ becomes trivial,
simply removing the Kronecker-$\delta$:
\be
	e^{-\bar{J} \cdot\varrho\cdot \delta/\delta \phi } e^{-\Sint_\star[\phi]}
	=
	\sum_{j=0}^\infty \frac{(-1)^j}{j!} 
	\Biggl\{
	\sum_{i_1=0}^\infty \frac{1}{i_1!}
	\biggl(
		-\bar{J} \cdot \varrho \cdot \fder{}{\phi}
	\biggr)^{i_1} \Sint_\star[\phi]
	\Biggr\}^j
	=
	\exp
	\Bigl(
		-e^{-\bar{J} \cdot \varrho \cdot \delta/\delta \phi} \Sint_\star[\phi]
	\Bigr),
\ee
thus demonstrating~\eq{exp-manip-reverse}. 

Using this result we find, from the definition of $\dualshift$, \eq{dual-J}, and our solution for $\Tact_\star$, \eq{T-solution}, that
\be
	-\dualshift_\star[\phi,J] = 
	-\hf \bar{J} \cdot \ep^{-1} \bigl(\varrho+1 \bigr) \varrho \cdot \bar{J}
	+
	\ln 
	\biggl\{
		e^{\op}
		e^{\bar{J} \cdot \ep^{-1} (\varrho+1) \cdot \phi}
		e^{-\bar{J} \cdot \varrho \cdot \delta/\delta \phi} e^{-\Sint_\star[\phi]}
	\biggr\}, \qquad \eta_\star <2.
\ee
To proceed, we exploit the fact that the logarithm generates \emph{connected} diagrams.
This is very restrictive. In particular, the $\bar{J} \cdot \ep^{-1} \bigl(\varrho+1 \bigr) \cdot \phi$
can appear in only three ways: on its own, two copies connected to each other with an
$\op$, or any number of copies spliced onto the $\phi$-legs of diagrams built out of the
$\Sint_\star$. From this we conclude that
\be
	\dualshift_\star[\phi,J] = -\bar{J} \cdot \ep^{-1} \bigl(\varrho+1 \bigr) \cdot \phi
		- \hf \bar{J} \cdot \ep^{-1} \bigl(\varrho+1 \bigr) \cdot \bar{J}
		-\ln 
		\biggl(
			e^{\op}
			e^{\bar{J} \cdot  \delta/\delta \phi} e^{-\Sint_\star[\phi]}
		\biggr), \qquad \eta_\star <2.
\label{eq:dual_J-solution}
\ee
Using the definition of the dual action~\eq{dual}, together with the appropriate variant of~\eq{exp-manip-reverse} yields:
\be
	\dualshift_\star[\phi,J]  - \dual_\star[\phi]
	=
	\Bigl(	
		e^{\bar{J} \cdot  \delta/\delta \phi}  -1
	\Bigr)
	\Bigl\{
		\dual_\star[\phi] - \hf \phi \cdot \ep^{-1} \bigl(\varrho+1 \bigr) \cdot \phi
	\Bigr\}, \qquad \eta_\star <2.
\ee
It is easy to check that, at a fixed-point, $\dualshift_\star[\phi,J]$ satisfies~\eq{dualshift-flow}. Recalling~\eq{correlations-dual} it is apparent that, in dimensionless variables,
\begin{multline}
	G(p_1,\ldots, p_n) \, \deltahat{p_1 + \cdots + p_n}
	=
\\
	\left.
	- \fder{}{J(p_1)} \cdots \fder{}{J(p_n)} 
	\Bigl(
		e^{\bar{J} \cdot \delta/\delta \phi} - 1
	\Bigr)
	\Bigl\{
		\dual_\star[\phi] - \hf \phi \cdot  \ep^{-1} \bigl(\varrho+1 \bigr) \cdot \phi
	\Bigr\}
	\right\vert_{J,\phi=0},
\label{eq:CorrFnsByDual}
\end{multline}
making clear the relationship between the correlation functions and the dual action. This is very natural: the $n>2$-point critical correlation functions are given by the vertices of the dual action---which we recall from~\eq{homog-solution} transform homogeneously with momenta---multiplied by $1/\mathrm{mom}^2$ on each leg. The exception is at the two-point level, where we need an extra term to subtract off the inhomogeneous part of $\dualv{2}$. Recalling~\eq{D2-solution}, \eqs{h}{c} it is straightforward to show 
that
\begin{align}
	G(p) & = \frac{\intconst_{\eta_\star} + \const_{\eta_\star}}{p^2},
\label{eq:G_2}
\\
	G(p_1,\ldots,p_n) & = (-1)^{n+1} 
	\dualv{n}(p_1,\ldots,p_n) \prod_{i=1}^{n} \frac{1}{p_i^2}.
\label{eq:G_n}
\end{align}
As we will see in the next section, correlation functions of this form are automatically dilatation covariant.
Notice that, finally, we see why we chose the sign the way that we did in~\eq{constants}: for $\eta_\star <2 ,\neq 0$ we have that $\const_{\eta_\star} = 0$ and so taking $\intconst_{\eta_\star}$ guarantees positivity of the connected two-point function. The slight difference between the $\eta_\star = 0$ and $\eta_\star \neq 0$ cases is important and can be used to show that the only fixed-point with $\eta_\star = 0$ [subject to positivity of $G(p)$] is the Gaussian one~\cite{OJR-Pohl}.

From the perspective of~\eqs{G_2}{G_n},
let us return to the issue of whether it might be possible to find other fixed-point solutions to the source-dependent flow equation which satisfy the boundary condition~\eq{source_bc}. This would seem unlikely for, should such solutions exist, they would not yield the expected form of the correlation functions at a fixed-point. Furthermore, at the two-point level  we can explicitly show that the only solution is~\eq{T-solution}.%
\footnote{%
To see this note that, for two-point solutions, 
$\dualshift_\star[\phi,J] = \hf \phi \cdot g \cdot \phi + \hf \phi \cdot A_0 \cdot \phi + 
 \bar{J} \cdot A_1\cdot \phi + \hf \bar{J} \cdot A_2 \cdot \bar{J}$, 
where $A_i(p^2) = a_i p^{2(1+\eta_\star/2)}$, for constants $a_i$. Without any loss of generality, we can perform a rescaling such that either $a_1 = a_2$ or $a_1 = -a_2$. Using the operator $e^{-\op}$ (which has a well defined action on two-point objects) it is easy to check that only if we choose $\dualshift_\star$ as in~\eq{dual_J-solution} do  we recover the correct boundary condition in dimensionful variables.}

Let us conclude this section by noting that the exact solution for $e^{-\dualshift_\star}$, given by~\eq{dual_J-solution}, is a non-trivial example of an object on which $e^{-\op}$ has a well defined action: it is easy to work backwards to show that $e^{-\op} e^{-\dualshift_\star} = e^{-\Tact_\star}$, with $\Tact_\star$ given by~\eq{T-solution}.

\subsection{Dilatation Covariance}
\label{sec:FP-scaling}

Let us conclude our discussion of the correlation functions
by showing that,
at a critical fixed-point, they are automatically covariant under dilatations. In other words, given the scaling factor, $a$, 
we would like to demonstrate that:
\be
	\eval{\phi(ax_1)\cdots \phi(ax_n)}_{\conn} = a^{-nd_\star} \eval{\phi(x_1)\cdots \phi(x_n)}_{\conn}.
\label{eq:CorrFnScaling}
\ee
To this end, let us recall that, for general
seed action, the correlation functions are related to $\dualshift[0,J]$
via~\eq{correlations-dual}. For dimensionful $J$, $\dualshift[0,J]$
satisfies~\eq{dual_J-flow}; in the dimensionless case we have
\be
	\left(
		\partial_t -\hat{D}^J
	\right) \dualshift[0,J] = 0.
\label{eq:D-flow-rescaled}
\ee
Although~\eq{D-flow-rescaled} can be read off from~\eq{dualshift-flow},
let us note that it holds more generally than this: \eqn{D-flow-rescaled}
follows directly from~\eq{dual_J-flow}---which is valid for any seed action---by rescaling $J$,
whereas~\eq{dualshift-flow} is true only for $\hSint=0$.

However, now the shortcomings of the analysis of the previous section
do force us to take $\hSint=0$. The point is that it is only in this case that
we have shown that each critical fixed-point can be used to generate a source-dependent
fixed-point which, in dimensionful variables reduces to $J \cdot \phi$ in 
the $\Lambda \rightarrow \infty$ limit. Therefore, it is only for $\hSint = 0$ that we have succeeded
in showing that we have
\be
	\partial_t \dualshift_\star[0,J] = 0,
\label{eq:dualshift-fp}
\ee
in dimensionless variables. Trivially, when~\eq{dualshift-fp} is satisfied we have that
\be
	\hat{D}^J_\star \dualshift_\star[0,J] = 0.
\label{eq:dualshift-fp-condition}
\ee

It is useful to write this equation out in position space. Examining~\eq{DJ}, observe that
\[
	 \int_p [p\cdot \partial_p J(p)]  \fder{}{J(p)} = \int_x J(x) x\cdot \partial_x \fder{}{J(x)}
\]
and so~\eq{dualshift-fp-condition} becomes:
\be
	\Int{x} J(x)
	\bigl(
		x_\mu \partial_\mu + d_\star
	\bigr)
	\fder{}{J(x)}
	 \dualshift_\star[0,J] = 0,
\label{eq:DilatationInvariance}
\ee
where we recall that $d_\star \equiv (\D-2+\eta_\star)/2$.
We now recognize $x_\mu \partial_\mu + d_\star$ as the generator of dilatations
(see e.g.~\cite{YellowPages}), and~\eq{DilatationInvariance} as the infinitesimal version
of~\eq{CorrFnScaling}.

Thus, we have proven that the correlation functions at a 
critical fixed-point are annihilated by the dilatation generator---and, therefore, that the correlation functions exhibit the expected dilatation covariance---even though the fixed-point action is
not, itself, dilatation-invariant. However, the action (at a fixed-point or otherwise) is Euclidean invariant and so the correlation functions automatically inherit covariance under translations and rotations.
A subset of fixed-points will additionally be covariant under special conformal transformations.
Such conformal fixed-points are expected to be critical and it would be nice to investigate this further. Note also that applying conformal covariance as a constraint on the correlation functions might
render the inverse problem of deducing the corresponding fixed-point action more tractable.

\section{Flow Equations for Other Theories}
\label{sec:other-theories}

Up to this point, we have dealt almost exclusively with theories of single scalar field, $\phi$.  We will now briefly describe ERGs for other theories. In \sect{Other-Overview} we will indicate, at the schematic level, how to modify some of the flow equations that we have encountered and will mention some applications. \Sect{mgierg} is devoted to outlining the key concepts of manifestly gauge invariant ERGS; some new insights are also presented.

\subsection{Overview}
\label{sec:Other-Overview}

In the context of the generalized flows of \sect{General_ERG},
incorporating multiple scalars is, as mentioned around~\eq{flow-N}, easy. The generalization to non-scalar theories follows the same pattern. In this section and the next,
we take $\varphi_i$ to represent some set of fields which are not necessarily scalars.
Thus we introduce a set of kernels labelled by the fields, $\dd^{\varphi_i \varphi_j}$. Note that 
$\dd^{\varphi_i \varphi_j}$ is not a function of the fields: the notation is just meant to read `the kernel for $\varphi_i$ and $\varphi_j$'.
The various kernels may very well be different from one another. 
The generalized flow equation~\eq{ProtoFlow} becomes:
\be
	-\flow  S 
	= \hf \fder{S}{\varphi_i} \cdot \dd^{\varphi_i \varphi_j} \cdot \fder{\Sigma}{\varphi_j} 
	- \hf \fder{}{\varphi_i} \cdot \dd^{\varphi_i \varphi_j}  \cdot \fder{\Sigma}{\varphi_j},
\label{eq:flow-multi}
\ee
where a sum over repeated indices is understood, and the dots sandwiched between
the functional derivatives and the kernels represent not only an integral over momentum,
but also sums over Lorentz indices, spinor indices and so-forth, as appropriate.
Including fermions and non-gauge vector fields is now easy: all that we must do is 
make sure that the $\varphi_i$ incorporate the necessary fields. 

Similarly modifying the effective average action approach is equally straightforward: returning to~\eq{Morris-1PI},
all we need to do is include the appropriate fields and interpret the trace appropriately. Further discussion of
fermionic systems, together with references, can  be found in~\cite{Wetterich-Rev,JMP-Review,Salmhofer+Honerkamp}. Let us note that there has been a recent focus on non-relativistic
systems~\cite{Birse-Pairing, Gies-Flow-BCS-BEC,Gies-Ultra-Cold,Krippa-Ultra-Cold,Diehl-3body,Birse-2body,Strack-Superfluids,Diehl-Particle-Hole,Moroz-trion,
Moroz-Efimov-Effect,Floerchinger-Lithium,Gies-FRG-BCE-BEC,Kopietz-Truncate,
Birse-3body,Krippa-Few+Many}, particularly in the context of the topical
subject of ultra-cold gases. In a different direction, the effective average action approach has been recently used, for the first time, to study the physics of polymerized membranes~\cite{Mouhanna-Membranes}.

Supersymmetry in either of the generalized ERG or effective average action approaches presents no particular problems, for which
the reader is referred to~\cite{Bonini+Vian,Falkenberg,Susy-Chiral,Sonoda-SUSY-Construct}. 
A fairly up-to-date list of references can be found in~\cite{Susy-Chiral} but it is worth
mentioning that there has been a recent increase in
activity in the investigation of supersymmetric flows~\cite{Higashi:2007tn,Higashi:2007dm,Synatschke:2008pv,Gies:2009az,Synatschke:2009nm,Synatschke:2009da,Sonoda-NonRenorm,Synatschke:2010ub}. Interestingly, supersymmetric theories are so constrained that just the existence 
of the Wilsonian effective action, together with a knowledge of  the non-renormalization theorem, allows one to essentially rule out an  asymptotic safety scenario for the Wess--Zumino model~\cite{Safety}.%
\footnote{Such a scenario would, amongst other things, require the associated UV  fixed-point to have negative anomalous dimension. Thus, even if such a fixed-point exists, we expect the theory to be non-unitary upon continuation to Minkowski space.}
The flow equation has also been adapted for
use in noncommutative scalar field theories~\cite{G+W-PC,G+W-2D,G+W-4D,RG+OJR}, as mentioned earlier.

Gauge fields, unsurprisingly, present their own problems. Below, we will sketch the construction of generalized ERGs for gauge theories which, quite remarkably,  can be done in a manifestly gauge invariant manner. Before doing so, however, we note that the overwhelming bulk of work into gauge theories using the ERG has been done using the
effective average action. This approach, which was initiated in~\cite{Reuter+Wetterich},
proceeds via the more conventional gauge-fixed route (several different gauges have been considered, in practice).
Since fixing the gauge anyway breaks manifest gauge invariance, additional breaking
due to a cutoff is perhaps not quite so severe and anyway one can hope to keep track of
the effects (which formally vanish in the limit that all fluctuations are integrated out, corresponding to $\Lambda \rightarrow 0$). 

As a practical tool, there is no question that this way of doing things is currently superior to the manifestly gauge invariant approach, and a considerable amount of work has been devoted to this subject. There are two recent reviews~\cite{JMP-Review,Gies-Rev} which, respectively, cover work done up to the
end of 2005 and 2006. Since then, there has been some very interesting work on Landau gauge Yang--Mills~\cite{JMP+CS-1,JMP+CS-2,Braun+Gies+JMP} and also QCD at finite temperature~\cite{Braun-0810.1727,Braun-0908.1543}; see~\cite{Blaizot-QGP} for a recent review focusing on
the quark-gluon plasma.

The effective average action approach is also the one used for ERG
studies into asymptotic safety in quantum gravity (a manifestly diffeomorphism
invariant approach has yet to be formulated). Inspired by the original
work of Weinberg~\cite{Weinberg-AS} (who has very recently returned to
this topic~\cite{Weinberg-ASI}), the idea received a new lease of
life following the pioneering work of Reuter~\cite{Reuter-Genesis}. Since
then, this has become an active field of research, for which reviews / papers with an
extensive guide to the literature can
be found in~\cite{Reuter-Living,Reuter+Saueressig,Percacci-Review,Litim-AS-Review,Christoph-Rev,Dario-Review}. 

Typically, the so-called `Einstein--Hilbert' truncation is employed, in which
all terms besides those in the Einstein--Hilbert action (including cosmological constant) 
are thrown away. (Of course, both Newton's `constant' and
the cosmological `constant' are allowed to run with energy.) Although this
truncation is rather crude, there are two 
noteworthy papers in which richer truncations are considered~\cite{Percacci-Rn,Dario-C2};
in both cases, the non-trivial fixed-point remains, providing
perhaps
the most compelling evidence to date that its existence is not illusory. 
Then again, it should be emphasised that much work remains to be done, particularly bearing in mind some of the lessons of scalar field theory: (i) certain truncations are known to generate spurious fixed-points~\cite{TRM-Truncations} (ii) as the analysis of \sect{Gen-TP} shows, non-unitary fixed-points seem to greatly outnumber their physical counterparts (certainly at the non-interacting level; whether this persists more generally is not known).

Doubtless, some of the future work on asymptotic safety in quantum gravity will focus on
the effects of including matter. Let us mention here that, 
building on~\cite{Scalar-Tensor-I},  it has
been shown (beyond the Einstein--Hilbert truncation of the gravitational sector) that the non-trivial fixed-point persists in the presence of
a minimally coupled scalar field~\cite{Scalar-Tensor-II}.

There has also been a recent series of works
drawing parallels between asymptotic safety in gravity and non-linear sigma models~\cite{Percacci-FP-NLS,Percacci-HigherDeriv-Sigma,Percacci-AS-Sigma},
as well as investigations into asymptotic safety in chiral Yukawa systems~\cite{Gies-AS-SimpleYukawa,Gies-AS-Towards,Gies-AS-ChiralYukawa}.

\subsection{Manifestly Gauge Invariant ERGs}
\label{sec:mgierg}

\subsubsection{The Pure Abelian Theory}

As a warmup for the non-Abelian case, we consider the pure Abelian case which, though straightforward, provides some useful lessons. 
The gauge field will be denoted by $A_\mu$, with gauge transformations taking the form
\be
	\delta A_\mu(x) = \partial_\mu \omega(x)
\label{eq:GT-Abelian}
\ee
for arbitrary $\omega(x)$.
In Abelian gauge theory, the functional derivative $\delta / \delta A_\mu$ is gauge invariant and so the analogue of the scalar Polchinski equation is trivial to write down and, moreover, is manifestly gauge invariant:
\be
	-\flow \Sint_\Lambda[A] = 
	\hf 		
	\fder{\Sint}{A_\mu} \cdot \dd_{\mu\nu} \cdot \fder{\Sint}{A_\nu}
	-
	\hf	
	\fder{}{A_\mu} \cdot \dd_{\mu\nu} \cdot \fder{\Sint}{A_\nu}
,
\label{eq:Pol-Abelian}
\ee
where
\be
	\ep_{\mu\nu}(p^2;\Lambda) = \frac{\cutoff(p^2/\Lambda^2)}{p} \delta_{\mu\nu},
\qquad
	\mathrm{with}
\qquad
	\dd_{\mu\nu}(p^2;\Lambda) = -\flow \ep_{\mu\nu}(p^2;\Lambda)
.
\ee
Note that, along the lines of~\eq{split}, we have split the total action according to
\be
	S_\Lambda[A] = \hf  A_\mu \cdot {\ctp}_{\mu\nu} \cdot A_\nu + \Sint_\Lambda[A],
\label{eq:split-Abelian}
\ee
where
\be
	{\ctp}_{\mu\nu}(p;\Lambda) = \cutoff^{-1}(p^2/\Lambda^2) \bigl(p^2 \delta_{\mu\nu} - p_\mu p_\nu\bigr)
.
\ee
The tensor structure of this vertex is dictated by manifest gauge invariance. Since the action is invariant under~\eq{GT-Abelian}, all vertices are transverse on all legs:
\be
	p_{\mu_i} S_{\mu_1\cdots \mu_i\cdots\mu_n}(p_1,\ldots,p_i,\ldots,p_n) = 0, \qquad \forall i,
	\qquad
	\Rightarrow
	\qquad
	p_\mu \fder{S[A]}{A_\mu} = 0
.
\label{eq:WID-Abelian}
\ee
An upshot of this is that, in contrast to the non-Abelian case, $\Sint$ is gauge invariant by itself.

Now, even in this simple context we find an interesting departure from a gauge-fixed theory. Consider multiplying together the (momentum space) effective propagator and the two-point vertex 
${\ctp}_{\mu\nu}$:
\be
	{\ctp}_{\mu\alpha}(p) \ep_{\alpha \nu}(p)
	= \delta_{\mu\nu} - \frac{p_\mu p_\nu}{p^2}
.
\label{eq:EPR-Abelian}
\ee
In a gauge-fixed setting, the second term on the \rhs\ would be absent (\ie\ the propagator is the inverse of the gauge-fixed two-point term). With this in mind, the $p_\mu p_\nu/p^2$ piece has been christened a `gauge remainder'~\cite{aprop} since its presence is forced by the manifest gauge invariance.

Let us now turn to the correlation functions. Our treatment will mirror that of \sect{CorrFns}. 
Taking $\Sint_\Lambda[A,J]$ such that $\Sint_\Lambda[A,0]$ is just the standard (interaction part of) the Wilsonian effective action we define
\be
	\dual[A,J]
	\equiv
	-\ln
	\bigl(
		e^{\op}
		e^{-\Sint_\Lambda[A,J]}
	\bigr)
,
\label{eq:dual-Abelian}
\ee
where 
$J$ couples to some operator at the bare scale and
\be
	\op \equiv 
	\hf \fder{}{A_\mu} \cdot \ep_{\mu\nu} \cdot \fder{}{A_\nu}
.
\ee
Precisely as for the scalar Polchinski equation it is easy to check, using the flow equation~\eq{Pol-Abelian}, that
\be
	-\flow \dual[A,J] = 0.
\ee
As before, let us consider evaluating the ERG invariant $\dual[A,J]$ by taking $\Lambda =0$ on the \rhs\ of~\eq{dual-Abelian}. Given the usual assumption of no IR pathologies, we therefore take the \naive\ limit
\be
	\lim_{\Lambda \rightarrow 0}\ln \Bigl( e^{\op} e^{-\Sint_\Lambda[A,J]} \Bigr) 
	=
	-\Sint_{\Lambda = 0}[A,J],
\ee
from which we conclude (according to the recipe of \sect{CorrFns}) that $\dual[0,J]$ generates the connected correlation function of whatever (gauge invariant) operator couples to $J$ at the bare scale. 

However, suppose that we decide to evaluate $\dual[0,J]$ by taking $\Lambda >0$ on the \rhs\ of~\eq{dual-Abelian}. Before doing anything else, let us exploit~\eq{WID-Abelian} to show that we can write
\be
\dual[A,J]
	=
	-\ln
	\bigl(
		e^{\op_\xi}
		e^{-\Sint_\Lambda[A,J]}
	\bigr)
,
\label{eq:dual-Abelian-gen}
\ee
where, in momentum space,
\be
	\op_\xi =\hf \int_p \fder{}{A_\mu(-p)} 
	\frac{\cutoff(p^2/\Lambda^2)}{p^2}
	\biggl(\delta_{\mu\nu} - (1-\xi) \frac{p_\mu p_\nu}{p^2}\biggr)
	\fder{}{A_\nu(p)}
.
\label{eq:GenCovGauge}
\ee
The object sandwiched between the functional derivatives obviously takes the form of a UV regularized propagator in general covariant gauge.%
\footnote{Note that we are not entitled to directly furnish the ERG kernel, as appearing in the flow equation, with a $p_\mu p_\nu/p^2$ piece since all ingredients of the flow equation must be quasi-local.}
Now, let us emphasise that no gauge fixing has been done! Rather, we have understood how to map calculations in a manifestly gauge invariant setting into a more standard form. The trick is to separate out a two-point piece from the total action and then to perform calculations of ERG invariant quantities at $\Lambda \neq 0$. As we will comment below, in the non-Abelian setting  we are able to \emph{generate} the Faddeev--Popov determinant starting from a manifestly gauge invariant setting without ever actually performing the standard gauge fixing operations.

\subsubsection{Non-Abelian Gauge Theory}

The formalism for treating non-Abelian gauge theory is much more involved and we will not discuss it in great depth, though will give more than just a cursory overview (see~\cite{qcd} for the most recent detailed description of the formalism and also the earlier works~\cite{ym1,aprop,qed,mgierg1,mgierg2,Conf}). We now take $A_\mu$ to denote a non-Abelian field, out of which the coupling, $g$, has been scaled, so that the covariant derivative is
\be
	\nabla_\mu = \partial_\mu - iA_\mu.
\ee
The field strength tensor is defined, as usual, to be
\be
	F_{\mu\nu} = i[\nabla_\mu, \nabla_\nu],
\ee
and gauge transformations are given by
\be
	\delta A_\mu = [\nabla_\mu, \omega].
\label{eq:GT-nonAbelian}
\ee

The first issue to be solved is how to reconcile a cutoff with gauge invariance. This requires two ingredients. First of all, the cutoff must be `covariantized'. 
We can see what this means by noting that our first stab at a regularized kinetic term
\[
	\tr \Int{x} \volume{y} F_{\mu\nu}(x) \cutoff^{-1}(x-y;\Lambda) F_{\mu\nu}(y)
\]
is not invariant under~\eq{GT-nonAbelian}.
This can be rectified by replacing the regularized kinetic term by
\be
	\tr \Int{x} \volume{y} F_{\mu\nu}(x) \{\cutoff^{-1}(x-y;\Lambda)\} F_{\mu\nu}(y),
\label{eq:cov}
\ee
where $\{\cutoff^{-1}\}$ is some covariantization of the kernel which renders the above expression gauge invariant~\cite{ym1,aprop}. An example of this would
be to take the momentum space kernel to depend on $\nabla^2/\Lambda^2$,
\viz\ $\cutoff (\nabla^2/\Lambda^2)$. This amounts to furnishing the cutoff function
with vertices. Just as the vertices of the action are subject to Ward identities, as
a consequence of gauge symmetry, so too are the vertices of $\{K\}$.

One might hope that this procedure is sufficient to regularize the theory, but a standard
perturbative analysis reveals that a 
set of one-loop divergences slip through~\cite{sunn}, corresponding to those diagrams with $\leq \D$
external legs. This is perhaps not
that surprising since although the UV behaviour of the propagator is improved by 
an insertion of the cutoff function, the behaviour of the higher-point vertices
is made correspondingly worse, as can be seen from~\eq{cov}.

The solution to this problem is to include a set of Pauli--Villars fields to kill the
remaining divergences. There is an elegant way of doing this:
the physical $\SU(N)$ gauge theory is embedded in a spontaneously 
broken%
\footnote{To adhere to convention, we blithely use the term `spontaneous symmetry breaking',
despite the fact that Elitzur's theorem~\cite{Elitzur} implies that this is nothing more than
a `convenient fiction'~\cite{Wilczek-convenient,Greensite-Rev} in the case of \emph{local} symmetries (which is particularly pertinent since, as we will see, we never fix the gauge). Thus, we do not encounter any phase boundary as
we go to high energies; rather, we find that the large-momentum behaviour of loop diagrams is smoothly
cutoff as a consequence of the underlying  $\SU(N|N)$ symmetry.} 
$\SU(N|N)$ gauge theory~\cite{sunn}.%
\footnote{The covariant higher derivative regularization is now understood to apply to the \emph{entire} spontaneously broken $\SU(N|N)$ gauge theory, thereby avoiding the problem of overlapping divergences. }
 The heavy fields resulting from the symmetry breaking (which are given a mass at the effective cutoff scale)
provide precisely the set of required Pauli--Villars fields!

 In a little more detail, the picture is as follows.
Considering the problematic one-loop diagrams mentioned above, focus first the planar ones.
For these diagrams (but not for the non-planar ones, as it turns out), there is now the option of either a physical field or a regulator field circulating in the loop, and the combination of the two is finite. Thus, in the planar limit, everything is now regularized at the scale $\Lambda$. Moving on, note that, essentially on account of the fact that $\tr A_\mu =0$, all non-planar diagrams with fewer than four external fields vanish. Recalling that potentially problematic diagrams have at most $\D$ legs it is apparent that only in $\D \geq 4$ are there still problematic diagrams in the non-planar sector. However, it turns out that gauge invariance lessens the superficial degree of divergence of these diagrams, guaranteeing finiteness in $\D<8$. 

Nevertheless, there is something not entirely satisfactory about this (which has not been pointed out before). In $\D\geq4$, it seems that there are (non-planar) diagrams which are essentially oblivious to the presence of the effective scale, being finite only as a consequence of gauge invariance.  Whilst we can agree that the theory is regularized by the scheme described above, it is not clear if this regularization truly corresponds to a cutoff, in the sense of all modes above a certain scale being suppressed (at least in the non-planar sector). This issue, which we will mention again below, needs further attention. (Also, whether the regularization works throughout theory space or just for the admittedly most interesting case of the trajectory emanating from the Gaussian fixed-point has never been addressed.)

It should be noted that, in the literature~\cite{sunn,aprop,mgierg1}, a massless, unphysical
gauge field remains in the particle spectrum. Previously, it has been argued that this particle decouples in
the $\Lambda\rightarrow\infty$ limit, at least if we are on the renormalized trajectory emanating from the Gaussian fixed-point, and so is harmless. However, there is a subtlety: this gauge field comes with a wrong-sign action and so the $\beta$ function of its coupling is positive rather than negative. This indicates that we cannot sit on the desired renormalized trajectory. A solution to this is to extend the symmetry breaking sector to ensure that all components of this gauge field are given a mass of order the cutoff.

This $\SU(N|N)$ scheme shares a common ideology with Slavnov's higher derivative scheme~\cite{Slavnov-HD,Fadeev-HD,Bakeyev-HD,Asorey-HD}; together with the lattice, these three approaches constitute the only known nonperturbative regularizations of QCD. Let us note in passing that, 
in the context of the AdS-CFT correspondence~\cite{Malda-AdS,Witten-AdSCFT,GKP-AdSCFT}, this scheme can be 
used to furnish an understanding of how the radial direction on the gravity side of the duality plays the role of a \emph{gauge invariant} cutoff~\cite{EMR}.

Having discussed the regularization, let us now turn to the flow equation. The essential idea is to covariantize the general form~\eq{flow-multi}:
\be
	-\flow  S 
	= \hf \fder{S}{\varphi_i} \{ \dd^{\varphi_i \varphi_j} \} \fder{\Sigma_g}{\varphi_j} 
	- \hf \fder{}{\varphi_i} \{ \dd^{\varphi_i \varphi_j}  \} \fder{\Sigma_g}{\varphi_j},
\label{eq:cov-flow}
\ee
where the $\varphi_i$ now include the complete spectrum of fields
present in spontaneously broken $\SU(N|N)$ gauge theory. Notice that
the $\Sigma$ of~\eq{flow-multi} has been replaced by\footnote{In many works
on this subject, $\hS$ is defined so that in $\Sigma_g$ it does not come with the
additional factor of $g^2$.}
\[
	\Sigma_g \equiv g^2 (S - 2\hS),
\]
which is appropriate for $g$ having been scaled out of the covariant derivative.
It is worth noting that if one does this rescaling carefully~\cite{ym1}, there
appears an additional, inconvenient term on the \lhs\ of the flow equation. One
of the beauties of the general approach to ERGs that we have take is that this
extra term can in fact be dropped, this procedure corresponding as it does to a
different---perfectly legal---choice of $\Psi$ [\cf~\eq{blocked}]; the result is~\eq{cov-flow}.
The seed action, $\hS$, is taken to be a (manifestly) gauge invariant functional of the fields
which, whilst it can be left largely arbitrary, is subject to certain constraints~\cite{aprop,mgierg2}.

One of the truly remarkable things about this flow equation is that it is
\emph{manifestly} gauge invariant: no gauge fixing has been---nor ever needs
to be---performed. This is very different from independence of the gauge in a gauge-fixed formalism.
In particular, we are restricted to computing correlation functions of manifestly gauge invariant operators.
Nevertheless, within this context it is now understood how to map calculations in this formalism onto standard gauge fixed ones. In the past~\cite{univ,evalues,mgiuc,univ}, it has been realized that the dual action in non-Abelian gauge theory must be supplemented by a new term; thus~\eq{dual-Abelian} becomes
\be
	\dual[\varphi,J]
	\equiv
	-\ln
	\bigl(
		e^{\op_g}
		e^{-\Sint_\Lambda[\varphi,J] - \mathcal{G}[\varphi]}
	\bigr)
,
\label{eq:dual-nonAbelian}
\ee
where
\be
	\op_g \equiv
	\frac{g^2}{2}
	\fder{}{\varphi^i} \cdot \ep^{\varphi^i \varphi^j} \cdot \fder{}{\varphi^j}
.
\ee
(Note that, previously, all expressions for $\dual$ were diagrammatic.) Momentarily suppressing our curiosity about $\mathcal{G}$, it turns out that it is this enhanced expression for the dual action which satisfies
\be
	\flow \dual[0,J] = 0
\ee
and, therefore, generates the correlation functions of whatever couples to $J$ in the ultraviolet. 

Interestingly, it has recently been determined~\cite{WIP} that $\mathcal{G}[\varphi]$ can (formally, at any rate) be  interpreted as nothing other than the Faddeev--Popov determinant! Thus, the picture is as follows. We start with a manifestly gauge invariant formalism, and at no point fix the gauge. However, consider performing the gauge-variant separation of the action into a two-point piece and interactions. Using the latter as a building block, construct the object $\dual[\varphi,J]$ according to~\eq{dual-nonAbelian}; whilst the entire thing is gauge invariant, it is made up of gauge-variant components. The crucial point about $\dual[\varphi,J]$ is that $\dual[0,J]$ is an ERG invariant.
Now, if we evaluate the \rhs\ of~\eq{dual-nonAbelian} at $\Lambda=0$---which should amount to simply computing $S_{\Lambda=0}[A,J]$---we never encounter anything gauge-variant (this object can be computed, in principle, directly from the manifestly gauge invariant flow equation). However, if instead we evaluate this ERG invariant quantity by working at $\Lambda \neq 0$ then we encounter the Faddeev--Popov determinant. Moreover, precisely as in~\eq{GenCovGauge}, we can (should we so desire) map ourselves into a general covariant gauge.

Let us pause, however, to flag a possible problem. As mentioned earlier in this section, it seems that, for the non-planar theory, there are diagrams which essentially do not feel the presence of the effective scale, $\Lambda$. With this in mind, it is not obvious that the $\Lambda \rightarrow 0$ limit of the \rhs\ of~\eq{dual-nonAbelian} (with $A=0$) actually yields just $\Sint_{\Lambda=0}[0,J]$; this needs to be checked.

Before concluding this short discussion of manifestly gauge invariant ERGs it is important to state that a heavy price has been paid in their construction. In addition to the greatly expanded field content, there is much more besides going on beneath the surface of the apparently placid~\eq{cov-flow}. Perhaps the biggest problem is that the simple structure of the Polchinski equation has been spoilt (much as the notation attempts to hide this)
by the necessary inclusion of a non-trivial seed action and covariantization of the kernels. The upshot of this is that there are major obstacles in the way of repeating even the simplest calculations performed in \sect{solve}. In particular, it has never been explicitly demonstrated that the equation possesses a Gaussian fixed-point with the expected set of eigenperturbations. Thus, much of the work done to date with this formalism has been rather implicit. It would be desirable to improve on this.

\section{Conclusion}
\label{sec:conc}

Of the various aspects pertaining to the ERG that have been discussed in this
paper it is worth asking, now that we are almost finished, whether any 
in particular can stake a claim to being the most profound. In part, the
answer to this rather subjective question is coloured by the angle at
which one approaches the subject and can be expected to contain a 
certain amount of personal prejudice. For example, suppose that we
are interested in studying the properties of some system with many 
degrees of freedom per correlation length. Then, from a pragmatic point
of view, we might view the fact that the ERG provides computational 
access to such problems as being of primary importance.
If, instead,  one prefers to demand that something profound should yield 
broad, intuitive understanding then there is no better candidate
than the picture of universality of systems approaching a second
order phase transition provided by the ERG. However, 
the focus of this review, if only implicitly, has been on QFT (mainly due to the
limitations of the author) and it is from this
perspective that I would like to put forward what I believe to be one of 
the  deepest insights that the ERG has to offer. As will become clear, it is closely
related to the notion of universality, though with a somewhat different emphasis.

The more standard approaches to QFT of text-book 
canonical or path integral quantization generally display a marked preference
for free field theories or small modifications thereof. The success and prevalence 
of this program are well justified and easy to understand. Much of 
the impetus for developing QFT has come from the field of high energy physics
and, to date, our best picture of nature at small scales---encoded by
the standard model of particle physics---deals with field theories constructed around a Gaussian
fixed-point. And yet even this last point is actually a subtle one. 

As discussed at
great length in \sect{Qualitative}, the $\SU(3)$ and $\SU(2)$ sectors of the standard model
are asymptotically free meaning that, as stand-alone theories,  they make sense down to arbitrarily small distances.
To be precise, both theories constitute a (marginally) relevant perturbation of their
associated Gaussian fixed-points. The same cannot be said of the $\mathrm{U}(1)$ and 
Higgs sectors of the standard model. In neither of these theories does the Gaussian fixed-point support an interacting
renormalized trajectory: the standard model as a whole only makes sense as a low
energy effective theory. (One might hope that coupling a scalar sector to a gauge sector, as in the standard model, could reverse the sign of the positive scalar $\beta$-function. Whilst such completely asymptotically free gauge-Higgs systems do exist---see \eg~\cite{Callaway} for a review---this mechanism sadly does not work for the standard model.)

Nevertheless, suppose that one chooses a bare action for the
standard model that
is near to the critical surface of the Gaussian fixed-point. Since both the U$(1)$ charge
and the Higgs' self coupling are only \emph{marginally} irrelevant, the low energy theory effective theory is, up to corrections going like inverse powers of the bare scale, precisely what is written down in the standard model.
The reason that it is uniquely (to leading order) the standard model that appears as the low energy effective theory is precisely the same one that lies behind the universality associated with second order
phase transitions. Indeed, having focused this discussion around QFT, we seem to have been ineluctably led back to the conclusion that it is universality that is the most important conceptual issue contained within the framework of the ERG.

However, there is an associated concept which has been somewhat masked by the fact that this discussion has centred around the Gaussian fixed-point. Suppose that we consider a set of fields for which the space of all allowed theories---`theory space'---supports a non-trivial fixed-point. Then, of course, this fixed-point provides (just like the Gaussian one) on the one hand the basis for constructing theories that make sense down to arbitrarily small distances and, on the other, universality of the IR dynamics of theories near to the critical surface.
But the real point to make is that this fixed-point is something which has been \emph{solved} for. 
Similarly, if we wish to use this fixed-point as the basis for a renormalized trajectory, then we
must solve for the relevant and marginally relevant perturbations. This should be compared to
the more usual way of constructing a QFT, where we write down some bare action and then do (perturbative) computations to determine the renormalizability of the correlation functions. Perhaps unfortunately, the fact that
there is often a focus on theories built around the Gaussian fixed-point means that the distinction
between these two methodologies is largely washed away by the comparative simplicity of the problem.

Nevertheless, the idea that renormalizable QFTs are things which should be solved for is a compelling one: in the entire space of allowed theories, we have an equation (the ERG equation, of course!) which can be solved for the very special set of fixed-points theories and associated renormalized trajectories. 

To conclude, I would like to advocate the idea that this procedure can in fact be taken one step further.
The main result of \sect{CorrFns} is that correlation functions are (nonperturbatively) renormalizable 
if they follow either from a source-dependent fixed-point or a relevant/exactly marginal source-dependent perturbation thereof. 
For correlation functions of scalar fields at different points,
$\eval{\phi(x_1)\cdots \phi(x_n)}_{\conn}$,
 this does not really tell us anything new; rather, it yields a different way of seeing why renormalizability of the Wilsonian effective action implies renormalizability of the aforementioned correlation functions. But in gauge theories, one can expect the picture to be very different.

Let us recall that, as sketched in \sect{other-theories}, it is possible to formulate manifestly gauge invariant ERGs. In this case the only correlation functions that are non-zero are built from objects which are themselves manifestly gauge invariant. Consequently, the `standard' correlation functions
$\eval{A(x_1)\cdots A(x_n)}_{\conn}$ have no role to play in such a formulation. In this case, it is a very important question to ask how one determines the nonperturbative renormalization properties of correlation functions of gauge invariant operators. I contend that the answer, in a similar vein to the above, is that this can be done (in principle) by \emph{solving} the appropriately modified ERG equation.

Again, it is worth comparing this to the standard way of doing this: having in mind what we think we should be computing, we fix the gauge and proceed as usual. But if we never fix the gauge then it becomes clear that we should \emph{determine} from the QFT in question those objects 
that we should be considering in the first place!%
\footnote{%
	It would be very interesting to try to link this with the program of constructing gauge invariant charges being carried out by Lavelle, McMullan and collaborators---see~\cite{L+M-Review,L+M-Construction,L+M-Renormalization,L+M-Physical} and other papers by the same authors. Note, though, that their procedure breaks down for non-Abelian gauge theories (this break down being identified with confinement), whereas the program advocated above is expected to work, on account of asymptotic freedom. On the other hand, for QED where gauge invariant charges can be constructed, the
 program advocated above is not suitable for $\D=4$ since QED 
is only a low energy effective theory and does not sit on a renormalized trajectory! 
Moreover, the composite operators corresponding to these charges are non-local, which presents
a challenge. Nevertheless
it might well be that the two approaches can be related in certain circumstances and it would be
worthwhile exploring this further. 
}
 Only by answering this question will we arrive at correlation functions which are \emph{guaranteed} to be nonperturbatively renormalizable. It is thus irresistible to speculate that perhaps we should be asking not what quantum field theory can compute for us, but what we can compute for quantum field theory.

\begin{acknowledgments}
I would like to extend a huge debt of gratitude to Roberto Percacci and, particularly, Hugh Osborn for
reading and commenting on substantial amounts of this long piece of work. I have also benefited from discussions with John Cardy, Mike Birse, Daniel Litim, Jan Pawlowski and Manfred Salmhofer. I am truly grateful to Holger Gies for providing some much needed encouragement. This work was supported by the Science and Technology Facilities Council [grant number ST/F008848/1].
\end{acknowledgments}

\appendix

\section{The Flow of the Dual Action}
\label{app:Dual-flow}

Recalling the definition of the dual action, \eq{dual},
\be
	-\dual[\phi] \equiv
	\ln \bigl( e^{\op} e^{-\Sint} \bigr)
	\equiv
	\ln
	\left[
		\exp
		\left(
			\hf \fder{}{\phi} \cdot \ep \cdot \fder{}{\phi}
		\right)
		e^{-\Sint}
	\right],
\label{eq:dual-recap}
\ee
in this section we will derive the flow of the dual
action given the flow equation with
general seed action~\eq{reduced-flow}
\be
	\left(-\flow + \frac{\eta}{2} \Count_\phi\right) \Sint
	=\hf \classical{\Sint}{\dd}{\Sigint}
	-\hf \quantum{\dd}{\Sigint}
	- \phi \cdot \ep^{-1} \dd \cdot \fder{\hSint}{\phi}
	-\frac{\eta}{2} \phi \cdot \ep^{-1} \cdot \phi.
\label{eq:flow-recap}
\ee

From~\eq{dual-recap} it is apparent that
\be
	\flow \dual
	=
	e^{\dual} e^{\op}
	\left[
		\hf \classical{\Sint}{\dd}{\Sint} -\hf \quantum{\dd}{\Sint}
		+\flow \Sint
	\right]
	e^{-\Sint},
\label{eq:dual-flow-a}
\ee
where the signs work out since $\dd \equiv -\flow \ep$. Recalling that $\Sigint \equiv \Sint - 2\hSint$,
we substitute~\eq{flow-recap} into~\eq{dual-flow-a} to yield:
\begin{multline}
	\flow \dual
	=
	e^{\dual} e^{\op}
	\biggl[
		 \classical{\Sint}{\dd}{\hSint} - \quantum{\dd}{\hSint}
		+ \phi \cdot \ep^{-1} \dd \cdot \fder{\hSint}{\phi}
		+ \frac{\eta}{2} \Count_\phi \Sint
		+\frac{\eta}{2} \phi \cdot \ep^{-1} \cdot \phi
	\biggr]
	e^{-\Sint}.
\label{eq:dual-flow-b}
\end{multline}
 
The game now is to commute any explicitly occurring $\phi$s through the $e^{\op}$.
To this end, we note that
\be
	\left[
		\op, \phi(p) 
	\right]
	=
	\fder{}{\phi(-p)} \ep(p^2),
	\qquad
	\Rightarrow
	\qquad
	\left[
		e^{\op}, \phi(p)
	\right]
	= 
	e^{\op}
	\fder{}{\phi(-p)} \ep(p^2).
\label{eq:commutator-phi}
\ee
Consequently, it is apparent that
\be
	e^{\dual} e^{\op}
	 \phi \cdot \ep^{-1} \dd \cdot \fder{\hSint}{\phi} e^{-\Sint}
	 =
	 e^{\dual} 
	  \phi \cdot \ep^{-1} \dd \cdot e^{\op} \, \fder{\hSint}{\phi} e^{-\Sint}
	+ 
	e^{\dual} e^{\op}
	\frac{\vec{\delta}}{\delta \phi}
	\cdot
	\dd
	\cdot
	\fder{\hSint}{\phi}
	e^{-\Sint},
\ee
where the arrow above the functional derivative is just to emphasise that it
hits all terms to its right. Therefore, the final term in this expression exactly
cancels the first and second  terms on the \rhs\ of~\eq{dual-flow-b}. Thus,
at this stage of the proceedings we have that
\be
	\flow \dual
	=
	-\frac{\eta}{2} e^{\dual} e^{\op}
	\Bigl(
	\Count_\phi
	-
	\phi \cdot \ep^{-1} \cdot \phi
	\Bigr)
	e^{-\Sint}
	+
	e^{\dual} 
	  \phi \cdot \ep^{-1} \dd \cdot e^{\op} \, \fder{\hSint}{\phi} e^{-\Sint}.
\label{eq:dual-flow-pen}
\ee
The first term on the \rhs\ can be processed by writing
\be
	\hf e^{\dual} e^{\op}
	\Bigl(
	\Count_\phi
	-
	\phi \cdot \ep^{-1} \cdot \phi
	\Bigr)
	e^{-\Sint}
	=
	\hf e^{\dual} 
	\Bigl(
	\Count_\phi
	-
	\phi \cdot \ep^{-1} \cdot \phi
	\Bigr)
	e^{\op}
	e^{-\Sint}
	+
	\hf e^{\dual} 
	\Bigl[
	e^{\op},
	\Count_\phi
	-
	\phi \cdot \ep^{-1} \cdot \phi
	\Bigr]
	e^{-\Sint}.
\label{eq:dual-flow-comm}
\ee

Focusing on the commutator term, 
the $\Count_\phi$ piece can be processed directly from~\eq{commutator-phi},
\be
	\left[
		e^{\op}, {\textstyle \hf} \Count_\phi
	\right]
	=
	e^{\op} \op.
\label{eq:commutator-Delta_phi}
\ee
Next we must commute the $\phi \cdot \ep^{-1} \cdot \phi$
to the right of the $e^{\op}$. To do this 
we note that, for some $X(p^2)$,
\begin{align}
	\left[
		\op, \phi \cdot X \cdot \phi
	\right]
	& =
	\deltahat{0} X \cdot \ep + 2 \phi \cdot \ep X \cdot \fder{}{\phi}
\\[2ex]
	\left[
		\op, 2 \phi \cdot \ep X \cdot \fder{}{\phi}
	\right]	
	& =
	2 \classical{}{\ep^2 X}{}
\label{eq:commutator-phiXphi}
\end{align}
In order to compute the commutator of $e^{-\op}$ with $\phi \cdot X \cdot \phi$, we now employ
a trick (see \eg\ section 2.7 of \cite{Georgi}):
\be
	\Bigl[
		e^{\op} , F[\phi]
	\Bigr]
	=
	\int_0^1 ds\, e^{s\op} \left[ \op, F\right] e^{(1-s)\op},
\ee
where $F$ is some functional of $\phi$. Note that, in the case where $[\op,[\op,F]] = 0$, then
the \rhs\ simply becomes $e^{\op} [\op,F]$. This is one way to derive the second
part of~\eq{commutator-phi}. Returning to the case in hand,
\be
	\Bigl[
		e^{\op} , \phi \cdot X \cdot \phi
	\Bigr]
	=
	\deltahat{0} \ep \cdot X
	e^{\op}
	+2
	\int_0^1 ds
	e^{s\op} 
	\phi \cdot C X \cdot \fder{}{\phi}
	e^{(1-s)\op}.
\label{eq:commutator-phiphi}
\ee
The second term can, using by now familiar techniques, be rewritten according to
\begin{align}
	2
	\int_0^1 ds
	e^{s\op} 
	\phi \cdot C X \cdot \fder{}{\phi}
	e^{(1-s)\op}
&	=
	2\phi \cdot C X \cdot \fder{}{\phi}
	e^{\op}
	+2
	\biggl( \int_0^1 s \, ds \biggr)  \classical{}{\ep^2 X}{} e^{\op}
\\[1ex]
& = 
	2\phi \cdot C X \cdot \fder{}{\phi}
	e^{\op}
	+  \classical{}{\ep^2 X}{} e^{\op}.
\end{align}
Substituting this back into~\eq{commutator-phiphi} yields
\be
	\Bigl[
		e^{\op} , \phi \cdot X \cdot \phi
	\Bigr]
	=
	\deltahat{0} \ep \cdot X
	e^{\op}
	+
	2\phi \cdot C X \cdot \fder{}{\phi}
	e^{\op}
	+  \classical{}{\ep^2 X}{} e^{\op}.
\label{eq:useful-commutator}
\ee
Before returning to the case in question, let us note that~\eq{useful-commutator} can be readily adapted for
$e^{-\op}$ by sending $\op \rightarrow -\op$ and $\ep \rightarrow -\ep$:
\be
	\Bigl[
		e^{-\op} , \phi \cdot X \cdot \phi
	\Bigr]
	=
	-\deltahat{0} \ep \cdot X
	e^{-\op}
	-
	2\phi \cdot C X \cdot \fder{}{\phi}
	e^{-\op}
	+  \classical{}{\ep^2 X}{} e^{-\op}.
\label{eq:useful-commutator-b}
\ee

Focusing our interest back on~\eq{dual-flow-pen}, we set $X=\ep^{-1}$, 
in~\eq{useful-commutator} to give:
\be
	\Bigl[
		e^{\op} , \hf \phi \cdot \ep^{-1} \cdot \phi
	\Bigr]
	=
	\hf \deltahat{0} \ep \cdot \ep^{-1}
	e^{\op}
	+
	\Count_\phi
	e^{\op}
	+  \op e^{\op}.
\ee
Combining this with~\eq{commutator-Delta_phi} we find that
\be
	\hf e^{\dual} 
	\Bigl[
	e^{\op},
	\Count_\phi
	-
	\phi \cdot \ep^{-1} \cdot \phi
	\Bigr]
	e^{-\Sint}
	=
	- e^{\dual}
	\Biggl[	
	\hf \deltahat{0} \ep \cdot \ep^{-1}
	+
	\Count_\phi
	\Biggr]
	e^{\op}
	e^{-\Sint}
	=
	-\hf \deltahat{0} \ep \cdot \ep^{-1}
	- 
	e^{\dual} \Count_\phi e^{\op} e^{-\Sint},
\ee
where we have used $e^{\dual}  e^{\op} e^{-\Sint} = 1$ (so long as there is nothing to the right of this expression on which $e^{\op}$ can act). Substituting this expression into~\eq{dual-flow-comm} gives the useful result
\begin{align}
	\hf e^{\dual} e^{\op}
	\Bigl(
	\Count_\phi
	-
	\phi \cdot \ep^{-1} \cdot \phi
	\Bigr)
	e^{-\Sint}
	& =
	-\hf e^{\dual} 
	\Bigl(
	\Count_\phi
	+
	\phi \cdot \ep^{-1} \cdot \phi
	\Bigr)
	e^{\op}
	e^{-\Sint}
	-\hf \deltahat{0} \ep \cdot \ep^{-1}
\nonumber
\\
	& = \hf \Count_\phi \dual -\hf \phi \cdot \ep^{-1} \cdot \phi -\hf \deltahat{0} \ep \cdot \ep^{-1},
\label{eq:useful-manip}
 \end{align}
where we have again used $e^{\dual}  e^{\op} e^{-\Sint} = 1$. The calculation
can now be finished by substituting this
expression into~\eq{dual-flow-pen} to yield:
\be
	\left(
		\flow + \frac{\eta}{2} \Count_\phi
	\right) \dual
	=
	\frac{\eta}{2} \phi \cdot \ep^{-1} \cdot \phi
	+
	e^{\dual} 
	\phi \cdot \ep^{-1} \dd \cdot e^{\op} \, \fder{\hSint}{\phi} e^{-\Sint}
	+\frac{\eta}{2}
	\deltahat{0} \ep \cdot \ep^{-1}.
\ee
Dropping the vacuum term gives~\eq{dualflow-Seed}.

\section{The Exactly Marginal, Redundant Operator}
\label{app:redundant}

In this section, we will show that the marginal operator
\be
	\marginal [\phi] =
	\left(\hf \Count_\phi + \Count_{\cutoff}\right) \Sint_\star[\phi]
	\equiv
	\hat{\Count} \Sint_\star[\phi]
\ee
is related to the marginal, redundant operator of O'Dwyer and Osborn via~\eq{simple}, 
for $\eta_\star<2, \ \neq 0 $ and via~\eq{simple-GFP} at the Gaussian fixed-point.
The first step
is to recall~\eq{O_mar-proj}:
\[
	\marginal[\phi] = e^{\Sint_\star} e^{-\op} e^{-\dual_\star} \hat{\Count} \dual_\star.
\]
Our aim now is to substitute for $\hat{\Count} \dual_\star$
using~\eqs{Count_K-=0}{Count_K-neq0}:
\[
	\hat{\Count} \dual_\star[\phi] =
	\left\{
		\begin{array}{ll}
			\ds
			\hf \Count_\phi \dual_\star[\phi] - \hf \phi \cdot \ep^{-1} \bigl(1+\varrho \bigr) \cdot \phi, 
			& \eta_\star <2, \ \neq 0,
		\\[2ex]
			\ds
			\hf \Count_\phi \dual_\star[\phi], & \eta_\star =0.
		\end{array}
	\right.
\]
Focusing on the common $\Count_\phi$ term, we utilize~\eq{commutator-Delta_phi} (with $\op \rightarrow -\op)$ to give
\be
	\hf e^{\Sint_\star} e^{-\op}  e^{-\dual_\star}\Count_\phi \dual_\star
	=
	-\hf e^{\Sint_\star} e^{-\op} \Count_\phi e^{-\dual_\star}
	=
	-
	\hf e^{\Sint_\star}
	\Count_\phi
	e^{-\op} e^{-\dual_\star}
	+
	e^{\Sint_\star} 
	\op 
	e^{-\op}
	e^{-\dual_\star}.
\label{eq:marginal-a}
\ee
To process the first term we note that
\be
	e^{\Sint_\star} \fder{}{\phi(p)} e^{-\op} e^{-\dual_\star} = 
	e^{\Sint_\star} \fder{}{\phi(p)} e^{-\Sint_\star}
	=-\fder{\Sint_\star}{\phi(p)},
\label{eq:dual-insert1}
\ee
where we have used the fact that $e^{-\op} e^{-\dual_\star} = e^{-\Sint_\star} $. 
The second term in~\eq{marginal-a} can be similarly dealt with:
\be
	e^{\Sint_\star} \fder{}{\phi(q)} \fder{}{\phi(p)} e^{-\op} e^{-\dual_\star}
	= -\fder{}{\phi(q)} \fder{\Sint_\star}{\phi(p)} + \fder{\Sint_\star}{\phi(q)} \fder{\Sint_\star}{\phi(p)}.
\label{eq:dual-insert2}
\ee
Combining these results we thus find that
\be
	\hf e^{\Sint_\star} e^{-\op}  e^{-\dual_\star}\Count_\phi \dual_\star
	= \hf \Count_\phi \Sint_\star + \hf \classical{\Sint_\star}{\ep}{\Sint_\star}
	- \hf \quantum{\ep}{\Sint_\star}.
\label{eq:O_mar-component1}
\ee

To complete the analysis, we use~\eq{useful-commutator-b} to show that 
\begin{multline}
	\left[
		e^{-\op}, \phi \cdot \ep^{-1} (\varrho +1) \cdot \phi
	\right]
	=
	-\deltahat{0}
	\int_p [\varrho(p^2)+1] e^{-\op}
\\
	-
	2 
	\phi \cdot
	 (\varrho+1) \cdot
	 \fder{}{\phi}
	 e^{-\op}
	+e^{-\op}
	\classical{}{\ep(\varrho+1)}{},
\end{multline}
from which it follows that
\begin{multline}
	-\hf e^{\Sint_\star} e^{-\op} e^{-\dual_\star}
 	\phi \cdot \ep^{-1}(\varrho +1) \cdot \phi
	=
	-\hf \phi \cdot \ep^{-1} (\varrho +1) \cdot \phi
\\
	+e^{\Sint_\star}
	\phi \cdot 
	 (\varrho+1) \cdot  
	 \fder{}{\phi}
	e^{-\op} e^{-\dual_\star}
	-
	\hf e^{\Sint_\star}  
		\classical{}{\ep(\varrho+1)}{} 
	e^{-\op}
	e^{-\dual_\star}
	+ \mbox{const}.
\label{eq:marginal-b}
\end{multline}
Dropping the constant, the second and third terms can be processed by using~\eqs{dual-insert1}{dual-insert2}:
\begin{multline}
	-\hf e^{\Sint_\star} e^{-\op} e^{-\dual_\star}
 	\phi \cdot \ep^{-1} (\varrho +1) \cdot \phi
	=
	-\hf \phi \cdot \ep^{-1} (\varrho +1) \cdot \phi
\\
	- \phi \cdot \bigl(1+\varrho \bigr) \cdot \fder{\Sint_\star}{\phi}
	-\hf \classical{\Sint_\star}{\ep \bigl(1+\varrho \bigr)}{\Sint_\star}
	+\hf \quantum{\ep \bigl(1+\varrho \bigr)}{\Sint_\star}.
\label{eq:O_mar-component2}
\end{multline}
Summing the contributions from~\eqs{O_mar-component1}{O_mar-component2} yields
\begin{multline}
	-2 \marginal[\phi]
	=
	\phi \cdot \ep^{-1} (\varrho +1) \cdot \phi
\\
	+
	\phi \cdot (2\varrho+1) \cdot \fder{\Sint_\star}{\phi}
	+
	\classical{\Sint_\star}{\ep \varrho}{\Sint_\star}
	-
	\quantum{\ep \varrho}{\Sint_\star}, 
	\qquad
	\eta_\star <2, \ \neq 0,
\end{multline}
whereas the result for vanishing $\eta_\star$ is
\be
	-2 \marginal[\phi]
	=- \Count_\phi \Sint_\star - \classical{\Sint_\star}{\ep}{\Sint_\star} + \quantum{\ep}{\Sint_\star},
	\qquad \eta_\star = 0.
\ee

The final step is to recognize that in the two cases the 
operator can be constructed from~\eq{redundant} by taking
\be
	-2 \Theta(p) = 
	\left\{
	\begin{array}{ll}	
	\ds
		\bigl[\varrho(p^2)+1\bigr] \phi(p) + \ep (p^2) \varrho(p^2) \fder{\Sint_\star}{\phi(-p)},
		& \eta_\star <2,\ \neq 0
	\\[2ex]
	\ds
		-\fder{\Sint_\star}{\phi(p)} \ep(p^2), & \eta_\star = 0,
	\end{array}
	\right.
\label{eq:theta-redundant}
\ee
as can be checked by direct substitution. In the first case, $\eta_\star <2,\neq 0$
our operator is therefore redundant, since the \rhs\ is quasi-local [the $1/p^2$ contained
in the $\ep$ is compensated for by the behaviour of $\varrho$, as is apparent from~\eq{HO-b}]. 
The operator constructed by O'Dwyer and Osborn~\cite{JOD+HO} corresponds to
\be
	\Theta'(p) = [\varrho(p^2) + 1] \phi(p) + 
	\ep (p^2) \varrho(p^2) \fder{\Sint_\star}{\phi(-p)},
\ee
and so we see that the two operators are the same, up to a factor of $-2$, at least for $\eta_\star \neq 0$.

For $\eta_\star =0$,  let us start by supposing that we are at the Gaussian fixed-point, in which case
we can use~\eq{Gen-Gauss}. Here we find that
\be
	2 \Theta(p) = \frac{\intconst \cutoff(p^2)}{1-\intconst \cutoff(p^2)}  \phi(p),
\ee
whereas
\be
	\Theta'(p) =
	\left\{
		1+ 
		\frac{\intconst \bigl[\cutoff(p^2) -1 \bigr] }{1-\intconst \cutoff(p^2)}  
	\right\}\cutoff(p^2)  \phi(p)
	=
	 \frac{(1-\intconst) \cutoff(p^2)}{1-\intconst \cutoff(p^2)}  \phi(p),
\ee
from which it is apparent that
\be
	\Theta'(p) =\frac{2\bigl(1-\intconst\bigr)}{B}  \Theta(p), \qquad \mbox{Gaussian fixed-point}.
\ee
Should it be the case that other fixed-points exist with $\eta_\star = 0$ (as mentioned earlier, this
certainly cannot happen in integer dimension) then it would appear that $\hat{\Count} \Sint$ is
unrelated to the marginal, redundant operator.

\section{A Menagerie of Redundant Operators}
\label{app:Menagerie}

In this appendix, we will explicitly verify that various (redundant) operators constructed using the crutch of the dual action are indeed solutions of the eigenvalue equation~\eq{full-eigen}. To start with, let us observe the following:
\begin{subequations}
\begin{align}
	\classifier \, \phi(p) &= 
		\biggl(\frac{d+2-\eta_\star}{2} + p \cdot \partial_p \biggr)\phi(p)
		+ 2\fder{\Sint_\star}{\phi(-p)} \cutoff'(p^2),
\label{eq:classify-phi}		
\\
	\classifier \, \fder{\Sint_\star}{\phi(-p)} & =
		\eta_\star \ep^{-1}(p^2) \phi(p)
		+\biggl(\frac{d-2+\eta_\star}{2} + p \cdot \partial_p \biggr)\fder{\Sint_\star}{\phi(-p)},
\label{eq:classify-dS/dphi}
\end{align}
\end{subequations}
where the latter equation is most readily derived simply by hitting the fixed-point equation~\eq{FP-eq} with $\delta / \delta \phi$. The other relationships that we will need are
\begin{subequations}
\begin{align}
	\biggl[
		\classifier, \fder{}{\phi(p)}
	\biggr]
	& =
	-2 \fder{}{\phi} \biggl( \fder{\Sint_\star}{\phi(p)}\biggr) \cdot \cutoff' \cdot \fder{}{\phi}
	+ \biggl(
		\frac{\D-2+\eta_\star}{2} + p \cdot \partial_p
	\biggr) \fder{}{\phi(p)},
\label{eq:classify-comm}
\\
	\classifier \, X Y & = \bigl(\classifier X \bigr) Y + X\, \classifier Y -2 \fder{X}{\phi} \cdot \cutoff' \cdot \fder{Y}{\phi}
\label{eq:classify-action}
,
\end{align}
\end{subequations}
where $X$ and $Y$ are arbitrary.

\subsection{The Marginal, Redundant Operator}
\label{app:Menagerie-mro}

In this section, we will explicitly demonstrate that the marginal, redundant operator---taken in the form of~\eq{HO-mro}---satisfies
\be
	\classifier \eop'^{\mathrm{R}}_\mathrm{mar} = 0.
\label{eq:HO-mro-eqn}
\ee
The first step is to rewrite
\be
	 \eop'^{\mathrm{R}}_\mathrm{mar} = 
	 \int_p 
	\biggl[
		A(p) - \fder{}{\phi(-p)} \cutoff(p^2)
	\biggr]
	B(-p),
\label{eq:HO-mro-Rewrite}
\ee
where
\begin{subequations}
\begin{align}
	A(p) & = p^2 \phi(p) + \fder{\Sint_\star}{\phi(-p)} \cutoff(p^2),
\label{eq:A}
\\
	B(-p) & = \frac{1+\varrho(p^2)}{\cutoff(p^2)}\phi(-p) + \frac{\varrho(p^2)}{p^2} \fder{\Sint_\star}{\phi(p)}.
\label{eq:B}
\end{align}
\end{subequations} 
Before moving on, let us observe from~\eqs{classify-phi}{classify-dS/dphi} that
\begin{subequations}
\begin{align}
	\classifier \, A(p) & = 
		\biggl(\frac{d-2+\eta_\star}{2} + p \cdot \partial_p \biggr) A(p),
\label{eq:classify-A}
\\
	\classifier \, B(q) & = 
		\biggl(\frac{d+2-\eta_\star}{2} + q \cdot \partial_q \biggr) B(q).
\label{eq:classify-B}
\end{align}
\end{subequations}

Now, utilizing~\eq{classify-action}, we have that
\begin{multline}
	\classifier \eop'^{\mathrm{R}}_\mathrm{mar} =
	\int_p
	\biggl\{
		B(-p) \, \classifier A(p) + A(p) \, \classifier B(-p) - 2 \fder{A(p)}{\phi} \cdot \cutoff' \cdot \fder{B(-p)}{\phi}
\\
		- \fder{}{\phi(-p)} \cutoff(p^2) \classifier B(-p)
		- 
		\biggl[
			\classifier, \fder{}{\phi(p)}
		\biggr] \cutoff(p^2) B(-p)
	\biggr\}
\label{eq:HO-mro-Process}
\end{multline}
Using~\eqs{classify-A}{classify-B}, it is immediately apparent that (after integration by parts) the first two terms cancel. To process the final two terms, notice that we can rewrite them as
\begin{multline*}
	-\int_p
	\biggl\{
		\fder{}{\phi(-p)} \cutoff(p^2) \biggl(\frac{d+2-\eta_\star}{2} + p \cdot \partial_p \biggr) B(-p)
		-
		2 \fder{}{\phi} \biggl( \fder{\Sint_\star}{\phi(-p)} \cutoff(p^2) \biggr) \cdot \cutoff' \cdot \fder{B(-p)}{\phi}
\\
		+
		\cutoff(p^2) \biggl[ \biggl(\frac{d-2+\eta_\star}{2} + p \cdot \partial_p \biggr) \fder{}{\phi(-p)} \biggr] B(-p)
	\biggr\}
\end{multline*}
Combining the first and last terms and integrating by parts, we have:
\[
	\int_p
	\biggl\{
		2 p^2 \cutoff'(p^2) B(-p) + 2 \fder{}{\phi} \biggl( \fder{\Sint_\star}{\phi(-p)} \cutoff(p^2) \biggr) \cdot \cutoff' 
			\cdot \fder{B(-p)}{\phi}
	\biggr\}
	=
	2 \int_p 
	\fder{A(p)}{\phi} \cdot \cutoff' \cdot \fder{B(-p)}{\phi},
\]
cancelling the remaining term in~\eq{HO-mro-Process}, thereby demonstrating~\eq{HO-mro-eqn}.

\subsection{Operators with a Two-Point Dual}
\label{app:Menagerie-D^2}

In this subsection, we will verify that the operators given in~\eq{redundant-family} satisfy
\be
	\classifier \eop_{2,r}[\phi] = (2+\eta_\star - r) \eop_{2,r}[\phi]
.
\ee
To this end, observe that we can write
\be
	\eop_{2,r}[\phi] = 
	\int_p
	\biggl[
		A(p) - \fder{}{\phi(-p)} \cutoff(p^2)
	\biggr] P_{r-4}
	A(-p),
\ee
where we recall that $P_r(p) \sim p^{2r/2}$. Following steps similar to those above, it is apparent that
\begin{align*}
	\classifier \eop_{2,r}[\phi] & = 
	\int_p  P_{r-4}
	\biggl\{
		(2+\eta_\star-r) A(p) A(-p) -2 p^2 \cutoff'(p^2) \fder{A(-p)}{\phi(-p)}
	\\	
	& 
	\qquad \qquad \qquad \qquad \qquad
		- \cutoff(p^2) \bigl(d-2+\eta_\star + p \cdot \partial_p \bigr) \fder{A(-p)}{\phi(-p)}
	\biggr\}
	\\
	&
	=
	(2+\eta_\star-r) \eop_{2,r}[\phi] ,
\end{align*}
as required.

\bibliography{../../Biblios/ERG.bib,../../Biblios/Renormalons.bib,../../Biblios/Books.bib,../../Biblios/CFT.bib,../../Biblios/NC.bib,../../Biblios/Misc.bib}

\end{document}